\documentclass[11pt]{book} 

\usepackage[utf8]{inputenc} 

\usepackage{geometry} 

\usepackage[numbers,square]{natbib}
\usepackage{tabularx}

\usepackage{graphicx} 

\usepackage{marginnote}
\usepackage{makeidx}
\usepackage{booktabs} 
\usepackage{array} 
\usepackage{paralist} 
\usepackage{verbatim} 
\usepackage{subfig} 
\usepackage{caption} 
\usepackage[formats]{listings}
\usepackage{tikz}
\usetikzlibrary{arrows,tikzmark,shadows}
\tikzset{
  comment/.style={
    draw,
    fill=blue!70,
    text=white,
    rounded corners,
    drop shadow,
    align=left,
  },
}

\usepackage{amsmath}
\usepackage{xspace}
\usepackage[draft]{pdfcomment}
\usepackage{marginnote}
\usepackage{framed}

\lstdefineformat{none}{}
\lstdefineformat{c}{%
\{=\string\newline\indent,%
\}=[;\ ]\newline\noindent\string\newline,%
\}\ =\noindent\string,%
\};=\newline\noindent\string\newline,%
;=[\ ]\string\space}

\usepackage{fancyhdr} 
\renewcommand{\sectionmark}[1]{}
\usepackage{sectsty}
\allsectionsfont{\sffamily\mdseries\upshape} 

\usepackage[nottoc,notlof,notlot]{tocbibind} 
\usepackage[titles,subfigure]{tocloft} 

\usepackage{url}
\usepackage{framed,color}
\usepackage{import}
\usepackage{todonotes}

\definecolor{shadecolor}{rgb}{.75,.9,.9}
\definecolor{asidecolor}{rgb}{.75,.75,.75}
\definecolor{exercisecolor}{rgb}{.75,.9,.9}

\newcommand{\executeiffilenewer}[3]{%
\ifnum\pdfstrcmp{\pdffilemoddate{#1}}{\pdffilemoddate{#2}}>0%
{\immediate\write18{#3}}\fi%

}

\newcommand{\note}[1]{}

\newcommand{\sym}[1]{\texttt{#1}}
\newcommand{\code}[1]{\texttt{#1}}
\newcommand{\term}[1]{{\color{blue}\textit{#1}}\index{#1}}  
\newcommand{\vivado}{Vivado\textsuperscript{\textregistered}}
\newcommand{\VHLS}{Vivado\textsuperscript{\textregistered} HLS\xspace}
\newcommand{\tabspace}{\vspace{1em}}
\newcommand{\stevecomment}[1]{\pdfmargincomment[color=red,icon=Insert,author={steve}]}

\newcommand{\unit}[2]{{\frac{\texttt{#1}}{\texttt{#2}}}}
\newenvironment{tabularpad}[1]{\parskip=0pt\par\noindent\begin{minipage}{\textwidth}\parskip=0.5em\hfill\par\hspace{2em}\begin{tabular}{#1}}{\end{tabular}\hfill\par\hfill\end{minipage}}
\newenvironment{padbox}[1]{\parskip=0pt\par\noindent\begin{minipage}{\textwidth}\parskip=0.5em\hfill\par\hspace{2em}\begin{minipage}{#1}}{\end{minipage}\hfill\par\hfill\end{minipage}}

\newlength{\boxwidth}
\newenvironment{coloredbox}[1]{
\def\FrameCommand{\fboxsep=\FrameSep \fboxrule=\FrameRule \fcolorbox{black}{#1}}%
  \MakeFramed {\setlength{\boxwidth}{\textwidth}
  \addtolength{\boxwidth}{-2\FrameSep}
  \addtolength{\boxwidth}{-2\FrameRule}
  \setlength{\hsize}{\boxwidth} \FrameRestore}}%
 {\endMakeFramed}

\newenvironment{aside}{\begin{coloredbox}{asidecolor}}{\end{coloredbox} }
\newenvironment{exercise}{\begin{coloredbox}{exercisecolor}}{\end{coloredbox} }

\usepackage[toc,acronym]{glossaries}

\makenoidxglossaries

\newglossaryentry{hlsg}
{
	name={HLS},
    	description={High-level synthesis is a hardware design process that translates an algorithmic description (which is decoupled from the cycle to cycle behavior) into a register transfer level (RTL) hardware description language which specifies the exact behavior of the circuit on a cycle-by-cycle basis}
}

\newglossaryentry{rtlg}
{
	name={RTL},
    	description={Register-transfer level (RTL) is a hardware design abstraction which models a synchronous digital circuit using logical operations that occur between between hardware registers. It is common design entry for modern digital design}
}

\newglossaryentry{asicg}
{
	name={ASIC},
    	description={An application-specific integrated circuit (ASIC) is a circuit created for a particular use, e.g., an audio decoder or a wireless modem. }
}

\newglossaryentry{edag}
{
	name={EDA},
    	description={Electronic design automation (EDA) are a set of software tools used to aid the hardware design process.  }
}

\newglossaryentry{fpgag}
{
	name={FPGA},
    	description={A field-programmable gate array (FPGA) is an integrated circuit that can be customized or programmed after it is manufactured (``in the field''). }
}

\newglossaryentry{lutg}
{
	name={LUT},
    	description={A lookup table (LUT) is a memory where the address signal are the inputs and the corresponding outputs are contained in the memory entries. It is a key computational component of modern \gls{fpga}s.}
}

\newglossaryentry{ffg}
{
	name={FF},
    	description={A flip-flop (FF) is a circuit that can store information. We typically think of it as storing one bit of data and are a fundamental building block for creating memories in digital circuits.}
}

\newglossaryentry{bramg}
{
	name={BRAM},
    	description={A block RAM is a configurable random access memory that is embedded throughout an FPGA for data storage and communication.}
}

\newglossaryentry{romg}
{
	name={ROM},
    	description={A Read-only Memory is a memory which is initialized to a particular value and then read but never written.  In many cases the storage for ROMs can be highly optimized because their value never changes.}
}
\newglossaryentry{crsg}
{
	name={Compressed Row Storage},
    	description={Compressed Row Storage is a technique for representing a sparse matrix.  It allows large matrices with a small number of elements to be stored and operated on efficiently.}
}
\newglossaryentry{ssag}
{
	name={Static Single Assignment},
    	description={Static Single Assignment is an intermediate representation in compilers where each variable is assigned only once.  This form makes many common optimizations simpler to write.}
}

\newglossaryentry{core}
{
	name={IP~core},
	description={An RTL-level component with well-defined interfaces enabling it to be incorporated into a larger design.  Often used as a way of hiding the `intellectual property' from another company, hence the name.}
} 

\newglossaryentry{synth}
{
	name={logic~synthesis},
	description={The process of converting an gls{rtl} design into a netlist of device-level primitives.}
}

\newglossaryentry{parg}
{
	name={place and route},
	description={The process of converting a netlist of device-level primitives into the configuration of a particular device.}
} 

\newglossaryentry{fftg}
{
	name={Fast Fourier Transform},
    	description={An optimized version of the \gls{dft} which requires fewer operations.}
}

\newglossaryentry{dftg}
{
	name={Discrete Fourier Transform},
    	description={An transformation that takes a discrete signal and converts it to a freqeuncy-domain representation.}
}

\newglossaryentry{netlist}
{
	name={netlist},
	description={An intermediate design artifact consisting of device-level primitive elements and the connections between them.  In FPGA designs, the primitive elements include \glspl{lut},\glspl{ff}, and \glspl{bram}.}
} 

\newglossaryentry{looppipelining}
{
	name={loop~pipelining},
	description={Enabling multiple iterations of a loop to run concurrently sharing the same functional units.}
} 

\newglossaryentry{arraypartitioning}
{
	name={array~partitioning},
	description={Dividing a single logical array into multiple physical memories.}
} 

\newglossaryentry{recurrence}
{
	name={recurrence},
	description={A code structure that results in a feedback loop when implemented in a circuit.  Recurrences limit the throughput of the circuit.}
} 
\newglossaryentry{loopinterchange}
{
	name={loop~interchange},
	description={A code transformation that changes the order of loop operations.   This transformation is often a useful approach to addressing recurrences in code.}
} 
\newglossaryentry{taskpipelining}
{
	name={task~pipelining},
	description={Being able to execute more than one task concurrently on the same accelerator in a pipelined fashion.}
} 
\newglossaryentry{peg}
{
	name={processing element},
	description={A coarse-grained concurrently executing component in a design.  In HLS, this is often used in the context of a dataflow design.}
} 
\newglossaryentry{routingchannel}
{
	name={routing channel},
	plural={routing channels},
	description={A routing channel provides a flexible set of connections between the FPGA programmable logic elements. }
}

\newglossaryentry{switchbox}
{
	name={switchbox},
	plural={switchboxes},
	description={A switchbox connects \gls{routingchannel}s to provide a flesible routing structure for data routed between the programmable logic and \gls{ioblock}. }
}

\newglossaryentry{ioblock}
{
	name={I/O block},
	plural={I/O blocks},
	description={An I/O block provides the interface between the FPGA fabric and the remainder of the system. I/O blocks can talk to memories (e.g., on-chip caches and off-chip DRAM, microprocessors (using AXI or other protocols), sensors, actuators, etc.. }
}

\newglossaryentry{bitstream}
{
	name={bitstream},
	plural={bitstreams},
	description={The configuration data used to program the functionality of an FPGA}
}

\newglossaryentry{slice}
{
	name={slice},
	plural={slices},
	description={A (typically small) set of \glspl{lut}, \glspl{ff} and multiplexors. These are often reported in \gls{fpga} resource utilization reports. }
}

\newglossaryentry{task}
{
	name={task},
	plural={tasks},
	description={A fundamental atomic unit of behavior or high-level synthesis computation; this corresponds to a function invocation in high-level synthesis}
}

\newglossaryentry{task-latency}
{
	name={task~latency},
	description={The time between when a task starts and when it finishes}
}

\newglossaryentry{task-interval} 
{
	name={task~interval},
	description={The time between when one task starts and the next starts or the difference between the start times of two consecutive tasks}
}

\newglossaryentry{data-rate} 
{
	name={data~rate},
	description={The frequency at which a task can process the input data. This is often expressed in bits/second and thus also depends on the size of the input data}
}

\newglossaryentry{fir} 
{
	name={finite impulse response},
	description={A common digital signal processing task that performs a convolution on the input signal with a fixed signal that is defined by its coefficients. The FIR is often performed in hardware and can be efficiently implemented}
}

\newglossaryentry{process} 
{
	name={process},
	plural={processes},
	description={An individual component in a dataflow architecture}
}

\newglossaryentry{partial_loop_unrolling} 
{
	name={partial loop unrolling},
	description={A transformation where the body of a loop is replicated multiple times.  This is often used in processor systems to reduce loop condition overhead or to provide opportunities for vectorization.  In HLS, it can have a similar effect, enabling more operations from the same loop nest to be considered in scheduling.  This can improve the performance of a design.}
}

\newglossaryentry{cosimulation} 
{
	name={C/RTL cosimulation},
	description={The process of verifying an \gls{rtl} design generated by HLS using testvectors captured from the C testbench.}
}

\newglossaryentry{stable_sort} 
{
	name={stable sort},
	description={A sorting algorithm that keeps different elements with the same sorting key in their original sequence after sorting.}
}

\newglossaryentry{sorting_cell} 
{
	name={sorting cell},
	description={An simple stateful component that forms part of a larger sorting network or algorithm.  Commonly a cell performs a compare-and-swap operation between two elements.}
}

\newglossaryentry{systolic_array} 
{
	name={systolic array},
	description={An array of processing elements that coordinate to perform a more complex algorithm.  Systolic arrays are usually designed so that each processing element encapsulates some local information and only communicates with its local neighbors.  This often enables systolic arrays to easily scale to large problem sizes by increasing the size of the array.}
}

\newglossaryentry{hls}
{
	type=\acronymtype, 
	name={HLS}, 
	description={high-level synthesis}, 
	first={high-level synthesis (HLS)\glsadd{hlsg}}, 
	see=[Glossary:]{hlsg}
}

\newglossaryentry{rtl}
{
	type=\acronymtype, 
	name={RTL}, 
	description={register-transfer level}, 
	first={register-transfer level (RTL)\glsadd{rtlg}}, 
	see=[Glossary:]{rtlg}
}

\newglossaryentry{asic}
{
	type=\acronymtype, 
	name={ASIC}, 
	description={application-specific integrated circuit}, 
	first={application-specific integrated circuit (ASIC)\glsadd{asicg}}, 
	see=[Glossary:]{asicg}
}

\newglossaryentry{eda}
{
	type=\acronymtype, 
	name={EDA}, 
	description={electronic design automation}, 
	first={electronic design automation (EDA)\glsadd{edag}}, 
	see=[Glossary:]{edag}
}

\newglossaryentry{fpga}
{
	type=\acronymtype, 
	name={FPGA}, 
	description={field-programmable gate array}, 
	first={field-programmable gate array (FPGA)\glsadd{fpgag}}, 
	plural={FPGAs},
	see=[Glossary:]{fpgag}
}

\newglossaryentry{lut}
{
	type=\acronymtype, 
	name={LUT}, 
	description={lookup table}, 
	first={lookup table (LUT)\glsadd{lutg}}, 
	plural={LUTs},
	see=[Glossary:]{lutg}
}

\newglossaryentry{par}
{
	type=\acronymtype, 
	name={PAR}, 
	description={place and route}, 
	first={place and route (PAR)\glsadd{parg}}, 
	see=[Glossary:]{parg}
}

\newglossaryentry{ff}
{
	type=\acronymtype, 
	name={FF}, 
	description={flip-flop}, 
	first={flip-flop (FF)\glsadd{ffg}}, 
	plural={FFs},
	see=[Glossary:]{ffg}
}

\newglossaryentry{bram}
{
	type=\acronymtype, 
	name={BRAM}, 
	description={block RAM}, 
	first={block RAM (BRAM)\glsadd{bramg}}, 
	plural={BRAMs},
	see=[Glossary:]{bramg}
}

\newglossaryentry{rom}
{
	type=\acronymtype, 
	name={ROM}, 
	description={Read-only Memory}, 
	first={Read-only Memory (ROM)\glsadd{romg}}, 
	plural={ROMs},
	see=[Glossary:]{romg}
}

\newglossaryentry{pe}
{
	type=\acronymtype, 
	name={PE}, 
	description={Processing Element}, 
	first={Processing Element (PE)\glsadd{peg}}, 
	plural={PEs},
	see=[Glossary:]{peg}
}

\newglossaryentry{fft}
{
	type=\acronymtype, 
	name={FFT}, 
	description={Fast Fourier Transform}, 
	first={Fast Fourier Transform (FFT)\glsadd{fftg}}, 
	plural={FFTs},
	see=[Glossary:]{fftg}
}
\newglossaryentry{dft}
{
	type=\acronymtype, 
	name={DFT}, 
	description={Discrete Fourier Transform}, 
	first={Discrete Fourier Transform (DFT)\glsadd{dftg}}, 
	plural={DFTs},
	see=[Glossary:]{dftg}
}
\newglossaryentry{crs}
{
	type=\acronymtype, 
	name={CRS}, 
	description={compressed row storage}, 
	first={compressed row storage (CRS)\glsadd{crsg}}, 
	see=[Glossary:]{crsg}
}
\newglossaryentry{ssa}
{
	type=\acronymtype, 
	name={SSA}, 
	description={static single assigment}, 
	first={static single assignment (SSA)\glsadd{ssag}}, 
	see=[Glossary:]{ssag}
}

\title{Parallel Programming for FPGAs}
\author{Ryan Kastner, Janarbek Matai, and Stephen Neuendorffer\\\note{Notes are enabled}}

\makeindex

\begin{document}
\setlength{\emergencystretch}{5em}
\setlength{\fboxrule}{1pt}
\setlength{\fboxsep}{9pt}

\setlength{\FrameRule}{\fboxrule}
\setlength{\FrameSep}{\fboxsep}

\maketitle
\vspace*{\stretch{1.0}} \ \\
\begin{center}
\setlength{\parskip}{0.5\baselineskip}
Copyright 2011-\the\year.
\end{center}
\vspace*{20mm} \ \\
\begin{minipage}{\linewidth}
This work is licensed under the Creative Commons Attribution 4.0 International License.

To view a copy of this license, visit \url{http://creativecommons.org/licenses/by/3.0/}. 
\end{minipage}
\vspace*{20mm} \ \\
\begin{minipage}{\linewidth}
The newest version of this book can be found at \url{http://hlsbook.ucsd.edu}.  The authors welcome your feedback and suggestions.
\end{minipage}
\vspace*{\stretch{2.0}} \ \\
\newpage
\tableofcontents
\newpage

\chapter*{Preface}
\addcontentsline{toc}{chapter}{Preface}
\begin{aside}
``When someone says, 'I want a programming language in which I need only say what I wish done', give him a lollipop.'' -Alan Perlis
\end{aside}

This book focuses on the use of algorithmic high-level synthesis (HLS) to build application-specific FPGA systems.  Our goal is to give the reader an appreciation of the process of creating an optimized hardware design using HLS. Although the details are, of necessity, different from parallel programming for multicore processors or GPUs, many of the fundamental concepts are similar.  For example, designers must understand memory hierarchy and bandwidth, spatial and temporal locality of reference, parallelism, and tradeoffs between computation and storage.

This book is a practical guide for anyone interested in building FPGA systems.  In a university environment, it is appropriate for advanced undergraduate and graduate courses.  At the same time, it is also useful for practicing system designers and embedded programmers.  The book assumes the reader has a working knowledge of C/C++ and includes a significant amount of sample code. In addition, we assume familiarity with basic computer architecture concepts (pipelining, speedup, Amdahl's Law, etc.).  A knowledge of the RTL-based FPGA design flow is helpful, although not required.

The book includes several features that make it particularly valuable in a classroom environment.  It includes questions within each chapter that will challenge the reader to solidify their understanding of the material.  It provides specific projects in the Appendix. These were developed and used in the HLS class taught at UCSD (CSE 237C). We will make the files for these projects available to instructors upon request. These projects teach concepts in HLS using examples in the domain of digital signal processing with a focus on developing wireless communication systems. Each project is more or less associated with one chapter in the book. The projects have reference designs targeting FPGA boards distributed through the Xilinx University Program (\url{http://www.xilinx.com/support/university.html}).   The FPGA boards are available for commercial purchase.  Any reader of the book is encouraged to request an evaluation license of \VHLS at \url{http://www.xilinx.com}.

This book is {\em not} primarily about HLS algorithms.  There are many excellent resources that provide details about the HLS process including algorithms for scheduling, resource allocation, and binding \cite{micheli1994synthesis, gupta2004spark, coussy2010high, gajski2012high}.  This book is valuable in a course that focuses on these concepts as supplementary material, giving students an idea of how the algorithms fit together in a coherent form, and providing concrete use cases of applications developed in a HLS language.  This book is also {\em not} primarily about the intricacies of FPGA architectures or RTL design techniques.  However, again it may be valuable as supplementary material for those looking to understand more about the system-level context.

This book focuses on using Xilinx tools for implementing designs, in particular \VHLS to perform the translation from C-like code to RTL.  C programming examples are given that are specific to the syntax used in \VHLS.  In general, the book explains not only \VHLS specifics, but also the underlying generic HLS concepts that are often found in other tools.  We encourage readers with access to other tools to understand how these concepts are interpreted in any HLS tool they may be using.

Good luck and happy programming!

\chapter*{Acknowledgements}
\addcontentsline{toc}{chapter}{Acknowledgements}

Many people have contributed to making this book happen.  Probably first and foremost are the many people who have done research in the area of High-Level Synthesis.  Underlying each of the applications in this book are many individual synthesis and mapping technologies which combine to result in high-quality implementations.

Many people have worked directly on the \VHLS tool over the years.  From the beginning in Jason Cong's research group at UCLA as the AutoPilot tool, to making a commercial product at AutoESL Inc., to widespread adoption at Xilinx, it's taken many hours of engineering to build an effective tool.  To Zhiru Zhang, Yiping Fan, Bin Liu, Guoling Han, Kecheng Hao, Peichen Pan, Devadas Varma, Chuck Song, and many others: your efforts are greatly appreciated.

The idea for this book originally arose out of a hallway conversation with Salil Raje shortly after Xilinx acquired the AutoESL technology.  Much thanks to Salil for his early encouragement and financial support.  Ivo Bolsens and Kees Vissers also had the trust that we could make something worth the effort.  Thanks to you all for having the patience to wait for good things to emerge.

This book would not have been possible without substantial support from the UCSD Wireless Embedded Systems Program. This book grew out of the need to make a hardware design class that could be broadly applicable to students coming from a mix of software and hardware backgrounds.  The program provided substantial resources in terms of lab instructors, teaching assistants, and supplies that were invaluable as we developed (and re-developed) the curriculum that eventually morphed into this book. Thanks to all the UCSD 237C students over the years for providing feedback on what made sense, what didn't, and generally acting as guinea pigs over many revisions of the class and this book. Your suggestions and feedback were extremely helpful. A special thanks to the TAs for these classes, notable Alireza Khodamoradi, Pingfan Meng, Dajung Lee, Quentin Gautier, and Armaiti Ardeshiricham; they certainly felt the growing pains a lot more than the instructor.

Various colleagues have been subjected to early drafts of the book, including Zhiru Zhang, Mike Wirthlin, Jonathan Corbett. We appreciate your feedback.  

\chapter{Introduction}
\glsresetall
\label{chapter:introduction}
\section{High-level Synthesis (HLS)}

The hardware design process has evolved significantly over the years. When the circuits were small, hardware designers could more easily specify every transistor, how they were wired together, and their physical layout. Everything was done manually. As our ability to manufacture more transistors increased, hardware designers began to rely on automated design tools to help them in the process of creating the circuits. These tools gradually become more and more sophisticated and allowed hardware designers to work at higher levels of abstraction and thus become more efficient. Rather than specify the layout of every transistor, a hardware designer could instead specify digital circuits and have \gls{eda} tools automatically translate these more abstract specifications into a physical layout. 

The Mead and Conway approach \cite{mead1980introduction} of using a programming language (e.g., Verilog or VHDL) that compiles a design into physical chips took hold in the 1980s. Since that time, the hardware complexity has continued to increase at an exponential rate, which forced hardware designers to move to even more abstract hardware programming languages. \gls{rtl} was one step in abstraction, enabling a designer to simply specify the registers and the operations performed on those registers, without considering how the registers and operations are eventually implementation. \gls{eda} tools can translate \gls{rtl} specifications into a digital circuit model and then subsequently into the detailed specification for a device that implements the digital circuit.  This specification might be the files necessary to manufacture a custom device or might be the files necessary to program an off-the-shelf device, such as an \gls{fpga}. Ultimately, the combination of these abstractions enables designers to build extraordinarily complex systems without getting lost in the details of how they are implemented.  A non-technical perspective on the value of these abstractions can be found in \cite{lee2017plato}.

 \gls{hls} is yet another step in abstraction that enables a designer to focus on larger architectural questions rather than individual registers and cycle-to-cycle operations.  Instead a designer captures behavior in a program that does not include specific registers or cycles and an \gls{hls} tool creates the detailed \gls{rtl} micro-architecture.  One of the first tools to implement such a flow was based on behavioral Verilog and generated an \gls{rtl}-level architecture also captured in Verilog\cite{knapp96bc}.  Many commercial tools now use C/C++ as the input language.  For the most part the language is unimportant, assuming that you have a tool that accepts the program you want to synthesize!

Fundamentally, algorithmic HLS does several things automatically that an RTL designer does manually:
\begin{itemize}
\item \gls{hls} analyzes and exploits the concurrency in an algorithm.
\item \gls{hls} inserts registers as necessary to limit critical paths and achieve a desired clock frequency.
\item \gls{hls} generates control logic that directs the data path.
\item \gls{hls} implements interfaces to connect to the rest of the system.
\item \gls{hls} maps data onto storage elements to balance resource usage and bandwidth.
\item \gls{hls} maps computation onto logic elements performing user specified and automatic optimizations to achieve the most efficient implementation.
\end{itemize}

Generally, the goal of \gls{hls} is to make these decisions automatically based upon user-provided input specification and design constraints.  However, \gls{hls} tools greatly differ in their ability to do this effectively.  Fortunately, there exist many mature HLS tools (e.g., Xilinx \VHLS, LegUp \cite{canis2011legup}, and Mentor Catapult HLS) that can make these decisions automatically for a wide range of applications. We will use \VHLS as an exemplar for this book; however, the general techniques are broadly applicable to most HLS tools though likely with some changes in input language syntax/semantics. 

In general, the designer is expected to supply the \gls{hls} tool a functional specification, describe the interface, provide a target computational device, and give optimization directives. More specifically, \VHLS requires the following inputs:
\begin{itemize}
\item A function specified in C, C++, or SystemC 
\item A design testbench that calls the function and verifies its correctness by checking the results.
\item A target FPGA device
\item The desired clock period
\item Directives guiding the implementation process
\end{itemize}

In general, \gls{hls} tools can not handle any arbitrary software code. Many concepts that are common in software programming are difficult to implement in hardware. Yet, a hardware description offers much more flexibility in terms of how to implement the computation. It typically requires additional information to be added by the designers (suggestions or \lstinline|#pragmas|) that provide hints to the tool about how to create the most efficient design. Thus, \gls{hls} tools simultaneously limit and enhance the expressiveness of the input language. For example, it is common to not be able to handle dynamic memory allocation. There is often limited support for standard libraries. System calls are typically avoided in hardware to reduce complexity. The ability to perform recursion is often limited. On the other hand, \gls{hls} tools can deal with a variety of different interfaces (direct memory access, streaming, on-chip memories). And these tools can perform advanced optimizations (pipelining, memory partitioning, bitwidth optimization) to create an efficient hardware implementation. 

We make the following assumptions about the input function specification, which generally adheres to the guidelines of the \VHLS tool:
\begin{itemize}
\item No dynamic memory allocation (no operators like \lstinline|malloc()|, \lstinline|free()|, \lstinline|new|, and \lstinline|delete()|)
\item Limited use of pointers-to-pointers (e.g., may not appear at the interface)
\item System calls are not supported (e.g., \lstinline|abort()|, \lstinline|exit()|, \lstinline|printf()|, etc. They can be used in the code, e.g., in the testbench, but they are ignored (removed) during synthesis.
\item Limited use of other standard libraries (e.g., common \lstinline|math.h| functions are supported, but uncommon ones are not)
\item Limited use of function pointers and virtual functions in C++ classes (function calls must be compile-time determined by the compiler).
\item No recursive function calls.
\item The interface must be precisely defined.
\end{itemize}

The primary output of an \gls{hls} tool is a \gls{rtl} hardware design that is capable of being synthesized through the rest of the hardware design flow. Additionally, the tool may output testbenches to aid in the verification process. Finally, the tool will provide some estimates on resource usage and performance.  \VHLS generates the following outputs:
\begin{itemize}
\item Synthesizable Verilog and VHDL
\item RTL simulations based on the design testbench
\item Static analysis of performance and resource usage
\item Metadata at the boundaries of a design, making it easier to integrate into a system.
\end{itemize}

Once an \gls{rtl}-level design is available, other tools are usually used in a standard \gls{rtl} design flow.  In the Xilinx \vivado Design Suite, \gls{synth} is performed, translating the \gls{rtl}-level design into a \gls{netlist} of primitive \gls{fpga} logical elements.  The \gls{netlist} (consisting of logical elements and the connections between them) is then associated with specific resources in a target device, a process called \gls{par}.  The resulting configuration of the \gls{fpga} resources is captured in a \gls{bitstream}, which can be loaded onto the FPGA to program its functionality.  The \gls{bitstream} contains a binary representation of the configuration of each \gls{fpga} resource, including logic elements, wire connections, and on-chip memories. A large Xilinx UltraScale FPGAs will have over 1 billion configuration bits and even the ``smaller'' devices have hundreds of millions of bits \cite{ultrascaleArchConfig}. 


\section{FPGA Architecture}

It is important to understand the modern FPGA architectures since many of the \gls{hls} optimizations specifically target these features. Over the decades, FPGAs have gone from small arrays of programmable logic and interconnect to massive arrays of programmable logic and interconnect with on-chip memories, custom data paths, high speed I/O, and microprocessor cores all co-located on the same chip. In this section, we discuss the  architectural features that are relevant to \gls{hls}. It is not our intention (nor do we think it is necessary) to provide substantial details of the FPGA architecture. Rather we aim to give the reader enough information to understand the \gls{hls} reports and successfully use and leverage the \gls{hls} directives, many of which very specifically target modern FPGA architectural features.

\glspl{fpga} are an array of programmable logic blocks and memory elements that are connected together using programmable interconnect. Typically these logic blocks are implemented as a \gls{lut} -- a memory where the address signal are the inputs and the outputs are stored in the memory entries. An $n$-bit \gls{lut} can be programmed to compute any $n$-input Boolean function by using the function's truth table as the values of the \gls{lut} memory. 

\begin{figure}
\centering
\includegraphics[width= \textwidth]{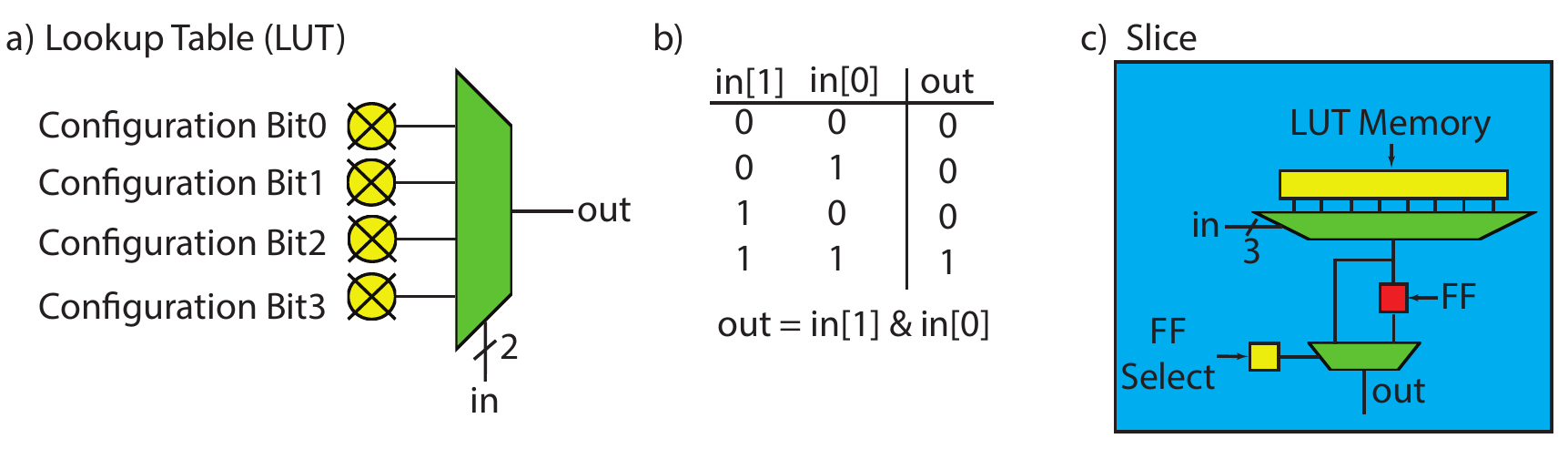}
\caption{Part a) shows a 2 input \gls{lut}, i.e., a 2-LUT. Each of the four configuration bits can be programmed to change the function of the 2-LUT making it a fully programmable 2 input logic gate. Part b) provides a sample programming to implement an AND gate.  The values in the ``out'' column from top to bottom correspond directly to configuration bits 0 through 3. Part c) shows a simple \gls{slice} that contains a slightly more complex 3-LUT with the possibility of storing the output into a \gls{ff}. Note that there are nine configuration bits: eight to program the 3-LUT and one to decide whether the output should be direct from the 3-LUT or the one stored in the \gls{ff}. More generally, a \gls{slice} is defined as a small number of \glspl{lut} and \glspl{ff} combined with routing logic (multiplexers) to move inputs, outputs, and internal values between the \glspl{lut} and \glspl{ff}.}
\label{fig:lut}
\end{figure}

Figure \ref{fig:lut} a) shows a 2 input \gls{lut}. It has $2^2 = 4$ configuration bits. These bits are the ones that are programmed to determine the functionality of the \gls{lut}.  Figure \ref{fig:lut} b) shows the truth table for a 2 input AND gate. By using the four values in the ``out'' column for configuration bits 0-3, we can program this 2 input LUT to be a 2 input AND gate. The functionality is easily changed by reprogramming the configuration bits. This is a flexible and fairly efficient method for encoding smaller Boolean logic functions. Most \glspl{fpga} use a \glspl{lut} with 4-6 input bits as their base element for computation. Larger \glspl{fpga} can have millions of these programmable logic elements.

\begin{exercise}
How would you program the 2-LUT from Figure \ref{fig:lut} to implement an XOR gate? An OR gate? How many programming bits does an $n$ input ($n$-LUT) require?
\end{exercise}

\begin{exercise}
How many unique functions can a 2-LUT be programmed to implement? How many unique functions can a $n$ input ($n$-LUT) implement? 
\end{exercise}

The \gls{ff} is the basic memory element for the \gls{fpga}. They are typically co-located with a \glspl{lut}. \gls{lut}s can be replicated and combined with \glspl{ff} and other specialized functions (e.g., a full adder) to create a more complex logic element called a configurable logic block (CLB), logic array block (LAB), or \gls{slice} depending on the vendor or design tool. We use the term \gls{slice} since it is the resource reported by the \VHLS tool.  A \gls{slice} is a small number of \glspl{lut}, \glspl{ff} and multiplexers combined to make a more powerful programmable logic element. The exact number and combination of  \glspl{lut}, \glspl{ff} and multiplexers varies by architecture, but generally a slice has only few of each of these elements. Figure \ref{fig:lut} c) shows a very simple slice with one 3-LUT and one FF. A \gls{slice} may also use some more complex logic functions. For example, it is common to embedded a full adder into a slice. This is an example of ``hardening'' the FPGA; this full adder is not programmable logic -- it can only be used as a full adder, but it is common to use full adders (to make addition operations) and it is more efficient to use the custom full adder as opposed to implementing a full adder on the programmable logic (which is also an option). And thus, overall it is beneficial to have a hard full adder in the slice.

Programmable interconnect is the other key element of an FPGA. It provides a flexible network of wires to create connections between the slices. The inputs and outputs of the \gls{slice} are connected to a \gls{routingchannel}. The \gls{routingchannel} contains a set configuration bits can be programmed to connect or disconnect the inputs/outputs of the \gls{slice} to the programmable interconnect. Routing channels are connected to \glspl{switchbox}. A \gls{switchbox} is a collection of switches that are implemented as pass transistors. These provide the ability to connect the routing tracks from one \glspl{routingchannel} to another. 

\begin{figure}
\centering
\includegraphics[width= 0.8\textwidth]{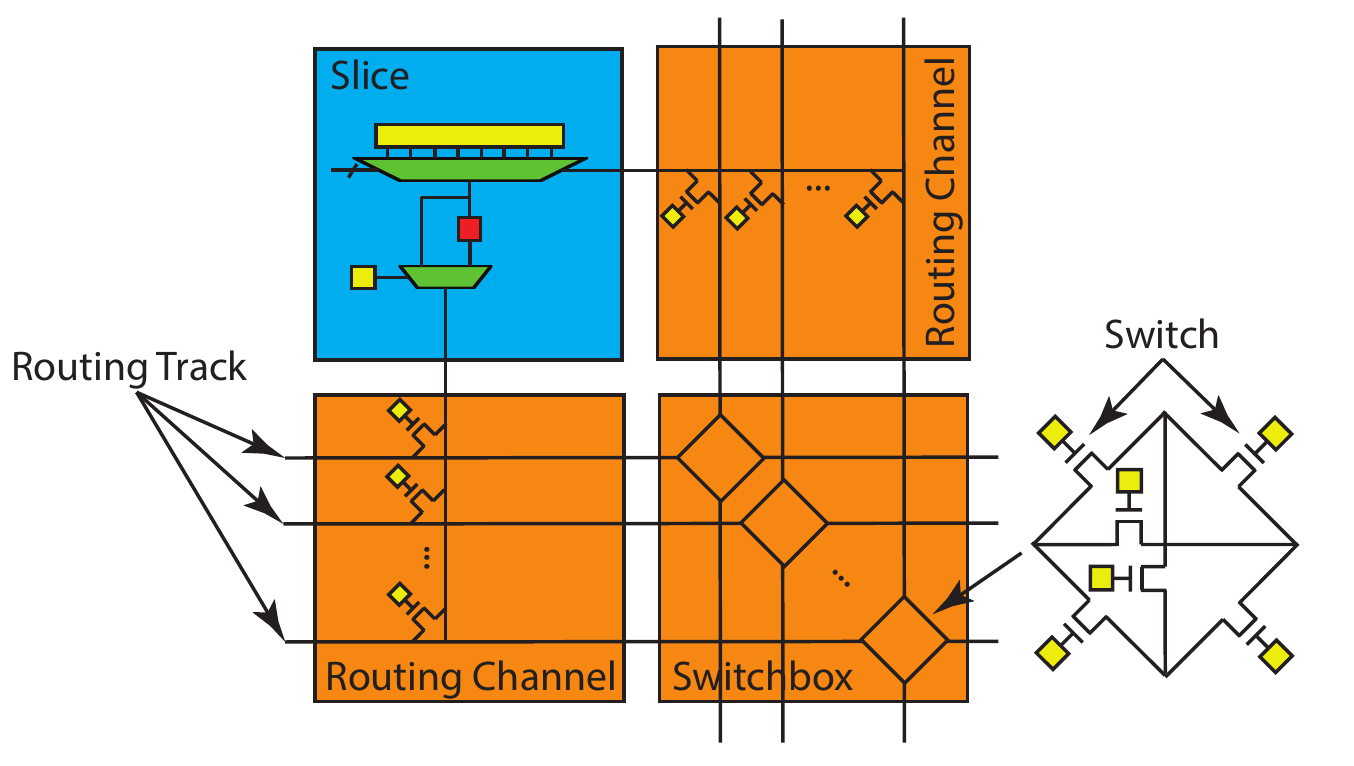}
\caption{ A \gls{slice} contains a small number of \glspl{lut} and \gls{ff}. We show a very simple \gls{slice} with one \gls{lut} and one \gls{ff} though generally these have two or more of each. Slices are connected to one another using a \gls{routingchannel} and \gls{switchbox}. These two provide a programmable interconnect that provide the data movement between the programmable logic elements (\glspl{slice}). The \gls{switchbox} has many switches (typically implemented as pass transistors) that allow for arbitrary wiring configurations between the different routing tracks in the routing tracks adjacent to the switchbox. }
\label{fig:slice_channel}
\end{figure}

Figure \ref{fig:slice_channel} provides an example of how a \gls{slice}, \gls{routingchannel}, and \gls{switchbox} are connected. Each input/output to the \gls{slice} can be connected to one of many routing tracks that exist in a routing channel. You can think of routing tracks as single bit wires. The physical connections between the slice and the routing tracks in the routing channel are configured using a pass transistor that is programmed to perform a connect or disconnect from the input/output of the slice and the programmable interconnect. 

The \glspl{switchbox} provides a connection matrix between routing tracks in adjacent \glspl{routingchannel}. Typically, an \gls{fpga} has a logical 2D representation. That is, the \gls{fpga} is designed in a manner that provides a 2D abstraction for computation. This is often called an ``island-style'' architecture where the \glspl{slice} represent ``logic islands'' that are connected using the \glspl{routingchannel} and \glspl{switchbox}. Here the \gls{switchbox} connects to four \glspl{routingchannel} to the north, east, south, and west directions. The exact programming of the switches in the \glspl{routingchannel} and \glspl{switchbox} determines how the inputs and outputs of the programmable logic blocks are connected. The number of channels, the connectivity of the switchboxes, the structure of the slice, and other logic and circuit level FPGA architectural techniques are very well studied; we refer the interested reader to the following books and surveys on this topic for more information \cite{brown1996fpga, betz1997vpr, hauck2010reconfigurable}. Generally, it is not necessary to understand all of the nitty-gritty details of the FPGA architecture in order to successfully use the \gls{hls} tools, rather it is more important to have a general understanding of the various FPGA resources and how the \gls{hls} optimizations effect the resource usage.

\begin{figure}
\centering
\includegraphics[width= 0.7\textwidth]{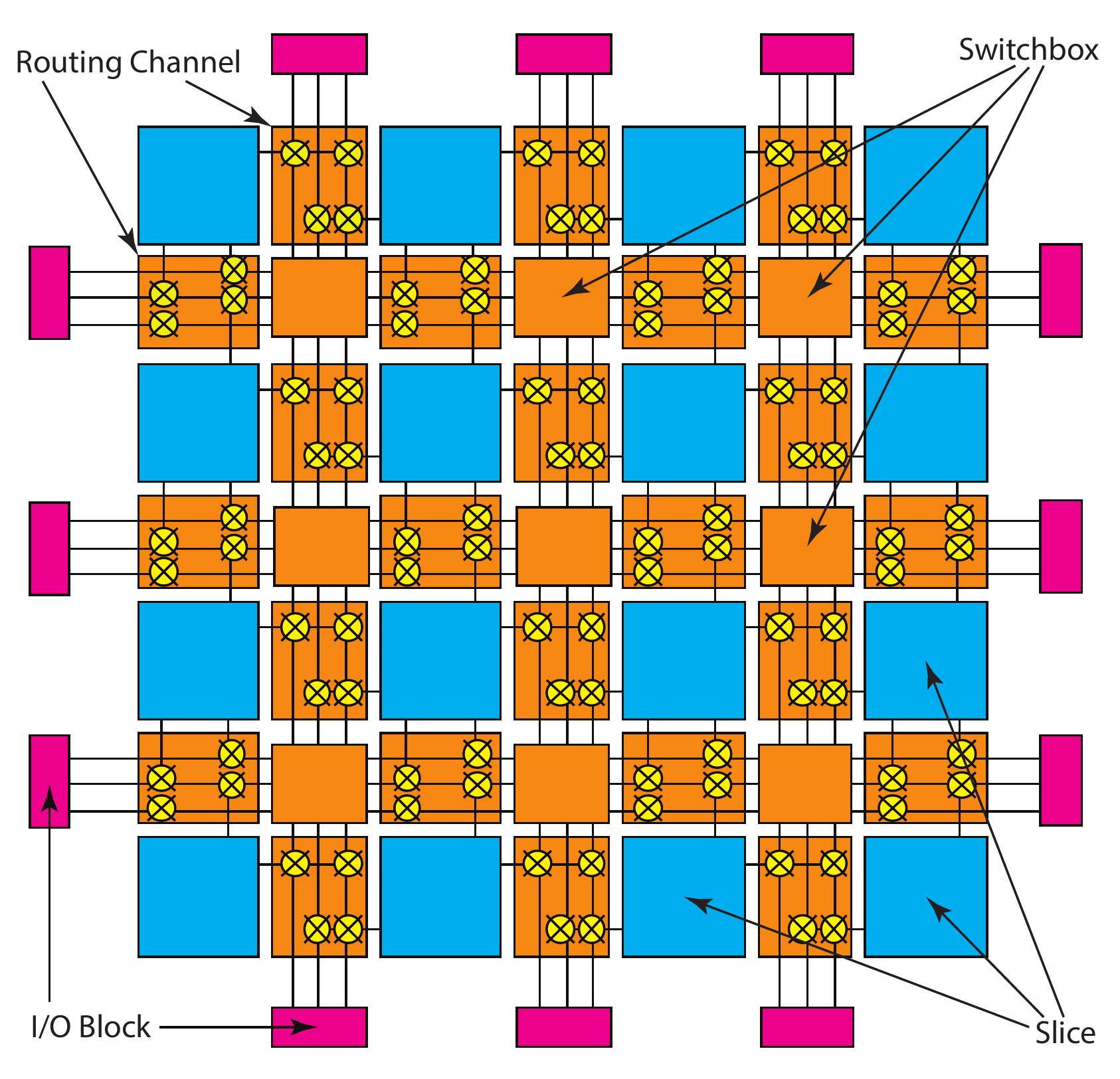}
\caption{ The two-dimensional structure of an \gls{fpga} showing an island style architecture. The logic and memory resources in the \glspl{slice} are interconnected using \glspl{routingchannel} and \glspl{switchbox}. The input/output (I/O) blocks provide an external interface, e.g., to a memory, microprocessor, sensor, and/or actuator. On some FPGAs, the I/O directly connects to the chip pins. Other FPGAs use the I/O to connect the programmable logic fabric to on-chip resources (e.g., a microprocessor bus or cache).   }
\label{fig:fpga}
\end{figure}

Figure \ref{fig:fpga} provides an even more abstract depiction of an \gls{fpga} programmable logic and interconnect. This provides a larger view of the 2D dimensional layout of the programmable logic (e.g., \glspl{slice}), \glspl{routingchannel}, and \glspl{switchbox}. The \gls{fpga} programmable logic uses \glspl{ioblock} to communicate with an external device. This may be a microcontroller (e.g., an on-chip ARM processor using an AXI interface), memory (e.g., an on-chip cache or an off-chip DRAM memory controller), a sensor (e.g., an antenna through an A/D interface), or a actuator (e.g., a motor through an D/A interface). More recently, \glspl{fpga} have integrated custom on-chip I/O handlers, e.g., memory controllers, transceivers, or analog-to-digital (and vice versa) controllers directly into the fabric in order to increase performance.

\begin{figure}
\centering
\includegraphics[width=\textwidth]{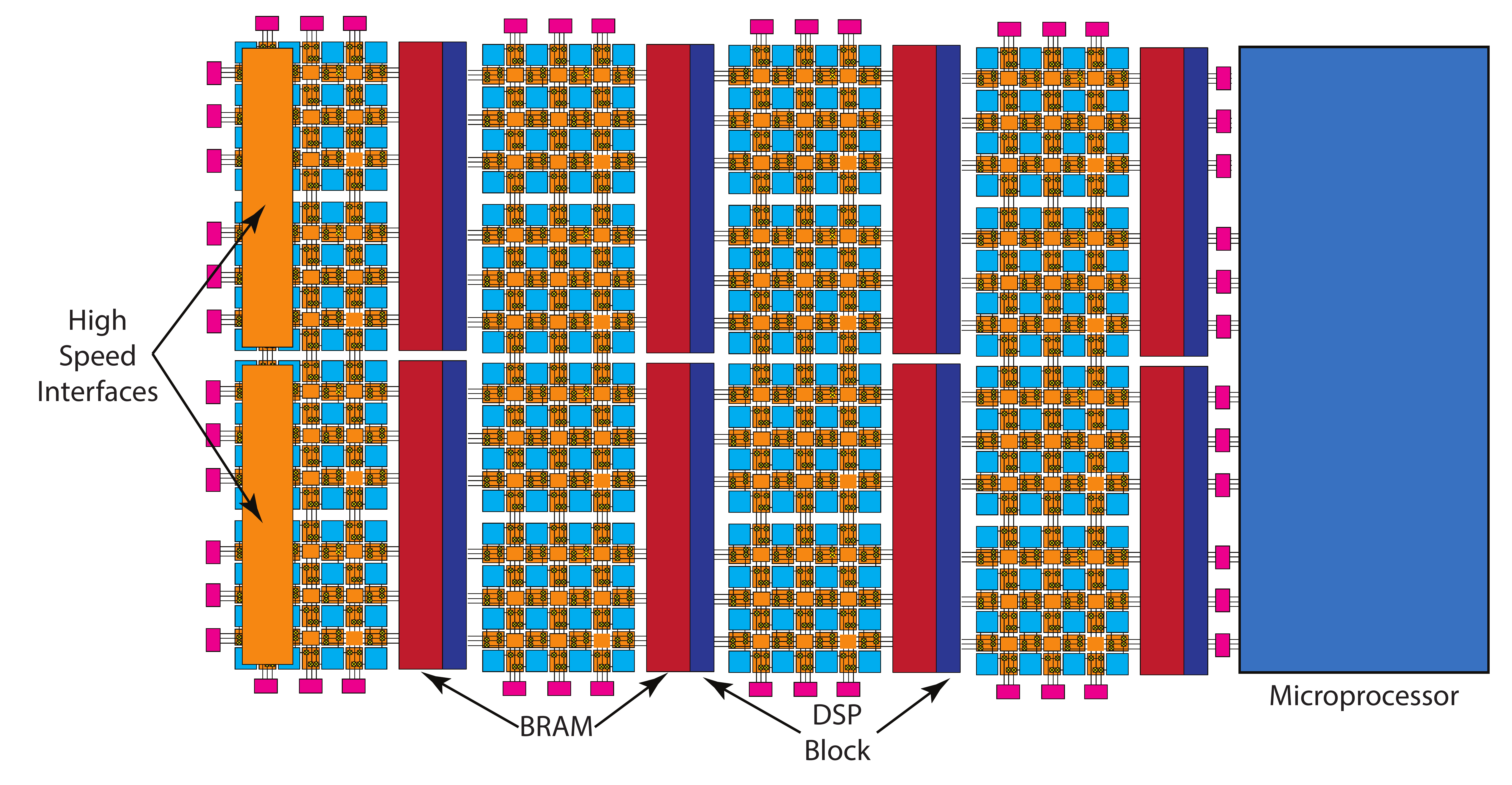}
\caption{ Modern \glspl{fpga} are becoming more heterogenous with a mix of programmable logic elements and ``hardened'' architectural elements like register files, custom datapaths, and high speed interconnect. The \gls{fpga} is often paired with one or more microprocessors, e.g., ARM or x86 cores, that coordinates the control of the system. }
\label{fig:heterogenous_fpga}
\end{figure}

As the number of transistors on the FPGA has continued to increase, FPGA architectures have begun to incorporate more and more ``hard'' resources. These are custom resources designed specifically to perform a task. For example, many applications make heavy use of addition and multiplication operations. Thus, the FPGA fabric added custom resources targeted at these operations. An example of this is the DSP48 custom datapaths, which efficiently implement a series of arithmetic operations including multiplication, addition, multiply-accumulate, and word level logical operations. These DSP48 blocks have some programmability, but are not as flexible as the programmable logic. Yet, implementing a multiply or MAC operation on these DSP48s is much more efficient than performing the same operation on the programmable logic. Thus, there is a fundamental tradeoff of efficiency versus flexibility. Modern FPGAs will have hundreds to thousands of these DSP48 distributed throughout the logic fabric as shown in Figure \ref{fig:heterogenous_fpga}.

\begin{exercise}
Compare the performance of a multiply accumulate operation using the programmable logic versus the DSP48s. What is the maximum frequency that one can obtain in both cases? How does the FPGA resource usage change?
\end{exercise} 

A \gls{bram} is another example of a hardened resource. \glspl{bram} are configurable random access memory modules that support different memory layouts and interfaces. For example, they can be changed to have byte, half-word, word, and double word transfers and connected to a variety of different interfaces including local on-chip buses (for talking to the programmable fabric) and processor buses (to communicate with on-chip processors). Generally, these are used to transfer data between on-chip resources (e.g., the FPGA fabric and microprocessor) and store large data sets on chip. We could choose to store the data set in the slices (using the \glspl{ff}) but this would incur overheads in performance and resource usage. 

A typical \gls{bram} has around 32 Kbit of memory storage which can be configured as 32K x 1 bit, 16K x 2 bits, 8K x 4 bits, etc. They can be cascaded together to create larger memories. All of this configuration is done by the Vivado tools; this is a major advantage of \VHLS: the designer does not need to worry about these low level details. The \glspl{bram} are typically co-located next to the DSP48. For HLS, it may be beneficial to think of the \glspl{bram} as configurable register files. These \glspl{bram} can directly feed the custom datapaths (DSP48s), talk to on-chip microprocessors, and transfer data to custom datapaths implemented on the programmable logic.

\begin{exercise}
Implement a design that stores a large (e.g., thousand item) array in BRAMs and programmable logic. How does the performance change? What about the resource usage?
\end{exercise}

Figure \ref{fig:FPGAmemories} provides a comparison between different on-chip and off-chip memory resources. There are millions of \glspl{ff} on-chip and these provide hundreds of Kbytes of bit level storage. These can be read to and written to on every cycle and thus provide a tremendous amount of total bandwidth. Unfortunately, they do not provide the most efficient storage capacity. \glspl{bram} provide a bit more storage density at the cost of limited total bandwidth. Only one or two entries of the \glspl{bram} can be accessed during every cycle which is the major limiting factor for the bandwidth. Going even further in this direction, we can use very high density off-chip external memory, but the bandwidth is even further reduced. The decision about where to place your application's data is crucial and one that we will consider extensively throughout this book. The \VHLS tool has many options that allow the designer to specify exactly where and how to store data.

\begin{figure}
\centering
 
\begin{tabular}{|r|c|c|c|}
\hline
 & External &  & \\
  & Memory & BRAM & FFs \\
\hline
count & 1-4 & thousands & millions \\
size & GBytes & KBytes & Bits \\
total size & GBytes & MBytes & 100s of KBytes \\
width & 8-64 & 1-16 & 1 \\
total bandwidth & GBytes/sec & TBytes/sec & 100s of TBytes/sec \\
\hline
\end{tabular}
\caption{A comparison of three different on- and off-chip memory storage options. External memory provides the most density but has limited total bandwidth. Moving on-chip there are two options: \glspl{ff} and \glspl{bram}. \glspl{ff} have the best total bandwidth but only a limited amount of total data storage capability. \glspl{bram} provide an intermediate value between external memory and \glspl{ff}.}
\label{fig:FPGAmemories}
\end{figure}

As on-chip transistors have become more plentiful, we have the ability to consider integrating even more complex hardened resources. On-chip microprocessors are a prime example of this. High-end modern FPGAs can include four or more on-chip microprocessors (e.g., ARM cores). While only the largest FPGAs include four microprocessors, it is very common to see one microprocessor included in even the smaller FPGA devices. This provides the ability to run an operating system (e.g., Linux) and quickly leverage all of its facilities, e.g., to communicate with devices through drivers, run larger software packages like OpenCV, and use more common high level programming languages like Python to get the system up and running quickly. The microprocessor is often the controller for the system. It orchestrates data movement between off-chip memories, sensors, and actuators  to on-chip resources like the \glspl{bram}. And the microprocessor can coordinate between the custom IP cores developed with the \gls{hls} tools, third party IP cores, and the board level resources.   

\section{FPGA Design Process}

\begin{figure}
\centering
\def\svgwidth{\columnwidth}
\executeiffilenewer{embedded_design_template.svg}{images/embedded_design_template.pdf}%
{inkscape -z -D --file=embedded_design_template.svg %
--export-pdf=images/embedded_design_template.pdf --export-latex}%
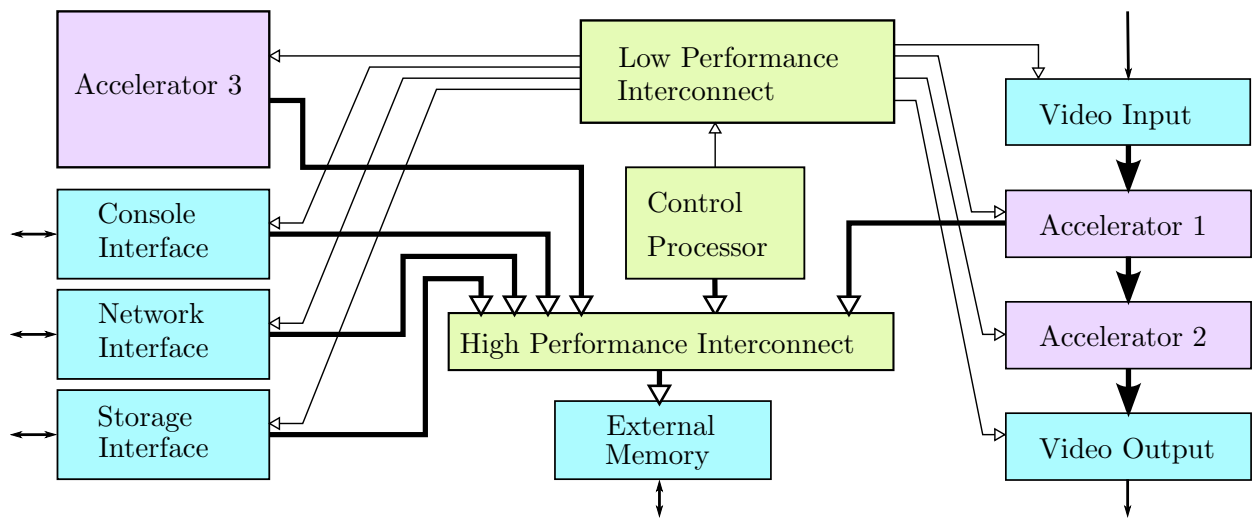%

\caption{A block diagram showing a hypothetical embedded FPGA design, consisting of I/O interface cores (shown in blue), standard cores (shown in green), and application specific accelerator cores (shown in purple).  Note that accelerator cores might have streaming interfaces (Accelerator 2), memory-mapped interfaces (Accelerator 3), or both (Accelerator 1). }
\label{fig:designTemplate}
\end{figure}

Given the complexity and size of modern FPGA devices, designers have looked to impose a higher-level structure on building designs.  As a result, FPGA designs are often composed of components or \glspl{core}, structured something like Figure \ref{fig:designTemplate}.  At the periphery of the design, close to the I/O pins, is a relatively small amount of logic that implements timing-critical I/O functions or protocols, such as a memory controller block, video interface core or analog-to-digital converter.  This logic, which we will refer to as an \term{I/O interface core}, is typically implemented as structural RTL, often with additional timing constraints that describe the timing relationships between signals and the variability of these signals.   These constraints must also take into account interference of signals propagating through traces in a circuit board and connectors outside of the FPGA.  In order to implement high speed interfaces, these cores typically make use of dedicated logic in the FPGA architecture that is close to the I/O pins for serializing and deserializing data, recovering and distributing clock signals, and delaying signals with picosecond accuracy in order to repeatably capture data in a register.  The implementation of I/O interface cores is often highly customized for a particular FPGA architecture and hence are typically provided by FPGA vendors as reference designs or off-the-shelf components, so for the purposes of this book, we can ignore the detailed implementation of these blocks.

Away from I/O pins, FPGA designs often contain \term{standard cores}, such as processor cores, on-chip memories, and interconnect switches.   Other standard cores include generic, fixed-function processing components, such as filters, FFTs, and codecs.  Although instances of these cores are often parameterized and assembled in a wide variety of ways in a different designs, they are not typically the differentiating element in a customers design.  Instead, they represent commodity, horizontal technology that can be leveraged in designs in many different application areas.   As a result, they are often provided by an FPGA vendor or component provider and only rarely implemented by a system designer.  Unlike interface I/O cores, standard cores are primarily synchronous circuits that require few constraints other than basic timing constraints specifying the clock period.  As a result, such cores are typically portable between FPGA families, although their circuit structure may still be highly optimized.

Lastly, FPGA designs typically contain customized, application-specific \term{accelerator cores}.  As with standard cores, accelerator cores are primarily synchronous circuits that can be characterized by a clock period timing constraint.  In contrast, however, they are almost inevitably constructed for a specific application by a system designer.  These are often the ``secret sauce'' and used to differentiate your system from others. Ideally a designer can quickly and easily generate high-performance custom cores, perform a design space exploration about feasible designs, and integrate these into their systems in a short timeframe. This book will focus on the development of custom cores using HLS in a fast, efficient and high performance manner.

When integrating a design as in Figure \ref{fig:designTemplate}, there are two common design methodologies.  One methodology is to treat HLS-generated accelerator cores just like any other cores.  After creating the appropriate accelerator cores using HLS, the cores are composed together (for instance, in a tool such as \vivado IP Integrator) along with I/O interface cores and standard cores to form a complete design.  This \term{core-based design methodology} is similar to the way that FPGA designs are developed without the use of HLS.  A newer methodology focuses on standard design templates or platforms, which combine a stable, verified composition of standard cores and I/O interface cores targeting a particular board. This \term{platform-based design methodology} enables a high-level programmer to quickly integrate different algorithms or \term{roles} within the interfaces provided by a single platform or \term{shell}.  It can also make it easier to port an accelerator from one platform to another as long as the shells support the same standardized interfaces.

\section{Design Optimization}
\label{sec:designOptimization}

\subsection{Performance Characterization}
\label{sec:designCharacterization}
Before we can talk about optimizing a design, we need to discuss the key criterion that are used to characterize a design.   The computation time is a particularly important metric for design quality.  When describing synchronous circuits, one often will use the number of clock cycles as a measure of performance.  However, this is not appropriate when comparing designs that have different clock rates, which is typically the case in HLS. For example, the clock frequency is specified as an input constraint to the \VHLS, and it is feasible to generate different architectures for the same exact code by simply changing the target clock frequency.  Thus, it is most appropriate to use seconds, which allows an apples-to-apples comparison between any HLS architecture. The \VHLS tool reports the number of cycles and the clock frequency. These can be used to calculate the exact amount of time that some piece of code requires to compute.

\begin{aside}
It is possible to optimize the design by changing the clock frequency. The \VHLS tool takes as input a target clock frequency, and changing this frequency target will likely result in the tool generating different implementations. We discuss this throughout the book. For example, Chapter \ref{sec:fir-performance} describes the constraints the are imposed on the \VHLS tool depending on the clock period. Chapter \ref{sec:fir-chaining} discusses how increasing the clock period can increase the throughput by employing operation chaining.
\end{aside}

We use the term \gls{task} to mean a fundamental atomic unit of behavior; this corresponds to a function invocation in \VHLS.  The \gls{task-latency} is the time between when a task starts and when it finishes.  The \gls{task-interval} is the time between when one task starts and the next starts or the difference between the start times of two consecutive tasks.  All task input, output, and computation is bounded by the task latency, but the start of a task may not coincide with the reading of inputs and the end of a task may not coincide with writing of outputs, particularly if the task has state that is passed from one task to the next.  In many designs, \gls{data-rate} is a key design goal, and depends on both the \gls{task-interval} and the size of the arguments to the function.

\begin{figure}
\centering
\def\svgwidth{\columnwidth}
\executeiffilenewer{intervalduration.svg}{images/intervalduration.pdf}%
{inkscape -z -D --file=intervalduration.svg %
--export-pdf=images/intervalduration.pdf --export-latex}%
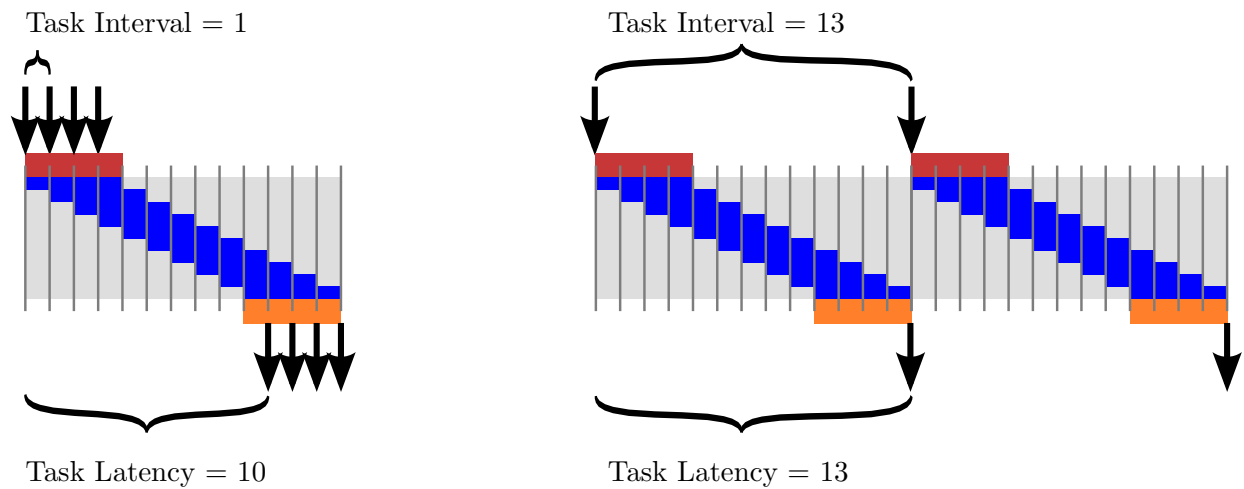%

\caption{The task interval and task latency for two different designs. The left design is pipelined while the right one uses a more sequential implementation.}
\label{fig:intervalDuration}
\end{figure}

Figure \ref{fig:intervalDuration} shows two different designs of some hypothetical application.   The horizontal axis represents time (increasing to the right) and the vertical axis represents different functional units in the design.  Input operations are shaded red and and output operations are shaded orange.  Active operators are represented in dark blue and inactive operators are represented in light blue.  Each incoming arrow represents the start of a task and each outgoing arrow represents the completion of a task.  The diagram on the left represents four executions of an architecture that starts a new task every cycle. This corresponds to a `fully-pipelined' implementation. On the right, there is task with a very different architecture, reading a block of four pieces of input data, processing it, and then producing a block of four output samples after some time.  This architecture has the same latency and interval (13 cycles) and can only process one task at a time.  This behavior is in contrast to the behavior of the pipelined design on the left, which has multiple tasks executing at any given instant of time.  Pipelining in HLS is very similar to the concept of instruction pipelines in a microprocessor.  However, instead of using a simple 5-stage pipeline where the results of an operation are usually written back to the register file before being executed on again, the \VHLS tool constructs a circuit that is customized to run a particular program on a particular FPGA device.  The tool optimizes the number of pipeline stages, the initiation interval (the time between successive data provided to the pipeline -- similar to the task interval), the number and types of functional units and their interconnection based on a particular program and the device that is being targeted.

The \VHLS tool counts cycles by determining the maximum number of registers between any input and output of a task. Thus, it is possible for a design to have a task latency of zero cycles, corresponding to a combinational circuit which has no registers in any path from input to output. Another convention is to count the input and/or output as a register and then find the maximum registers on any path. This would result in a larger number of cycles. We use the \VHLS convention throughout this book.

\begin{aside}
Note that many tools report the task interval as ``throughput''. This terminology is somewhat counterintuitive since a longer task interval almost inevitably results in fewer tasks being completed in a given time and thus lower data rates at the interfaces.  Similarly, many tools use ``latency'' to describe a relationship between reading inputs and writing outputs.  Unfortunately, in designs with complex behavior and multiple interfaces, it is hard to characterize tasks solely in terms of inputs and outputs, e.g., a task may require multiple reads or writes from a memory interface. 
\end{aside}


\subsection{Area/Throughput Tradeoffs}
\label{sec:filterThroughputTradeoffs}

In order to better discuss some of the challenges that one faces when using an HLS tool, let's consider a simple yet common hardware function -- the \gls{fir} (FIR) filter. An FIR performs a convolution on an input sequence with a fixed set of coefficients. An FIR is quite general -- it can be used to perform different types of filter (high pass filter, low pass, band pass, etc.).  Perhaps the most simple example of an FIR is a moving average filter. We will talk more background on FIR in Chapter \ref{chapter:fir} and describe many specific optimizations that can be done using HLS. But in the meantime just consider its implementation at a high level. 

The C code in Figure \ref{fig:FIR} provides a functional or task description for HLS; this can be directly used as input to the \VHLS tool, which will analyze it and  produce a functionally equivalent RTL circuit. This is a complex process, and we will not get into much detail about this now, but think of it as a compiler like gcc. Yet instead of outputting assembly code, the HLS ``compiler'' creates an RTL hardware description. In both cases, it is not necessary to understand exactly how the compiler works. This is exactly why we have the compiler in the first place -- to automate the design process and allow the programmer/designer to work at a higher level of abstraction. Yet at the same time, someone that knows more about how the compiler works will often be able to write more efficient code. This is particularly important for writing HLS code since there are many options for synthesizing the design that are not typically obvious to one that only knows the ``software'' flow. For example, ideas like custom memory layout, pipelining, and different I/O interfaces are important for HLS, but not for a ``software compiler''. These are the concepts that we focus on in this book.

A key question to understand is: ``What circuit is generated from this code?''.  Depending on your assumptions and the capabilities of a particular HLS tool, the answer can vary widely. There are multiple ways that this could be synthesized by an HLS tool. 

\begin{figure}
{
\begin{lstlisting}[format=none]
#define NUM_TAPS 4

void fir(int input, int *output, int taps[NUM_TAPS])
{
	static int delay_line[NUM_TAPS] = {};

	int result = 0;
	for (int i = NUM_TAPS - 1; i > 0; i--) {
		delay_line[i] = delay_line[i - 1];
	}
	delay_line[0] = input;

	for (int i = 0; i < NUM_TAPS; i++) {
		result += delay_line[i] * taps[i];
	}

	*output = result;
}
\end{lstlisting}
}
\caption{Code for a four tap FIR filter.
}\label{fig:FIR}
\end{figure}

\begin{figure}
{
\hbox{
\begin{lstlisting}
fir:
        .frame  r1,0,r15                # vars= 0, regs= 0, args= 0
        .mask   0x00000000
        addik   r3,r0,delay_line.1450
        lwi     r4,r3,8           # Unrolled loop to shift the delay line
        swi     r4,r3,12
        lwi     r4,r3,4
        swi     r4,r3,8
        lwi     r4,r3,0
        swi     r4,r3,4
        swi     r5,r3,0           # Store the new input sample into the delay line
        addik   r5,r0,4 	  # Initialize the loop counter
        addk    r8,r0,r0          # Initialize accumulator to zero
        addk    r4,r8,r0          # Initialize index expression to zero
$L2:
        muli    r3,r4,4           # Compute a byte offset into the delay_line array
        addik   r9,r3,delay_line.1450
        lw      r3,r3,r7          # Load filter tap
        lwi     r9,r9,0           # Load value from delay line
        mul     r3,r3,r9	  # Filter Multiply
        addk    r8,r8,r3	  # Filter Accumulate
        addik   r5,r5,-1          # update the loop counter
        bneid   r5,$L2
        addik   r4,r4,1           # branch delay slot, update index expression

        rtsd    r15, 8		  
        swi     r8,r6,0		  # branch delay slot, store the output
        .end    fir
\end{lstlisting}}\hbox{%
\executeiffilenewer{filter_asm_behavior.svg}{images/filter_asm_behavior.pdf}%
{inkscape -z -D --file=filter_asm_behavior.svg %
--export-pdf=images/filter_asm_behavior.pdf --export-latex}%
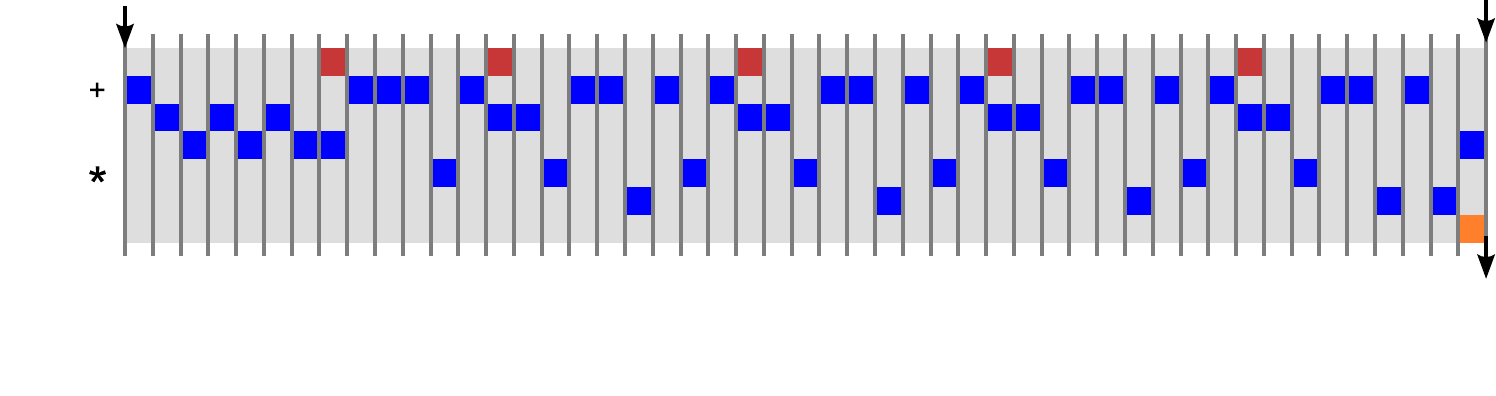%
}
}
\caption{RISC-style assembly generated from the C code in Figure \ref{fig:FIR}, targetting the Xilinx Microblaze processor.  This code is generated using \texttt{microblazeel-xilinx-linux-gnu-gcc -O1 -mno-xl-soft-mul -S fir.c}}\label{fig:FIR_microblaze}
\end{figure}

One possible circuit would execute the code sequentially, as would a simple RISC microprocessor.  Figure \ref{fig:FIR_microblaze} shows assembly code for the Xilinx Microblaze processor which implements the C code in Figure \ref{fig:FIR}.  Although this code has been optimized, many instructions must still be executed to compute array index expressions and loop control operations.  If we assume that a new instruction can be issued each clock cycle, then this code will take approximately 49 clock cycles to compute one output sample of the filter.  Without going into the details of how the code works, we can see that one important barrier to performance in this code is how many instructions can be executed in each clock cycle.  Many improvements in computer architecture are fundamentally attempts to execute more complex instructions that do more useful work more often.  One characteristic of HLS is that architectural tradeoffs can be made without needing to fit in the constraints of an instruction set architecture.  It is common in HLS designs to generate architectures that issue hundreds or thousands of RISC-equivalent instructions per clock with pipelines that are hundreds of cycles deep.

By default, the \VHLS tool will generate an optimized, but largely sequential architecture.   In a sequential architecture, loops and branches are transformed into control logic that enables the registers, functional units, and the rest of the data path.  Conceptually, this is similar to the execution of a RISC processor, except that the program to be executed is converted into a finite state machine in the generated RTL rather than being fetched from the program memory.   A sequential architecture tends to limit the number of functional units in a design with a focus on resource sharing over massive parallelism.  Since such an implementation can be generated from most programs with minimal optimization or transformation, this makes it easy for users to get started with HLS.  One disadvantage of a sequential architecture is that analyzing and understanding data rates is often difficult since the control logic can be complex.  The control logic dictates the number of cycles for the task interval and task latencies. The control logic can be quite complex, making it difficult to analyze. In addition, the behavior of the control logic may also depend on the data being processed, resulting in performance that is data-dependent.

\begin{figure}
\centering
\mbox{%
\executeiffilenewer{filter_one_sample.svg}{images/filter_one_sample.pdf}%
{inkscape -z -D --file=filter_one_sample.svg %
--export-pdf=images/filter_one_sample.pdf --export-latex}%
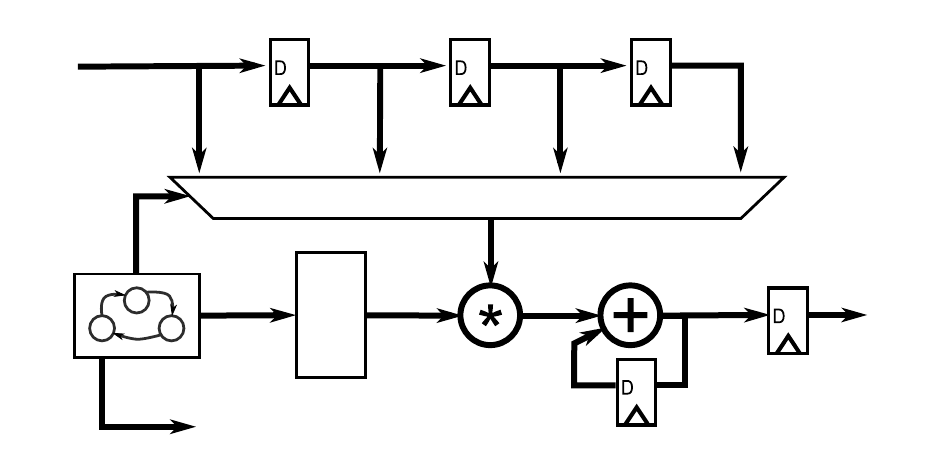%
}\mbox{%
\executeiffilenewer{filter_one_sample_behavior.svg}{images/filter_one_sample_behavior.pdf}%
{inkscape -z -D --file=filter_one_sample_behavior.svg %
--export-pdf=images/filter_one_sample_behavior.pdf --export-latex}%
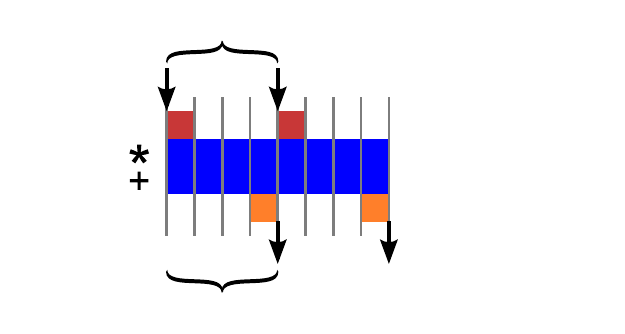%
}
\caption{A ``one tap per clock'' architecture for an FIR filter.  This architecture can be implemented from the code in Figure \ref{fig:FIR}.}
\label{fig:FIR_sequential}
\end{figure}

\begin{figure}
\centering
\mbox{%
\executeiffilenewer{filter_one_tap.svg}{images/filter_one_tap.pdf}%
{inkscape -z -D --file=filter_one_tap.svg %
--export-pdf=images/filter_one_tap.pdf --export-latex}%
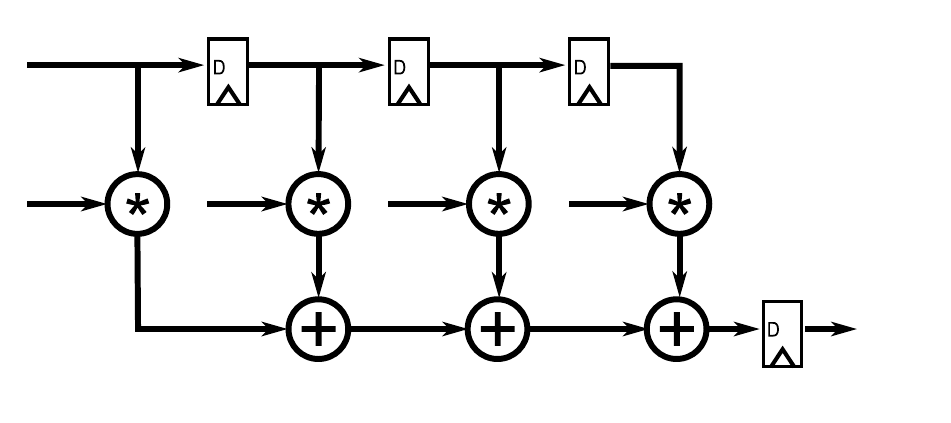%
}\mbox{%
\executeiffilenewer{filter_one_tap_behavior.svg}{images/filter_one_tap_behavior.pdf}%
{inkscape -z -D --file=filter_one_tap_behavior.svg %
--export-pdf=images/filter_one_tap_behavior.pdf --export-latex}%
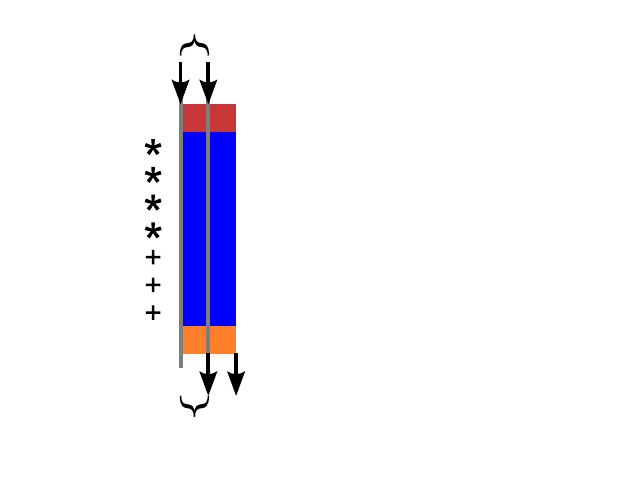%
}
\caption{A ``one sample per clock'' architecture for an FIR filter.  This architecture can be implemented from the code in Figure \ref{fig:FIR} using function pipeline.}\label{fig:FIR_function_pipeline}
\end{figure}

However, the \VHLS tool can also generate higher performance pipelined and parallel architectures.  One important class of architectures is called a \term{function pipeline}.  A function pipeline architecture is derived by considering the code within the function to be entirely part of a computational data path, with little control logic.  Loops and branches in the code are converted into unconditional constructs.  As a result, such architectures are relatively simple to characterize, analyze, and understand and are often used for simple, high data rate designs where data is processed continuously.  Function pipeline architectures are beneficial as components in a larger design since their simple behavior allows them to be resource shared like primitive functional units.  One disadvantage of a function pipeline architecture is that not all code can be effectively parallelized.

The \VHLS tool can be directed to generate a function pipeline by placing the \lstinline|#pragma HLS pipeline| directive in the body of a function.  This directive takes a parameter that can be used to specify the initiation interval of the the pipeline, which is the same as a task interval for a function pipeline.  Figure \ref{fig:FIR_sequential} shows one potential design -- a ``one tap per clock'' architecture, consisting of a single multiplier and single adder to compute the filter.  This implementation has a task interval of 4 and a task latency of 4.  This architecture can take a new sample to filter once every 4 cycles and will generate a new output 4 cycles after the input is consumed. The implementation in Figure \ref{fig:FIR_function_pipeline} shows a ``one sample per clock'' architecture, consisting of 4 multipliers and 3 adders.  This implementation has a task interval of 1 and a task latency of 1, which in this case means that the implementation accept a new input value every cycle. Other implementations are also possible, such as architectures with ``two taps per clock'' or ``two samples per clock'', which might be more appropriate for a particular application.   We discuss more ways to optimize the FIR function in depth in Chapter \ref{chapter:fir}. 

In practice, complex designs often include complicated tradeoffs between sequential architectures and parallel architectures, in order to achieve the best overall design.  In \VHLS, these tradeoffs are largely controlled by the user, through various tool options and code annotations, such as \lstinline|#pragma| directive.

\begin{exercise}
What would the task interval of a ``two taps per clock'' architecture for a 4 tap filter be?  What about for a ``two samples per clock'' architecture?
\end{exercise}

\subsection{Restrictions on Processing Rate}

As we have seen, the task interval of a design can be changed by selecting different kinds of architectures, often resulting in a corresponding increase in the processing rate.  However, it is important to realize that the task interval of any processing architecture is fundamentally limited in several ways.  The most important limit arises from \glspl{recurrence} or feedback loops in a design.  The other key limitation arises from \term{resource limits}.


A \term{recurrence} is any case where a computation by a component depends on a previous computation by the same component.  A key concept is that {\em recurrences fundamentally limit the throughput of a design}, even in the presence of pipelining \cite{papaefthymiou91,leiserson93}.  As a result, analyzing recurrences in algorithms and generating hardware that is guaranteed to be correct is a key function of an HLS tool.  Similarly, understanding algorithms and selecting those without tight recurrences is a important part of using HLS (and, in fact, parallel programming in general).

Recurrences can arrive in different coding constructs, such as static variables (Figure \ref{fig:FIR}), sequential loops (Figure \ref{fig:FIR_sequential}).  Recurrences can also appear in a sequential architecture and disappear when pipelining is applied (as in Figures \ref{fig:FIR_sequential} and \ref{fig:FIR_function_pipeline}).  In other cases, recurrences can exist without limiting the throughput of a sequential architecture, but can become problematic when the design is pipelined.


%


Another key factor that can limit processing rate are resource limitations.  One form of resource limitation is associated with the wires at the boundary of a design, since a synchronous circuit can only capture or transmit one bit per wire per clock cycle.  As a result, if a function with the signature \lstinline|int32_t f(int32_t x);| is implemented as a single block at 100 MHz with a task interval of 1, then the most data that it can process is 3.2 Gbits/sec. Another form of resource limitation arises from memories since most memories only support a limited number of accesses per clock cycle.  Yet another form of resource limitation comes from user constraints.  If a designer limits the number of operators that can instantiated during synthesis, then this places a limit on the processing rate of the resulting design.

%

\subsection{Coding Style}

Another key question you should ask yourself is, ``Is this code the best way to capture the algorithm?''.  In many cases, the goal is not only the highest quality of results, but maintainable and modifiable code.  Although this is somewhat a stylistic preference, coding style can sometimes limit the architectures that a HLS tool can generate from a particular piece of code.

For instance, while a tool might be able to generate either architecture in Figure \ref{fig:FIR_function_pipeline} or \ref{fig:FIR_sequential} from the code in Figure \ref{fig:FIR}, additing additional directives as shown in Figure \ref{fig:block_FIR} would result in a specific architecture. In this case the delay line is explicitly unrolled, and the multiply-accumulate \lstinline|for| loop is stated to be implemented in a pipelined manner. This would direct the HLS tool to produce an architecture that looks more like the pipelined one in Figure \ref{fig:FIR_function_pipeline}.

\begin{figure}
\begin{lstlisting}

#define NUM_TAPS 4

void block_fir(int input[256], int output[256], int taps[NUM_TAPS],
							 int delay_line[NUM_TAPS]) {
	int i, j;
	for (j = 0; j < 256; j++) {
		int result = 0;
		for (i = NUM_TAPS - 1; i > 0; i--) {
#pragma HLS unroll
			delay_line[i] = delay_line[i - 1];
		}
		delay_line[0] = input;

		for (i = 0; i < NUM_TAPS; i++) {
#pragma HLS pipeline
			result += delay_line[i] * taps[i];
		}
		output[j] = result;
	}
}
\end{lstlisting}
\caption{Alternative code implementing an FIR filter.}\label{fig:block_FIR}
\end{figure}


\begin{exercise}
The chapter described how to build filters with a range of different processing rates, up to one sample per clock cycle.  However, many designs may require processing data at a higher rate, perhaps several samples per clock cycle.  How would you code such a design?  Implement a 4 samples per clock cycle FIR filter.  How many resources does this architecture require (e.g., number of multipliers and adders)? Which do you think will use more FPGA resources: the 4 samples per clock or the 1 sample per clock cycle design?  
\end{exercise}

We look further into the optimization of an FIR function in more detail in Chapter \ref{chapter:fir}. We discuss how to employ different optimizations (pipelining, unrolling, bitwidth optimization, etc.), and describe their effects on the performance and resource utilization of the resulting generated architectures.

\section{Restructured Code}

Writing highly optimized synthesizable HLS code is often not a straightforward process. It involves a deep understanding of the application at hand, the ability to change the code such that the \VHLS tool creates optimized hardware structures and utilizes the directives in an effective manner. 

Throughout the rest of the book, we walk through the synthesis of a number of different application domains -- including digital signal processing, sorting, compression, matrix operations, and video processing. In order to get the most efficient architecture, it is important to have a deep understanding of the algorithm. This enables optimizations that require rewriting the code rather than just adding synthesis directives -- a processing that we call code restructuring. 


\term{Restructured code} maps well into hardware, and often represents the eccentricities of the tool chain, which requires deep understanding of micro-architectural constructs in addition to the algorithmic functionality. Standard, off-the-shelf code typically yields very poor quality of results that are orders of magnitude slower than CPU designs, even when using HLS directives such as pipelining, unrolling, and data partitioning. Thus, it is important to understand how to write code that the \VHLS will synthesize in an optimal manner.

Restructured code typically differs substantially from a software implementation -- even one that is highly optimized. A number of studies suggest that restructuring code is an essential step to generate an efficient FPGA
design \cite{mataidesigning, matai2energy, cong2011high, chen2012fpga, lee250high}. Thus, in order to get an efficient hardware design, the user must write restructured code with the underlying
hardware architecture in mind. Writing restructured code requires significant
hardware design expertise and domain specific knowledge. 

Throughout the rest of this book, we go through a number of different applications, and show how to restructure the code for a more efficient hardware design. We present applications such as finite impulse response (FIR), discrete Fourier transform (DFT), fast Fourier transform (FFT), sparse matrix vector multiply (SpMV), matrix multiplication, sorting, and Huffman encoding. We discuss the impact of restructured code on the final hardware generated from high-level synthesis. And we propose a restructuring techniques based on best practices. In each chapter, we aim to:
\begin{enumerate}
	\item Highlight the importance of restructuring code to obtain FPGA
          designs with good quality of result, i.e., a design that has high performance and low area usage;
	\item Provide restructured code for common applications;
	\item Discuss the impact of the restructuring on the underlying hardware; and
	\item Perform the necessary HLS directives to achieve the best design
\end{enumerate}
Throughout the book, we use example applications to show how to move from a baseline implementation and restructure the code to provide more efficient hardware design. We believe that the optimization process is best understood through example. Each chapter performs a different set of optimization directives including pipelining, dataflow, array partitioning, loop optimizations, and bitwidth optimization.  Additionally, we provide insight on the skills and knowledge necessary to perform the code restructuring process.

\section{Book Organization}
We organized this book to teach by example. Each chapter presents an application (or application domain) and walks through its implementation using  different HLS optimizations. Generally, each chapter focuses on a limited subset of optimization techniques. And the application complexity generally increases in the later chapters. We start with a relatively simple to understand finite impulse response (FIR) filter in Chapter \ref{chapter:fir} and move on to implement complete video processing systems in Chapter \ref{chapter:video}. 

There are of course benefits and drawbacks to this approach. The benefits are: 1) the reader can see how the optimizations are directly applicable to an application, 2) each application provides an exemplar of how to write HLS code, and 3) while simple to explain, toy examples can be hard to generalize and thus do not always make the best examples. 

The drawbacks to the teach by example approach are: 1) most applications requires some background to give the reader a better understanding of the computation being performed. Truly understanding the computation often necessitates an extensive discussion on the mathematical background on the application. For example, implementing the best architecture for the fast Fourier transform (FFT) requires that the designer have deep understanding of the mathematical concepts behind a discrete Fourier transform (DFT) and FFT. Thus, some chapters, e.g., Chapter \ref{chapter:dft} (DFT) and Chapter \ref{chapter:fft} (FFT), start with a non-trivial amount of mathematical discussion. This may be off-putting to a reader simply looking to understand the basics of HLS, but we believe that such a deep understanding is necessary to understand the code restructuring that is necessary to achieve the best design. 2) some times a concept could be better explained by a toy example that abstracts away some of the non-important application details. 

The organization for each chapter follows the same general pattern. A chapter begins by providing a background on the application(s) under consideration. In many cases, this is straightforward, e.g., it is not too difficult to explain matrix multiplication (as in Chapter \ref{chapter:matrix_multiplication}) while the DFT requires a substantial amount of discussion (Chapter \ref{chapter:dft}). Then, we provide a \emph{baseline implementation} -- a functionally correct but unoptimized implementation of the application using \VHLS. After that, we perform a number of different optimizations. Some of the chapters focus on a small number of optimizations (e.g., Chapter \ref{chapter:cordic} emphasizes bitwidth optimizations) while others look at a broad range of optimizations (e.g., Chapter \ref{chapter:fir} on FIR filters). The key optimizations and design methods are typically introduced in-depth in one chapter and then used repeatedly in the subsequent chapters. 

The book is made to be read sequentially. For example, Chapter \ref{chapter:fir} introduces most of the optimizations and the later chapters provide more depth on using these optimizations. The applications generally get more complex throughout the book. However, each chapter is relatively self-contained. Thus, a more advanced HLS user can read an individual chapter if she only cares to understand a particular application domain. For example, a reader interested in generating a hardware accelerated sorting engine can look at Chapter \ref{chapter:sorting} without necessarily have to read all of the previous chapters. This is another benefit of our teach by example approach.

\begin{table}[htp]
\caption{A map describing the types of HLS optimizations and the chapters that discuss the concepts beyond them.}
\begin{center}
\begin{tabular}{c||c|c|c|c|c|c|c|c|c|c|}
Chapter & \rotatebox{90}{FIR} & \rotatebox{90}{CORDIC} & \rotatebox{90}{DFT} & \rotatebox{90}{FFT} & \rotatebox{90}{SPMV} & \rotatebox{90}{MatMul} & \rotatebox{90}{Histogram} & \rotatebox{90}{Video} & \rotatebox{90}{Sorting} & \rotatebox{90}{Huffman} \\
 & 2 & 3 & 4 & 5 & 6 & 7 & 8 & 9 & 10 & 11 \\
\hline \hline
Loop Unrolling & X & & X & X & X & & X & & X &  \\
Loop Pipelining & X & & X & X & X & & X & X & X & X \\
Bitwidth Optimization & X & X & & & & & & & & X  \\
Function Inlining & X & & & & & & & & & X  \\
Hierarchy & X & & & X & & & X & X & X & X  \\
Array Optimizations &  & & X & X & X & X & X & X & X & X  \\
Task Pipelining &  & &  & X & & & X & X & X & X  \\
Testbench &  & & & & X & X & & & X & X  \\
Co-simulation &  & & & & X & & & & &  \\
Streaming &  & & & &  & X & & X & X &  \\
Interfacing &  & & & &  & & & X & &  \\

\end{tabular}
\end{center}
\label{table:optimizations}
\end{table}%

Table \ref{table:optimizations} provides an overview of the types of optimization and the chapters where they are covered in at least some level of detail. Chapter \ref{chapter:fir} provides a gentle introduction the HLS design process. It overviews several different optimizations, and shows how they can be used in the optimization of a FIR filter. The later chapters go into much more detail on the benefits and usage of these optimizations. 

The next set of chapters (Chapters \ref{chapter:cordic} through \ref{chapter:fft}) build digital signal processing blocks (CORDIC, DFT, and FFT). Each of these chapters generally focuses on one optimization: bitwidth optimization (Chapter \ref{chapter:cordic}), array optimizations (Chapter \ref{chapter:dft}, and task pipelining (Chapter \ref{chapter:fft}). For example, Chapter \ref{chapter:cordic} gives an in-depth discussion on how to perform bitwidth optimizations on the CORDIC application.  Chapter \ref{chapter:dft} provides an introduction to array optimizations, in particular, how to perform array partitioning in order to increase the on-chip memory bandwidth. This chapter also talks about loop unrolling and pipelining, and the need to perform these three optimizations in concert. Chapter \ref{chapter:fft} describes the optimizations of the FFT, which itself is a major code restructuring of the DFT. Thus, the chapter gives a background on how the FFT works. The FFT is a staged algorithm, and thus highly amenable to task pipelining. The final optimized FFT code requires a number of other optimizations including loop pipelining, unrolling, and array optimizations. Each of these chapters is paired with a project from the Appendix. These projects lead the design and optimization of the blocks, and the integration of these blocks into wireless communications systems. 

Chapters \ref{chapter:spmv} through \ref{chapter:huffman} provide a discussion on the optimization of more applications. Chapter \ref{chapter:spmv} describes how to use a testbench and how to perform RTL co-simulation. It also describes array and loop optimizations; these optimizations are common and thus are used in the optimization of many applications. Chapter \ref{chapter:matrix_multiplication} introduces the streaming optimization for data flow between tasks. Chapter \ref{chapter:prefixsum} presents two applications (prefix sum and histogram) that are relatively simple, but requires careful code restructuring in order to create the optimal architecture. Chapter \ref{chapter:video} talks extensively about how to perform different types interfacing, e.g., with a video stream using different bus and memory interfaces. As the name implies, the video streaming requires the use of the stream primitive, and extensive usage of loop and array optimizations. Chapter \ref{chapter:sorting} goes through a couple of sorting algorithms. These require a large number of different optimizations. The final chapter creates a complex data compression architecture. It has a large number of complex blocks that work on a more complex data structure (trees). 

%
%
%
%
%
%
%
%


\chapter{Finite Impulse Response (FIR) Filters}
\glsresetall
\label{chapter:fir}

\section{Overview}
Finite Impulse Response (FIR) filters are commonplace in digital signal processing (DSP) applications -- they are perhaps the most widely used operation in this domain. They are well suited for hardware implementation since they can be implemented as a highly optimized architecture. A key property is that they are a linear transform on contiguous elements of a signal. This maps well to a data structures (e.g., FIFOs or tap delay lines) that can be implemented efficiently in hardware. In general, streaming applications tend to map well to FPGAs, e.g., most of the examples that we present throughout the book have some sort of streaming behavior. 

Two fundamental uses for a filter are signal restoration and signal separation. Signal separation is perhaps the more common use case: here one tries to isolate the input signal into different parts. Typically, we think of these as different frequency ranges, e.g., we may want perform a low pass filter in order remove high frequencies that are not of interest. Or we may wish to perform a band pass filter to determine the presence of a particular frequency in order to demodulate it, e.g., for isolating tones during frequency shift keying demodulation.  Signal restoration relates to removing noise and other common distortion artifacts that may have been introduced into the signal, e.g., as data is being transmitted across the wireless channel. This includes smoothing the signal and removing the DC component.

Digital FIR filters often deal with a discrete signal generated by sampling a continuous signal. The most familiar sampling is performed in time, i.e., the values from a signal are taken at discrete instances. These are most often sampled at regular intervals. For instance, we might sample the voltage across an antenna at a regular interval with an analog-to-digital converter. Alternatively we might sample the current created by a photo-diode to determine the light intensity.  Alternatively, samples may be taken in space.  For instance, we might sample the value of different locations in an image sensor consisting of an array of photo-diodes to create a digital image.  More in-depth descriptions of signals and sampling can be found in \cite{lee2011signalsandsystems}.

The format of the data in a sample changes depending upon the application. Digital communications often uses complex numbers (in-phase and quadrature or I/Q values) to represent a sample. Later in this chapter we will describe how to design a complex FIR filter to handle such data. In image processing we often think of a pixel as a sample. A pixel can have multiple fields, e.g., red, green, and blue (RGB) color channels.  We may wish to filter each of these channels in a different way again depending upon the application.

The goal of this chapter is to provide a basic understanding of the process of taking an algorithm and creating a good hardware design using high-level synthesis. The first step in this process is always to have a deep understanding of the algorithm itself. This allows us to make design optimizations like code restructuring much more easily. The next section provides an understanding of the FIR filter theory and computation. The remainder of the chapter introduces various HLS optimizations on the FIR filter. These are meant to provide an overview of these optimizations. Each of them will described in more depth in subsequent chapters. 

\section{Background}

The output signal of a filter given an impulse input signal is its \term{impulse response}. The impulse response of a linear, time invariant filter contains the complete information about the filter. As the name implies, the impulse response of an FIR filter (a restricted type of linear, time invariant filter) is finite, i.e., it is always zero far away from zero.  Given the impulse response of an FIR filter, we can compute the output signal for any input signal through the process of \term{convolution}.  This process combines samples of the impulse response (also called \term{coefficients} or \term{taps}) with samples of the input signal to compute samples of the output signal.  The output of filter can be computed in other ways (for instance, in the frequency domain), but for the purposes of this chapter we will focus on computing in the time domain.

The convolution of an N-tap FIR filter with coefficients $h[]$ with an input signal $x[]$ is described by the general difference equation:
\begin{equation}
y[i] =  \displaystyle\sum\limits_{j=0}^{N-1} h[j] \cdot x[i-j]
\end{equation}
Note that to compute a single value of the output of an N-tap filter requires N multiplies and N-1 additions.

\term{Moving average filters} are a simple form of lowpass FIR filter where all the coefficients are identical and sum to one. For instance in the case of the three point moving filter, the coefficients are $h = [\frac{1}{3}, \frac{1}{3}, \frac{1}{3}]$.  It is also called a \term{box car filter} due to the shape of its convolution kernel.  Alternatively, you can think of a moving average filter as taking the average of several adjacent samples of the input signal and averaging them together. We can see that this equivalence by substituting $1/N$ for $h[j]$ in the convolution equation above and rearranging to arrive at the familiar equation for an average of $N$ elements:
\begin{equation}
y[i] = \frac{1}{N} \displaystyle\sum\limits_{j=0}^{N} x[i-j]
\end{equation}

Each sample in the output signal can be computer by the above equation using $N-1$ additions and one final multiplication by $1/N$.  Even the final multiplication can often be regrouped and merged with other operations.  As a result, moving average filters are simpler to compute than a general FIR filter.  Specifically, when $N = 3$ we perform this operation to calculate $y[12]$:
\begin{equation}
y[12] = \frac{1}{3} \cdot (x[12] + x[11] + x[10])
\end{equation}
This filter is \term{causal}, meaning that the output is a function of no future values of the input. It is possible and common to change this, for example, so that the average is centered on the current sample, i.e., $y[12] = \frac{1}{3} \cdot (x[11] + x[12] + x[13])$. While fundamentally causality is an important property for system analysis, it is less important for a hardware implementation as a finite non-causal filter can be made causal with buffering and/or reindexing of the data.

Moving average filters can be used to smooth out a signal, for example to remove random (mostly high frequency) noise.   As the number of taps $N$ gets larger, we average over a larger number of samples, and we correspondingly must perform more computations. For a moving average filter, larger values of $N$ correspond to reducing the bandwidth of the output signal. In essence, it is acting like a low pass filter (though not a very optimal one). Intuitively, this should make sense. As we average over larger and larger number of samples, we are eliminating higher frequency variations in the input signal. That is, ``smoothing'' is equivalent to reducing higher frequencies. The moving average filter is optimal for reducing white noise while keeping the sharpest step response, i.e., it creates the lowest noise for a given edge sharpness.
\note{The previous paragraph has alot of concepts.  Can we add a reference here?}

Note that in general, filter coefficients can be crafted to create many different kinds of filters: low pass, high pass, band pass, etc.. In general, a larger value of number of taps provides more degrees of freedom when designing a filter, generally resulting in filters with better characteristics. There is substantial amount of literature devoted to generating filter coefficients with particular characteristics for a given application. When implementing a filter, the actual values of these coefficients are largely irrelevant and we can ignore how the coefficients themselves were arrived at.  However, as we saw with the moving average filter, the structure of the filter, or the particular coefficients can have a large impact on the number of operations that need to be performed.  For instance, symmetric filters have multiple taps with exactly the same value which can be grouped to reduce the number of multiplications.  In other cases, it is possible to convert the multiplication by a known constant filter coefficient into shift and add operations \cite{kastner2010arithmetic}. In that case, the values of the coefficients can drastically change the performance and area of the filter implementation \cite{mirzaei2007fpga}. But we will ignore that for the time being, and focus on generating architectures that have constant coefficients, but do not take advantage of the values of the constants.

\section{Base FIR Architecture}
\label{sec:base_fir}

Consider the code for an 11 tap FIR filter in Figure \ref{fig:fir11_initial}. The function takes two arguments, an input sample \lstinline{x}, and the output sample \lstinline{y}. This function must be called multiple times to compute an entire output signal, since each time that we execute the function we provide one input sample and receive one output sample. This code is convenient for modeling a streaming architecture, since we can call it as many times as needed as more data becomes available.

\begin{figure}
\begin{lstlisting}
#define N 11
#include "ap_int.h"

typedef int coef_t;
typedef int data_t;
typedef int acc_t;

void fir(data_t *y, data_t x) {
	coef_t c[N] = {53, 0, -91, 0, 313, 500, 313, 0, -91, 0, 53};
	static data_t shift_reg[N];
	acc_t acc;
	int i;

	acc = 0;
Shift_Accum_Loop:
	for (i = N - 1; i >= 0; i--) {
		if (i == 0) {
			acc += x * c[0];
			shift_reg[0] = x;
		} else {
			shift_reg[i] = shift_reg[i - 1];
			acc += shift_reg[i] * c[i];
		}
	}
	*y = acc;
}
\end{lstlisting}
\caption{A functionally correct, but highly unoptimized, implementation of an 11 tap FIR filter. }
\label{fig:fir11_initial}
\end{figure}

The coefficients for the filter are stored in the \lstinline{c[]} array declared inside of the function. These are statically defined constants. Note that the coefficients are symmetric. i.e., they are mirrored around the center value \lstinline{c[5] = 500}. Many FIR filter have this type of symmetry. We could take advantage of it in order to reduce the amount of storage that is required for the \lstinline{c[]} array.

The code uses \lstinline{typedef} for the different variables. While this is not necessary, it is convenient for changing the types of data. As we discuss later, bit width optimization -- specifically setting the number of integer and fraction bits for each variable -- can provide significant benefits in terms of performance and area.  

\begin{exercise}
Rewrite the code so that it takes advantage of the symmetry found in the coefficients. That is, change \lstinline{c[]} so that it has six elements (\lstinline{c[0]} through \lstinline{c[5]}). What changes are necessary in the rest of the code? How does this effect the number of resources? How does it change the performance?
\end{exercise}

The code is written as a streaming function. It receives one sample at a time, and therefore it must store the previous samples. Since this is an 11 tap filter, we must keep the previous 10 samples. This is the purpose of the \lstinline{shift_reg[]} array. This array is declared \lstinline{static} since the data must be persistent across multiple calls to the function. 

The \lstinline{for} loop is doing two fundamental tasks in each iteration. First, it performs the multiply and accumulate operation on the input samples (the current input sample \lstinline{x} and the previous input samples stored in \lstinline{shift_reg[]}). Each iteration of the loop performs a multiplication of one of the constants with one of the sample, and stores the running sum in the variable \lstinline{acc}. The loop is also shifting values through \lstinline{shift_array}, which works as a FIFO. It stores the input sample \lstinline{x} into  \lstinline{shift_array[0]}, and moves the previous elements ``up'' through the \lstinline{shift_array}:

\begin{padbox}{.5\textwidth}
\noindent\lstinline{shift_array[10] = shift_array[9]} \\
\lstinline{shift_array[9]  =  shift_array[8]} \\
\lstinline{shift_array[8]  =  shift_array[7]} \\
$\cdots$ \\
\lstinline{shift_array[2]  =  shift_array[1]} \\
\lstinline{shift_array[1]  =  shift_array[0]} \\
\lstinline{shift_array[0]  =  x} \\
\end{padbox}
\begin{aside}
The label \lstinline{Shift_Accum_Loop:} is not necessary. However it can be useful for debugging. The \VHLS tool adds these labels into the views of the code. 
\end{aside}

After the for loop completes, the \lstinline{acc} variable has the complete result of the convolution of the input samples with the FIR coefficient array. The final result is written into the function argument \lstinline{y} which acts as the output port from this \lstinline{fir} function. This completes the streaming process for computing one output value of an FIR.

This function does not provide an efficient implementation of a FIR filter. It is largely sequential, and employs a significant amount of unnecessary control logic. The following sections describe a number of different optimizations that improve its performance.

\section{Calculating Performance}
\label{sec:fir-performance}

Before we get into the optimizations, it is necessary to define precise metrics. When deriving the performance of a design, it is important to carefully state the metric. For instance, there are many different ways of specifying how ``fast'' your design runs. For example, you could say that it operates at $X$ bits/second. Or that it can perform $Y$ operations/sec. Other common performance metrics specifically for FIR filters talk about the number of filter operations/second. Yet another metric is multiply accumulate operations: MACs/second. Each of these are related to one another, in some manner, but when comparing different implementations it is important to compare apples to apples. For example, directly comparing one design using bits/second to another using filter operations/second can be misleading; fully understanding the relative benefits of the designs requires that we compare them using the same metric. And this may require additional information, e.g., going from filter operations/second to bits/second requires information about the size of the input and output data. 

All of the aforementioned metrics use seconds. High-level synthesis tools talk about the designs in terms of number of cycles, and the frequency of the clock. The frequency is inversely proportional to the time it takes to complete one clock cycle. Using them both gives us the amount of time in seconds to perform some operation. The number of cycles and the clock frequency are both important: a design that takes one cycle, but with a very low frequency is not necessary better than another design that takes 10 clock cycles but operates at a much higher frequency. 

The clock frequency is a complicated function that the \VHLS tool attempts to optimize alongside the number of cycles. Note that it is possible to specify a target frequency to the \VHLS tool. This is done using the \lstinline{create_clock} tcl command.  For example, the command \lstinline{create_clock -period 5} directs the tool to target a clock period of 5 ns and equivalently a clock frequency of 200 MHz.  Note that this is only a target clock frequency only and primarily affects how much operation chaining is performed by the tool.  After generating RTL, the \VHLS tool provides an initial timing estimate relative to this clock target.  However, some uncertainty in the performance of the circuit remains which is only resolved once the design is fully place and routed.

While achieving higher frequencies are often critical for reaching higher performance, increasing the target clock frequency  is not necessarily optimal in terms of an overall system. Lower frequencies give more leeway for the tool to combine multiple dependent operations in a single cycle, a process called \term{operation chaining}. This can sometimes allow higher performance by enabling improved logic synthesis optimizations and increasing the amount of code that can fit in a device. Improved operation chaining can also improve (i.e., lower) the initiation interval of pipelines with recurrences.   In general providing a constrained, but not over constrained target clock latency is a good option. Something in the range of $5 - 10$ ns is typically a good starting option. Once you optimize your design, you can vary the clock period and observe the results. We will describe operation chaining in more detail in the next section.

Because \VHLS deals with clock frequency estimates, it does include some margin to account for the fact that there is some error in the estimate.  The goal of this margin is to ensure enough timing slack in the design that the generated RTL can be successfully placed and routed.  This margin can be directly controlled using the \lstinline{set_clock_uncertainty} TCL command.  Note that this command only affects the HLS generated RTL and is different from the concept of clock uncertainty in RTL-level timing constraints.  Timing constraints generated by \VHLS for the RTL implementation flow are solely based on the target clock period.

It is also necessary to put the task that you are performing in context with the performance metric that you are calculating. In our example, each execution of the \lstinline{fir} function results in one output sample. But we are performing $N = 11$ multiply accumulate operations for each execution of \lstinline{fir}. Therefore, if your metric of interest is MACs/second, you should calculate the task latency for \lstinline{fir} in terms of seconds, and then divide this by $11$ to get the time that it takes to perform the equivalent of one MAC operation. 

Calculating performance becomes even more complicated as we perform pipelining and other optimizations. In this case, it is important to understand the difference between task interval and task latency. It is a good time to refresh your understanding of these two metrics of performance. This was discussed in Chapter \ref{sec:designOptimization}. And we will continue to discuss how different optimizations effect different performance metrics.

\section{Operation Chaining}
\label{sec:fir-chaining}

\begin{figure}
\centering
\includegraphics[width=5.5in]{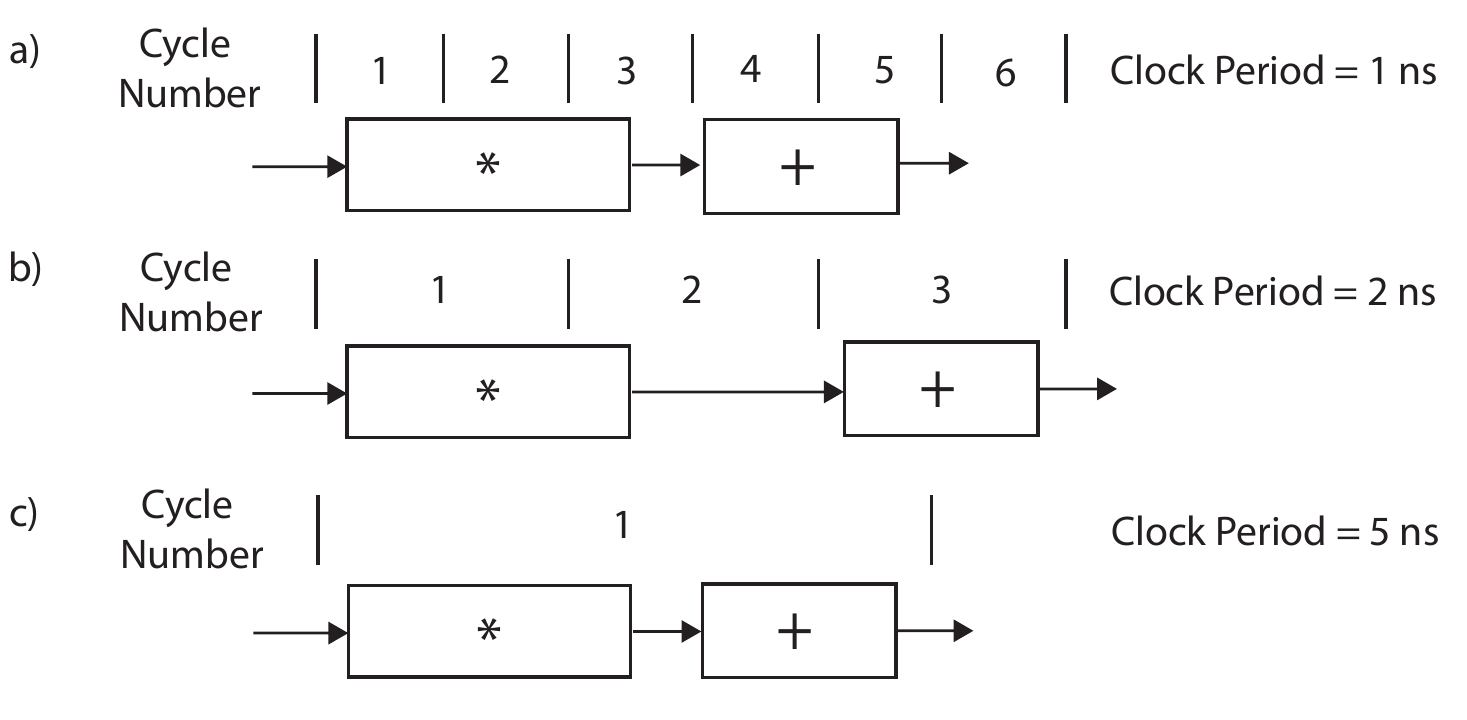}
\caption{The performance of multiply accumulate operation changes depending upon the target clock period. Assume the \lstinline{multiply} operation takes 3 ns and \lstinline{add} operation takes 2 ns. Part a) has a clock period of 1 ns, and one MAC operation takes 5 cycles. Thus the performance is 200 million MACs/sec. Part b) has a clock period of 2 ns, and the MAC takes 3 cycles resulting in approximately 167 million MACs/sec.  Part c) has a clock period of 5 ns. By using operation chaining, a MAC operation takes 1 cycle for a clock period of 200 million MACs/sec. }
\label{fig:mac}
\end{figure}

\term{Operation chaining} is an important optimization that the \VHLS performs in order to optimize the final design. It is not something that a designer has much control over, but it is important that the designer understands how this works especially with respect to performance. Consider the multiply accumulate operation that is done in a FIR filter tap. Assume that the \lstinline{add} operation takes 2 ns to complete, and a \lstinline{multiply} operation takes 3 ns. If we set the clock period to 1 ns (or equivalently a clock frequency of 1 GHz), then it would take 5 cycles for the MAC operation to complete. This is depicted in Figure \ref{fig:mac} a). The \lstinline{multiply} operation is executed over 3 cycles, and the \lstinline{add} operation is executed across 2 cycles. The total time for the MAC operation is 5 cycles $\times$ 1 ns per cycle $=$ 5 ns. Thus we can perform 1/5 ns $=$ 200 million MACs/second. 

If we increase the clock period to 2 ns, the \lstinline{multiply} operation now spans over two cycles, and the \lstinline{add} operation must wait until cycle 3 to start. It can complete in one cycle. Thus the MAC operation requires 3 cycles, so 6 ns total to complete. This allows us to perform approximately 167 million MACs/second. This result is lower than the previous result with a clock period of 1 ns. This can be explained by the ``dead time'' in cycle 2 where no operation is being performed.

However, it is not always true that increasing the clock period results in worse performance. For example, if we set the clock period to 5 ns, we can perform both the \lstinline{multiply} and \lstinline{add} operation in the same cycle using operation chaining. This is shown in Figure \ref{fig:mac} c). Thus the MAC operation takes 1 cycle where each cycle is 5 ns, so we can perform 200 million MACs/second. This is the same performance as Figure \ref{fig:mac} a) where the clock period is faster (1 ns). 

So far we have performed chaining of only two operations in one cycle. It is possible to chain multiple operations in one cycle. For example, if the clock period is 10 ns, we could perform 5 \lstinline{add} operations in a sequential manner. Or we could do two sequential MAC operations. 

It should start to become apparent that the clock period plays an important role in how the \VHLS tool optimizes the design. This becomes even more complicated with all of the other optimizations that the \VHLS tool performs. It is not that important to fully understand the entire process of the \VHLS tool. This is especially true since the tool is constantly being improved with each new release. However, it is important to have a good idea about how the tool may work. This will allow you to better comprehend the results, and even allow you to write more optimized code. 

Although the \VHLS can generate different different hardware for different target clock periods, overall performance optimization and determining the optimal target clock period still requires some creativity on the part of the user. For the most part, we advocate sticking within a small subset of clock periods. For example, in the projects we suggest that you set the clock period to 10 ns and focus on understanding how other optimizations, such as pipelining, can be used to create different architectures. This 100 MHz clock frequency is relatively easy to achieve, yet it is provides a good first order result. It is certainly possible to create designs that run at a faster clock rate.  200 MHz and faster designs are possible but often require more careful balance between clock frequency targets and other optimization goals.  You can change the target clock period and observe the differences in the performance. Unfortunately, there is no good rule to pick the optimal frequency. 

\begin{exercise}
Vary the clock period for the base FIR architecture (Figure \ref{fig:fir11_initial}) from 10 ns to 1 ns in increments of 1 ns. Which clock period provides the best performance? Which gives the best area? Why do you think this is the case? Do you see any trends? 
\end{exercise}

\section{Code Hoisting}

The \lstinline{if/else} statement inside of the \lstinline{for} loop is inefficient.  For every control structure in the code, the \VHLS tool creates logical hardware that checks if the condition is met, which is executed in every iteration of the loop. Furthermore, this conditional structure limits the execution of the statements in either the \lstinline{if} or \lstinline{else} branches; these statements can only be executed after the \lstinline{if} condition statement is resolved.

The \lstinline{if} statement checks when \lstinline{x == 0}, which happens only on the last iteration. Therefore, the statements within the \lstinline{if} branch can be ``hoisted'' out of the loop. That is we can execute these statements after the loop ends, and then remove the \lstinline{if/else} control flow in the loop. Finally, we must change the loop bounds from executing the ``0th'' iteration. This transform is shown in Figure \ref{fig:fir11_ifelse}. This shows just the changes that are required to the \lstinline{for} loop. 

\begin{figure}
\begin{lstlisting}
Shift_Accum_Loop:
for (i = N - 1; i > 0; i--) {
	shift_reg[i] = shift_reg[i - 1];
	acc += shift_reg[i] * c[i];
}

acc += x * c[0];
shift_reg[0] = x;
\end{lstlisting}
\caption{ Removing the conditional statement from the \lstinline{for} loop creates a more efficient hardware implementation. }
\label{fig:fir11_ifelse}
\end{figure}

The end results is a much more compact implementation that is ripe for further loop optimizations, e.g., unrolling and pipelining. We discuss those optimizations later.

\begin{exercise}
Compare the implementations before and after the removal of the \lstinline{if/else} condition done through loop hoisting. What is the difference in performance? How do the number of resources change?
\end{exercise}


\section{Loop Fission}

We are doing two fundamental operations within the \lstinline{for} loop. The first part shifts the data through the \lstinline{shift_reg} array. The second part performs the multiply and accumulate operations in order to calculate the output sample. \term{Loop fission} takes these two operations and implements each of them in their own loop. While it may not intuitively seem like a good idea, it allows us to perform optimizations separately on each loop. This can be advantageous especially in cases when the resulting optimizations on the split loops are different. 

\begin{figure}
\begin{lstlisting}
TDL:
for (i = N - 1; i > 0; i--) {
	shift_reg[i] = shift_reg[i - 1];
}
shift_reg[0] = x;

acc = 0;
MAC:
for (i = N - 1; i >= 0; i--) {
	acc += shift_reg[i] * c[i];
}
\end{lstlisting}
\caption{ A code snippet corresponding to splitting the \lstinline{for} loop into two separate loops. }
\label{fig:fir11_partition}
\end{figure}

The code in Figure \ref{fig:fir11_partition} shows the result of manual loop fission optimization.  The code snippet splits the loop from Figure \ref{fig:fir11_ifelse} into two loops. Note the label names for the two loops. The first is \lstinline{TDL} and the second is \lstinline{MAC}. \term{Tapped delay line (TDL)} is a common DSP term for the FIFO operation; MAC is short-hand for ``multiply accumulate''.

\begin{exercise}
Compare the implementations before and after loop fission. What is the difference in performance? How do the number of resources change?
\end{exercise}

Loop fission alone often does not provide a more efficient hardware implementation. However, it allows each of the loops to be optimized independently, which could lead to better results than optimizing the single, original \lstinline{for} loop. The reverse is also true; merging two (or more) \lstinline{for} loops into one \lstinline{for} loop may yield the best results. This is highly dependent upon the application, which is true for most optimizations. In general, there is not a single `rule of thumb' for how to optimize your code. There are many tricks of the trade, and your mileage may vary. Thus, it is important to have many tricks at your disposal, and even better, have a deep understanding of how the optimizations work. Only then will you be able to create the best hardware implementation. Let us continue to learn some additional tricks...

\section{Loop Unrolling}
By default, the \VHLS tool synthesizes \lstinline{for} loops in a sequential manner. The tool creates a data path that implements one execution of the statements in the body of the loop. The data path executes sequentially for each iteration of the loop. This creates an area efficient architecture; however, it limits the ability to exploit parallelism that may be present across loop iterations. 

\term{Loop unrolling} replicates the body of the loop by some number of times (called the \term{factor}). And it reduces the number of iterations of the loop by the same factor. In the best case, when none of the statements in the loop depend upon any of the data generated in the previous iterations, this can substantially increase the available parallelism, and thus enables an architecture that runs much faster. 

The first \lstinline{for} loop (with the label \lstinline{TDL}) in Figure \ref{fig:fir11_partition} shifts the values up through the \lstinline{shift_reg} array. The loop iterates from largest value (\lstinline{N-1}) to the smallest value (\lstinline{i = 1}). By unrolling this loop, we can create a data path that executes a number of these shift operations in parallel. 

\begin{figure}
\begin{lstlisting}
TDL:
for (i = N - 1; i > 1; i = i - 2) {
	shift_reg[i] = shift_reg[i - 1];
	shift_reg[i - 1] = shift_reg[i - 2];
}
if (i == 1) {
	shift_reg[1] = shift_reg[0];
}
shift_reg[0] = x;
\end{lstlisting}
\caption{ Manually unrolling the \lstinline{TDL} loop in the \lstinline{fir11} function.  }
\label{fig:fir11_unrollTDL}
\end{figure}

Figure \ref{fig:fir11_unrollTDL} shows the result of unrolling the loop by a factor of two. This code replicates the loop body twice. Each iteration of the loop now performs two shift operations. Correspondingly, we must perform half of the number of iterations.

Note that there is an additional \lstinline{if} condition after the \lstinline{for} loop. This is required in the case when the loop does not have an even number of iterations. In this case, we must perform the last ``half'' iteration by itself. The code in the \lstinline{if} statement performs this last ``half'' iteration, i.e., moving the data from \lstinline{shift_reg[0]} into \lstinline{shift_reg[1]}.

Also note the effect of the loop unrolling on the \lstinline{for} loop header. The decrement operation changes from \lstinline{i--} to \lstinline{i=i-2}. This is due to the fact that we are doing two times the ``work'' in each iteration, thus we should decrement by 2 instead of 1.

Finally, the condition for terminating the \lstinline{for} loop changes from \lstinline{i > 0} to \lstinline{i > 1}. This is related to the fact that we should make sure that the ``last'' iteration can fully complete without causing an error. If the last iteration of the \lstinline{for} loop executes when \lstinline{i = 1}, then the second statement would try to read from \lstinline{shift_reg[-1]}. Rather than perform this illegal operation, we do the final shift in the \lstinline{if} statement after the \lstinline{for} loop. 

\begin{exercise}
Write the code corresponding to manually unrolling this \lstinline{TDL for} loop by a factor of three. How does this change the loop body? What changes are necessary to the loop header? Is the additional code in the \lstinline{if} statement after the \lstinline{for} loop still necessary? If so, how is it different?
\end{exercise}

Loop unrolling can increase the overall performance provided that we have the ability to execute some (or all) of the statements in parallel. In the unrolled code, each iteration requires that we read two values from the \lstinline{shift_reg} array; and we write two values to the same array. Thus, if we wish to execute both statements in parallel, we must be able to perform two read operations and two write operations from the \lstinline{shift_reg} array in the same cycle. 

Assume that we store the \lstinline{shift_reg} array in one BRAM, and that BRAM has two read ports and one write port. Thus we can perform two read operations in one cycle. But we must sequentialize the write operations across two consecutive cycles. 

There are ways to execute these two statements in one cycle. For example, we could store all of the values of the \lstinline{shift_reg} array in separate registers. It is possible to read and write to each individual register on every cycle. In this case, we can perform both of the statements in this unrolled \lstinline{for} loop in one cycle. You can tell the \VHLS tool to put all of the values in the \lstinline{shift_reg} array into registers using the directive \lstinline{#pragma HLS array_partition  variable=shift_reg complete}. This is an important optimization, thus we discuss the \lstinline{array_partition} directive in more detail later. 

A user can tell the \VHLS tool to automatically unroll the loop using the \lstinline{unroll} directive. To automatically perform the unrolling done manually in Figure \ref{fig:fir11_unrollTDL}, we should put the directive \lstinline{#pragma HLS unroll factor=2} into the body of the code, right after the \lstinline{for} loop header. While we can always manually perform loop unrolling, it is much easier to allow the tool to do it for us. It makes the code easier to read; and it will result in fewer coding errors. 

\begin{exercise}
Unroll the \lstinline{TDL for} loop automatically using the \lstinline{unroll} directive. As you increase the unroll factor, how does this change the number of resources (FFs, LUTs, BRAMs, DSP48s, etc.)? How does it effect the throughput? What happens when you use the \lstinline{array_partition} directive in conjunction with the \lstinline{unroll} directive?  What happens if you do not use the \lstinline{unroll} directive?
\end{exercise}

Now, consider the second \lstinline{for} loop (with the label \lstinline{MAC}) in Figure \ref{fig:fir11_partition}. This loop multiplies a value from the array \lstinline{c[]} with a value from the array \lstinline{shift_array[]}. In each iteration it accesses the \lstinline{i}th value from both arrays. And then it adds the result of that multiplication into the \lstinline{acc} variable. 

Each iteration of this loop performs one multiply and one add operation. Each iteration performs one read operation from array \lstinline{shift_reg[]} and array \lstinline{c[]}. The result of the multiplication of these two values is accumulated into the variable \lstinline{acc}. 

The load and multiplication operations are independent across all of the iterations of the for loop. The addition operation, depending on how it is implemented, may depend upon the values of the previous iterations. However, it is possible to unroll this loop and remove this dependency.

\begin{figure}
\begin{lstlisting}
acc = 0;
MAC:
for (i = N - 1; i >= 3; i -= 4) {
	acc += shift_reg[i] * c[i] + shift_reg[i - 1] * c[i - 1] +
				 shift_reg[i - 2] * c[i - 2] + shift_reg[i - 3] * c[i - 3];
}

for (; i >= 0; i--) {
	acc += shift_reg[i] * c[i];
}
\end{lstlisting}
\caption{ Manually unrolling the \lstinline{MAC} loop in the \lstinline{fir11} function by a factor of four.  }
\label{fig:fir11_unrollMAC}
\end{figure}

Figure \ref{fig:fir11_unrollMAC} shows the code corresponding to unrolling the \lstinline{MAC for} loop by a factor of four.  The first \lstinline{for} loop is the unrolled loop. The \lstinline{for} loop header is modified in a similar manner to when we unrolled the  \lstinline{TDL} loop. The bound is changed to \lstinline{i>=3}, and \lstinline{i} is decremented by a factor of $4$ for each iteration of the unrolled loop.

While there was loop carried dependency in the original, unrolled \lstinline{for}, it is no longer present in the unrolled loop. The loop carried dependency came due to the \lstinline{acc} variable; since the result of the multiply accumulate is written to this variable ever iteration, and we read from this register in every iteration (to perform the running sum), it creates a read-after-write (RAW) dependency across iterations.  Note that there is not a dependency on the \lstinline{acc} variable in the unrolled \lstinline{for} loop due to the way this is written. Thus we are free to parallelize the four individual MAC operations in the unrolled \lstinline{for} loop. 

There is an additional \lstinline{for} loop after the unrolled \lstinline{for} loop. This is necessary to perform any partial iterations. Just like we required the \lstinline{if} statement in the \lstinline{TDL}, this performs any computations on a potential last iteration. This occurs when the number of iterations in the original, unrolled \lstinline{for} loop is not an even multiple of $4$. 

Once again, we can tell the \VHLS tool to automatically unroll the loop by a factor of $4$ by inserting the code \lstinline{#pragma HLS unroll factor=4} into the \lstinline{MAC} loop body. 

By specifying the optional argument \lstinline{skip_exit_check} in that directive, the \VHLS tool will not add the final \lstinline{for} loop to check for partial iterations. This is useful in the case when you know that the loop will never require these final partial iterations. Or perhaps performing this last few iterations does not have an (major) effect on the results, and thus it can be skipped. By using this option, the \VHLS tool does not have to create that additional \lstinline{for} loop. Thus the resulting hardware is simpler, and more area efficient.

The \lstinline{for} loop is completely unrolled when no factor argument is specified. This is equivalent to unrolling by the maximum number of iterations; in this case a complete unrolling and unrolling by a factor of $11$ is equivalent. In both cases, the loop body is replicated 11 times. And the loop header is unnecessary; there is no need to keep a counter or check if the loop exit condition is met. In order to perform a complete unrolling, the bounds of the loop must be statically determined, i.e., the \VHLS tool must be able to know the number of iterations for the \lstinline{for} loop at compile time.

Complete loop unrolling exposes a maximal amount of parallelism at the cost of creating an implementation that requires a significant amount of resources. Thus, it ok to perform a complete loop unroll on ``smaller'' for loops. But completely unrolling a loop with a large number of iterations (e.g., one that iterates a million times) is typically infeasible. Often times, the \VHLS tool will run for a very long time (and many times fail to complete after hours of synthesis) if the resulting loop unrolling creates code that is very large. 

\begin{aside}
If you design does not synthesize in under 15 minutes, you should carefully consider the effect of your optimizations. It is certainly possible that large designs can take a significant amount for the \VHLS tool to synthesize them. But as a beginning user, your designs should synthesize relatively quickly.  If they take a long time, that most likely means that you used some directives that significantly expanded the code, perhaps in a way that you did not intend.
\end{aside}

\begin{exercise}
Synthesize a number of designs by varying the unroll factor for the \lstinline{MAC} loop. How does the performance change? How does the unroll factor number affect the number of resources? Compare these results with the trends that you found by unrolling the \lstinline{TDL}.   
\end{exercise}

\section{Loop Pipelining}

By default, the \VHLS tool synthesizes \lstinline{for} loops in a sequential manner. For example, the \lstinline{for} loop in Figure \ref{fig:fir11_initial} will perform each iteration of the loop one after the other. That is, all of the statements in the second iteration happen only when all of the statements from the first iteration are complete; the same is true for the subsequent iterations. This happens even in cases when it is possible to perform statements from the iterations in parallel. In other cases, it is possible to start some of the statements in a later iteration before all of the statements in a former iteration are complete. This does not happen unless the designer specifically states that it should. This motivates the idea of \term{loop pipelining}, which allows for multiple iterations of the loop to execute concurrently.

Consider the \lstinline{MAC for} loop from Figure \ref{fig:fir11_partition}. This performs one multiply accumulate (MAC) operation per iteration.  This \lstinline{MAC for} loop has four operations in the loop body:
\begin{itemize}
\item \lstinline{Read c[]}: Load the specified data from the \lstinline{C} array.
\item \lstinline{Read shift_reg[]}: Load the specified data from the \lstinline{shift_reg} array.
\item \lstinline{*}: Multiply the values from the arrays \lstinline{c[]} and \lstinline{shift_reg[]}.
\item \lstinline{+}: Accumulate this multiplied result into the \lstinline{acc} variable.
\end{itemize}

\begin{figure}
\centering
\includegraphics[width=6in]{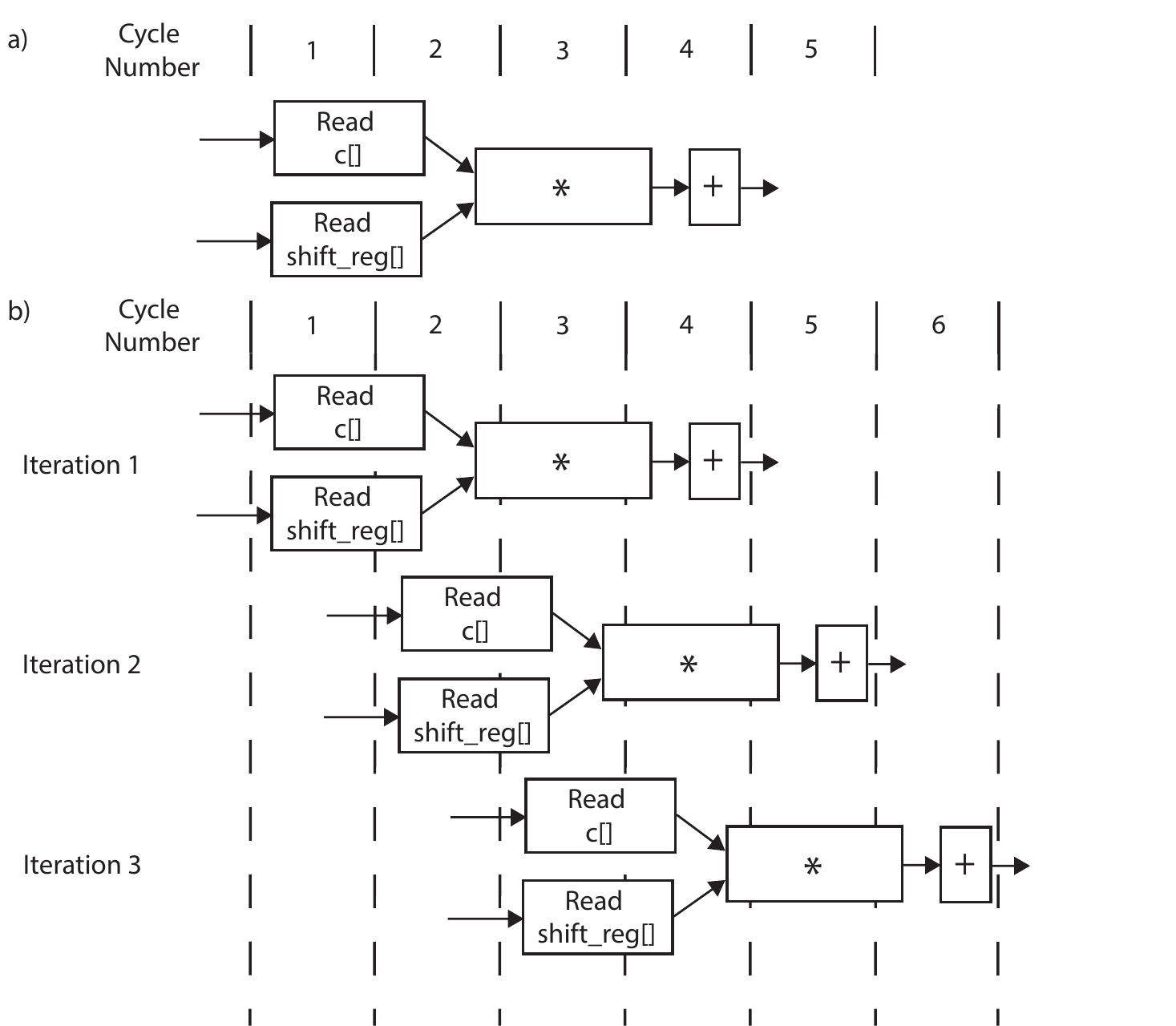}
\caption{Part a) shows a schedule for the body of the \lstinline{MAC for} loop. Part b) shows the schedule for three iterations of a pipelined version of the \lstinline{MAC for} loop.}
\label{fig:pipeline_mac}
\end{figure}

A schedule corresponding to one iteration of the \lstinline{MAC for} loop is shown in Figure \ref{fig:pipeline_mac} a). The \lstinline{Read} operations each require 2 cycles. This is due to the fact that the first cycle provides the address to the memory, and the data from the memory is delivered during the second cycle.  These two \lstinline{Read} operations can be done in parallel since there are no dependencies between them. The \lstinline{*} operation can begin in Cycle 2; assume that it takes three cycles to complete, i.e., it is finished is Cycle 4. The \lstinline{+} operation is chained to start and complete during Cycle 4. The entire body of the \lstinline{MAC for} loop takes 4 cycles to complete.

There are a number of performance metrics associated with a \lstinline{for} loop. The \term{iteration latency} is the number of cycles that it takes to perform one iteration of the loop body. The iteration latency for this \lstinline{MAC for} loop is 4 cycles. The \term{\lstinline{for} loop latency} is the number of cycles required to complete the entire execution of the loop. This includes time to calculate the initialization statement (e.g., \lstinline{i = 0}), the condition statement (e.g., \lstinline{i >= 0}), and the increment statement (e.g., \lstinline{i--}). Assuming that these three header statements can be done in parallel with the loop body execution, the \VHLS tool reports the latency of this \lstinline{MAC for} loop as 44 cycles. This is the  number of iterations (11) multiplied by the iteration latency (4 cycles) plus one additional cycle to determine that the loop should stop iterating. And then you subtract one. Perhaps the only strange thing here is the ``subtract 1''. We will get to that in a second. But first, there is one additional cycle that is required at the beginning of the next iteration, which checks if the condition statement is satisfied (it is not) and then exits the loop. Now the ``subtract 1'': \VHLS determines the latency as the cycle in which the output data is ready. In this case, the final data is ready during Cycle 43. This would be written into a register at the end of Cycle 43 and correspondingly the beginning of Cycle 44.   Another way to think of this is that the latency is equal to the maximum number of registers between the input data and the output data. 

Loop pipelining is an optimization that overlaps multiple iterations of a \lstinline{for} loop. Figure \ref{fig:pipeline_mac} b) provides an example of pipelining for the \lstinline{MAC for} loop. The figure shows three iterations of the \lstinline{for} which are executed simultaneously. The first iteration is equivalent the the non-pipelined version as depicted in Figure \ref{fig:pipeline_mac} a). The difference is the start times of the subsequent iterations. In the non-pipelined version, the second iteration begins after the first iteration is completed, i.e., in Cycle 5. However, the pipelined version can start the subsequent iteration before the previous iterations complete. In the figure, Iteration 2 starts at Cycle 2, and Iteration 3 starts at Cycle 3. The remaining iterations start every consecutive cycle. Thus, the final iteration, Iteration 11, would start at Cycle 11 and it would complete during Cycle 14. Thus, the loop latency is 14.

The \term{loop initiation interval (II)} is another important performance metric. It is defined as the number of clock cycles until the next iteration of the loop can start. In our example, the loop II is 1, which means that we start a new iteration of the loop every cycle. This is graphically depicted in Figure \ref{fig:pipeline_mac} b). The II can be explicitly set using the directive. For example, the directive \lstinline{#pragma HLS pipeline II=2} informs the \VHLS tool to attempt to set the \lstinline{II=2}. Note that this may not always be possible due to resource constraints and/or dependencies in the code. The output reports will tell you exact what the \VHLS tool was able to achieve. 

\begin{exercise}
Explicitly set the loop initiation interval starting at 1 and increasing in increments of 1 cycle. How does increasing the II effect the loop latency? What are the trends?  At some point setting the II to a larger value does not make sense. What is that value in this example? How do you describe that value for a general \lstinline{for} loop?
\end{exercise}

Any \lstinline{for} loop can be pipelined, so let us now consider the \lstinline{TDL for} loop. This \lstinline{for} loop as a similar header to the \lstinline{MAC for} loop. The body of the loop performs an element by element shift of data through the array as described in Section \ref{sec:base_fir}. There are two operations: one \lstinline{Read} and one \lstinline{Write} to the \lstinline{shift_reg} array. The iteration latency of this loop is 2 cycles. The \lstinline{Read} operation takes two cycles, and the \lstinline{Write} operation is performed at the end of Cycle 2. The \lstinline{for} loop latency for this non-pipelined loop is 20 cycles.

\begin{figure}
\centering
\includegraphics[width=6in]{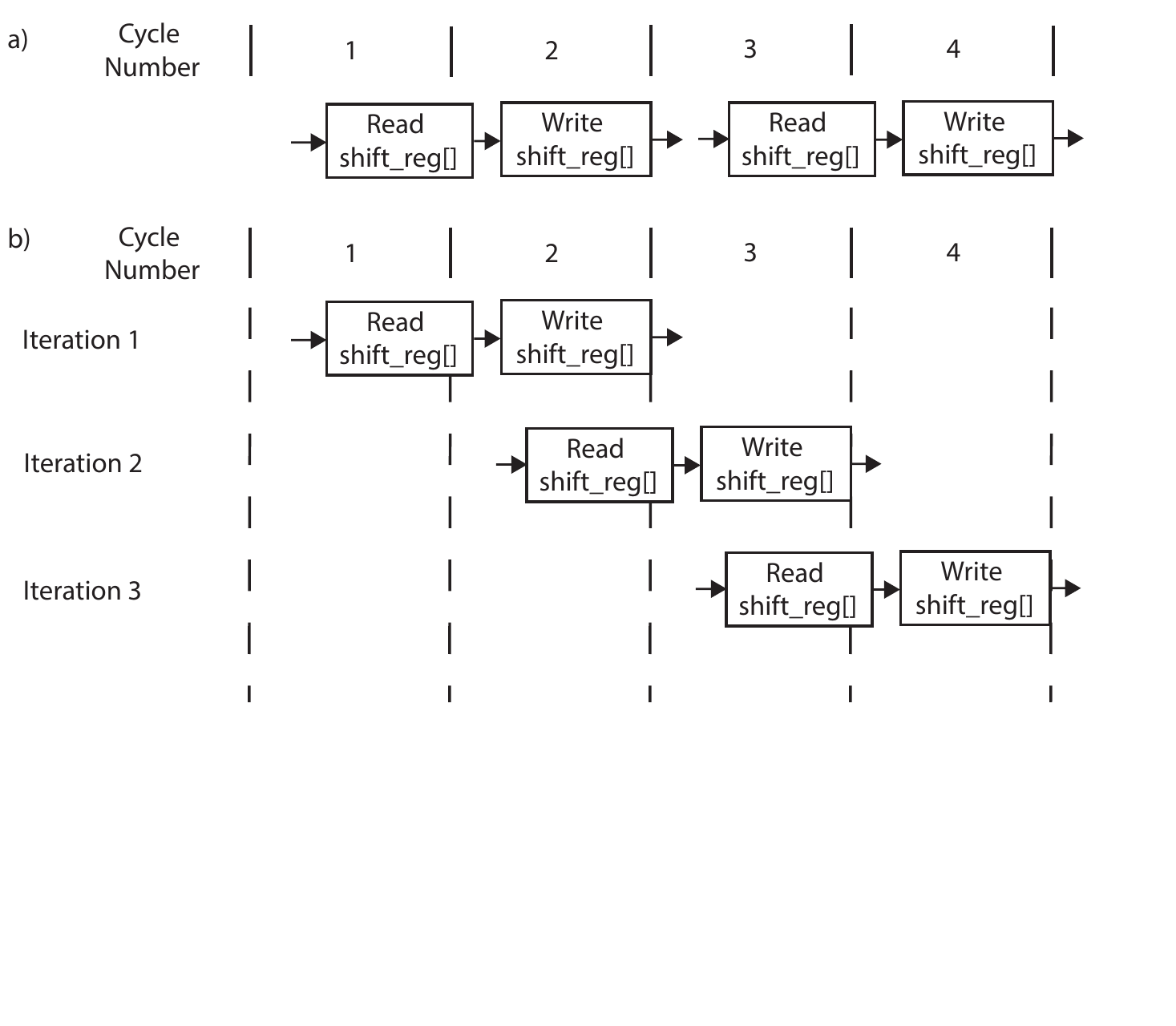}
\caption{Part a) shows a schedule for two iterations of the body of the \lstinline{TDL for} loop. Part b) shows the schedule for three iterations of a pipelined version of the \lstinline{TDL for} loop with II=1.}
\label{fig:pipeline_tdl}
\end{figure}

We can pipeline this loop by inserting the directive \lstinline{#pragma HLS pipeline II=1} after the loop header. The result of the synthesis is a loop initiation interval equal to 1 cycle.  This means that we can start a loop iteration every cycle. 

By modifying the example slightly, we can demonstrate a scenario where the resource constraints do not allow the \VHLS tool to achieve an II=1. To do this, we explicitly set the type of memory for the \lstinline{shift_reg} array. By not specifying the resource, we leave it up to the \VHLS tool to decide. But we can specify the memory using a directive, e.g., the directive \lstinline{#pragma HLS resource variable=shift_reg core=RAM_1P} forces the \VHLS tool to use a single port RAM. When using this directive in conjunction with the loop pipelining objective, the \VHLS tool will fail to pipeline this loop with an II=1. This is due to the fact that a pipelined version of this code requires both a \lstinline{Read} and \lstinline{Write} operation in the same cycle. This is not possible using a single port RAM. This is evident in Figure \ref{fig:pipeline_tdl} b). Looking at Cycle 2, we require a \lstinline{Write} operation to the array \lstinline{shift_reg} in Iteration 1, and a \lstinline{Read} operation to the same array in Iteration 2.  We can modify the directive to allow HLS more scheduling freedom by removing the explicit request for II=1, e.g. \lstinline{#pragma HLS pipeline}.  In this case, HLS will automatically increase the initiation interval until it can find a feasible schedule.

\begin{aside}
The \lstinline{RESOURCE} directive allows the user to force the \VHLS tool to map an operation to a hardware core. This can be done on arrays (as shown above) and also for variables. Consider the code \lstinline{a = b + c;}. We can use the \lstinline{RESOURCE} directive \lstinline{#pragma HLS RESOURCE variable=a core=AddSub_DSP} to tell the \VHLS tool that the \lstinline{add} operation is implemented using a DSP48. There are a wide variety of cores described in the \VHLS documentation \cite{ug902}. In general, it is advised to let the \VHLS decide the resources. If these are not satisfactory, then the designer can use directives.
\end{aside}


\section{Bitwidth Optimization}

The C language provides many different data types to describe different kinds of behavior.  Up until this point, we have focused on the \lstinline{int} type, which \VHLS treats as a 32-bit signed integer.   The C language also provides floating point data types, such as \lstinline{float} and \lstinline{double}, and integer data types, such as \lstinline{char}, \lstinline{short}, \lstinline{long}, and \lstinline{long long}. Integer datatypes may be \lstinline{unsigned}. All of these data types have a size which is a power of 2.

The actual number of bits for these C data types may vary depending upon the processor architecture. For example, an \lstinline{int} can be 16 bits on a micro controller and 32 bits on a general purpose processor. The C standard dictates minimum bit widths (e.g., an \lstinline{int} is at least 16 bits) and relations between the types (e.g., a \lstinline{long} is not smaller than an \lstinline{int} which is not smaller than a \lstinline{short}).  The C99 language standard eliminated this ambiguity with types such as \lstinline{int8_t}, \lstinline{int16_t}, \lstinline{int32_t}, and \lstinline{int64_t}.  

The primary benefits of using these different data types in software revolve around the amount of storage that the data type requires. For large arrays, using 8 bit values instead of 16 bit values can cut memory usage in half. The drawback is that the range of values that you can represent is reduced. An 8 bit signed value allows numbers in the range [-128,127] while the 16 bit signed data type has a range of [-32,768, 32,767].  Smaller operations may also require fewer clock cycles to execute, or may allow more instructions to be executed in parallel.

The same benefits are seen in an FPGA implementation, but they are even more pronounced. Since the \VHLS tool generates a custom data path, it will create an implementation matched to the specified data types. For example, the statement \lstinline{a = b * c} will have different latency and resource usage depending upon the data type. If all of the variables are 32 bits wide, then more primitive boolean operations need to be performed than if the variables are only 8 bits wide.   As a result, more FPGA resources (or more complex resources) must be used.  Additionally, more complex logic typically requires more pipelining in order to achieve the same frequency.  A 32-bit multiplication might require 5 internal registers to meet the same frequency that an 8 bit multiplication can achieve with only one internal register. As a result, the latency of the operation will be larger (5 cycles instead of 1 cycle) and HLS must take this into account.

\begin{exercise}
Create a simple design that implements the code \lstinline{a = b * c}. Change the data type of the variables to \lstinline{char}, \lstinline{short}, \lstinline{int}, \lstinline{long}, and \lstinline{long long}. How many cycles does the multiply operation take in each case? How many resources are used for the different data types?
\end{exercise}

\begin{exercise}
What primitive boolean operations are needed to implement the multiplication of 8-bit numbers?  How does this change when implementing multiplication of 32-bit numbers?  Hint: How many primitive decimal operations are needed to implement multiplication of two 8 digit decimal numbers? \end{exercise}

In many cases to implement optimized hardware, it is necessary to process data where the bitwidth is not a power of two. For example, analog to digital converters often output results in 10 bits, 12 bits, or 14 bits.  We could map these to 16 bit values, but this would likely reduce performance and increase resource usage.  To more accurately describe such values, \VHLS provides \term{arbitrary precision data types} which allow for signed and unsigned data types of any bitwidth. 

There are two separate classes for unsigned and signed data types:
\begin{itemize}
\item Unsigned: \lstinline{ap_uint<width>}
\item Signed: \lstinline{ap_int<width>}
\end{itemize}
where the \lstinline{width} variable is an integer between 1 and 1024\footnote{1024 is the default maximum value, and this can be changed if needed. See the \VHLS user manuals for more information on how to do this.}.  For example, \lstinline{ap_int<8>} is an 8 bit signed value (same as \lstinline{char}), and \lstinline{ap_uint<32>} is a 32 bit unsigned value (same as \lstinline{unsigned int}). This provides a more powerful data type since it can do any bitwidth, e.g., \lstinline{ap_uint<4>} or \lstinline{ap_int<537>}. To use these data types you must use C++ and include the file \lstinline{ap_int.h}, i.e., add the code \lstinline{#include "ap_int.h"} in your project and use a filename ending in `.cpp'\footnote{Similar capabilities are also available in C using the file \lstinline{ap_cint.h}}.

Consider coefficients array \lstinline{c[]} from the \lstinline{fir} filter code in Figure \ref{fig:fir11_initial}. It is reprinted here for you convenience: \lstinline|coef_t c[N] = {53, 0, -91, 0, 313, 500, 313, 0, -91, 0, 53};|. The data type \lstinline{coef_t} is defined as an \lstinline{int} meaning that we have 32 bits of precision. This is unnecessary for these constants since they range from -91 to 500. Thus we could use a smaller data type for this. We will need a signed data type since we have positive and negative values. And the maximum absolute value for any of these 11 entries is 500, which requires $\lceil \log_2 500 \rceil = 9$ bits. Since we need negative numbers, we add an additional bit. Thus \lstinline{coef_t} can be declared as \lstinline{ap_int<10>}.

\begin{exercise}
What is the appropriate data type for the variable \lstinline{i} in the \lstinline{fir} function (see Figure \ref{fig:fir11_initial})?
\end{exercise}

We can also more accurately define the data types for the other variables in the \lstinline{fir} function, e.g., \lstinline{acc} and \lstinline{shift_reg}. Consider the \lstinline{shift_reg} array first. This is storing the last 11 values of the input variable \lstinline{x}. So we know that the \lstinline{shift_reg} values can safely have the same data type as \lstinline{x}. By ``safe'', we mean that there will be no loss in precision, i.e., if \lstinline{shift_reg} had a data type with a smaller bitwidth, then some significant bits of \lstinline{x} would need to be eliminated to fit them into a value in \lstinline{shift_reg}. For example, if \lstinline{x} was defined as 16 bits (\lstinline{ap_uint<16>}) and \lstinline{shift_reg} was defined as 12 bits (\lstinline{ap_uint<12>}), then we would cut off the 4 most significant bits of \lstinline{x} when we stored it into \lstinline{shift_reg}. 

Defining the appropriate data type for \lstinline{acc} is a more difficult task. The \lstinline{acc} variable stores the multiply and accumulated sum over the \lstinline{shift_reg} and the coefficient array \lstinline{c[]} i.e., the output value of the filter. If we wish to be safe, then we calculate the largest possible value that could be stored in \lstinline{acc}, and set the bitwidth as that. 

To accomplish this, we must understand how the bitwidth increases as we perform arithmetic operations. Consider the operation \lstinline{a = b + c} where \lstinline{ap_uint<10> b} and \lstinline{ap_uint<10> c}. What is the data type for the variable \lstinline{a}? We can perform a worst case analysis here, and assume both \lstinline{a} and \lstinline{b} are the largest possible value $2^{10} = 1024$. Adding them together results in $a = 2024$ which can be represented as an 11 bit unsigned number, i.e., \lstinline{ap_uint<11>}. In general we will need one more bit than the largest bitwidth of the two number being added. That is, when  \lstinline{ap_uint<x> b} and \lstinline{ap_uint<y> c}, the data type for \lstinline{a} is \lstinline{ap_uint<z>} where $z = \max(x,y) + 1$. This same is also true when adding signed integers. 

That handles one part of the question for assigning the appropriate data type to \lstinline{acc}, but we must also deal with the multiply operation. Using the same terminology, we wish to determine the value the bitwidth $z$ given the bitwidths $x$ and $y$ (i.e., \lstinline{ap_int<z> a}, \lstinline{ap_int<x> b}, \lstinline{ap_int<y> c}) for the operation \lstinline{a = b * c}. While we will not go into the details, the formula is $z = x + y$. 

\begin{exercise}
Given these two formulas, determine the bitwidth of \lstinline{acc} such that it is safe.
\end{exercise}

Ultimately we are storing \lstinline{acc} into the variable \lstinline{y} which is an output port of the function. Therefore, if the bitwidth of \lstinline{acc} is larger than the bitwidth of \lstinline{c}, the value in \lstinline{acc} will be truncated to be stored into \lstinline{y}. Thus, is it even important to insure that the bitwidth of \lstinline{acc} is large enough so that it can handle the full precision of the multiply and accumulate operations? 

The answer to the question lies in the tolerance of the application. In many digital signal processing applications, the data itself is noisy, meaning that the lower several bits may not have any significance. In addition, in signal processing applications, we often perform numerical approximations when processing data which can introduce additional error. Thus, it may not be important that we insure that the \lstinline{acc} variable has enough precision to return a completely precise result.  On the other hand, it may be desirable to keep more bits in the accumulation and then round the final answer again to reduce the overall rounding error in the computation. Other applications, such as scientific computing, more dynamic range is often required, which may lead to the use of floating point numbers instead of integer or fixed-point arithmetic. So what is the correct answer? Ultimately it is up to the tolerance of the application designer.


\section{Complex FIR Filter}


To this point, we have solely looked at filtering real numbers. Many digital wireless communication systems deal with complex numbers using in-phase (I) and quadrature (Q) components (see Chapter \ref{chap:Wireless} for more details). Fortunately, it is possible to create a complex FIR filter using real FIR filter as we describe in the following.

To understand how to build a complex FIR filter from real FIR filters consider Equation \ref{eq:complex_fir}. Assume that $(I_{in}, Q_{in})$ is one sample of the input data that we wish to filter.  And one of the complex FIR filter coefficients is denoted as $(I_{fir}, Q_{fir})$. There will be more than one input sample and complex coefficient, but let us not worry about that for now.

\begin{equation}
(I_{in} + j  Q_{in})(I_{fir} + j  Q_{fir}) = (I_{in} I_{fir} - Q_{in}  Q_{fir}) + j  (Q_{in}  I_{fir} + I_{in} Q_{fir})
\label{eq:complex_fir}
\end{equation}

\begin{figure}
\centering
\includegraphics[width=6in]{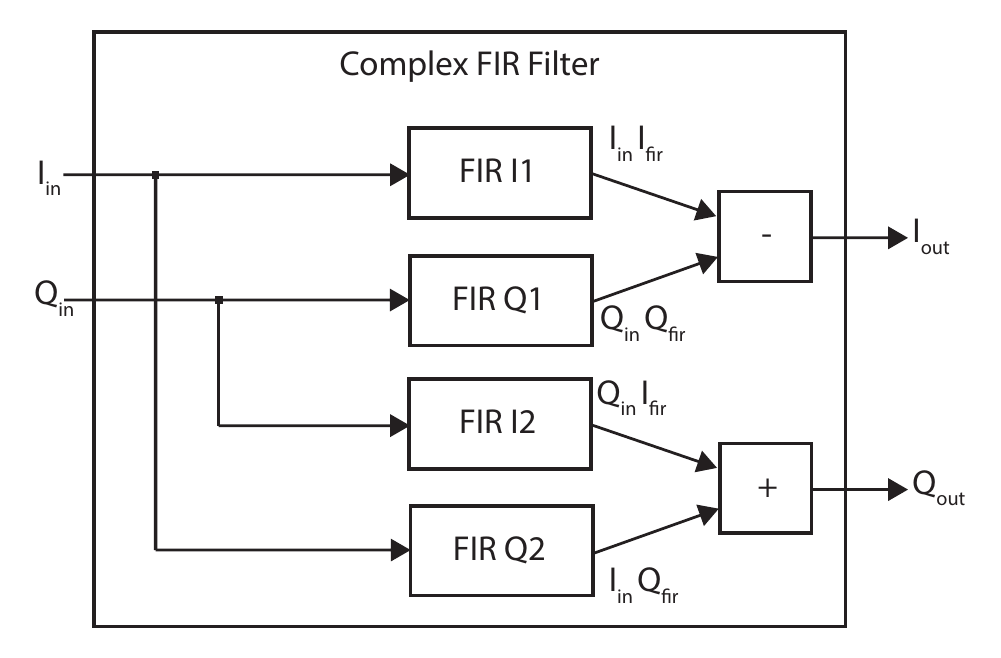}
\caption{A complex FIR filter built from four real FIR filters. The input I and Q samples are feed into four different real FIR filters. The FIR filters hold the in-phase (FIR I) and quadrature (FIR Q) complex coefficients. }
\label{fig:complex_fir}
\end{figure}

Equation \ref{eq:complex_fir} shows the multiplication of the input complex number by one coefficient of the complex FIR filter. The right side of the equation shows that the real portion of the output of complex input filtered by a complex FIR filter is $I_{in} I_{fir} - Q_{in}  Q_{fir}$ and the imaginary output is $Q_{in}  I_{fir} + I_{in} Q_{fir}$. This implies that we can separate the complex FIR filter operation into four real filters as shown in Figure \ref{fig:complex_fir}.

A complex FIR filter takes as input a complex number $(I_{in},Q_{in})$ and outputs a complex filtered value $(I_{out},Q_{out})$. Figure \ref{fig:complex_fir} provides a block diagram of this complex filter using four real FIR filters (FIR I1, FIR Q1, FIR I2, FIR Q2). The filters FIR I1 and FIR I2 are equivalent, i.e., they have the exact same coefficients. FIR Q1 and FIR Q2 are also equivalent. The output of each of these filters corresponds to a term from Equation \ref{eq:complex_fir}. These output are then added or subtracted to provide the final filtered complex output $(I_{out}, Q_{out})$. 

We used a hierarchical structure to define this complex FIR filter. \VHLS implements hierarchy using functions. Taking the previous real FIR function \lstinline{void fir (data_t *y, data_t x)} we can create another function that encapsulates four versions of this \lstinline{fir} function to create the complex FIR filter. This code is shown in Figure \ref{fig:complex_fir_code}.

\begin{figure}
\begin{lstlisting}
typedef int	data_t;
void firI1(data_t *y, data_t x);
void firQ1(data_t *y, data_t x);
void firI2(data_t *y, data_t x);
void firQ2(data_t *y, data_t x);

void complexFIR(data_t Iin, data_t Qin, data_t *Iout, data_t *Qout) {

  data_t IinIfir, QinQfir, QinIfir, IinQfir;

  firI1(&IinIfir, Iin);
  firQ1(&QinQfir, Qin);
  firI2(&QinIfir, Qin);
  firQ2(&IinQfir, Iin);

  *Iout = IinIfir + QinQfir;
  *Qout = QinIfir - IinQfir;
}
\end{lstlisting}
\caption{The \VHLS code to hierarchically implement a complex FIR filter using four real FIR filters.}
\label{fig:complex_fir_code}
\end{figure}

The code defines four functions \lstinline{firI1}, \lstinline{firQ1}, \lstinline{firI2}, and \lstinline{firQ2}. Each of these functions has the exact same code, e.g., that of the \lstinline{fir} function from Figure \ref{fig:fir11_initial}. Typically, we would not need to replicate the function; we could simply call the same function four times. However, this is not possible in this case due to the \lstinline{static} keyword used within the \lstinline{fir} function for the \lstinline{shift_reg}. \note{A good solution to this is to convert the function to a class.}

The function calls act as interfaces. The \VHLS tool does not optimize across function boundaries. That is, each \lstinline{fir} function synthesized independently, and treated more or less as a black box in the \lstinline{complexFIR} function. You can use the \lstinline{inline} directive if you want the \VHLS tool to co-optimize a particular function within its parent function. This will add the code from that function into the parent function and eliminate the hierarchical structure. While this can increase the potential for benefits in performance and area, it also creates a large amount of code that the tool must synthesize. That may take a long time, even fail to synthesize, or may result in a non-optimal design. Therefore, use the \lstinline{inline} directive carefully. Also note that the \VHLS tool may choose to inline functions on its own. These are typically functions with a small amount of code.

\begin{figure}
\begin{lstlisting}
float mul(int x, int y) { return x * y; }

float top_function(float a, float b, float c, float d) {
	return mul(a, b) + mul(c, d) + mul(b, c) + mul(a, d);
}

float inlined_top_function(float a, float b, float c, float d) {
	return a * b + c * d + b * c + a * d;
}
\end{lstlisting}
\caption{ A simple and trivial example to demonstrate the \lstinline{inline} directive. The \lstinline{top_function} has four function calls to the function \lstinline{mul}. If we placed an \lstinline{inline} directive on the \lstinline{mul} function, the result is similar to what you see in the function \lstinline{inlined_top_function}. }
\label{fig:inline}
\end{figure}

\begin{aside}
The \lstinline{inline} directive removes function boundaries, which may enable additional opportunities for the \VHLS tool at the cost of increasing the complexity of the synthesis problem, i.e., it will likely make the synthesis time longer.  It also eliminates any overhead associated with performing the function call. It allows for different implementations while maintaining the structure of the code, and making it hierarchical and more readable.

The code in Figure \ref{fig:inline} provides an example of how the inline directive works. The function \lstinline{inlined_top_function} is the result of using the \lstinline{inline} directive on the \lstinline{mul} function.

The \VHLS tool will sometimes choose to inline functions automatically. For example, it will very likely choose to inline the \lstinline{mul} function from Figure \ref{fig:inline} since it is small. You can force the tool to keep the function hierarchy by placing an \lstinline{inline} directive in the function with the \lstinline{off} argument.

The \lstinline{inline} directive also has a \lstinline{recursive} argument that inlines all functions called within the inlined function to also be inlined. That is, it will recursively add the code into the parent functions from every child function. This could create a substantial code base, so use this function carefully.

An inlined function will not have separate entries in the report since all of the logic will be associated with the parent function.
\end{aside}


\section{Conclusion}
This chapter describes the specification and optimization of a FIR filter using the \VHLS tool. The goal is to provide an overview of the HLS process. The first step in this process is understanding the basic concepts behind the computation of the FIR filter. This does not require a deep mathematical understanding, but certainly enough knowledge to write it in a manner that is synthesizable by the \VHLS tool. This may require translating the code from a different language (e.g., MATLAB, Java, C++, Python, etc.). Many times it requires rewriting to use simpler data structures, e.g., one that is explicitly implemented in an array. And it often involves removing system calls and other code not supported by the HLS tool. 

Creating an optimum architecture requires a basic understanding about how the HLS tool performs its synthesis and optimization process to RTL code. It is certainly not unnecessary to understand the exact HLS algorithms for schedule, binding, resource allocation, etc. (and many times these are proprietary). But having a general idea of the process does aid the designer in writing code that maps well to hardware. Throughout the chapter, we talked about some of the key features of the HLS synthesis process that are necessary to understand when performing various optimizations. It is especially important to understand the way that the HLS tool reports performance, which we describe in Chapter \ref{sec:fir-performance}. 

Additionally, we presented some basic HLS optimizations (including loop and bitwidth optimizations). We highlighted their benefits and potential drawbacks using the FIR filter as an example. These are common optimizations that can be applied across a wide range of applications. We provide more details about these optimizations in subsequent chapters as we walk through the implementation of other more complex applications. 

Finally, there is an entire project devoted to further optimizing the FIR filter. This is located in the Appendix in Chapter \ref{chapter:FIR_Project}. 

\chapter{CORDIC}
\glsresetall
\label{chapter:cordic}

\section{Overview}
\label{subsec:CORDIC_Overview}


CORDIC (Coordinate Rotation DIgital Computer) is an efficient technique to calculate trigonometric, hyperbolic, and other mathematical functions. It is a digit-by-digit algorithm that produces one output digit per iteration. This allows us to tune the accuracy of the algorithm to the application requirements; additional iterations produce a more precise output result. Accuracy is another common design evaluation metric alongside performance and resource usage. CORDIC performs simple computations using only addition, subtraction, bit shifting, and table lookups, which are efficient to implement in FPGAs and more generally in hardware. 

\begin{aside}
The CORDIC method was developed by Jack Volder in the 1950's as a digital solution to replace an analog resolver for real-time navigation on a B-58 bomber. A resolver measures degrees of rotation. At that time hardware implementations of multiply operations were prohibitively expense and CPUs had very limited amount of state. Thus the algorithm needed to have low complexity and use simple operations.  Over the years, it has been used in math co-processors \cite{duprat1993cordic}, linear systems \cite{ahmed1982highly}, radar signal processing \cite{andraka1996building}, Fourier transforms \cite{despain1974fourier}, and many other digital signal processing algorithms. It is now commonly used in FPGA designs. \VHLS uses a CORDIC core for calculating trigonometric functions and it is a common element of modern FPGA IP core libraries.
\end{aside}


The goal of this chapter is to demonstrate how to create an optimized CORDIC core using high-level synthesis. We are gradually increasing the complexity of the types of hardware cores that we are developing as we progress through the book. The CORDIC method is an iterative algorithm; thus most of the computation is performed within a single \lstinline|for| loop. The code itself is not all that complex. However, understanding the code such that we can create an optimal hardware implementation requires deep insight. And a good HLS designers must always understand the computation if they wish to create the optimal design. Thus, we spend the early part of this chapter giving the mathematical and computational background of the CORDIC method.  

The major HLS optimization that we wish to highlight in this chapter is choosing the correct number representation for the variables. As we discuss later in the chapter, the designer must carefully tradeoff between the accuracy of the results, the performance, and resource utilization of the design. Number representation is one big factor in this tradeoff -- ``larger'' numbers (i.e., those with more bits) generally provide more precision at the cost of increased resource usage (more FFs and logic blocks) and reduced performance. We provide a background on number representation and arbitrary data types in Chapter \ref{sec:arbitrary_precision}. 

This chapter is coupled with the project described in Chapter \ref{chapter:phase_detector} that allows more in-depth experimentation with the tradeoffs between precision (accuracy of the computation), resource usage, and performance. The aim of this chapter is to provide enough insight so that one can perform the exercises from that project, i.e., this chapter and that project are meant to complement each other. The goal of the project is to build a phase detector which uses a CORDIC and a complex matched filter which we have conveniently covered in this and the previous chapter. 

\section{Background}
\label{subsec:CORDIC_Basics}

The core idea behind the CORDIC is to efficiently perform a set of vector rotations in a two-dimensional plane. By overlaying these rotations with some simple control decisions, we can perform a variety of fundamental operations, e.g., trigonometric, hyperbolic, and logarithmic functions, real and complex multiplication, and matrix decompositions and factorizations.  CORDIC has been used in a wide range of applications including signal processing, robotics, communications, and many scientific computations. CORDIC is commonly used in FPGA design since it has a small resource usage. 

In the following, we walk through the process of how a CORDIC performs the sine and cosine of a given an input angle $\theta$. This is done using a series of vector rotations using only simple operations which are very efficient to implement in hardware. At the high level, the algorithm works using a series of rotations with the goal of reaching the target input angle $\theta$. The key innovation that makes this efficient is that the rotations can be done in a manner that requires minimal computation. In particular, we perform the rotations using multiplications by constant powers of two. This translates to simply moving bits around in hardware which is extremely efficient as it does not require any sort of logic.  

\begin{figure}
\centering
\includegraphics[width=.5\textwidth]{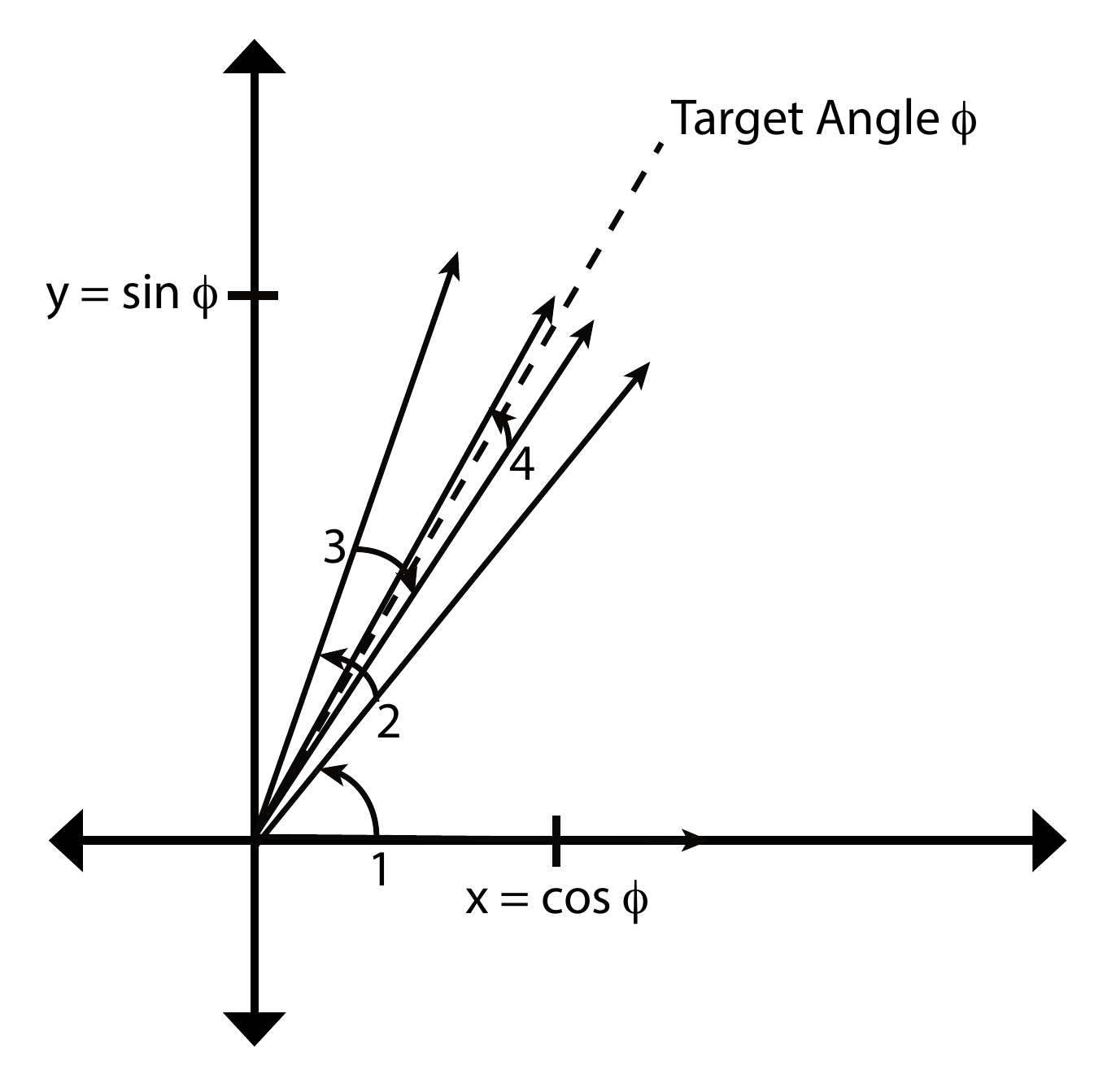}
\caption{ Using the CORDIC to calculate the functions $\sin \phi$ and $\cos \phi$. Here, the CORDIC starts at the x-axis with a corresponding $0^\circ$ angle. It then performs four iterative positive/negative rotations in increasingly smaller rotation angle with the ultimate goal of reaching the target angle $\phi$. Once we finish our rotations we are close to the target angle. We take the corresponding $x$ and $y$ values of the final vector which correspond to  $\cos \phi$ and $\sin \phi$ (respectively) assuming the length of the vector is $1$. The key to the CORDIC is doing all of this in a computationally efficient manner. }
\label{fig:cordic_overview}
\end{figure}

Figure \ref{fig:cordic_overview} provides a high level overview of the CORDIC procedure for calculating $\cos \phi$ and $\sin \phi$. In this case, we start our initial rotating vector on the x-axis, i.e, at a $0^\circ$ angle. Then, we perform an iterative series of rotations; in this example we only perform four rotations, but generally this is on the order of 40 rotations. Each of the subsequent rotation uses an increasingly smaller angle, which means that every iteration adds a bit more precision to the output value. At each iteration, we decide between doing a positive or negative rotation by that smaller angle. The angle values that we rotate are fixed a priori; thus, we can easily store their values in a small memory and keep a running sum of the cumulative angle that we have rotated so far. If this cumulative angle is larger than our target angle $\phi$, then we perform a negative rotation. If it is smaller, then the rotation is positive. Once we have completed a sufficient number of rotations, we can determine the $\cos \phi$ and $\sin \phi$ by directly reading the $x$ and $y$ values from the final rotated vector. If our final vector has a magnitude of $1$, then $x = \cos \phi$ and $y = \sin \phi$.

We start with some terminology. The goal is to refresh your memory about some basic trigonometric and vector concepts. Feel free to skim this if it is familiar. But keep in mind that one of the most important aspects of creating an efficient hardware design is to truly understand the application; only then can the designer effectively utilize the optimization directives and perform code refactoring which is required to get the most efficient designs. 

The fundamental goal of the CORDIC algorithm is to perform a series of rotations in an efficient manner. Let us start by thinking about how to generally perform a rotation. In two dimensions, the rotation matrix is:
\begin{equation}
R(\theta) = \begin{bmatrix}
\cos \theta & -\sin \theta \\
\sin \theta & \cos \theta \\
\end{bmatrix}
\label{eq:rotation_matrix}
\end{equation}
The CORDIC uses an iterative algorithm that rotates a vector $v$ to some target of angle which depends on the function that the CORDIC is performing. One rotation is a matrix vector multiplications in the form of $v_{i} = R_{i} \cdot v_{i-1}$. Thus in each iteration of the CORDIC we perform the following operations to perform one rotation which is the matrix vector multiply:
\begin{equation}
\begin{bmatrix}
\cos \theta & -\sin \theta \\
\sin \theta & \cos \theta \\
\end{bmatrix}\begin{bmatrix}
x_{i-1} \\
y_{i-1} \\
\end{bmatrix}
= \begin{bmatrix}
x_i \\
y_i \\
\end{bmatrix} 
\end{equation}
Writing out the linear equations, the coordinates of the newly rotated vector are: 
\begin{equation}
x_i = x_{i-1}  \cos \theta - y_{i-1}  \sin \theta\
\end{equation} and 
\begin{equation}
y_i = x_{i-1} \sin \theta + y_{i-1} \cos \theta\
\end{equation}
This is precisely the operation that we need to simplify. We want to perform these rotations without having to perform any multiplications.

Consider first a $90^\circ$ rotation. In this case the rotation matrix is:
\begin{equation}
R(90^\circ) = \begin{bmatrix}
\cos 90^\circ & -\sin 90^\circ \\
\sin 90^\circ & \cos 90^\circ \\
\end{bmatrix} = \begin{bmatrix}
0 & -1 \\
1 & 0 \\
\end{bmatrix} 
\end{equation} and thus we only have to perform the operations:
\begin{align}
x_i &= x_{i-1}  \cos 90^\circ - y_{i-1} \sin 90^\circ \nonumber \\
& = x_{i-1} \cdot 0 - y_{i-1} \cdot 1 \nonumber \\
& = -y_{i-1}\
\end{align}
and 
\begin{align}
y_i &= x_{i-1} \sin  90^\circ + y_{i-1} \cos 90^\circ \nonumber \\
& = x_{i-1} \cdot 1 + y_{i-1} \cdot 0 \nonumber \\
&=  x_{i-1}\
\end{align}
Putting this altogether we get
\begin{equation}
\begin{bmatrix}
0 & -1 \\
1 & 0 \\
\end{bmatrix}\begin{bmatrix}
x \\
y \\
\end{bmatrix}
= \begin{bmatrix}
-y \\
x \\
\end{bmatrix} 
\end{equation}
You can see that this is requires a very minimal amount of calculation; the rotated vector simply negates the $y$ value, and then swaps the $x$ and $y$ values. A two's complement negation requires the hardware equivalent to an adder. Thus, we have achieved our goal of performing a $90^\circ$ rotation efficiently.

\begin{exercise}
What if you wanted to rotation by $-90^\circ$? What is the rotation matrix $R(-90^\circ)$? What type of calculation is required for this rotation? How would one design the most efficient circuit that could perform a positive and negative rotation by $-90^\circ$, i.e., the direction of rotation is an input to the circuit?
\end{exercise}

While it is great that we can rotate by $\pm 90^\circ$, we also need to rotate by smaller angles if we wish to have any sort of good resolution in moving to the target angle. Perhaps the next natural angle that we might wish to rotate would be $\pm 45^\circ$. Using the rotation matrix from Equation \ref{eq:rotation_matrix}, we get
\begin{equation}
R(45^\circ) = \begin{bmatrix}
\cos 45^\circ & -\sin 45^\circ \\
\sin 45^\circ & \cos 45^\circ \\
\end{bmatrix} = \begin{bmatrix}
\sqrt 2/2 & -\sqrt 2/2 \\
\sqrt 2/2 & \sqrt 2/2 \\
\end{bmatrix}
\end{equation} Calculating out the computation for performing the rotation, we get
\begin{align}
x_i &= x_{i-1}  \cos 45^\circ - y_{i-1} \sin 45^\circ \nonumber \\
& = x_{i-1} \cdot \sqrt 2/2  - y_{i-1} \cdot \sqrt 2/2
\end{align} and 
\begin{align}
y_i &= x_{i-1} \sin  45^\circ + y_{i-1} \cos 45^\circ \nonumber \\
&= x_{i-1} \cdot \sqrt 2/2 + y_{i-1} \cdot \sqrt 2/2
\end{align} which when put back into matrix vector notation is
\begin{equation}
\begin{bmatrix}
\sqrt 2/2 & -\sqrt 2/2 \\
\sqrt 2/2 & \sqrt 2/2 \\
\end{bmatrix}
\begin{bmatrix}
x \\
y \\
\end{bmatrix}
= \begin{bmatrix}
\sqrt 2/2 x - \sqrt 2/2 y \\
\sqrt 2/2 x + \sqrt 2/2 y \\
\end{bmatrix} 
\end{equation}
This certainly is not as efficient of a computation as compared to rotating by $\pm 90^\circ$. The $\pm 90^\circ$ rotation was ideal because the multiplication were by very simple constants (in this case $0$, $1$, and $-1$). The key to the CORDIC is doing these rotations in an efficient manner, i.e., defining the rotation matrix in a way that their multiplication is trivial to compute. That is, we wish to be more like the previous $\pm 90^\circ$ and less like the much more difficult computation required for the $\pm 45^\circ$ rotation that we just described.

What if we ``forced'' the rotation matrix to be constants that were easy to multiply? For example, a multiplication by any power of two turns into a shift operation. If we set the constants in the rotation matrix to be powers of two, we could very easily perform rotations without multiplication. This is the key idea behind the CORDIC -- finding rotations that are very efficient to compute while minimizing any side effects. We will discuss these ``side effects'' in more detail, but there is an engineering decision that is being made here. In order to get efficient computation, we have to give up something; in this case we have to deal with the fact that the rotation also performs scaling, i.e., it changes the magnitude of the rotated vector  -- more on that later.

To further explore the idea of ``simple'' rotation matrices, consider the matrix 
\begin{equation}
R() = \begin{bmatrix}
1 & -1 \\
1 & 1 \\
\end{bmatrix} 
\end{equation} with the corresponding computation for the transformation
\begin{equation}
x_i = x_{i-1}  - y_{i-1}\
\end{equation} and 
\begin{equation}
y_i = x_{i-1}  + y_{i-1}\
\end{equation} with the matrix vector form of
\begin{equation}
\begin{bmatrix}
1 & -1 \\ 
1 & 1 \\
\end{bmatrix}
\begin{bmatrix}
x \\
y \\
\end{bmatrix}
= \begin{bmatrix}
x - y \\
x + y \\
\end{bmatrix} 
\end{equation}

This is certainly easy to compute and does not require any ``difficult'' multiplications. But what is the consequence of this operation? It turns out that this performs a rotation by $45^\circ$ which is perfect; we now have an efficient way to perform a $45^\circ$ rotation. But, this transform also scales the vector by a factor of  $\sqrt 2$.  The square root of the determinant of this matrix tells us how much the transformation scales the vector, i.e., how the length of the vector has changed. The determinant of this matrix here is $1 \cdot 1 -  (-1) \cdot 1 = 2$. Thus, this operation rotates by $45^\circ$ and scales by $\sqrt 2$. This is the tradeoff that the CORDIC makes; we can make the computation for the rotation easy to compute but it has the side effect that scales the length of the vector. This may or may not be a problem depending on the application. But for now, we put aside the scaling issue and focus on how to generalize the idea of performing rotations that are computationally efficient to perform.

Now we generalize the notion of performing efficient matrix rotations, i.e., performing rotations by only performing addition/subtraction and multiplication by a power of two (i.e., by shift operations). Consider again the rotation matrix 

\begin{equation}
R_{i}(\theta) = \begin{bmatrix} \cos(\theta_{i}) & -\sin(\theta_{i}) \\ \sin(\theta_{i}) & \cos(\theta_{i})\end{bmatrix}
\end{equation}
By using the following trigonometric identities,
\begin{equation}
\cos(\theta_{i}) =  {\frac{1}{\sqrt{1 + \tan^2(\theta_{i})}}}
\end{equation}
\begin{equation}
\sin(\theta_{i})  =  \frac{\tan(\theta_{i})}{\sqrt{1 + \tan^2(\theta_{i})}}
\end{equation}
we can rewrite the rotation matrix as
\begin{equation}
R_i = \frac{1}{\sqrt{1 + \tan^2(\theta_i)}} \begin{bmatrix} 1 & -\tan(\theta_i) \\ \tan(\theta_i) & 1 \end{bmatrix}
\end{equation}
If we restrict the values of $\tan(\theta_i)$ to be a multiplication by a factor of two, the rotation can be performed using a shifts (for the multiplication) and additions. More specifically, we use let $\tan(\theta_i) = 2^{-i}$. The rotation then becomes 
\begin{equation}
v_i = K_i \begin{bmatrix} 1 & - 2^{-i} \\  2^{-i} & 1 \end{bmatrix} \begin{bmatrix} x_{i-1} \\ y_{i-1} \end{bmatrix}
\end{equation}
where
\begin{equation}
K_i = \frac{1}{\sqrt{1 + 2^{-2i}}}
\end{equation}

A few things to note here. The $2^{-i}$ is equivalent to a right shift by $i$ bits, i.e., a division by a power of two. This is essentially just a simple rewiring which does not require any sort of logical resources, i.e., it is essentially ``free'' to compute in hardware. This is a huge benefit, but it does not come without some drawbacks. First, we are limited to rotate by angles $\theta$ such that $\tan(\theta_i) = 2^{-i}$. We will show that this is not much of a problem. Second, we are only showing rotation in one direction; the CORDIC requires the ability to rotation by $\pm \theta$. This is simple to correct by adding in $\sigma$ which can have a value of $1$ or $-1$, which corresponds to performing a positive or negative rotation. We can have a different $\sigma_i$ at every iteration/rotation. Thus the rotation operation generalizes to
\begin{equation}
v_i = K_i \begin{bmatrix} 1 & -\sigma_i 2^{-i} \\ \sigma_i 2^{-i} & 1 \end{bmatrix} \begin{bmatrix} x_{i-1} \\ y_{i-1} \end{bmatrix}
\end{equation}
Finally, the rotation requires a multiplication by $K_i$.  $K_i$ is typically ignored in the iterative process and then adjusted for after the series of rotations is completed. The cumulative scaling factor is
\begin{equation}
K(n) = \prod_{i=0}^{n-1} K_i  = \prod_{i=0}^{n-1}\frac {1}{\sqrt{1 + 2^{-2i}}}
\end{equation} and  
\begin{equation}
K = \lim_{n \to \infty}K(n) \approx 0.6072529350088812561694
\end{equation}
The scaling factors for different iterations can be calculated in advance and stored in a table. If we always perform a fixed number of rotations, this is simply one constant. This correction could also be made in advance by scaling $v_0$ appropriately before performing the rotations. Sometimes it is ok to ignore this scaling, which results in a processing gain
\begin{equation}
A = \frac{1}{K} = \lim_{n \to \infty} \prod_{i=0}^{n-1} {\sqrt{1 + 2^{-2i}}}\approx 1.64676025812107
\label{eq:cordicgain}
\end{equation}

At each iteration, we need to know the angle $\theta_i$ of the rotation that was just performed. This is derived as $\theta_i = \arctan 2^{-i}$. We can precompute these values for each value of $i$ and store them in an on-chip memory and use them as a lookup table. Additionally, we have a control decision that determines whether the rotation is clockwise or counterclockwise, i.e., we must determine if $\sigma$ is $1$ or $-1$. This decision depends on the desired CORDIC mode. For example, for calculating $\cos \phi$ and $\sin \phi$, we keep a running sum of the cumulative angle that we have rotated so far. We compare this to the target angle $\phi$ and perform a positive rotation if our current angle is less than $\phi$ and a negative rotation is our current angle is greater than $\phi$. 

Table \ref{table:cordic} provides the statistics for the first seven iterations of a CORDIC. The first row is the ``zeroth'' rotation (i.e., when $i=0$), which is a $45^{\circ}$ rotation. It performs a scaling of the vector by a factor of $1.41421$. The second row is the does a rotation by $2^{-1} = 0.5$. This results in a rotation by $\theta = \arctan 2^{-1} = 26.565^{\circ}$. This rotation scales the vector by $1.11803$. The CORDIC gain is the overall scaling of the vector. In this case, it is the scaling factor of the first two rotations, i.e., $1.58114 = 1.41421 \cdot 1.11803$. This process continues by incrementing $i$ which results in smaller and smaller rotating angles and scaling factors. Note that the CORDIC gain starts to stabilize to $\approx 1.64676025812107$ as described in Equation \ref{eq:cordicgain}. Also, note as the angles get smaller, they have less effect on the most significant digits. 

\begin{exercise}
Describe the effect if the $i$th iteration on the precision of the results? That is, what bits does it change? How does more iterations change the precision of the final result, i.e., how do the values of $\sin \phi$ and $\cos \phi$ change as the CORDIC performs more iterations?
\end{exercise}

\begin{table}[htbp]
\caption{The rotating angle, scaling factor, and CORDIC gain for the first seven iterations of a CORDIC. Note that the angle decreases by approximately half each time. The scaling factor indicates how much the length the the vector increases during that rotation. The CORDIC gain is the overall increase in the length of the vector which is the product of all of the scaling factors for the current and previous rotations.}
\begin{center}
\begin{tabular}{|c|c|c|c|c|c|}
\hline
i & $2^{-i}$ 	& Rotating Angle  	& Scaling Factor 	& CORDIC Gain 	\\ \hline \hline
0 & 1.0 		& $45.000^{\circ}$	& 1.41421			& 1.41421		\\ \hline
1 & 0.5 		& $26.565^{\circ}$	& 1.11803			& 1.58114		\\ \hline
2 & 0.25 		& $14.036^{\circ}$	& 1.03078			& 1.62980		\\ \hline
3 & 0.125 		& $7.125^{\circ}$	& 1.00778			& 1.64248		\\ \hline
4 & 0.0625 	& $3.576^{\circ}$	& 1.00195			& 1.64569		\\ \hline
5 & 0.03125 	& $1.790^{\circ}$	& 1.00049			& 1.64649		\\ \hline
6 & 0.015625 	& $0.895^{\circ}$	& 1.00012			& 1.64669		\\ \hline
\end{tabular}
\end{center}
\label{table:cordic}
\end{table}%

\section{Calculating Sine and Cosine}

Now we describe more precisely our running example of using a CORDIC to calculate the sine and cosine of some given angle $\phi$. In order to do this, we start with a vector on the positive $x$-axis (i.e., with an initial angle of $0^{\circ}$) and perform a series of rotations until we are approximately at the given angle $\phi$. Then we can simply read the $x$ and $y$ values of the resulting rotated vector to get the values $\cos \phi$ and $\sin \phi$, respectively. This assumes that the amplitude of the final vector is equal to $1$, which as you will see is not too difficult to achieve.

\begin{figure}
\centering
\includegraphics[width=.5\textwidth]{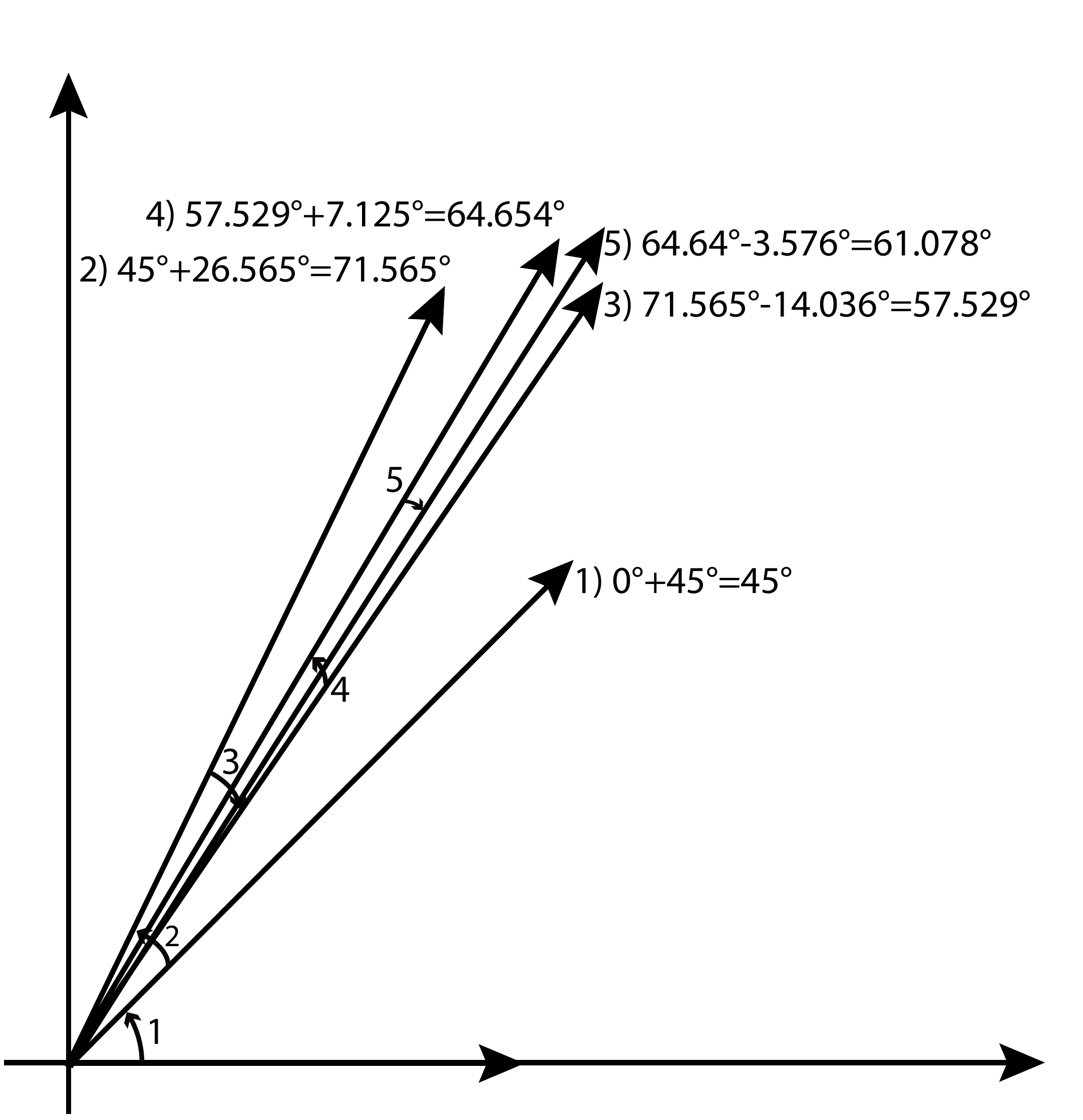}
\caption{Calculating $\cos 60^{\circ}$ and $\sin 60^{\circ}$ using the CORDIC algorithm. Five rotations are performed using incrementally larger $i$ values (0,1,2,3,4). The result is a vector with an angle of $61.078^{\circ}$. The corresponding $x$ and $y$ values of that vector give the approximate desired cosine and sine values. }
\label{fig:cordic_rotations}
\end{figure}

Let us illustrate this with an example: calculating $ \cos 60^{\circ}$ and $\sin 60^{\circ}$, i.e., $\phi = 60^{\circ}$. This process is depicted graphically in Figure \ref{fig:cordic_rotations}. Here we perform five rotations in order to give a final vector with an angle approximately equal to $60^{\circ}$. Our initial vector has a $0^{\circ}$ angle, i.e., it starts on the positive $x$-axis. The first rotation corresponds to $i=0$ which has a $45^{\circ}$ angle (see Table \ref{table:cordic}). Since we want to get to $60^{\circ}$, we rotate in the positive direction. The resulting rotated vector has a $45^{\circ}$ angle; also note that its amplitude is scaled by approximately 1.414. Now, we move on to $i=1$. As we wish to get to a $60^{\circ}$ angle, we rotate again in the positive direction. This rotation results in a vector that has an angle of $45^{\circ} + 26.565^{\circ} = 71.565^{\circ}$ and is scaled by a factor of 1.118; the total scaling resulting from the two rotations is $1.414 \times 1.118 = 1.581$. This is the CORDIC gain. Moving on to $i=2$, we now determine that our current angle is larger than the $60^{\circ}$ target, so we rotate by a negative angle resulting in a vector with a $57.529^{\circ}$ angle and scaled by a factor of $1.630$. This process continues by rotating the vector with incrementally larger $i$ values, resulting in smaller and smaller rotations that will eventually (approximately) reach the desired angle.  Also, note that the CORDIC gain begins to stabilize as the number of rotation increases.

After we perform a sufficient number of rotations, which is a function of the desired accuracy, we get a vector with an angle close to the desired input angle. The $x$ and $y$ values of that vector correspond to approximately $A_R \cos 60^{\circ}$ and $A_R \sin 60^{\circ}$, which is exactly what we want if $A_R = 1$. Since we typically know a priori the number of rotations that we will perform, we can insure that $A_R = 1$ by setting the magnitude of the initial vector to the reciprocal of the CORDIC gain. In the case of our example, assuming that we perform five rotations as shown in Figure \ref{fig:cordic_rotations}, this value is $1.64649^{-1} = 0.60735$ (the reciprocal of the CORDIC gain when $i=5$; see Table \ref{table:cordic}). We can easily set the amplitude of the initial vector by starting at a vector $(0.60735, 0)$. 

\begin{exercise}
How would the answer change if we performed one more rotation? How about two (three, four, etc.) more rotations? What is the accuracy (e.g., compared to MATLAB implementation) as we perform more rotations? How many rotations do you think is sufficient in the general case?
\end{exercise}

\begin{exercise}
Is it possible to get worse accuracy by performing more rotations? Provide an example when this would occur.
\end{exercise}

\begin{figure}
\begin{lstlisting}
// The file cordic.h holds definitions for the data types and constant values
#include "cordic.h"

// The cordic_phase array holds the angle for the current rotation
THETA_TYPE cordic_phase[NUM_ITERATIONS] = {45, 26.565, 14.036, 7.125,
                                           3.576, 1.790, 0.895, ...};

void cordic(THETA_TYPE theta, COS_SIN_TYPE &s, COS_SIN_TYPE &c)
{
  // Set the initial vector that we will rotate
  // current_cos = I; current_sin = Q
  COS_SIN_TYPE current_cos = 0.60735;
  COS_SIN_TYPE current_sin = 0.0;

  // Factor is the 2^(-L) value
  COS_SIN_TYPE factor = 1.0;

  // This loop iteratively rotates the initial vector to find the
  // sine and cosine values corresponding to the input theta angle
  for (int j = 0; j < NUM_ITERATIONS; j++) {
    // Determine if we are rotating by a positive or negative angle
    int sigma = (theta < 0) ? -1 : 1;

    // Save the current_cos, so that it can be used in the sine calculation
    COS_SIN_TYPE temp_cos = current_cos;

    // Perform the rotation
    current_cos = current_cos - current_sin * sigma * factor;
    current_sin = temp_cos * sigma * factor + current_sin;

    // Determine the new theta
    theta = theta - sigma * cordic_phase[j];

    // Calculate next 2^(-L) value
    factor = factor >> 1;
  }

  // Set the final sine and cosine values
  s = current_sin;  c = current_cos;
}
\end{lstlisting}
\caption{CORDIC code implementing the sine and cosine of a given angle.}
\label{fig:cordic_code}
\end{figure}

Figure \ref{fig:cordic_code} provides code that implements sine and cosine calculation using the CORDIC algorithm. It takes as input a target angle, and outputs the sine and cosine values corresponding to that angle. The code uses an array \lstinline{cordic_phase} as a lookup table that holds the angle of rotation for each iteration. This corresponds to the values in the ``Rotating Angle'' column in Table \ref{table:cordic}. We assume that the \lstinline{cordic.h} file defines the different data types (i.e., \lstinline{COS_SIN_TYPE} and \lstinline{THETA_TYPE}) and sets \lstinline{NUM_ITERATIONS} to some constant value. The data types can be changed to different fixed or floating point types, and \lstinline{NUM_ITERATIONS} set depending on our desired accuracy, area, and throughput. 
\begin{aside}
Notice that the variable \lstinline{sigma} is set as a two bit integer. Since we know that this will only take the value of $\pm 1$ we can change its data type which will result in smaller area and better performance than if we were to use the current \lstinline{int} data type. We discuss data types and how to specify them in \VHLS shortly.
\end{aside}  

This code is close to a ``software'' version. It can be optimized in many ways to increase its performance and reduce its area. We will discuss how to optimize this code later in the chapter.

\section{Cartesian to Polar Conversion}

With some modifications, the CORDIC can perform other functions. For example, it can convert between Cartesian and polar representations; we describe that in more detail in this section. The CORDIC can also do many other functions, which we will leave as an exercise to the reader. 

A two-dimensional vector $v$ can be represented using a Cartesian coordinate system $(x, y)$ or in the polar coordinate system $(r, \theta)$ where  $r$ is the radial coordinate (length of the vector) and $\theta$ is the angular coordinate. Both of these coordinate systems have their benefits and drawbacks. For example, if we want to do a rotation, then it is easier to think about the polar form while a linear transform is more easily described using the Cartesian system. 

\begin{figure}
\centering
\includegraphics[width=.55\textwidth]{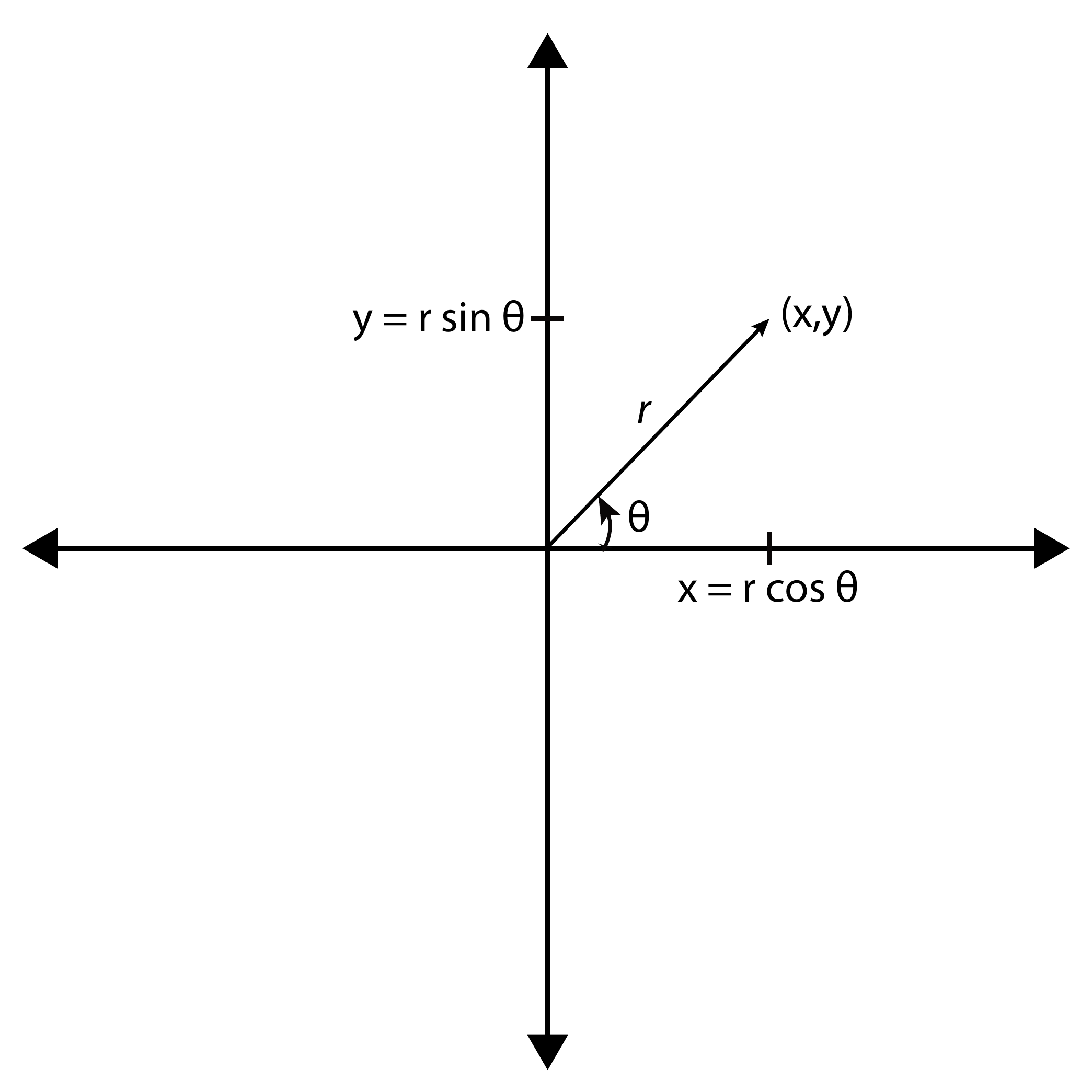}
\caption{ The figure shows a two-dimensional plane and a vector represented in both the Cartesian form $(x,y)$ and the polar form $(r, \theta)$ and provides the relationship between those two coordinate systems.  }
\label{fig:cordic_polar}
\end{figure}

The relationship between these coordinates is shown in the following equations:

\begin{equation}
x = r \cos \theta
\end{equation}

\begin{equation}
y = r \sin \theta
\end{equation}

\begin{equation}
r =\sqrt{x^2 + y^2}
\end{equation}

\begin{equation}
\theta = \operatorname{atan2}(y, x)
\end{equation}
where atan2 is a common variation on the arctangent function defined as
\begin{equation}
\operatorname{atan2}(y, x) =
\begin{cases}
\arctan(\frac{y}{x}) & \mbox{if } x > 0\\
\arctan(\frac{y}{x}) + \pi & \mbox{if } x < 0 \mbox{ and } y \ge 0\\
\arctan(\frac{y}{x}) - \pi & \mbox{if } x < 0 \mbox{ and } y < 0\\
\frac{\pi}{2} & \mbox{if } x = 0 \mbox{ and } y > 0\\
-\frac{\pi}{2} & \mbox{if } x = 0 \mbox{ and } y < 0\\
\text{undefined} & \mbox{if } x = 0 \mbox{ and } y = 0
\end{cases}
\end{equation}

This provides a way to translate between the two coordinate systems. However, these operations are not easy to implement in hardware. For example, sine, cosine, square root, and arctan are not simple operations and they require significant amount of resources. But we can use the CORDIC to perform these operations using a series of simple iterative rotation operations. 

Given a number in Cartesian form $(x,y)$, we can calculates its radial and amplitude coordinate (i.e., convert it to polar form) using the CORDIC. To do this, we rotate the given Cartesian number to $0^{\circ}$. Once this rotation is complete, the amplitude is the $x$ value of the final rotated vector. To determine the radial coordinate, we simply keep track of the cumulative angle of the rotations that the CORDIC performs. The angles of the rotating vector (for $i = 0,1,2,3, \dots$) are known and can be stored in a lookup table as done for calculating sine/cosine. Therefore, we simply need to keep track of the total rotation angle by performing an addition or subtraction of these angles, which depends on the direction of rotation.

The algorithm is similar to that of calculating the sine and cosine of a given angle. We perform a set of rotations with increasing values of $i$ such that the final vector resides on (close to) the positive $x$-axis (i.e., an angle of $0^{\circ}$). This can be done using positive or negative rotations  which is predicated on the $y$ value of the vector whose amplitude and phase we wish to determine. 

\begin{figure}
\centering
\includegraphics[width=\textwidth]{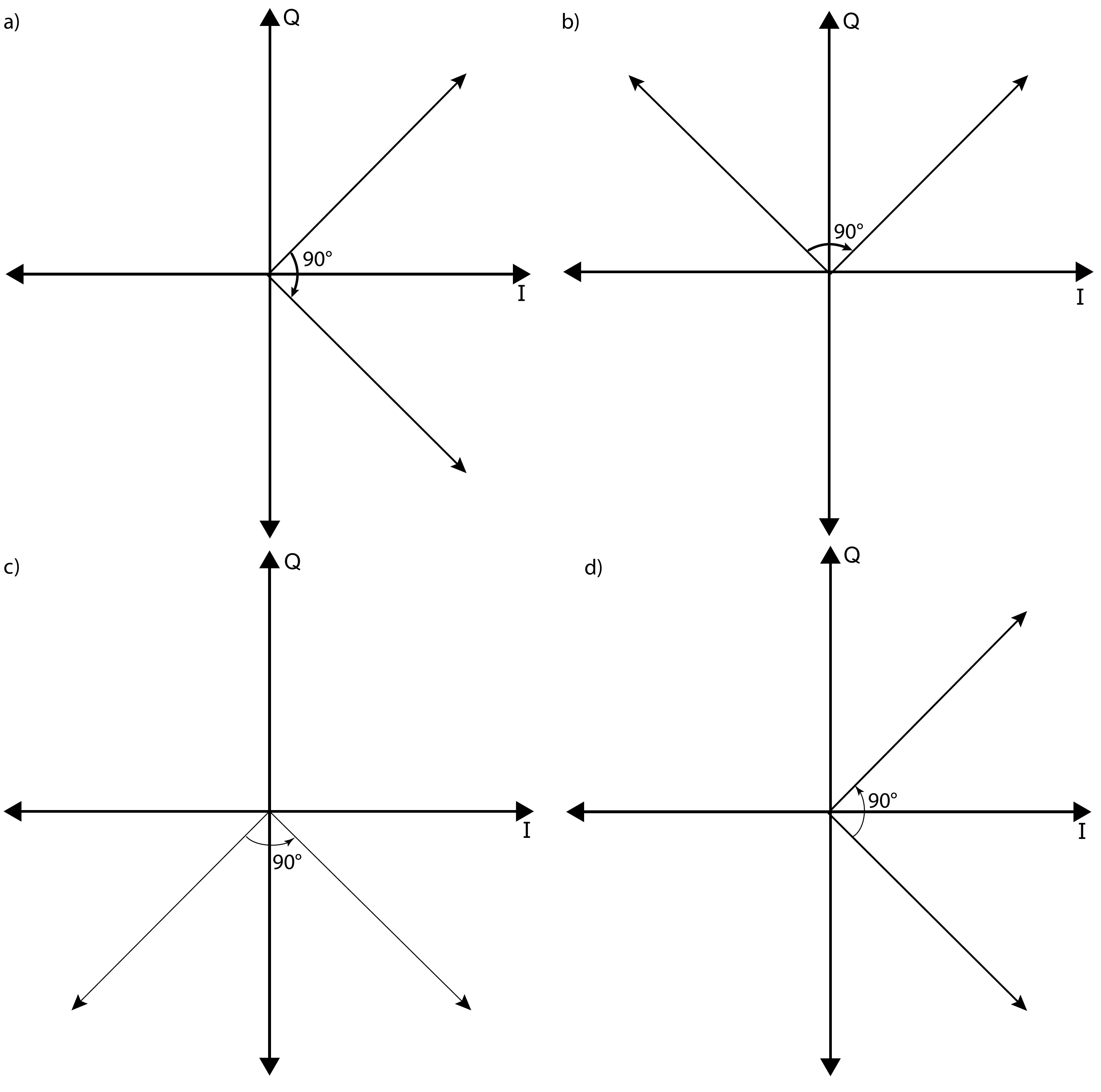}
\caption{The first step in performing a Cartesian to polar conversion is to perform a rotation by $\pm 90^{\circ}$ in order to get the initial vector into either Quadrant I or IV. Once it is in either of these two quadrants, subsequent rotations will allow the vector to reach a final angle of $0^{\circ}$. At this point, the radial value of the initial vector is the $x$ value of the final rotated vector and the phase of the initial vector is the summation of all the angles that the CORDIC performed. Parts a) and b) show an example with the initial $y$ value is positive, which means that the vector resides in either Quadrant I or II. Rotating by $-90^{\circ}$ puts them into the appropriate quadrant. Parts c) and d) show a similar situation when the $y$ value of the initial vector is negative. Here we wish to rotate by $90^{\circ}$ to get the vector into Quadrant I or IV. }
\label{fig:rotate90}
\end{figure}

The first step of the algorithm performs a rotation to get the initial vector into either Quadrant I or IV. This rotates the vector by $\pm 90^{\circ}$ depending on the sign of the $y$ value of the initial vector. If the $y$ value is positive, we know that we are in either Quadrant I or II. A rotation by $-90^{\circ}$ will put us into Quadrant IV or I, respectively. Once we are in either of those quadrants, we can guarantee that we will be able to asymptotically approach the target $0^{\circ}$ angle. If we are in Quadrant III or IV, the $y$ value of the initial vector will be negative. And a rotation by $90^{\circ}$ will put us into Quadrant IV or I, respectively. Recall that a $\pm 90^{\circ}$ rotation is done by negating either the $x$ or $y$ values of the vector and then swapping those values (see Equations \ref{eq:plus90} and \ref{eq:minus90}). The concept of these $\pm 90^{\circ}$ is shown in Figure \ref{fig:rotate90}.

There is an issue with the final radial value of the rotated vector; its magnitude is not the same as the initial magnitude before the rotations; it is scaled by the CORDIC gain. Or course, one could calculate the precise radial value of the vector by multiplying by the reciprocal of the appropriate CORDIC gain (approximately $1/1.647 = 0.607$)\footnote{Recall that the CORDIC gain is a function of the number of rotations as show in Table \ref{table:cordic}.}. However, this defeats the purpose of having a CORDIC, which eliminates the need for costly multiplication. And unfortunately this multiplication cannot be performed trivially using shifts and adds. Fortunately, this factor is often not important. e.g., in amplitude shift keying used in modulation in wireless communications, you only need to know a relative magnitude.  Or in other times, this amplitude gain can be compensated by other parts of the system.

\section{Number Representation}
\label{sec:number_representation}

The \lstinline{cordic} function uses currently uses common types for the variables. For example, the variable \lstinline{sigma} is defined as an \lstinline{int} and other variables use custom data types (e.g., \lstinline{THETA_TYPE} and \lstinline{COS_SIN_TYPE}). In many cases, HLS tools are able to further optimize the representation of these values to simplify the generated hardware.  For instance, in Figure \ref{fig:cordic_code}, the variable \lstinline{sigma} is restricted to be either \lstinline{1} or \lstinline{-1}.  Even though the variable is declared as an \lstinline{int} type of at least 32 bits, many fewer bits can be used to implement the variable without changing the behavior of the program. In other cases, particularly function inputs, memories, and variables that appear in recurrences, the representation cannot be automatically optimized.  In these cases, modifying the code to use smaller datatypes is a key optimization to avoid unnecessary resource usage.

Although reducing the size of variables is generally a good idea, this optimization can change the behavior of the program. A data type with fewer number of bits will not be able to express as much information as a data type with more bits and no finite binary representation can represent all real numbers with infinite accuracy. Fortunately, as designers we can pick numeric representations that are tuned to accuracy requirements of particular applications and tradeoff between accuracy, resource usage, and performance.

Before discussing these number representation optimizations further using our \lstinline{cordic} function, we first give a background on number representation. We provide the basics, as this is important in understand the data type specific representations provided by \VHLS. The next section starts with a fundamental background on number representation, and then proceeds to discuss the arbitrary precision variables available in \VHLS.

\subsection{Binary and Hexadecimal Numbers}

Computers and FPGAs typically represent numbers using \term{binary representation}, which enables numbers to be efficiently represented using on-off signals called binary digits, or simply \term{bits}.  Binary numbers work in most ways like normal decimal numbers, but can often be the cause of confusing errors if you are not familiar with how they work.  This is particularly true in many embedded systems and FPGAs where minimizing the number of bits used to represent variables can greatly increase the overall performance or efficiency of a system.  In this section, we will summarize binary arithmetic and the basic ways that computers represent numbers. 

Many readers may already be familiar with these ideas. In that case, you may skim these sections or skip them entirely. We do suggest that look at Section \ref{sec:arbitrary_precision} as this provides information specific to \VHLS on how to declare arbitrary data types. This is a key idea for optimizing the number representation of the \lstinline{cordic} function and any HLS code. 

When we write a normal integer, such as $4062$, what we really mean is implicitly $(4 * 1000) + (0 * 100) + (6* 10) + (2 * 1) = 4062$, or written in columns:
\begin{tabularpad}{*{4}{c}|l}
$10^3$ & $10^2$ & $10^1$ & $10^0$ & unsigned \\
\hline 
4&0&6&2 & = 4062 \\
\end{tabularpad}

A binary number is similar, except instead of using digits from zero to nine and powers of ten, we use numbers from zero to one and powers of 2:
\begin{tabularpad}{*{4}{c}|l}
$2^3$ & $2^2$ & $2^1$ & $2^0$ & unsigned \\
\hline 
1&0&1&1 &= 11\\
\end{tabularpad}
since $(1*8) + (0*4) + (1*2) + (1*1) = 11$.  To avoid ambiguity, binary numbers are often prefixed with "0b".  This makes it obvious that \lstinline|0b1011| is the number decimal 11 and not the number 1011.  The bit associated with the highest power of two is the \term{most significant bit}, and the bit associated with the lowest power of two is the \term{least significant bit}. 

Hexadecimal numbers use the digits representing numbers from zero to 15 and powers of 16:
\begin{tabularpad}{*{4}{c}|l}
$16^3$ & $16^2$ & $16^1$ & $16^0$ & unsigned\\
\hline 
8&0&3&15 &= 32831 \\
\end{tabularpad}
In order to avoid ambiguity, the digits from 10 to 15 are represented by the letters "A" through "F", and hexadecimal numbers are prefixed with "0x".  So the number above would normally be written in C code as \lstinline|0x803F|.

Note that binary representation can also represent fractional numbers, usually called \term{fixed-point} numbers, by simply extending the pattern to include negative exponents, so that "0b1011.01" is equivalent to:
\begin{tabularpad}{*{6}{c}|l}
$2^3$ & $2^2$ & $2^1$ & $2^0$ & $2^{-1}$ & $2^{-2}$ & unsigned \\
\hline 
1&0&1&1&0&1&= 11.25 \\
\end{tabularpad}
since $8 + 2 + 1 + \frac{1}{4} = 11.25$.  Unfortunately, the C standard doesn't provide a way of specifying constants in binary representation, although gcc and many other compilers allow integer constants (without a decimal point) to be specified  with the "0b" prefix.  The C99 standard does provide a way to describe floating-point constants with hexadecimal digits and a decimal exponent, however.  Note that the decimal exponent is required, even if it is zero.
\begin{lstlisting}
float p1 = 0xB.4p0; // Initialize p1 to "11.25"
float p2 = 0xB4p-4; // Initialize p2 to "11.25"
\end{lstlisting}

Notice that in general, it is only necessary to write non-zero digits and any digits not shown can be assumed to be zero without changing the represented value of an unsigned number.  As a result, it is easy to represent the same value with more digits: simply add as many zero digits as necessary.  This process is often called \term{zero-extension}.  Note that each additional digit increases the amount of numbers that can be represented.  Adding an additional bit to a binary number doubles the amount of numbers that can be represented, while an additional hexadecimal digit increases the amount of numbers by a factor of 16.
\begin{tabularpad}{*{12}{c}|l}
$2^7$ & $2^6$ & $2^5$ & $2^4$ & $2^3$ & $2^2$ & $2^1$ & $2^0$ & $2^{-1}$ & $2^{-2}$ & $2^{-3}$ & $2^{-4} $ & unsigned \\
\hline 
0&0&0&0&1&0&1&1&0&1&0&0&= 11.25 \\
\end{tabularpad}

\begin{aside}
Note that it is possible to have any number of bits in a binary number, not just 8, 16, or 32.   SystemC \cite{systemc}, for instance, defines several template classes for handling arbitrary precision integers and fixed-point numbers (including \lstinline|sc_int<>|, \lstinline|sc_uint<>|, \lstinline|sc_bigint<>|,  \lstinline|sc_ubigint<>|, \lstinline|sc_fixed<>|, and \lstinline|sc_ufixed<>|).  These classes can be commonly used in HLS tools, although they were originally defined for system modeling and not necessarily synthesis.  \VHLS, for instance, includes similar template classes (\lstinline|ap_int<>|, \lstinline|ap_uint<>|, \lstinline|ap_fixed<>|, and \lstinline|ap_ufixed<>|) that typically work better than the SystemC template classes, both in simulation and synthesis. \end{aside}

\begin{exercise}
Arbitrary precision numbers are even well defined (although not terribly useful) with zero digits.  List all the numbers that are representable with zero digits.
\end{exercise}

\subsection{Negative numbers}
Negative numbers are slightly more complicated than positive numbers, partly because there are several common ways to do it.  One simple way is represent negative numbers with a sign bit, often called \term{signed-magnitude} representation.  This representation just includes an additional bit to the front of the number to indicate whether it is signed or not.  One somewhat odd thing about signed-magnitude representation is that there is more than one way to represent zero.  This tends to make even apparently simple operations, like \lstinline|operator ==()|, more complex to implement.
\begin{tabularpad}{*{3}{c}|l}
+/-  & $2^1$ & $2^0$ & signed magnitude  \\
\hline 
0& 1&1& $=3$\\
0& 1&0& $=2$\\
0& 0&1& $=1$\\
0& 0&0& $=0$\\
1& 0&0& $=-0$\\
1& 0&1& $=-1$\\
1& 1&0& $=-2$\\
1& 1&1& $=-3$\\
\end{tabularpad} 

Another way to represent negative numbers is with \term{biased} representation.  This representation adds a constant offset (usually equal in magnitude to the value of the largest bit) to the value, which are otherwise treated as positive numbers:
\begin{tabularpad}{*{3}{c}|l}
$2^2$  & $2^1$ & $2^0$ & biased \\
\hline 
1& 1&1& $=3$\\
1& 1&0& $=2$\\
1& 0&1& $=1$\\
1& 0&0& $=0$\\
0& 1&1& $=-1$\\
0& 1&0& $=-2$\\
0& 0&1& $=-3$\\
0& 0&0& $=-4$\\
\end{tabularpad} 

However by far the most common technique for implementing negative numbers is known as \term{two's complement}.  In two's complement representation, the most significant bit represents the sign of the number (as in signed-magnitude representation), and \emph{also} whether or not an offset is applied.  One way of thinking about this situation is that the high order bit represents a negative contribution to the overall number.
\begin{tabularpad}{*{3}{c}|l}
$-2^2$  & $2^1$ & $2^0$ & two's complement \\
\hline 
0& 1&1& $=3$\\
0& 1&0& $=2$\\
0& 0&1& $=1$\\
0& 0&0& $=0$\\
1& 1&1& $=-1$\\
1& 1&0& $=-2$\\
1& 0&1& $=-3$\\
1& 0&0& $=-4$\\
\end{tabularpad}
\begin{tabularpad}{*{5}{c}|l}
$-2^4$  & $2^3$ & $2^2$  & $2^1$ & $2^0$ & two's complement \\
\hline 
0&0&0& 1&1& $=3$\\
0&0&0& 1&0& $=2$\\
0&0&0& 0&1& $=1$\\
0&0&0& 0&0& $=0$\\
1& 1&1& 1&1& $=-1$\\
1& 1&1& 1&0& $=-2$\\
1& 1&1& 0&1& $=-3$\\
1& 1&1& 0&0& $=-4$\\
\end{tabularpad}

One significant difference between unsigned numbers and two's complement numbers is that we need to know exactly how many bits are used to represent the number, since the most significant bit is treated differently than the remaining bits.  Furthermore, when widening a signed two's complement number with more bits, the sign bit is replicated to all the new most significant bits. This process is normally called \term{sign-extension}. For the rest of the book, we will generally assume that all signed numbers are represented in two's complement unless otherwise mentioned.

\begin{exercise}
What is the largest positive number representable with N bits in two's complement?  What is the largest negative number?
\end{exercise}
\begin{exercise}
Given a positive number $x$, how can you find the two's complement representation of $-x$?  
What is $-0$ in two's complement?  if $x$ is the largest negative number representable with N bits in two's complement, what is $-x$? 
\end{exercise}

\subsection{Overflow, Underflow, and Rounding}

\term{Overflow} occurs when a number is larger than the largest number that can be represented in a given number of bits.  Similarly, \term{underflow} occurs when a number is smaller than the smallest number that can be represented.  One common way of handling overflow or underflow is to simply drop the most significant bits of the original number, often called \term{wrapping}.
\begin{tabularpad}{*{10}{c}|l}
$2^5$ & $2^4$ & $2^3$ & $2^2$ & $2^1$ & $2^0$ & $2^{-1}$ & $2^{-2}$ & $2^{-3}$ & $2^{-4} $ & \\
\hline 
0&0&1&0&1&1&0&1&0&0&$=11.25$ \\
&0&1&0&1&1&0&1&0&0&$=11.25$ \\
&&1&0&1&1&0&1&0&0&$=11.25$ \\
&&&0&1&1&0&1&0&0&$=3.25$ \\
\end{tabularpad}

Handling overflow and underflow by wrapping two's complement numbers can even cause a positive number to become negative, or a negative number to become positive.
\begin{tabularpad}{*{8}{c}|l}
$-2^3$ & $2^2$ & $2^1$ & $2^0$ & $2^{-1}$ & $2^{-2}$ & $2^{-3}$ & $2^{-4} $ & two's complement \\
\hline 
1&0&1&1&0&1&0&0&$=-4.75$ \hfill\tabspace \\
\hfill& $-2^2$ & $2^1$ & $2^0$ & $2^{-1}$ & $2^{-2}$ & $2^{-3}$ & $2^{-4} $ & two's complement \\
\hline 
\hfill&0&1&1&0&1&0&0&$=3.25$ \hfill
\end{tabularpad}

Similarly, when a number cannot be represented precisely in a given number of fractional bits, it is necessary to apply \term{rounding}.  Again, there are several common ways to round numbers.  The simplest way is to just drop the extra fractional bits, which tends to result in numbers that are more negative.  This method of rounding is often called \term{rounding down} or \term{rounding to negative infinity}.  When rounding down to the nearest integer, this corresponds to the \lstinline|floor()| function, although it's possible to round to other bit positions as well.
\tabspace\\\makebox{\begin{tabular}{rl}
0b0100.00&$=4.0$   \\
0b0011.11&$=3.75$ \\
0b0011.10&$=3.5$ \\
0b0011.01&$=3.25$ \\
0b0011.00&$=3.0$ \\
0b1100.00&$=-4.0$\\
0b1011.11&$=-4.25$\\
0b1011.10&$=-4.5$\\
0b1011.01&$=-4.75$\\
0b1011.00&$=-5.0$\\
\end{tabular}
$\rightarrow$ \parbox{2cm}{Round to\\Negative\\Infinity} $\rightarrow$
\begin{tabular}{rl}
0b0100.0&$=4.0$   \\
0b0011.1&$=3.5$ \\
0b0011.1&$=3.5$ \\
0b0011.0&$=3.0$ \\
0b0011.0&$=3.0$ \\
0b1100.0&$=-4.0$\\
0b1011.1&$=-4.5$\\
0b1011.1&$=-4.5$\\
0b1011.0&$=-5.0$\\
0b1011.0&$=-5.0$\\
\end{tabular}}

It is also possible to handle rounding in other similar ways which force rounding to a more positive numbers (called \term{rounding up} or \term{rounding to positive infinity} and corresponding to the \lstinline|ceil()| function), to smaller absolute values (called \term{rounding to zero} and corresponding to the \lstinline|trunc()| function), or to larger absolute values (called \term{rounding away from zero} or \term{rounding to infinity} and corresponding to the \lstinline|round()| function)).  None of these operations always minimizes the error caused by rounding, however.  A better approach is called \term{rounding to nearest even}, \term{convergent rounding}, or \term{banker's rounding} and is implemented in the \lstinline|lrint()| function. As you might expect, this approach to rounding always picks the nearest representable number.  In addition, If there are two numbers equally distant, then the \emph{even} one is always picked.  An arbitrary-precision number is even if the last digit is zero.  This approach is the default handling of rounding with IEEE floating point, as it not only minimizes rounding errors but also ensures that the rounding error tends to cancel out when computing sums of random numbers.
\tabspace\\\makebox{\begin{tabular}{rl}
0b0100.00&$=4.0$   \\
0b0011.11&$=3.75$ \\
0b0011.10&$=3.5$ \\
0b0011.01&$=3.25$ \\
0b0011.00&$=3.0$ \\
0b1100.00&$=-4.0$\\
0b1011.11&$=-4.25$\\
0b1011.10&$=-4.5$\\
0b1011.01&$=-4.75$\\
0b1011.00&$=-5.0$\\
\end{tabular}
$\rightarrow$ \parbox{2cm}{Round to\\Nearest\\Even} $\rightarrow$
\begin{tabular}{rl}
0b0100.0&$=4.0$   \\
0b0100.0&$=4.0$ \\
0b0011.1&$=3.5$ \\
0b0011.0&$=3.0$ \\
0b0011.0&$=3.0$ \\
0b1100.0&$=-4.0$\\
0b1100.0&$=-4.0$\\
0b1011.1&$=-4.5$\\
0b1011.0&$=-5.0$\\
0b1011.0&$=-5.0$\\
\end{tabular}}
\tabspace\\

\subsection{Binary arithmetic} 
\label{sec:arithmetic}

Binary addition is very similar to decimal addition, simply align the binary points and add digits, taking care to correctly handle bits carried from one column to the next.  Note that the result of adding or subtracting two N-bit numbers generally takes N+1 bits to represent correctly without overflow.  The added bit is always an additional most significant bit for fractional numbers
\begin{tabularpad}{*{6}{c}|l}
  &$2^3$ & $2^2$  & $2^1$ & $2^0$ && unsigned \\
\hline 
&&0& 1&1&& $=3$\\
+&&0& 1&1&& $=3$\\
\hline
=&0&1& 1& 0&&$=6$\tabspace\\
 & $2^3$ & $2^2$  & $2^1$ & $2^0$ & $2^{-1}$ & unsigned \\
\hline 
&&1&1& 1&1& $=7.5$\\
+&&1&1& 1&1& $=7.5$\\
\hline
=&1&1&1& 1& 0&$=15$\\
\end{tabularpad}

Note that since the result of subtraction can be negative, the 'extra bit' becomes the sign-bit of a two's complement number.
\begin{tabularpad}{*{6}{c}|l}
 & &$2^3$ & $2^2$  & $2^1$ & $2^0$ & unsigned \\
\hline 
&&0&0& 1&1& $=3$\\
-&&0&0& 1&1& $=3$\\
\hline
=&&0&0& 0& 0&$=0$\tabspace\\
 &-$2^4$  & $2^3$ & $2^2$  & $2^1$ & $2^0$ & unsigned \\
\hline 
&&0&0& 1&1& $=3$\\
-&&1&1& 1&1& $=15$\\
\hline
=&1&0&1& 0& 0&$=-12$ (two's complement)\\
\end{tabularpad}

Multiplication for binary numbers also works similarly to familiar decimal multiplication.  In general, multiplying 2 N-bit numbers results in a 2*N bit result.
\begin{tabularpad}{*{8}{c}|l}
 &$2^6$  &$2^5$ &$2^4$ &$2^3$ & $2^2$  & $2^1$ & $2^0$ & two's complement \\
\hline 
&&&&1&0& 0&1& $=9$\\
*&&&&1&0& 0&1& $=9$\\
\hline
&&&&1&0& 0& 1&$=9$\\
&&&0&0& 0& 0&&$=0$\\
&&0&0& 0& 0&&&$=0$\\
+&1&0& 0& 1&&&&$=72$\\
\hline
&1&0&1&0&0&0&1&$=81$\\
\end{tabularpad}

Operations on signed numbers are somewhat more complex because of the sign-bit handling and won't be covered in detail.  However, the observations regarding the width of the result still applies: adding or subtracting two N-bit signed numbers results in an N+1-bit result, and Multiplying two N-bit signed numbers results in an 2*N-bit result.

\begin{exercise}
What about division?  Can the number of bits necessary to exactly represent the result a division operation of 2 N-bit numbers be computed?
\end{exercise}

\subsection{Representing Arbitrary Precision Integers in C and C++}
\label{sec:arbitrary_precision}

According to the C99 language standard, the precision of many standard types, such as \lstinline|int| and \lstinline|long| are implementation defined.  Although many programs can be written with these types in a way that does not have implementation-defined behavior, many cannot.  One small improvement is the \lstinline{inttypes.h} header in C99, which defines the types \lstinline|int8_t|,\lstinline|int16_t|,\lstinline|int32_t|, and \lstinline|int64_t| representing signed numbers of a given width and the corresponding types \lstinline|uint8_t|,\lstinline|uint16_t|,\lstinline|uint32_t|, and \lstinline|uint64_t| representing unsigned numbers.  Although these types are defined to have exactly the given bitwidths, they can still be somewhat awkward to use.  For instance, even relatively simple programs like the code below can have unexpected behavior.
\begin{lstlisting}
#include "inttypes.h"
uint16_t a =0x4000;
uint16_t b = 0x4000;
// Danger! p depends on sizeof(int)
uint32_t p = a*b;  
\end{lstlisting}
Although the values of \lstinline{a} and \lstinline{b} can be represented in 16 bits and their product (\lstinline{0x10000000}) can be represented exactly in 32 bits, the behavior of this code by the conversion rules in C99 is to first convert \lstinline|a| and \lstinline|b| to type \lstinline|int|, compute an integer result, and then to extend the result to 32 bits.  Although uncommon, it is correct for a C99 compiler to only have integers with only 16 bits of precision.  Furthermore, the C99 standard only defines 4 bitwidths for integer numbers, while FPGA systems often use a wide variety of bitwidths for arithmetic.  Also, printing these datatypes using \lstinline{printf()} is awkward, requiring the use of additional macros to write portable code.  The situation is even worse if we consider a fixed-point arithmetic example.  In the code below, we consider \lstinline{a} and \lstinline{b} to be fixed point numbers, and perform normalization correctly to generate a result in the same format.
\begin{lstlisting}
#include "inttypes.h"
// 4.0 represented with 12 fractional bits
uint16_t a =0x4000; 
// 4.0 represented with 12 fractional bits.
uint16_t b = 0x4000; 
// Danger! p depends on sizeof(int)
uint32_t p = (a*b) >> 12; 
\end{lstlisting}

The correct code in both cases requires casting the input variables to the width of the result before multiplying.
\begin{lstlisting}
#include "inttypes.h"
uint16_t a = 0x4000;
uint16_t b = 0x4000;
// p is assigned to 0x10000000
uint32_t p = (uint32_t) a*(uint32_t) b; 
\end{lstlisting}
\begin{lstlisting}
#include "inttypes.h"
// 4.0 represented with 12 fractional bits.
uint16_t a =0x4000; 
// 4.0 represented with 12 fractional bits.
uint16_t b = 0x4000; 
// p assigned to 16.0 represented with 12 fractional bits
uint32_t p = ( (uint32_t) a*(uint32_t) b ) >> 12; 
\end{lstlisting}

\begin{aside}
When using integers to represent fixed-point numbers, it is very important to document the fixed point format used, so that normalization can be performed correctly after multiplication.  Usually this is described using "Q" formats that give the number of fractional bits.  For instance, "Q15" format uses 15 fractional bits and usually applies to 16 bit signed variables.  Such a variable has values in the interval $[-1,1)$.  Similarly "Q31" format uses 31 fractional bits.
\end{aside}

For these reasons, it's usually preferable to use C++ and the \VHLS template classes \lstinline|ap_int<>|, \lstinline|ap_uint<>|, \lstinline|ap_fixed<>|, and \lstinline|ap_ufixed<>| to represent arbitrary precision numbers.  The \lstinline|ap_int<>| and \lstinline|ap_uint<>| template classes require a single integer template parameter that defines their width. Arithmetic functions generally produce a result that is wide enough to contain a correct result, following the rules in section \ref{sec:arithmetic}.
Only if the result is assigned to a narrower bitwidth does overflow or underflow occur.
\begin{lstlisting}
#include "ap_int.h"
ap_uint<15> a =0x4000;
ap_uint<15> b = 0x4000;
// p is assigned to 0x10000000.
ap_uint<30> p = a*b; 
\end{lstlisting}

The \lstinline|ap_fixed<>| and \lstinline|ap_ufixed<>| template classes are similar, except that they require two integer template arguments that define the overall width (the total number of bits) and the number of integer bits.
\begin{lstlisting}
#include "ap_fixed.h"
// 4.0 represented with 12 fractional bits.
ap_ufixed<15,12> a = 4.0; 
// 4.0 represented with 12 fractional bits.
ap_ufixed<15,12> b = 4.0; 
// p is assigned to 16.0 represented with 12 fractional bits
ap_ufixed<18,12> p = a*b; 
\end{lstlisting}

\begin{exercise}
Note that the \lstinline|ap_fixed<>| and \lstinline|ap_ufixed<>| template classes require the overall width of the number to be positive, but the number of integer bits can be arbitrary.  In particular, the number of integer bits can be 0 (indicating a number that is purely fractional) or can be the same as the overall width (indicating a number that has no fractional part).  However, the number of integer bits can also be negative or greater than the overall width!  What do such formats describe?  What are the largest and smallest numbers that can be represented by an \lstinline|ap_fixed<8,-3>|? \lstinline|ap_fixed<8,12>|?
\end{exercise}

\subsection{Floating Point}
\label{sec:floating_point}

\VHLS can also synthesize floating point calculations. Floating point numbers provide a large amount of precision, but this comes at a cost; it requires significant amount of computation which in turn translates to a large amount of resource usage and many cycles of latency. Thus, floating point numbers should be avoided unless absolutely necessary as dictated by the accuracy requirements application. In fact, the primary goal of this chapter is to allow the reader to understand how to effectively move from floating point to fixed point representations. Unfortunately, this is often a non-trivial task and there  are not many good standard methods to automatically perform this translation. This is partially due to the fact that moving to fixed point will reduce the accuracy of the application and this tradeoff is best left to the designer. 

The standard technique for high-level synthesis starts with a floating point representation during the initial development of the application. This allows the designer to focus on getting a functionally correct implementation. Once that is achieved, then she can move optimizing the number representation in order to reduce the resource usage and/or increase the performance. 

\begin{exercise}
Change all of the variables in the CORDIC from \lstinline{float} to \lstinline{int}. How does this effect the resource usage? How does it change the latency? How about the throughput? Does the accuracy change?
\end{exercise} 

\section{Further Optimizations}

In this section, we provide some brief thoughts and suggestions on the best way to optimize the CORDIC function. We focus on how the different optimizations change the precision of the result while providing the ability tradeoff between throughput, precision, and area. 

Ultimately, CORDIC produces an approximation. The error on that approximation generally decreases as the number of iterations increases. This corresponds to the number of times that we execute the \lstinline{for} loop in the \lstinline{cordic} function, which is set by \lstinline{NUM_ITERATIONS}.  Even if we perform a very large number of iterations, we may still have an approximation. One reason for this is that we may approach but never exactly match the desired target angle. We can, however, tune precision by choosing to perform greater or fewer iterations. All that needs to change in the algorithm is to modify the value of \lstinline{NUM_ITERATIONS}. The choice of \lstinline{NUM_ITERATIONS} depends on the number of digits of precision required by application using this CORDIC core.

\begin{exercise}
How do the area, throughput, and precision of the sine and cosine results change as you vary the data type?
\end{exercise}

\begin{exercise}
How does the constant \lstinline{NUM_ITERATIONS} affect the area, throughput, and precision? How does this affect the initial values of \lstinline{current_cos} and \lstinline{current_sin}? Do you need to modify the array \lstinline{cordic_phase}? Can you optimize the data types depending on the value of \lstinline{NUM_ITERATIONS}?
\end{exercise}

\begin{exercise}
The computations in the $for$ loop occupy most of the overall time. How do you best perform code transforms and/or use pragmas to optimize it?
\end{exercise}

\begin{exercise}
Setting the variable $sigma$ can be efficient in hardware using a two input multiplexer. Can you transform the code so that the high level synthesis tool implements it in this manner?
\end{exercise}

\begin{exercise}
The current code assumes that the given angle is between $\pm 90^{\circ}$. Can you add code to allow it to handle any angle between $\pm 180^{\circ}$?
\end{exercise}

\section{Conclusion}
In this chapter, we looked the Coordinate Rotation DIgital Computer (CORDIC) method for calculating trigonometric and hyperbolic functions based on vector rotations. We start with a background on the computation being performed by the CORDIC method. In particular, we focus on how to use the CORDIC method to calculate the sine and cosine values for a given angle. Additionally, we discuss how the same CORDIC method can be used to determine the amplitude and phase of a given complex number.

After this, we focus on the optimizations that can be done on the CORDIC method. Since it is an iterative method, there are fundamental tradeoffs between the number of iterations that are performed and the precision and accuracy of the resulting computation. We discuss how to reduce the precision/accuracy and get savings in FPGA resource usage and increases in performance.

We introduce the notion of using custom arbitrary data types for the variables in our \lstinline{cordic} function. This provides another method to reduce the latency, increase the throughput, and minimize the area while changing the precision of the intermediate and final results. \VHLS provides a method to specifically generate a large number of data types. We provide a background on number representation and introduce these custom data types. 

In general, there is a complex relationship between precision, resource utilization, and performance. We touch on some of these tradeoffs, and provide some insights on how to best optimize the \lstinline{cordic} function. We leave many of the optimization as well as the analysis of these tradeoffs, as an exercise to the reader. The CORDIC method is an integral part of the Phase Detector project described in Chapter \ref{chapter:phase_detector} -- a lab provided in the Appendix.

\chapter{Discrete Fourier Transform}
\glsresetall
\label{chapter:dft}


The \gls{dft} plays a fundamental role in digital signal processing systems. It is a method to change a discrete signal in the time domain to the same signal in the frequency domain.  By describing the signal as the sum of sinusoids, we can more easily compute some functions on the signal, e.g., filtering and other linear time invariant functions.  Therefore, it plays an important role in many wireless communications, image processing, and other digital signal processing applications.

This chapter provides an introduction to the \gls{dft} with a focus on its optimization for an FPGA implementation. At its core, the \gls{dft} performs a matrix-vector multiplication where the matrix is a fixed set of coefficients. The initial optimizations in Chapter \ref{subsec:dft_implementation} treat the \gls{dft} operation as a simplified matrix-vector multiplication. Then, Chapter \ref{subsec:dft_implementation} introduces a complete implementation of the \gls{dft} in \VHLS code. Additionally, we describe how to best optimize the \gls{dft} computation to increase the throughput. We focus our optimization efforts on array partitioning optimizations in Chapter \ref{subsec:dft_array_partitioning}.


There is a lot of math in the first two sections of this chapter. This may seem superfluous, but is necessary to fully comprehend the code restructuring optimizations, particularly for understanding the computational symmetries that are utilized by the \gls{fft} in the next chapter. That being said, if you are more interested in the HLS optimizations, you can skip to Chapter \ref{subsec:dft_implementation}. 

\section{Fourier Series}

In order to explain the discrete Fourier transform, we must first understand the \term{Fourier series}. The Fourier series provides an alternative way to look at a real valued, continuous, periodic signal where the signal runs over one period from $-\pi$ to $\pi$. The seminal result from Jean Baptiste Joseph Fourier states that any continuous, periodic signal over a period of $2 \pi$ can be represented by a sum of cosines and sines with a period of $2 \pi$. Formally, the Fourier Series is given as

\begin{equation}
\begin{array} {lcl} 
f(t) & \sim &  \frac{a_0}{2} + a_1 \cos (t) + a_2 \cos (2t) + a_3 \cos (3t) + \dots \\
& & + b_1 \sin(t) + b_2 \sin(2t) + b_3 \sin(3t) + \dots \\
& \sim & \frac{a_0}{2} + \displaystyle\sum\limits_{n=1}^{\infty} (a_n \cos (nt) + b_n \sin(nt))
\end{array}
\label{eq:fourier_series}
\end{equation}
where the coefficients $a_0, a_1, \dots$ and $b_1, b_2, \dots$ are computed as

\begin{equation}
\begin{array} {lcl} 
a_0 & = & \frac{1}{\pi} \int_{-\pi}^\pi f(t)\,\mathrm{d}t \\
a_n & = & \frac{1}{\pi} \int_{-\pi}^\pi f(t) \cos(nt)\,\mathrm{d}t \\
b_n & = & \frac{1}{\pi} \int_{-\pi}^\pi f(t) \sin(nt)\,\mathrm{d}t \\
\end{array}
\label{eq:fourier_coefficients}
\end{equation}

There are several things to note. First, the coefficients $a_0, a_1, a_2, \dots, b_1, b_2, \dots$ in Equation \ref{eq:fourier_coefficients} are called the Fourier coefficients. The coefficient $a_0$ is often called the \term{direct current (DC)} term (a reference to early electrical current analysis), the $n=1$ frequency is called the fundamental, while the other frequencies ($n \ge 2$) are called higher harmonics. The notions of fundamental and harmonic frequencies originate from acoustics and music. Second, the function $f$, and the $\cos()$ and $\sin()$ functions all have a period of $2 \pi$; changing this period to some other value is straightforward as we will show shortly. The DC value $a_0$ is equivalent to the coefficient of $\cos (0 \cdot t) = 1$, hence the use of symbol $a$. The $b_0$ value is not needed since $\sin (0 \cdot t) = 0$. Finally, the relation between the function $f$ and its Fourier series is approximate in some cases when there are discontinuities in $f$ (known as Gibbs phenomenon). This is a minor issue, and only relevant for the Fourier series, and not other Fourier Transforms. Therefore, going forward we will disregard this ``approximation'' ($\sim$) for ``equality'' ($=$). 

Representing functions that are periodic on something other than $\pi$ requires a simple change in variables. Assume a function is periodic on $[-L, L]$ rather than $[-\pi, \pi]$. Let 
\begin{equation}
t \equiv \frac{\pi t'}{L}
\end{equation} and 
\begin{equation}
\mathrm{d}t = \frac{\pi \mathrm{d}t'}{L}
\end{equation} which is a simple linear translation from the old $[-\pi, \pi]$ interval to the desired $[-L, L]$ interval.
Solving for $t'$ and substituting $t' = \frac{L t}{\pi}$ into Equation \ref{eq:fourier_series} gives
\begin{equation}
f(t') = \frac{a_0}{2} + \displaystyle\sum\limits_{n=1}^{\infty} (a_n \cos (\frac{n \pi t'}{L}) + b_n \sin(\frac{n \pi t'}{L}))
\end{equation} Solving for the $a$ and $b$ coefficients is similar:
\begin{equation}
\begin{array} {lcl} 
a_0 & = & \frac{1}{L} \int_{-L}^L f(t')\,\mathrm{d}t' \\
a_n & = & \frac{1}{L} \int_{-L}^L f(t') \cos(\frac{n \pi t'}{L})\,\mathrm{d}t' \\
b_n & = & \frac{1}{L} \int_{-L}^L f(t') \sin(\frac{n \pi t'}{L})\,\mathrm{d}t' \\
\end{array}
\end{equation}

We can use Euler's formula $e^{j n t} = \cos (n t) + j \sin (n t)$ to give a more concise formulation 
\begin{equation}
f(t) = \displaystyle\sum\limits_{n=-\infty}^{\infty} c_n e^{j n t}.
\end{equation} In this case, the Fourier coefficients $c_n$ are a complex exponential given by
\begin{equation}
c_n = \frac{1}{2 \pi} \int_{-\pi}^{\pi} f(t) e^{-j n t} \mathrm{d}t
\end{equation} which assumes that $f(t)$ is a periodic function with a period of $2\pi$, i.e., this equation is equivalent to Equation \ref{eq:fourier_series}. 

The Fourier coefficients $a_n$, $b_n$, and $c_n$ are related as
\begin{equation}
\begin{array} {lcl} 
a_n = c_n + c_{-n} \text{ for } n = 0,1,2, \dots \\
b_n = j(c_n - c_{-n}) \text{ for } n = 1,2, \dots \\
c_n = \left\{ 
  \begin{array}{l l }
  	\frac{1}{2} (a_n - j b_n) & n > 0 \\
	
	\frac{1}{2} a_0 & n = 0 \\
	\frac{1}{2} (a_{-n} + j b_{-n}) & n < 0 \\
  \end{array} \right .\\
\end{array}
\end{equation}

Note that the equations for deriving $a_n$, $b_n$, and $c_n$ introduce the notion of a ``negative'' frequency. While this physically does not make much sense, mathematically we can think about as a ``negative'' rotation on the complex plane. A ``positive'' frequency indicates that the complex number rotates in a counterclockwise direction in the complex plane. A negative frequency simply means that we are rotating in the opposite (clockwise) direction on the complex plane.

This idea is further illustrated by the relationship of cosine, sine, and the complex exponential.  Cosine can be viewed as the real part of the complex exponential and it can also be derived as the sum of two complex exponentials -- one with a positive frequency and the other with a negative frequency as shown in Equation \ref{eq:cos_exp}.
\begin{equation}
\cos(x) = \operatorname{Re} \{ e^{jx} \} = \frac{e^{jx} + e^{-jx}}{2}
\label{eq:cos_exp}
\end{equation} 
The relationship between sine and the complex exponential is similar as shown in Equation \ref{eq:sin_exp}. Here we subtract the negative frequency and divide by $2j$.
\begin{equation}
\sin(x) = \operatorname{Im} \{ e^{jx} \} = \frac{e^{jx} - e^{-jx}}{2j} 
\label{eq:sin_exp}
\end{equation}

\begin{figure}
\centering
\includegraphics[width= \textwidth]{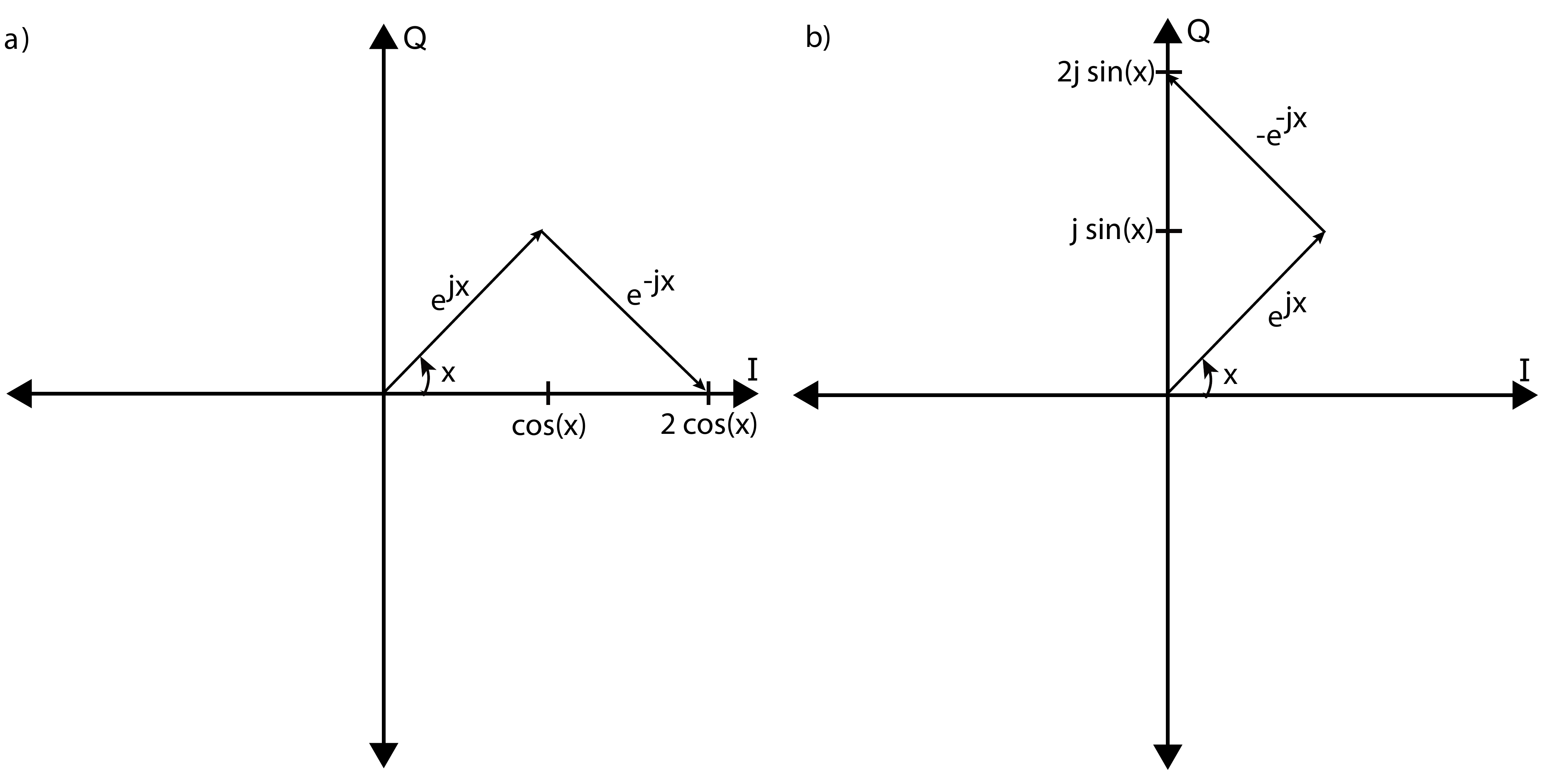}
\caption{A visualization of the relationship between the cosine, sine, and the complex exponential. Part a) shows the sum of two complex vectors, $e^{jx}$ and $e^{-jx}$. The result of this summation lands exactly on the real axis with the value $2 \cos (x)$. Part b) shows a similar summation except this time summing the vectors $e^{jx}$ and $-e^{-jx}$. This summation lands on the imaginary axis with the value $2 \sin (x)$.}
\label{fig:sin_cos_exp}
\end{figure}

Both of these relationships can be visualized as vectors in the complex plane as shown in Figure \ref{fig:sin_cos_exp}. Part a) shows the cosine derivation. Here we add the two complex vectors $e^{jx}$ and $e^{-jx}$. Note that the sum of these two vectors results in a vector on the real (in-phase or I) axis. The magnitude of that vector is $2 \cos(x)$. Thus, by dividing the sum of these two complex exponentials by $2$, we get the value $\cos (x)$ as shown in Equation \ref{eq:cos_exp}.  Figure \ref{fig:sin_cos_exp} b) shows the similar derivation for sine. Here we are adding the complex vectors $e^{jx}$ and $-e^{-jx}$. The result of this is a vector on the imaginary (quadrature or Q) axis with a magnitude of $2 \sin (x)$. Therefore, we must divide by $2j$ in order to get $\sin (x)$. Therefore, this validates the relationship as described in Equation \ref{eq:sin_exp}.

\section{\gls{dft} Background}
\label{sec:DFTbackground}

The previous section provided a mathematical foundation for the Fourier series, which works on signals that are continuous and periodic. The Discrete Fourier Transform requires {\it discrete} periodic signals. The \gls{dft} converts a finite number of equally spaced samples into a finite number of complex sinusoids. In other words, it converts a sampled function from one domain (most often the time domain) to the frequency domain. The frequencies of the complex sinusoids are integer multiples of the \term{fundamental frequency} which is defined as the frequency related to the sampling period of the input function. Perhaps the most important consequence of the discrete and periodic signal is that it can be represented by a finite set of numbers. Thus, a digital system can be used to implement the \gls{dft}.  

The \gls{dft} works on input functions that uses both real and complex numbers. Intuitively, it is easier to first understand how the real \gls{dft} works, so we will ignore complex numbers for the time being and start with real signals in order to gain ease into the mathematics a bit. 
\begin{aside}
A quick note on terminology: We use lower case function variables to denote signals in the time domain. Upper case function variables are signals in the frequency domain. We use $( )$ for continuous functions and $[ ]$ for discrete functions. For example, $f( )$ is a continuous time domain function and $F( )$ is its continuous frequency domain representation. Similarly $g[ ]$ is a discrete function in the time domain and $G[ ]$ is that function transformed into the frequency domain. 
\end{aside}

\begin{figure}
\centering
\includegraphics[width=\textwidth]{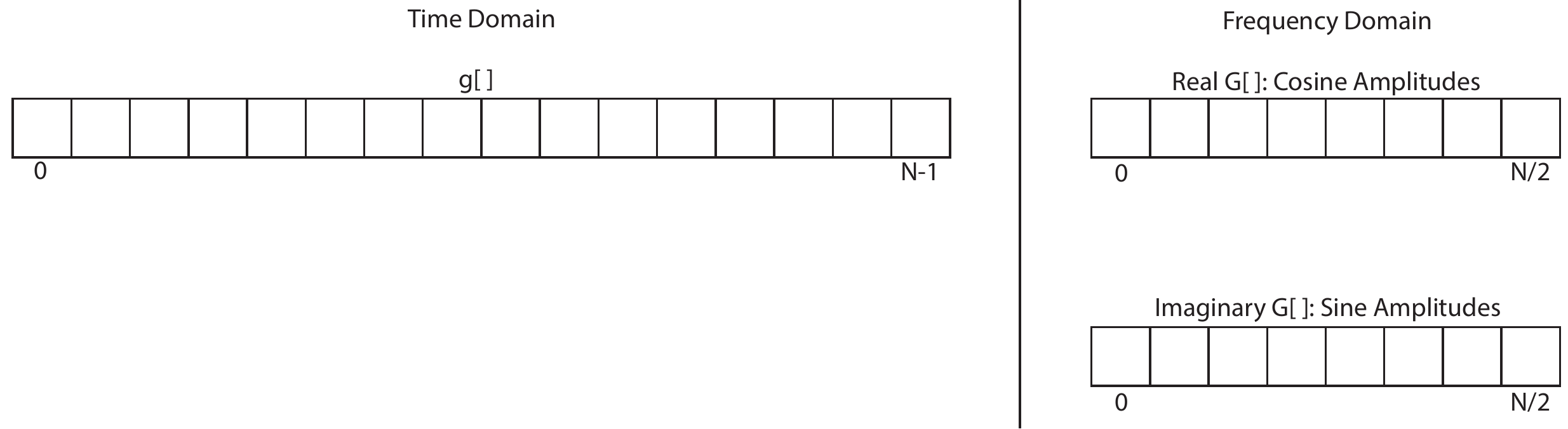}
\caption{ A real valued discrete function $g[ ]$ in the time domain with $N$ points has a frequency domain representation with $N/2 + 1$ samples. Each of these frequency domain samples has one cosine and one sine amplitude value. Collectively these two amplitude values can be represented by a complex number with the cosine amplitude representing the real part and the sine amplitude the imaginary part. }
\label{fig:basic-DFT}
\end{figure}

To start consider Figure \ref{fig:basic-DFT}. The figure shows on the left a real valued time domain signal $g[ ]$ with $N$ samples or points running from $0$ to $N-1$. The \gls{dft} is performed resulting in the frequency domain signals corresponding to the cosine and sine amplitudes for the various frequencies. These can be viewed as a complex number with the cosine amplitudes corresponding to the real value of the complex number and the sine amplitudes providing the imaginary portion of the complex number. There are $N/2 + 1$ cosine (real) and $N/2 + 1$ sine (imaginary) values. We will call this resulting complex valued frequency domain function $G[ ]$. Note that the number of samples in frequency domain ($N/2 + 1$) is due to the fact that we are considering a real valued time domain signal; a complex valued time domain signal results in a frequency domain signal with $N$ samples.

\begin{figure}
\centering
\includegraphics[width= 0.8 \textwidth]{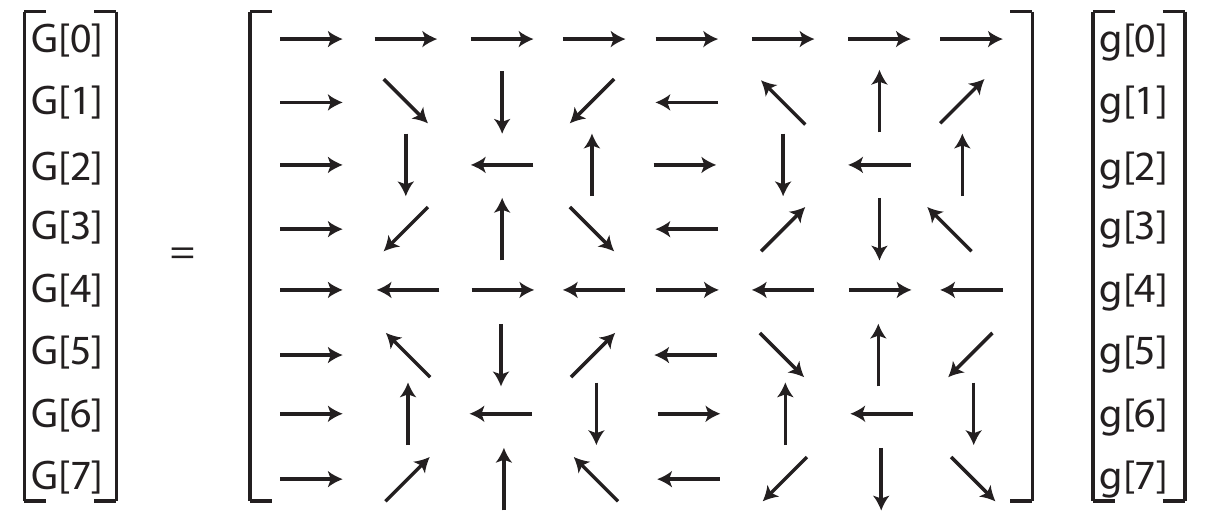}
\caption{ The elements of the $S$ shown as a complex vectors.  }
\label{fig:dft_visualization}
\end{figure}

An $N$ point \gls{dft} can be determined through a $N \times N$ matrix multiplied by a vector of size $N$, $G = S \cdot g$ where
\begin{equation}
S =
 \begin{bmatrix}
 \label{eq:Smatrix}
  1 & 1 & 1 & \cdots & 1 \\
  1 & s & s^2 & \cdots & s^{N-1} \\
  1 & s^2 & s^4 & \cdots & s^{2(N-1)} \\
  1 & s^3 & s^6 & \cdots & s^{3(N-1)} \\
  \vdots  & \vdots  & \vdots &\ddots & \vdots  \\
  1 & s^{N-1} & s^{2(N-1)}&\cdots & s^{(N-1)(N-1)}
 \end{bmatrix}
\end{equation} and $s = e^{\frac{-j 2 \pi}{N}}$.   Thus the samples in frequency domain are derived as 
\begin{equation}
G[k] = \displaystyle\sum\limits_{n=0}^{N-1} g[n] s^{kn} \text{ for } k = 0,\dots, N-1
\end{equation}
Figure \ref{fig:dft_visualization} provides a visualization of the \gls{dft} coefficients for an 8 point \gls{dft} operation. The eight frequency domain samples are derived by multiplying the 8 time domain samples with the corresponding rows of the $S$ matrix. Row 0 of the $S$ matrix corresponds to the DC component which is proportional to the average of the time domain samples. Multiplying Row 1 of the $S$ matrix with $g$ provides the cosine and sine amplitudes values for when there is one rotation around the unit circle. Since this is an 8 point \gls{dft}, this means that each phasor is offset by $45^{\circ}$. Performing eight $45^{\circ}$ rotations does one full rotation around the unit circle. Row 2 is similar except is performs two rotations around the unit circle, i.e., each rotation is $90^{\circ}$. This is a higher frequency. Row 3 does three rotations; Row 4 four rotations and so on. Each of these row times column multiplications gives the appropriate frequency domain sample. 

Notice that the $S$ matrix is diagonally symmetric, that is $S[i][j] = S[j][i]$.  In addition, $S[i][j] = s^i*s^j = s^{(i+j)}$.  There is also interesting symmetry around Row 4. The phasors in Rows 3 and 5 are complex conjugates of each other, i.e., $S[3][j] = S[5][j]^*$. Similarly, Rows 2 and 6 ($S[2][j] = S[6][j]^*$), and Rows 1 and 7 ($S[1][j] = S[7][j]^*$) are each related by the complex conjugate operation. It is for this reason that the \gls{dft} of a real valued input signal with $N$ samples has only $N/2 + 1$ cosine and sine values in the frequency domain. The remaining $N/2$ frequency domain values provide redundant information so they are not needed. However, this is not true when the input signal is complex. In this case, the frequency domain will have $N + 1$ cosine and sine values. 

\section{Matrix-Vector Multiplication Optimizations}
\label{subsec:mvmul_implementation}

Matrix-vector multiplication is the core computation of a \gls{dft}.  The input time domain vector is multiplied by a matrix with fixed special values. The result is a vector that corresponds to the frequency domain representation of the input time domain signal.  

In this section, we look at the hardware implementation of matrix-vector multiplication. We break this operation down into its most basic form (see Figure \ref{fig:matrix_vector_base}). This allows us to better focus the discussion on the optimizations rather than deal with all the complexities of using functionally correct \gls{dft} code. We will build a \gls{dft} core in the next section. 

\begin{figure}
\begin{lstlisting}
#define SIZE 8
typedef int BaseType;

void matrix_vector(BaseType M[SIZE][SIZE], BaseType V_In[SIZE], BaseType V_Out[SIZE]) {
	BaseType i, j;
data_loop:
	for (i = 0; i < SIZE; i++) {
		BaseType sum = 0;
	dot_product_loop:
		for (j = 0; j < SIZE; j++) {
			sum += V_In[j] * M[i][j];
		}
		V_Out[i] = sum;
	}
}
\end{lstlisting}
\caption{Simple code implementing a matrix-vector multiplication.}\label{fig:matrix_vector_base}
\end{figure}

The code in Figure \ref{fig:matrix_vector_base} provides an initial starting point for synthesizing this operation into hardware. We use a custom data type called \lstinline|BaseType| that is currently mapped as a \lstinline|float|. This may seem superfluous at the time, but this will allow us in the future to easily experiment with different number representations for our variables (e.g., signed or unsigned fixed point with different precision). The \lstinline|matrix_vector| function has three arguments. The first two arguments \lstinline|BaseType M[SIZE][SIZE]| and \lstinline|BaseType V_In[SIZE]| are the input matrix and vector to be multiplied. The third argument \lstinline|BaseType V_Out[SIZE]| is the resultant vector. By setting \lstinline|M = S|  and \lstinline|V_In| to a sampled time domain signal, the \lstinline|V_Out| will contain the \gls{dft}. \lstinline|SIZE| is a constant that determines the number of samples in the input signal and correspondingly the size of the \gls{dft}.

The algorithm itself is simply a nested \lstinline|for| loop. The inner loop (\lstinline|dot_product_loop|) computes the \gls{dft} coefficients starting from \lstinline|0| and going to \lstinline|SIZE - 1|. However, this relatively simple code has many design choices that can be performed when mapping to hardware. 

Whenever you perform HLS, you should think about the architecture that you wish to synthesize. Memory organization is one of the more important decisions. The question boils down to \emph{where do you store the data from your code?} There are a number of options when mapping variables to hardware. The variable could simply be a set of wires (if its value never needs saved across a cycle), a register, RAM or FIFO. All of these options provide tradeoffs between performance and area. 

Another major factor is the amount of parallelism that is available within the code. Purely sequential code has few options for implementation. On the other hand, code with a significant amount of parallelism has implementation options that range from purely sequentially to fully parallel. These options obviously have different area and performance. We will look at how both memory configurations and parallelism effect the hardware implementation for the matrix-vector implementation of the \gls{dft}.

\begin{figure}
\centering
\includegraphics[width= 0.4 \textwidth]{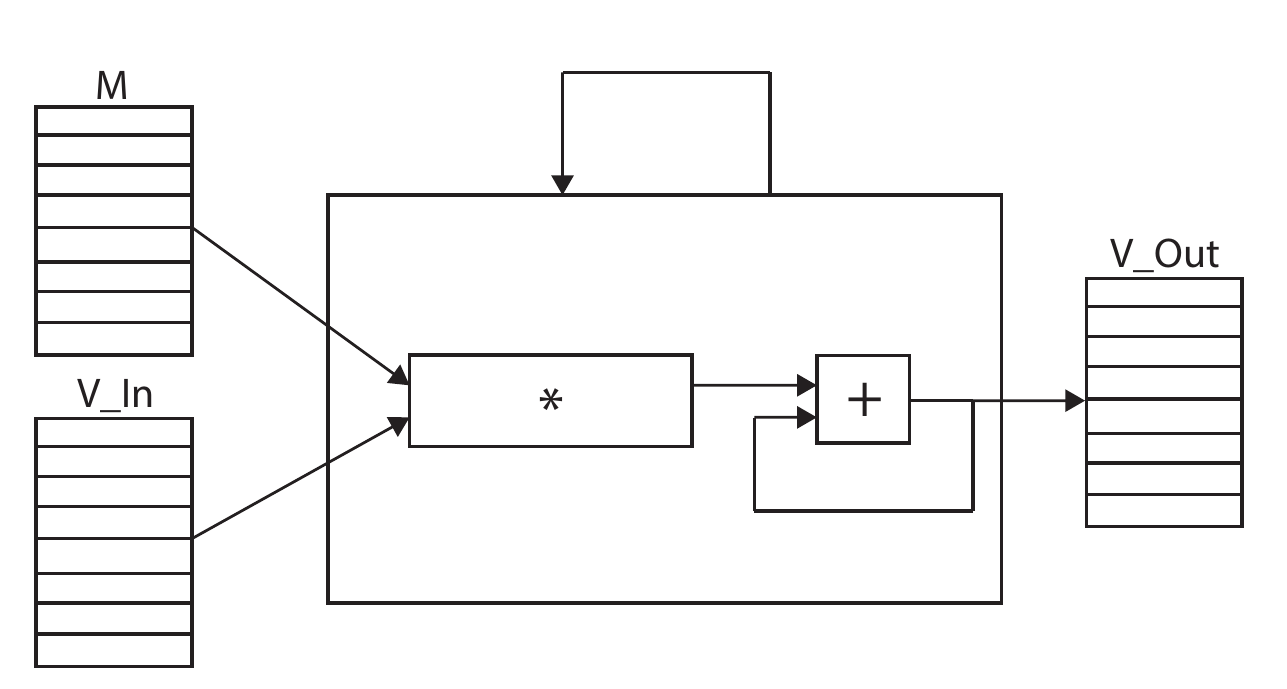}
\executeiffilenewer{dft_behavior_loop_sequential.svg}{images/dft_behavior_loop_sequential.pdf}%
{inkscape -z -D --file=dft_behavior_loop_sequential.svg %
--export-pdf=images/dft_behavior_loop_sequential.pdf --export-latex}%
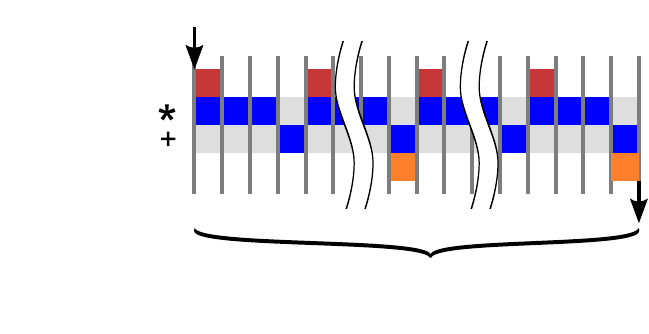%

\caption{A possible implementation of matrix-vector multiplication from the code in Figure \ref{fig:matrix_vector_base}.}
\label{fig:matrix_vector_sequential}
\end{figure}

Figure \ref{fig:matrix_vector_sequential} shows a sequential architecture for matrix-vector multiplication with one multiply and one addition operator.  Logic is created to access the \lstinline|V_In| and \lstinline|M| arrays which are stored in BRAMs. Each element of \lstinline|V_Out| is computed and stored into the BRAM. This architecture is essentially what will result from synthesizing the code from Figure \ref{fig:matrix_vector_base} with no directives. It does not consume a lot of area, but the task latency and task interval are relatively large.

\section{Pipelining and Parallelism}

There is substantial opportunity to exploit parallelism in the matrix-multiplication example. We start by focusing on the inner loop. The expression \lstinline|sum += V_In[j] * M[i][j];| is executed in each iteration of the loop. The variable \lstinline|sum|, which is keeping a running tally of the multiplications, is being reused in each iteration and takes on a new value. This inner loop can be rewritten as shown Figure \ref{fig:matrix_vector_base_unroll_inner}.  In this case, the sum variable has been completely eliminated and replaced with multiple intermediate values in the larger expression.

\begin{figure}
\begin{lstlisting}
#define SIZE 8
typedef int BaseType;

void matrix_vector(BaseType M[SIZE][SIZE], BaseType V_In[SIZE], BaseType V_Out[SIZE]) {
	BaseType i, j;
data_loop:
	for (i = 0; i < SIZE; i++) {
		BaseType sum = 0;
		V_Out[i] = V_In[0] * M[i][0] + V_In[1] * M[i][1] + V_In[2] * M[i][2] +
							 V_In[3] * M[i][3] + V_In[4] * M[i][4] + V_In[5] * M[i][5] +
							 V_In[6] * M[i][6] + V_In[7] * M[i][7];
	}
}
\end{lstlisting}
\caption{The matrix-vector multiplication example with a manually unrolled inner loop.}
\label{fig:matrix_vector_base_unroll_inner}
\end{figure}

\begin{aside}
Loop unrolling is performed automatically by \VHLS in a pipelined context.  Loop unrolling can also be requested by using \lstinline|#pragma HLS unroll| or the equivalent directive outside of a pipelined context.  
\end{aside} 

\begin{figure}
\centering
\executeiffilenewer{matrix_vector_unroll_inner.svg}{images/matrix_vector_unroll_inner.pdf}%
{inkscape -z -D --file=matrix_vector_unroll_inner.svg %
--export-pdf=images/matrix_vector_unroll_inner.pdf --export-latex}%
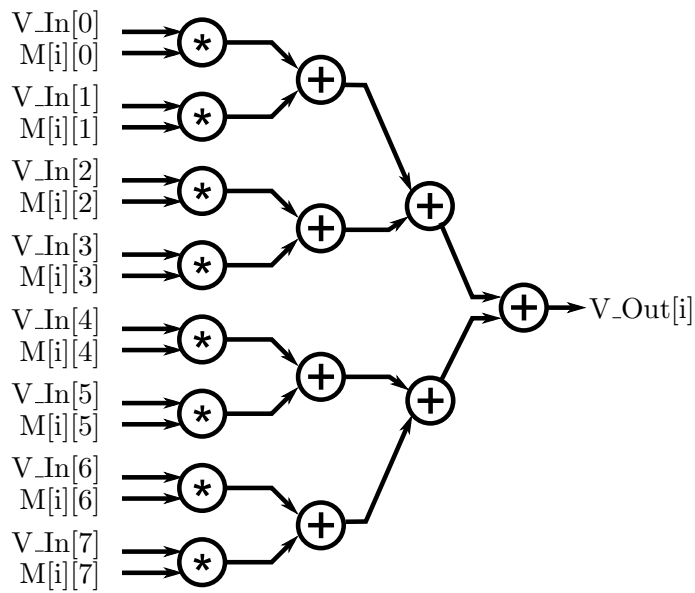%

\caption{A data flow graph of the expression resulting from the unrolled inner loop from Figure \ref{fig:matrix_vector_base_unroll_inner}.}\label{fig:matrix_vector_unroll_inner_dfg}
\end{figure}

It should be clear that the new expression replacing the inner loop has significant amount of parallelism. Each one of the multiplications can be performed simultaneously, and the summation can be performed using an adder tree. The data flow graph of this computation is shown in Figure \ref{fig:matrix_vector_unroll_inner_dfg}. 

\begin{figure}
\centering
\executeiffilenewer{dft_behavior1.svg}{images/dft_behavior1.pdf}%
{inkscape -z -D --file=dft_behavior1.svg %
--export-pdf=images/dft_behavior1.pdf --export-latex}%
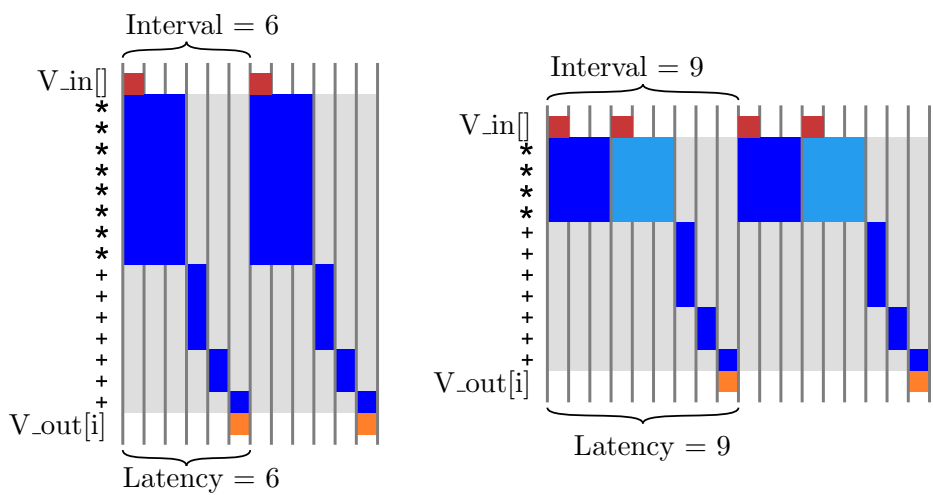%

\caption{Possible sequential implementations resulting from the unrolled inner loop from Figure \ref{fig:matrix_vector_base_unroll_inner}.}\label{fig:dft_behavior1}
\end{figure}

If we wish to achieve the minimum task latency for the expression resulting from the unrolled inner loop, all eight of the multiplications should be executed in parallel. Assuming that the multiplication has a latency of 3 cycles and addition has a latency of 1 cycle, then all of the \lstinline|V_In[j] * M[i][j]| operations are completed by the third time step. The summation of these eight intermediate results using an adder tree takes $\log 8 = 3$ cycles. Hence, the body of \lstinline|data_loop| now has a latency of 6 cycles for each iteration and requires 8 multipliers and 7 adders.  This behavior is shown in the left side of  Figure \ref{fig:dft_behavior1}.  Note that the adders could be reused across Cycle 4-6, which would reduce the number of adders to 4. However, adders are typically not shared when targeting FPGAs since an adder and a multiplexer require the same amount of FPGA resources (approximately 1 LUT per bit for a 2-input operator). 

If we are not willing to use 8 multipliers, there is an opportunity to reduce resource usage in exchange for increasing the number of cycles to execute the function. For example, using 4 multipliers would result in a latency of 6 cycles for the multiplication of the eight \lstinline|V_In[j] * M[i][j]| operations, and an overall latency of 9 cycles to finish the body of \lstinline|data_loop|. This behavior is shown in the right side of Figure \ref{fig:dft_behavior1}.  You could even use fewer multipliers at the cost of taking more cycles to complete the inner loop.

\begin{figure}
\centering
\executeiffilenewer{dft_behavior2.svg}{images/dft_behavior2.pdf}%
{inkscape -z -D --file=dft_behavior2.svg %
--export-pdf=images/dft_behavior2.pdf --export-latex}%
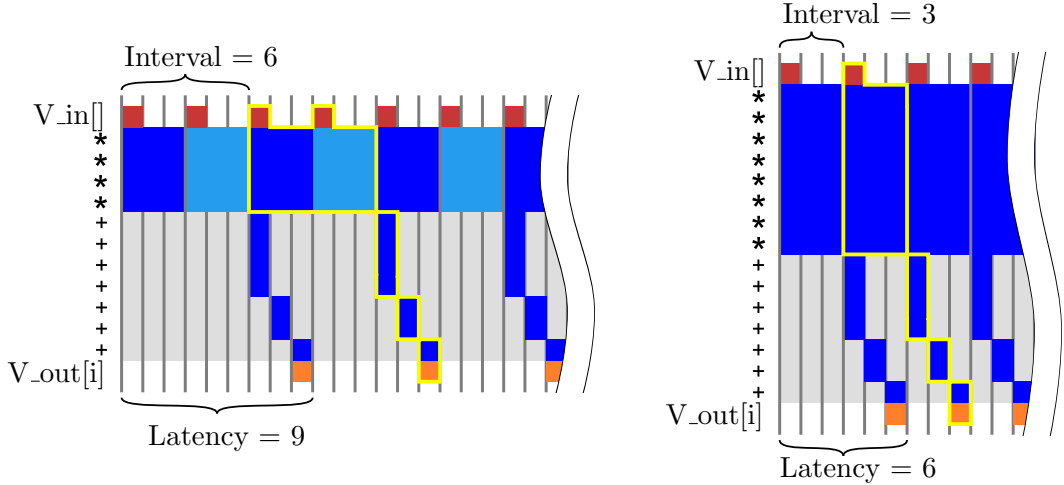%

\caption{Possible pipelined implementations resulting from the unrolled inner loop from Figure \ref{fig:matrix_vector_base_unroll_inner}.}\label{fig:dft_behavior2}
\end{figure}

Looking at Figure \ref{fig:dft_behavior1}, it is apparent that there are significant periods where the operators are not performing useful work, reducing the overall efficiency of the design.  It would be nice if we could reduce these periods.  In this case we can observe that each iteration of \lstinline|data_loop| is, in fact completely independent, which means that they can be executed concurrently.  Just as we unrolled \lstinline{dot_product_loop}, it's also possible to unroll \lstinline{data_loop} and perform all of the multiplications concurrently.  However, this would require a very large amount of FPGA resources.  A better choice is to enable each iteration of the loop to start as soon as possible, while the previous execution of the loop is still executing.  This process is called \gls{looppipelining} and is achieved in \VHLS using \lstinline|#pragma HLS pipeline|.  In most cases, loop pipelining reduces the interval of a loop to be reduced, but does not affect the latency.  Loop pipelined behavior of this design is shown in Figure \ref{fig:dft_behavior2}. 

Until now, we have only focused on operator latency.  It is common for functional units to be also be pipelined and most functional units in \VHLS are fully pipelined with an interval of one. Even though it might take 3 cycles for a single multiply operation to complete, a new multiply operation could start every clock cycle on a pipelined multiplier.  In this way, a single functional unit may be able to simultaneously execute many multiply operations at the same time.  For instance, a multiplier with a latency of 3 and an interval of 1 could be simultaneously executing three multiply operations. 

\begin{figure}
\centering
\executeiffilenewer{dft_behavior3.svg}{images/dft_behavior3.pdf}%
{inkscape -z -D --file=dft_behavior3.svg %
--export-pdf=images/dft_behavior3.pdf --export-latex}%
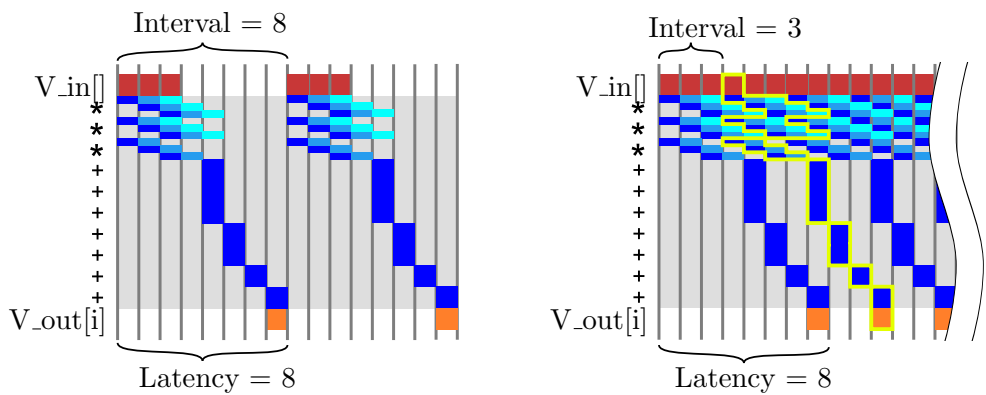%

\caption{Possible implementations resulting from the unrolled inner loop from Figure \ref{fig:matrix_vector_base_unroll_inner} using pipelined multipliers.}\label{fig:dft_behavior_pipelined}
\end{figure}

By taking advantage of pipelined multipliers, we can reduce the latency of the unrolled inner loop without adding additional operators.  One possible implementation using three pipelined multipliers is shown on the left in Figure \ref{fig:dft_behavior_pipelined}.  In this case, the multiplication operations can execute concurrently (because they have no data dependencies), while the addition operations cannot begin until the first multiplication has completed.  In the figure on the right, a pipelined version of this design is shown, with an interval of 3, which is similar to the results of \VHLS if \lstinline|#pragma  HLS pipeline II=3| is applied to the \lstinline|data_loop|.  In this case, not only are individual operations executing concurrently on the same operators, but those operations may come from different iterations of \lstinline|data_loop|.  

At this point you may have observed that pipelining is possible at different levels of hierarchy, including the operator level, loop level, and function level.  Furthermore, pipelining at different levels are largely independent!  We can use pipelined operators in a sequential loop, or we can use sequential operators to build a pipelined loop.  It's also possible to build pipelined implementations of large functions which can be shared in \VHLS just like primitive operators.  In the end, what matters most is how many operators are being instantiated, their individual costs, and how often they are used.



\section{Storage Tradeoffs and Array Partitioning}
\label{subsec:dft_array_partitioning}

Up until this point, we have assumed that the data in arrays (\lstinline|V_In[]|, \lstinline|M[][]|, and \lstinline|V_Out[]| are accessible at anytime.  In practice, however, the placement of the data plays a crucial role in the performance and resource usage. In most processor systems, the memory architecture is fixed and we can only adapt the program to attempt to best make use of the available memory hierarchy, taking care to minimize register spills and cache misses, for instance.  In HLS designs, we can also explore and leverage different memory structures and often try to find the memory structure that best matches a particular algorithm.  Typically large amounts of data are stored in off-chip memory, such as DRAM, flash, or even network-attached storage. However, data access times are typically long, on the order of tens to hundreds (or more) of cycles. Off-chip storage also relatively large amounts of energy to access, because large amounts of current must flow through long wires.  On-chip storage, in contrast can be accessed quickly and is much lower power.  I contrast it is more limited in the amount of data that can be stored.  A common pattern is to load data into on-chip memory in a block, where it can then be operated on repeatedly.  This is similar to the effect of caches in the memory hierarchy of general purpose CPUs.

The primary choices for on-chip storage on in embedded memories (e.g., block RAMs) or in flip-flops (FFs). These two options have their own tradeoffs. Flip-flop based memories allow for multiple reads at different addresses in a single clock.  It is also possible to read, modify, and write a Flip-flop based memory in a single clock cycle.  However, the number of FFs is typically limited to around 100 Kbytes, even in the largest devices. In practice, most flip-flop based memories should be much smaller in order to make effective use of other FPGA resources.  Block RAMs (BRAMs) offer higher capacity, on the order Mbytes of storage, at the cost of limited accessibility. For example, a single BRAM can store more than 1-4 Kbytes of data, but access to that data is limited to two different addresses each clock cycle. Furthermore, BRAMs are required to have a minimum amount of pipelining (i.e. the read operation must have a latency of at least one cycle).  Therefore, the fundamental tradeoff boils down to the required bandwidth versus the capacity. 

If throughput is the number one concern, all of the data would be stored in FFs. This would allow any element to be accessed as many times as it is needed each clock cycle. However, as the size of arrays grows large, this is not feasible.  In the case of matrix-vector multiplication, storing a 1024 by 1024 matrix of 32-bit integers would require about 4 MBytes of memory.   Even using BRAM, this storage would require about 1024 BRAM blocks, since each BRAM stores around 4KBytes.  On the other hand, using a single large BRAM-based memory means that we can only access two elements at a time.  This obviously prevents higher performance implementations, such as in Figure \ref{fig:matrix_vector_unroll_inner_dfg}, which require accessing multiple array elements each clock cycle (all eight elements of \lstinline|V_In[]| along with 8 elements of \lstinline|M[][]|).  In practice, most designs require larger arrays to be strategically divided into smaller BRAM memories, a process called \gls{arraypartitioning}.  Smaller arrays (often used for indexing into larger arrays) can be partitioned completely into individual scalar variables and mapped into FFs.  Matching pipelining choices and array partitioning to maximize the efficiency of operator usage and memory usage is an important aspect of design space exploration in HLS.

\begin{aside}
\VHLS will perform some array partitioning automatically, but as array partitioning tends to be rather design-specific it is often necessary to guide the tool for best results.  Global configuration of array partitioning is available in the \lstinline|config_array_partition| project option.  Individual arrays can be explicitly partitioned using the \lstinline|array_partition| directive. The directive \lstinline|array_partition complete| will split each element of an array into its own register, resulting in a flip-flop based memory.  As with many other directive-based optimizations, the same result can also be achieved by rewriting the code manually.  In general, it is preferred to use the tool directives since it avoids introducing bugs and keeps the code easy to maintain.
\end{aside}



\begin{figure}
\executeiffilenewer{matrix_vector_optimized.svg}{images/matrix_vector_optimized.pdf}%
{inkscape -z -D --file=matrix_vector_optimized.svg %
--export-pdf=images/matrix_vector_optimized.pdf --export-latex}%
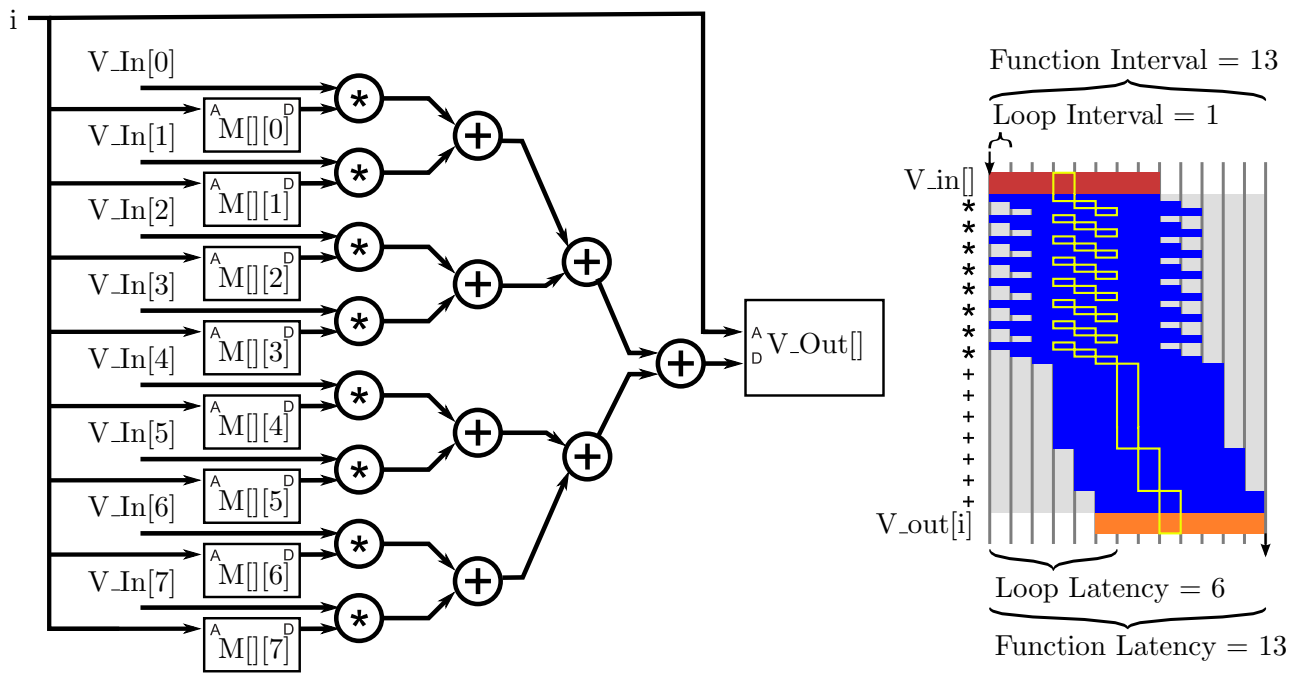%

\caption{Matrix-vector multiplication architecture with a particular choice of array partitioning and pipelining.  The pipelining registers have been elided and the behavior is shown at right.}
\label{fig:matrix_vector_optimized_behavior}
\end{figure}

Returning to the matrix-vector multiplication code in Figure \ref{fig:matrix_vector_base}, we can achieve a highly parallel implementation with the addition of only a few directives, as shown in Figure \ref{fig:matrix_vector_optimized}.  The resulting architecture is shown in Figure \ref{fig:matrix_vector_optimized_behavior}.  Notice that the inner \lstinline|j| loop is automatically unrolled by \VHLS and hence every use of \lstinline|j| is replaced with constants in the implementation.   This design demonstrates the most common use of array partitioning where the array dimensions that are partitioned (in this case, \lstinline|V_In[]| and the second dimension of \lstinline|M[][]|) are indexed with the constants (in this case the loop index \lstinline|j| of the unrolled loop).  This enables an architecture where multiplexers are not required to access the partitioned arrays.

\begin{figure}
\begin{lstlisting}
#define SIZE 8
typedef int BaseType;

void matrix_vector(BaseType M[SIZE][SIZE], BaseType V_In[SIZE], BaseType V_Out[SIZE]) {
#pragma HLS array_partition variable=M dim=2 complete
#pragma HLS array_partition variable=V_In complete
	BaseType i, j;
data_loop:
	for (i = 0; i < SIZE; i++) {
#pragma HLS pipeline II=1
		BaseType sum = 0;
	dot_product_loop:
		for (j = 0; j < SIZE; j++) {
			sum += V_In[j] * M[i][j];
		}
		V_Out[i] = sum;
	}
}
\end{lstlisting}
\caption{Matrix-vector multiplication with a particular choice of array partitioning and pipelining. }
\label{fig:matrix_vector_optimized}
\end{figure}

\begin{figure}
\executeiffilenewer{matrix_vector_partition_factor.svg}{images/matrix_vector_partition_factor.pdf}%
{inkscape -z -D --file=matrix_vector_partition_factor.svg %
--export-pdf=images/matrix_vector_partition_factor.pdf --export-latex}%
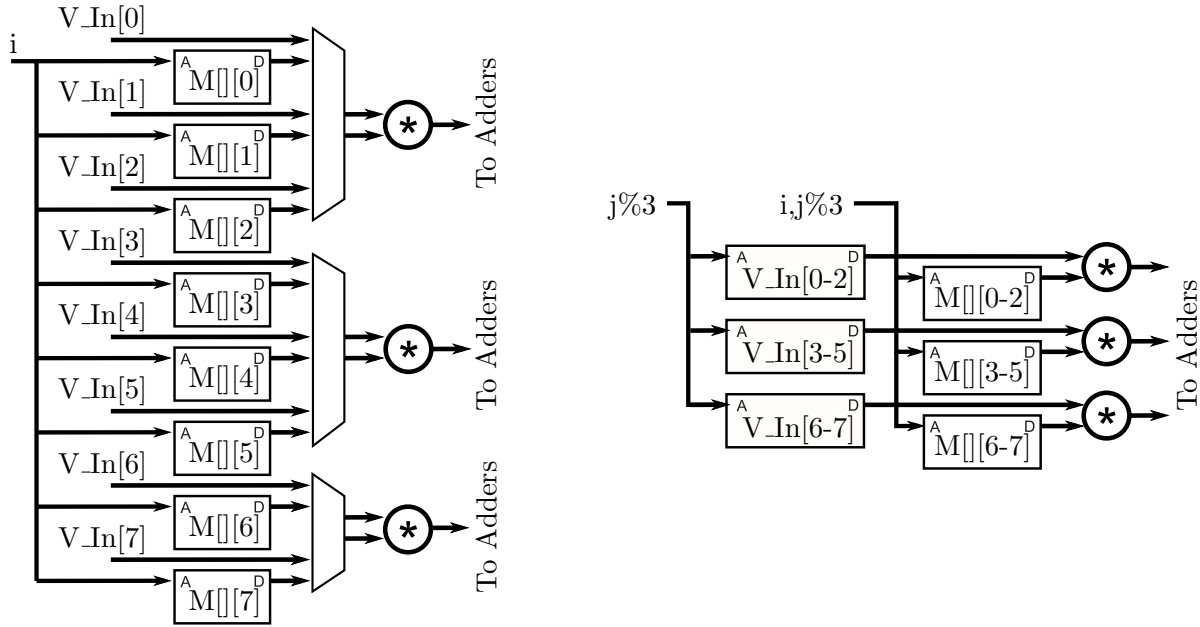%

\caption{Matrix-vector multiplication architectures at II=3 with a particular choices of array partitioning.  On the left, the arrays have been partitioned more than necessary, resulting in multiplexers.  On the right, the arrays are partitioned with factor=3. In this case, multiplexing has been reduced, but the \lstinline|j| loop index becomes a part of the address computations.}
\label{fig:matrix_vector_partition_factor}
\end{figure}

It's also possible to achieve other designs which use fewer multipliers and have lower performance.   For instance, in Figure \ref{fig:dft_behavior_pipelined}, these designs use only three multipliers, hence we need only need to read three elements of matrix \lstinline|M[][]| and vector \lstinline|V_in[]| each clock cycle. Completely partitioning these arrays would result in extra multiplexing as shown in Figure \ref{fig:matrix_vector_partition_factor}.  In actuality the arrays only need to be partitioned into three physical memories.  Again, this partitioning could be implemented manually by rewriting code or in \VHLS using the \lstinline|array_partition cyclic| directive.

\begin{aside}
Beginning with an array \lstinline|x| containing the values \[ 
\begin{bmatrix} 
1 & 2 & 3 & 4 & 5 & 6 & 7 & 8 & 9\\
\end{bmatrix}
\]
The directive \lstinline{array_partition variable=x factor=2 cyclic} on the array would split it into two arrays which are 
\[\begin{bmatrix}
1 & 3 & 5 & 7 & 9\\
\end{bmatrix} \text{and} 
\begin{bmatrix}
2 & 4 & 6 & 8 \\
\end{bmatrix}
\]
Similarly, the directive \lstinline{array_partition variable=x factor=2 block} would split it into two arrays
\[\begin{bmatrix}
1 & 2 & 3 & 4 & 5\\
\end{bmatrix} \text{and} 
\begin{bmatrix}
6 & 7 & 8 & 9 \\
\end{bmatrix}
\]
\end{aside}

\begin{exercise}
Study the effects of varying pipeline II and array partitioning on the performance and area. Plot the performance in terms of number of matrix vector multiply operations per second (throughput) versus the unroll and array partitioning factor. Plot the same trend for area (showing LUTs, FFs, DSP blocks, BRAMs). What is the general trend in both cases? Which design would you select? Why?
\end{exercise} 

Alternatively, similar results can be achieved by pipelining and applying \gls{partial_loop_unrolling} to the inner \lstinline|dot_product_loop|.   Figure \ref{fig:matrix_vector_unroll_inner2} shows the result of unrolling the inner loop of the matrix-vector multiplication code by a factor of 2. You can see that the loop bounds now increment by 2. Each loop iteration requires 2 elements of matrix \lstinline|M[][]| and vector \lstinline|V_in[]| each iteration and perform two multiplies instead of one. In this case after loop unrolling \VHLS can implement the operations in both expressions in parallel, corresponding to two iterations of the original loop. Note that without appropriate array partitioning, unrolling the inner loop may offer no increase in performance, as the number of concurrent read operations is limited by the number of ports to the memory.  In this case, we can store the data from the even columns in one BRAM and the data from the odd columns in the other. This is due to the fact that the unrolled loop is always performing one even iteration and one odd iteration. 

\begin{aside}
The HLS tool can automatically unroll loops using the \texttt{unroll} directive. The directive takes a \texttt{factor} argument which is a positive integer denoting the number of times that the loop body should be unrolled. 
\end{aside}

\begin{figure}
\begin{lstlisting}
#define SIZE 8
typedef int BaseType;

void matrix_vector(BaseType M[SIZE][SIZE], BaseType V_In[SIZE], BaseType V_Out[SIZE]) {
#pragma HLS array_partition variable=M dim=2 cyclic factor=2
#pragma HLS array_partition variable=V_In cyclic factor=2
	BaseType i, j;
data_loop:
	for (i = 0; i < SIZE; i++) {
		BaseType sum = 0;
	dot_product_loop:
		for (j = 0; j < SIZE; j+=2) {
#pragma HLS pipeline II=1
			sum += V_In[j] * M[i][j];
			sum += V_In[j+1] * M[i][j+1];
		}
		V_Out[i] = sum;
	}
}
\end{lstlisting}
\caption{The inner loop of matrix-vector multiply manually unrolled by a factor of two. }
\label{fig:matrix_vector_unroll_inner2}
\end{figure}

\begin{exercise}
Manually divide \lstinline|M[][]| and vector \lstinline|V_in[]| into separate arrays in the same manner as the directive \lstinline|array_partition cyclic factor=2|. How do you have to modify the code in order to change the access patterns? Now manually unroll the loop by a factor of two. How do the performance results vary between the original code (no array partitioning and no unrolling), only performing array partitioning, and performing array partitioning and loop unrolling? Finally, use the directives to perform array partitioning and loop unrolling. How do those results compare to your manual results?
\end{exercise} 

In this code, we see that array partitioning often goes hand in hand with our choices of pipelining. Array partitioning by a factor of 2 enables an increase in performance by a factor of 2, which can be achieved either by partially unrolling the inner loop by a factor of 2 or by reducing the II of the outer loop by a factor of 2.  Increasing performance requires a corresponding amount of array partitioning.  In the case of matrix vector multiplication, this relationship is relatively straightforward since there is only one access to each variable in the inner loop.  In other code, the relationship might be more complicated.  Regardless, the goal of a designer is usually to ensure that the instantiated FPGA resources are used efficiently.  Increasing performance by a factor of 2 should use approximately twice as many resources.  Decreasing performance by a factor of 2 should use approximately half as many resources.

\begin{exercise}
Study the effects of loop unrolling and array partitioning on the performance and area. Plot the performance in terms of number of matrix vector multiply operations per second (throughput) versus the unroll and array partitioning factor. Plot the same trend for area (showing LUTs, FFs, DSP blocks, BRAMs). What is the general trend in both cases? Which design would you select? Why?
\end{exercise}

\section{Baseline Implementation}
\label{subsec:dft_implementation}

We just discussed some optimizations for matrix-vector multiplication. This is a core computation in performing a \gls{dft}. However, there are some additionally intricacies that we must consider to move from the matrix-vector multiplication in the previous section to a functionally complete \gls{dft} hardware implementation. We move our focus to the \gls{dft} in this section, and describe how to optimize it to make it execute most efficiently.  

One significant change that is required is that we must be able to handle complex numbers.  As noted in Section \ref{sec:DFTbackground} because the elements of the $S$ matrix are complex numbers, the \gls{dft} of a real-valued signal is almost always a complex-valued signal.   It is also common to perform the \gls{dft} of a complex-valued signal, to produce a complex-valued result.  Additionally, we need to handle fractional or possibly floating point data, rather than integers.  This can increase the cost of the implementation, particularly if floating point operations need to be performed.  In addition, floating point operators, particularly addition, have much larger latency than integer addition.  This can make it more difficult to achieve II=1 loops.  A second change is that we'd like to be able to scale our design up to large input vector sizes, perhaps N=1024 input samples.  Unfortunately, if we directly use matrix-vector multiplication, then we must store the entire $S$ matrix.  Since this matrix is the square of the input size, it becomes prohibitive to store for large input sizes.  In the following sections, we'll discuss techniques to address both of these complexities. 

As is typical when creating a hardware implementation using high level synthesis, we start with a straightforward or naive implementation. This provides us with a baseline code that we can insure has the correct functionality. Typically, this code runs in a very sequential manner; it is not highly optimized and therefore may not meet the desired performance metrics. However, it is a necessary step to insure that the designer understand the functionality of the algorithm, and it serves as starting point for future optimizations.

\begin{figure}
\begin{lstlisting}
#include <math.h>					//Required for cos and sin functions
typedef double IN_TYPE;		// Data type for the input signal
typedef double TEMP_TYPE; // Data type for the temporary variables
#define N 256							// DFT Size

void dft(IN_TYPE sample_real[N], IN_TYPE sample_imag[N]) {
	int i, j;
	TEMP_TYPE w;
	TEMP_TYPE c, s;

	// Temporary arrays to hold the intermediate frequency domain results
	TEMP_TYPE temp_real[N];
	TEMP_TYPE temp_imag[N];

	// Calculate each frequency domain sample iteratively
	for (i = 0; i < N; i += 1) {
		temp_real[i] = 0;
		temp_imag[i] = 0;

		// (2 * pi * i)/N
		w = (2.0 * 3.141592653589  / N) * (TEMP_TYPE)i;

		// Calculate the jth frequency sample sequentially
		for (j = 0; j < N; j += 1) {
			// Utilize HLS tool to calculate sine and cosine values
			c = cos(j * w);
			s = sin(j * w);

			// Multiply the current phasor with the appropriate input sample and keep
			// running sum
			temp_real[i] += (sample_real[j] * c - sample_imag[j] * s);
			temp_imag[i] += (sample_real[j] * s + sample_imag[j] * c);
		}
	}

	// Perform an inplace DFT, i.e., copy result into the input arrays
	for (i = 0; i < N; i += 1) {
		sample_real[i] = temp_real[i];
		sample_imag[i] = temp_imag[i];
	}
}
\end{lstlisting}
\caption{Baseline code for the \gls{dft}.}
\label{fig:dft_code}
\end{figure}

Figure \ref{fig:dft_code} shows a baseline implementation of the \gls{dft}. This uses a doubly nested \lstinline|for| loop. The inner loop multiplies one row of the $S$ matrix with the input signal. Instead of reading the $S$ matrix as an input, this code computes an element of $S$ in each each iteration of the inner loop, based on the current loop indices.  This phasor is converted to Cartesian coordinates (a real part and an imaginary part) using the \lstinline|cos()| and \lstinline|sin()| functions. The code then performs a complex multiplication of the phasor with the appropriate sample of the input signal and accumulates the result. After $N$ iterations of this inner loop, one for each column of $S$, one frequency domain sample is calculated. The outer loop also iterates $N$ times, once for each row of $S$.  As a result, the code computes an expression for \lstinline|w| $N$ times, but computes the \lstinline|cos()| and \lstinline|sin()| functions and a complex multiply-add $N^2$ times.

\begin{figure}
\centering
\executeiffilenewer{dft_behavior_baseline.svg}{images/dft_behavior_baseline.pdf}%
{inkscape -z -D --file=dft_behavior_baseline.svg %
--export-pdf=images/dft_behavior_baseline.pdf --export-latex}%
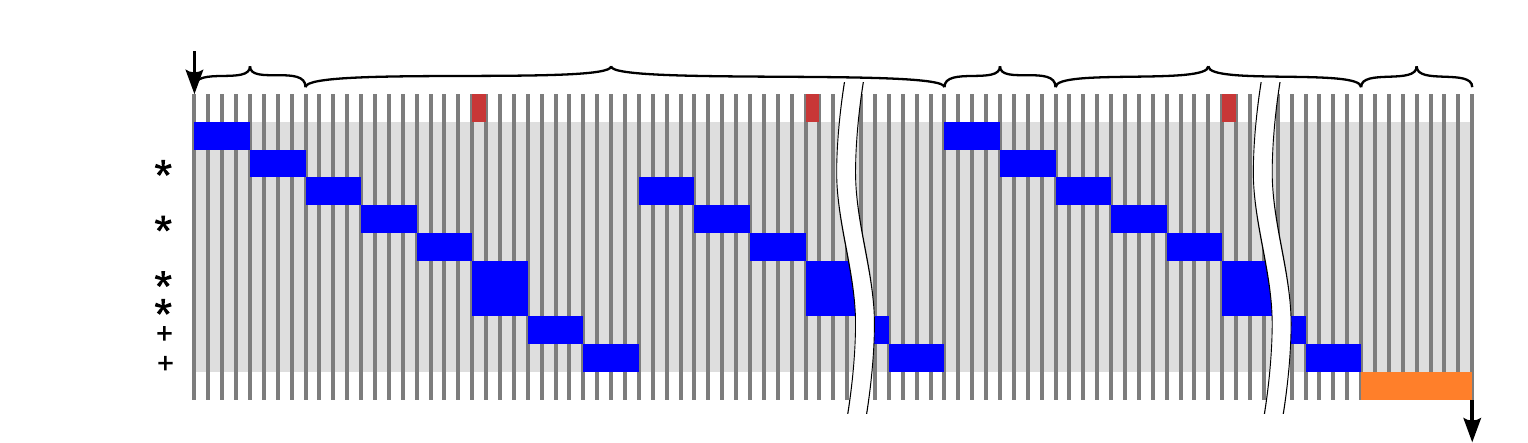%

\includegraphics[width= 0.7 \textwidth]{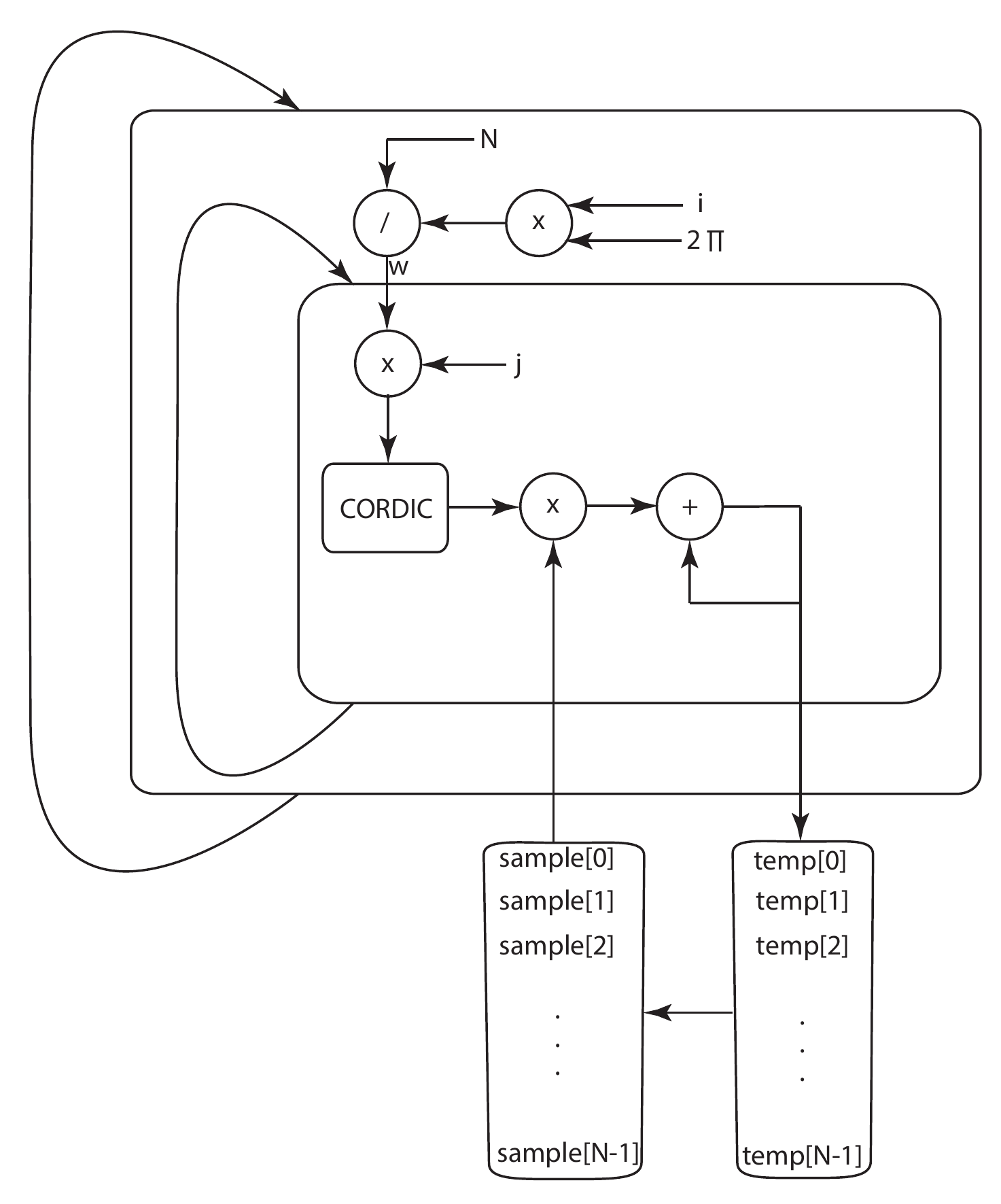}
\caption{ A high level architectural diagram of the \gls{dft} as specified in the code from Figure \ref{fig:dft_code}. This is not a comprehensive view of the architecture, e.g., it is missing components related to updating the loop counters \lstinline|i| and \lstinline|j|. It is meant to provide an approximate notion of how this architecture will be synthesized.  Here we've assumed that floating point operators take 4 clock cycles.}
\label{fig:dft_sequential_arch}
\end{figure}

This code uses a function call to calculate \lstinline|cos()| and \lstinline|sin()| values. \VHLS is capable of synthesizing these functions using its built-in math library. There are several possible algorithms \cite{detrey07hotbm} for implementing trigonometric functions including CORDIC, covered in Chapter \ref{chapter:cordic}.  However, generating precise results for these functions can be expensive.  There are several possibilities for eliminating these function calls, since the inputs aren't arbitrary.  We will discuss these tradeoffs in more detail later.    A sequential implementation of this code is show in Figure \ref{fig:dft_sequential_arch}. 


\begin{exercise}
What changes would this code require if you were to use a CORDIC that you designed, for example, from Chapter \ref{chapter:cordic}? Would changing the accuracy of the CORDIC core make the \gls{dft} hardware resource usage change? How would it effect the performance? 
\end{exercise}

\begin{exercise}
Implement the baseline code for the \gls{dft} using HLS.  Looking at the reports, what is the relative cost of the implementation of the trignometric functions, compared to multiplication and addition?  Which operations does it make more sense to try to optimize?  What performance can be achieved by pipelining the inner loop?
\end{exercise}

\section{\gls{dft} optimization}
\label{subsec:dft_optimization}

\begin{figure}
\centering
\executeiffilenewer{dft_recurrence_behavior.svg}{images/dft_recurrence_behavior.pdf}%
{inkscape -z -D --file=dft_recurrence_behavior.svg %
--export-pdf=images/dft_recurrence_behavior.pdf --export-latex}%
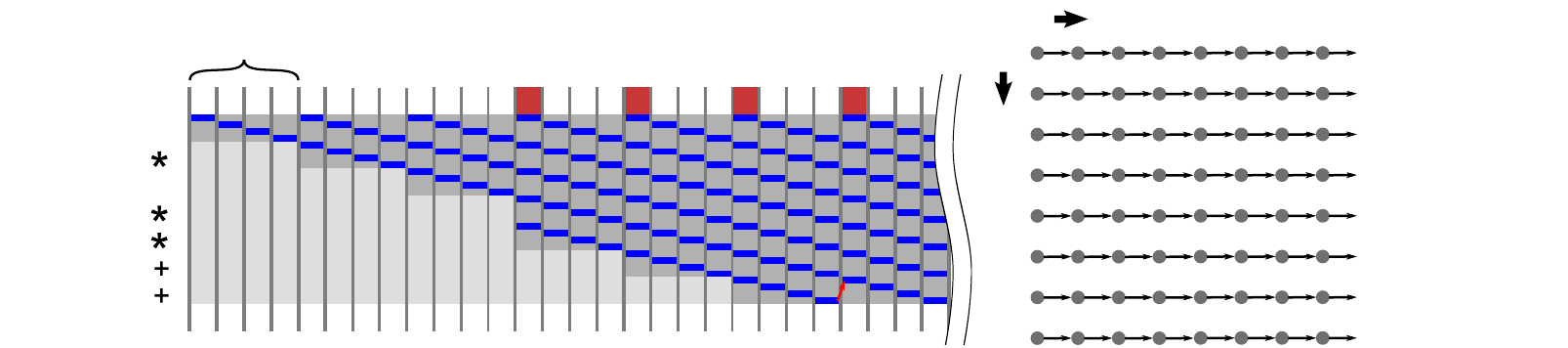%

\caption{Pipelined version of the behavior in Figure \ref{fig:dft_sequential_arch}.  In this case, the initiation interval of the loop is limited to 4, since each floating point addition takes 4 clock cycles to complete and the result is required before the next loop iteration begins (the dependence shown in red).  The dependencies  for all iterations are summarized in the diagram on the right.}
\label{fig:dft_recurrence_behavior}
\end{figure}

The baseline \gls{dft} implementation of the previous section uses relatively high precision \lstinline|double| datatypes.  Implementing floating point operations is typically very expensive and requires many pipeline stages, particularly for double precision.  We can see in Figure \ref{fig:dft_sequential_arch} that this significantly affects the performance of the loop.  With pipelining, the affect of these high-latency operations is less critical, since multiple executions of the loop can execute concurrently.  The exception in this code are the \lstinline|temp_real[]| and \lstinline|temp_imag[]| variables, which are used to accumulate the result.  This accumulation is a \gls{recurrence} and limits the achievable II in this design when pipelining the inner loop.   This operator dependence is shown in Figure \ref{fig:dft_recurrence_behavior}.

One possible solution is to reduce the precision of the computation.  This is always a valuable technique when it can be applied, since it reduces the resources required for each operation, the memory required to store any values, and often reduces the latency of operations as well.  For instance we could use the 32-bit \lstinline|float| type or the 16-bit \lstinline|half| types rather than double.  Many signal processing systems avoid floating point data types entirely and use fixed point data types\ref{sec:number_representation}.  For commonly used integer and fixed-point precisions, each addition can be completed in a single cycle, enabling the loop to be pipelined at II=1.

\begin{exercise}
What happens to the synthesis result of the code in Figure \ref{fig:dft_code} if you change all of the data types from $double$ to $float$? Or from $double$ to $half$? Or to a fixed point value? How does this change the performance (interval and latency) and the resource usage? Does it change the values of the output frequency domain samples?
\end{exercise}

A more general solution to achieve II=1 with floating point accumulations is to process the data in a different order.  Looking at Figure \ref{fig:dft_recurrence_behavior} we see that the recurrence exists (represented by the arrow) because the \lstinline|j| loop is the inner loop.  If the inner loop were the \lstinline|i| loop instead, then we wouldn't need the result of the accumulation before the next iteration starts.  We can achieve this in the code by interchanging the order of the two loops.  This optimization is often called \gls{loopinterchange} or pipeline-interleaved processing\cite{lee87sdfArchitecture}.  In this case, it may not be obvious that we can rearrange the loops because of the extra code inside the outer \lstinline|i| loop.  Fortunately, the $S$ matrix is diagonally symmetric, and hence \lstinline|i| and \lstinline|j| can be exchanged in the computation of \lstinline|w|.  The result is that we can now achieve an II of 1 for the inner loop.  The tradeoff is additional storage for the \lstinline|temp_real| and \lstinline|temp_imag| arrays to store the intermediate values until they are needed again.

\begin{exercise}
Reorder the loops of the code in Figure \ref{fig:dft_code} and show that you can pipeline the inner loop with an II of 1.
\end{exercise}

There are other optimizations that we can apply based on the structure of the $S$ matrix in the \gls{dft} to eliminate the trigonometric operations entirely.  Recall that the complex vectors for each element of the $S$ matrix are calculated based upon a fixed integer rotation around the unit circle.  Row $S[0][]$ of the $S$ matrix corresponds to zero rotations around the unit circle, row $S[1][]$ corresponds to a single rotation, and the following rows correspond to more rotations around the unit circle. It turns out that the vectors corresponding to the second row $S[1][]$, which is one rotation around the unit circle (divided into $360/8 = 45^{\circ}$ individual rotations), cover all of the vectors from every other row.  This can be visually confirmed by studying Figure \ref{fig:dft-visualization}.   Thus it is possible to store only the sine and cosine values from this one rotation, and then index into this memory to calculate the requisite values for the corresponding rows. This requires only $2 \times N = \mathcal{O}(N)$ elements of storage. This results in a $\mathcal{O}(N)$ reduction in storage, which for the 1024 point \gls{dft} would reduce the memory storage requirements to $1024 \times 2$ entries. Assuming 32 bit fixed or floating point values, this would require only 8 KB of on-chip memory. Obviously, this is a significant reduction compared to storing the entire $S$ matrix explicitly.   We denote this one dimensional storage of the matrix $S$ as $S'$ where

\begin{equation}
\label{eq:1DS}
S' = S[1][\cdot] = (1 \hspace{4mm} s \hspace{4mm} s^2 \hspace{4mm} \cdots \hspace{4mm} s^{N-1})
\end{equation}

\begin{exercise}
Derive a formula for the access pattern for the 1D array $S'$ given as input the row number $i$ and column element $j$ corresponding to the array $S$. That is, how do we index into the 1D $S$ array to access element $S(i,j)$ from the 2D $S$ array.
\end{exercise}

To increase performance further we can apply techniques that are very similar to the matrix-vector multiply.  Previously, we observed that increasing performance of matrix-vector multiply required partitioning the \lstinline|M[][]| array.  Unfortunately, representing the $S$ matrix using the $S'$ means that there is no longer an effective way to partition $S'$ to increase the amount of data that we can read on each clock. Every odd row and column of $S$ includes every element of $S'$.  As a result, there is no way to partition the values of $S'$ like were able to do with $S$.  The only way to increase the number of read ports from the memory that stores $S'$ is to replicate the storage.  Fortunately, unlike with a memory that must be read and written, it is relatively easy to replicate the storage for an array that is only read.  In fact, \VHLS will perform this optimization automatically when instantiates a \gls{rom} for an array which is initialized and then never modified.  One advantage of this capability is that we can simply move the $sin()$ and $cos()$ calls into an array initialization.  In most cases, if this code is at the beginning of a function and only initializes the array, then \VHLS is able to optimize away the trigonometric computation entirely and compute the contents of the ROM automatically.

\begin{exercise}
Devise an architecture that utilizes $S'$ -- the 1D version of the $S$ matrix. How does this affect the required storage space? Does this change the logic utilization compared to an implementation using the 2D $S$ matrix? 
\end{exercise}

In order to effectively optimize the design, we must consider every part of the code. The performance can only as good as the ``weakest link'' meaning that if there is a bottleneck the performance will take a significant hit. The current version of the \gls{dft} function performs an in-place operation on the input and output data, i.e., it stores the results in the same array as the input data. The input array arguments \lstinline|sample_real| and \lstinline|sample_imag| effectively act as a memory port. That is, you can think of these arguments arrays as stored in the same memory location. Thus, we can only grab one piece of data from each of these arrays on any given cycle. This can create a bottleneck in terms of parallelizing the multiplication and summation operations within the function. This also explains the reason why we must store all of the output results in a temporary array, and then copy all of those results into the ``sample'' arrays at the end of the function. We would not have to do this if we did not perform an in-place operation. 

\begin{exercise}
Modify the \gls{dft} function interface so that the input and outputs are stored in separate arrays. How does this effect the optimizations that you can perform? How does it change the performance? What about the area results?
\end{exercise}


\section{Conclusion}
In this chapter, we looked at the hardware implementation and optimization of the Discrete Fourier Transform (\gls{dft}). The \gls{dft} is a fundamental operation in digital signal processing. It takes a signal sampled in the time domain and converts it into the frequency domain. At the beginning of this chapter, we describe the mathematical background for the \gls{dft}. This is important for understanding the optimizations done in the next chapter (\gls{fft}). The remainder of the chapter was focused on specifying and optimizing the \gls{dft} for an efficient implementation on an FPGA.  

At its core, the \gls{dft} performs a matrix-vector multiplication. Thus, we spend some time initially to describe instruction level optimizations on a simplified code performing matrix-vector multiplication. These instruction level optimizations are done by the HLS tool. We use this as an opportunity to shed some light into the process that the HLS tool performs in the hopes that it will provide some better intuition about the results the tool outputs.

After that, we provide an functionally correct implementation for the \gls{dft}. We discuss a number of optimizations that can be done to improve the performance. In particular, we focus on the problem of dividing the coefficient array into different memories in order to increase the throughput. Array partitioning optimization are often key to achieving the highest performing architectures.


\chapter{Fast Fourier Transform}
\glsresetall
\label{chapter:fft}

Performing the \gls{dft} directly using matrix-vector multiply requires $\mathcal{O}(n^2)$ multiply and add operations, for an input signal with $n$ samples.  It is possible to reduce the complexity by exploiting the structure of the constant coefficients in the matrix.  This $S$ matrix encodes the coefficients of the \gls{dft}; each row of this matrix corresponds to a fixed number of rotations around the complex unit circle (please refer to Chapter \ref{sec:dft_background} for more detailed information). These values have a significant amount of redundancy, and that can be exploited to reduce the complexity of the algorithm. 

\begin{aside}
The 'Big O' notation used here describes the general order of complexity of an algorithm based on the size of the input data.  For a complete description of Big O notation and its use in analyzing algorithms, see \cite{CLR}.
\end{aside}

The \gls{fft} uses a divide-and-conquer approach based on the symmetry of the $S$ matrix.  The \gls{fft} was made popular by the Cooley-Tukey algorithm \cite{cooley65}, which requires $\mathcal{O}(n \log n)$ operations to compute the same function as the \gls{dft}.  This can provide a substantial speedup, especially when performing the Fourier transform on large signals. 

\begin{aside}
The divide-and-conquer approach to computing the \gls{dft} was initially developed by Karl Friedrich Gauss in the early 19th century. However, since Gauss' work on this was not published during his lifetime and only appeared in a collected works after his death, it was relegated to obscurity.  Heideman et al.~\cite{heideman84} provide a nice background on the history of the \gls{fft}.
\end{aside}

The focus of this chapter is to provide the reader with a good understanding of the \gls{fft} algorithm since that is an important part of creating an optimized hardware design. Thus, we start by giving a mathematical treatment of the \gls{fft}. This discussion focuses on small \gls{fft} sizes to give some basic intuition on the core ideas. After that, we focus on different hardware implementation strategies.

\section{Background}

The \gls{fft} brings about a reduction in complexity by taking advantage of symmetries in the \gls{dft} calculation. To better understand how to do this, let us look at \gls{dft} with a small number of points, starting with the 2 point \gls{dft}.  Recall that the \gls{dft} performs a matrix vector multiplication, i.e., $G[] = S[][] \cdot g[]$, where $g[]$ is the input data, $G[]$ is the frequency domain output data, and $S[][]$ are the \gls{dft} coefficients. We follow the same notation for the coefficient matrix, and the input and output vectors as described in Chapter \ref{sec:dft_background}.

For a 2 point \gls{dft}, the values of $S$ are:
\begin{equation}
S =
 \begin{bmatrix}
  W^{0 0}_2 & W^{0 1}_2  \\
  W^{1 0}_2 & W^{1 1}_2 \\
 \end{bmatrix}
 \end{equation}
Here we use the notation $W = e^{-j 2 \pi}$. The superscript on $W$ denotes values that are added to the numerator and the subscript on the $W$ indicates those values added in the denominator of the complex exponential. For example, $W^{2 3}_4 = e^{\frac{-j 2 \pi \cdot 2 \cdot 3}{4}}$. This is similar to the $s$ value used in the \gls{dft} discussion (Chapter \ref{sec:dft_background}) where $s = e^{\frac{-j 2 \pi}{N}}$. The relationship between $s$ and $W$ is $s = W_N$.

\begin{aside}
The $e^{-j 2 \pi}$ or $W$ terms are often called \term{twiddle factors}. This term has its origin in the 1966 paper by Gentleman and Sande \cite{gentleman1966fast}.
\end{aside}

\begin{equation}
\begin{bmatrix} 
G[0] \\ 
G[1] \\
\end{bmatrix} = 
 \begin{bmatrix}
  W^{0 0}_2 & W^{0 1}_2  \\
  W^{1 0}_2 & W^{1 1}_2 \\
 \end{bmatrix}
 \cdot
  \begin{bmatrix}
  g[0] \\
  g[1]\\
\end{bmatrix}
\end{equation}

Expanding the two equations for a 2 point \gls{dft} gives us:
\begin{equation}
\begin{array} {lll} 
G[0] & = & g[0] \cdot e^{\frac{-j 2 \pi \cdot 0 \cdot 0}{2}} + g[1] \cdot e^{\frac{-j 2 \pi \cdot 0 \cdot 1}{2}} \\
 & = & g[0] + g[1] \\
\end{array}
\label{eq:2ptlower}
\end{equation} due to the fact that since  $e^{0}  =  1$. The second frequency term
\begin{equation}
\begin{array} {lll} 
G[1] & = & g[0] \cdot e^{\frac{-j 2 \pi \cdot 1 \cdot 0}{2}} + g[1] \cdot e^{\frac{-j 2 \pi \cdot 1 \cdot 1}{2}} \\
 & = & g[0] - g[1] \\
\end{array}
\label{eq:2pthigher}
\end{equation} since  $e^{\frac{-j 2 \pi \cdot 1 \cdot 1}{2}}  = e^{-j \pi } = -1$.

\begin{figure}
\centering
\includegraphics[width= 0.8 \textwidth]{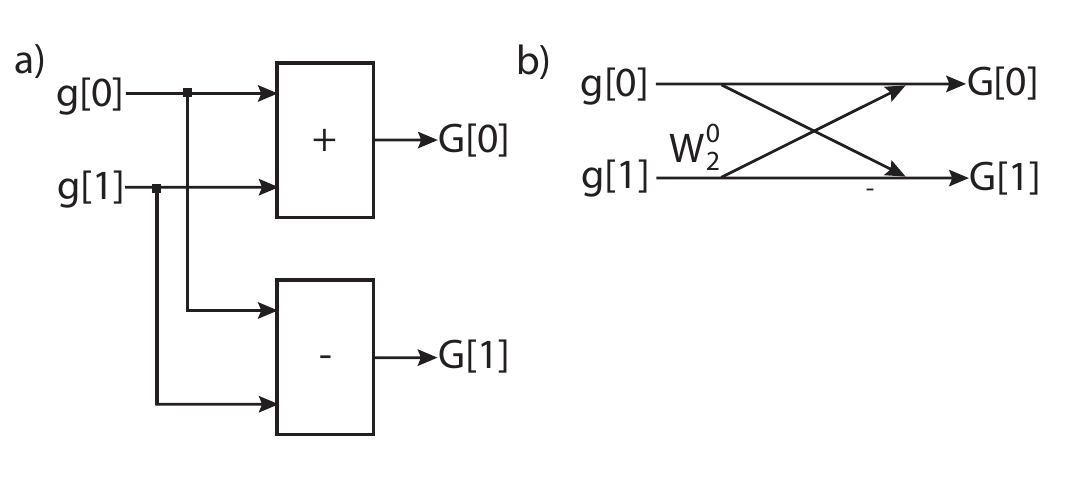}
\caption{Part a) is a data flow graph for a 2 point \gls{dft}/\gls{fft}. Part b) shows the same computation, but viewed as a butterfly structure. This is a common representation for the computation of an \gls{fft} in the digital signal processing domain.}
\label{fig:2pointFFT}
\end{figure}

Figure \ref{fig:2pointFFT} provides two different representations for this computation.  Part a) is the data flow graph for the 2 point \gls{dft}. It is the familiar view that we have used to represent computation throughout this book. Part b) shows a butterfly structure for the same computation. This is a typically structure used in digital signal processing, in particular, to represent the computations in an \gls{fft}. 

The butterfly structure is a more compact representation that is useful to represent large data flow graphs. When two lines come together this indicates an addition operation. Any label on the line itself indicates a multiplication of that label by the value on that line. There are two labels in this figure. The `$-$' sign on the bottom horizontal line indicates that this value should be negated. This followed by the addition denoted by the two lines intersecting is the same as subtraction. The second label is $W^0_2$. While this is a multiplication is unnecessary (since $W^0_2 = 1$ this means it is multiplying by the value `$1$'), we show it here since it is a common structure that appears in higher point \gls{fft}s.

Now let us consider a slightly larger \gls{dft} -- a 4 point \gls{dft}, i.e., one that has 4 inputs, 4 outputs, and a $4 \times 4$ $S$ matrix. The values of $S$ for a 4 point \gls{dft} are:
\begin{equation}
S =
 \begin{bmatrix}
  W^{0 0}_4 & W^{0 1}_4 & W^{0 2}_4 & W^{0 3}_4 \\
  W^{1 0}_4 & W^{1 1}_4 & W^{1 2}_4 & W^{1 3}_4 \\
  W^{2 0}_4 & W^{2 1}_4 & W^{2 2}_4 & W^{2 3}_4 \\
  W^{3 0}_4 & W^{3 1}_4 & W^{3 2}_4 & W^{3 3}_4 \\
 \end{bmatrix}
 \end{equation}
And the \gls{dft} equation to compute the frequency output terms are:
\begin{equation}
\begin{bmatrix} 
G[0] \\
G[1] \\
G[2] \\
G[3] \\
\end{bmatrix} = 
 \begin{bmatrix}
  W^{0 0}_4 & W^{0 1}_4 & W^{0 2}_4 & W^{0 3}_4 \\
  W^{1 0}_4 & W^{1 1}_4 & W^{1 2}_4 & W^{1 3}_4 \\
  W^{2 0}_4 & W^{2 1}_4 & W^{2 2}_4 & W^{2 3}_4 \\
  W^{3 0}_4 & W^{3 1}_4 & W^{3 2}_4 & W^{3 3}_4 \\
 \end{bmatrix}
 \cdot
  \begin{bmatrix}
  g[0] \\
  g[1]\\
  g[2]\\
  g[3]\\
\end{bmatrix}
\end{equation}

Now we write out the equations for each of the frequency domain values in $G[]$ one-by-one. The equation for G[0] is:
\begin{equation}
\begin{array} {lll} 
G[0] & = & g[0] \cdot e^{\frac{-j 2 \pi \cdot 0 \cdot 0}{4}} + g[1] \cdot e^{\frac{-j 2 \pi \cdot 0 \cdot 1}{4}} + g[2] \cdot e^{\frac{-j 2 \pi \cdot 0 \cdot 2}{4}} + g[3] \cdot e^{\frac{-j 2 \pi \cdot 0 \cdot 3}{4}}\\
 & = & g[0] + g[1] + g[2] + g[3] \\
\end{array}
\end{equation} since $e^0 = 1$. 

The equation for $G[1]$ is:
\begin{equation}
\begin{array} {lll} 
G[1] & = & g[0] \cdot e^{\frac{-j 2 \pi \cdot 1 \cdot 0}{4}} + g[1] \cdot e^{\frac{-j 2 \pi \cdot 1 \cdot 1}{4}} + g[2] \cdot e^{\frac{-j 2 \pi \cdot 1 \cdot 2}{4}} + g[3] \cdot e^{\frac{-j 2 \pi \cdot 1 \cdot 3}{4}}\\
 & = & g[0] + g[1] \cdot e^{\frac{-j 2 \pi}{4}} + g[2] \cdot e^{\frac{-j 4 \pi}{4}} + g[3] \cdot e^{\frac{-j 6 \pi}{4}}\\
 & = & g[0] + g[1] \cdot e^{\frac{-j 2 \pi}{4}} + g[2] \cdot e^{-j \pi}   + g[3] \cdot e^{\frac{-j 2 \pi}{4}} e^{-j \pi} \\
 & = & g[0] + g[1] \cdot e^{\frac{-j 2 \pi}{4}} - g[2] - g[3] \cdot e^{\frac{-j 2 \pi}{4}}\\
\end{array} 
\end{equation} The reductions were done based upon the fact that $e^{-j \pi} = -1$. 

The equation for $G[2]$ is:
\begin{equation}
\begin{array} {lll} 
G[2] & = & g[0] \cdot e^{\frac{-j 2 \pi \cdot 2 \cdot 0}{4}} + g[1] \cdot e^{\frac{-j 2 \pi \cdot 2 \cdot 1}{4}} + g[2] \cdot e^{\frac{-j 2 \pi \cdot 2 \cdot 2}{4}} + g[3] \cdot e^{\frac{-j 2 \pi \cdot 2 \cdot 3}{4}}\\
 & = & g[0] + g[1] \cdot e^{\frac{-j 4 \pi}{4}} + g[2] \cdot e^{\frac{-j 8 \pi}{4}} + g[3] \cdot e^{\frac{-j 12 \pi}{4}}\\
 & = & g[0] - g[1]  + g[2] -  g[3] \\
\end{array} 
\end{equation} The reductions were done by simplifications based upon rotations. E.g., $e^{\frac{-j 8 \pi}{4}} = 1$ and $e^{\frac{-12 j \pi}{4}} = -1$ since in both cases use the fact that $e^{-j 2\pi}$ is equal to $1$. In other words, any complex exponential with a rotation by $2 \pi$ is equal.

Finally, the equation for $G[3]$ is:
\begin{equation}
\begin{array} {lll} 
G[3] & = & g[0] \cdot e^{\frac{-j 2 \pi \cdot 3 \cdot 0}{4}} + g[1] \cdot e^{\frac{-j 2 \pi \cdot 3 \cdot 1}{4}} + g[2] \cdot e^{\frac{-j 2 \pi \cdot 3 \cdot 2}{4}} + g[3] \cdot e^{\frac{-j 2 \pi \cdot 3 \cdot 3}{4}}\\
 & = & g[0] + g[1] \cdot e^{\frac{-j  6 \pi}{4}} + g[2] \cdot e^{\frac{-j  12 \pi}{4}} + g[3] \cdot e^{\frac{-j 18 \pi}{4}}\\
 & = & g[0] + g[1] \cdot e^{\frac{-j 6  \pi }{4}}  - g[2] +  g[3] \cdot e^{\frac{-j 10  \pi}{4}}\\
  & = & g[0] + g[1] \cdot e^{\frac{-j 6  \pi }{4}}  - g[2] -  g[3] \cdot e^{\frac{-j 6  \pi}{4}}\\
\end{array} 
\end{equation} Most of the reductions that we have not seen yet deal with the last term. It starts out as $e^{\frac{-j 18 \pi}{4}}$. It is reduced to $e^{\frac{-j 10 \pi}{4}}$ since these are equivalent based upon a $2 \pi$ rotation, or, equivalently, $e^{\frac{-j 10 \pi}{4}} \cdot e^{\frac{-j 8 \pi}{4}}$ and the second term $e^{\frac{-j 8 \pi}{4}} = 1$. Finally, a rotation of $\pi$, which is equal to $-1$, brings it to $e^{\frac{-j 6  \pi}{4}}$. Another way of viewing this is $e^{\frac{-j 6 \pi}{4}} \cdot e^{\frac{-j 4 \pi}{4}}$ and $e^{\frac{-j 4 \pi}{4}} = -1$. We leave this term in this unreduced state in order to demonstrate symmetries in the following equations.

With a bit of reordering, we can view these four equations as:
\begin{equation}
\begin{array} {lll} 
G[0] & = & (g[0] + g[2]) + e^{\frac{-j 2 \pi 0}{4}} (g[1] + g[3])\\
G[1] & = & (g[0] - g[2]) + e^{\frac{-j 2 \pi 1}{4}} (g[1] - g[3])\\
G[2] & = & (g[0] + g[2]) + e^{\frac{-j 2 \pi 2}{4}} (g[1] + g[3])\\
G[3] & = & (g[0] - g[2]) + e^{\frac{-j 2 \pi 3}{4}} (g[1] - g[3])\\
\end{array}
\end{equation}

Several different symmetries are starting to emerge. First, the input data can be partitioned into even and odd elements, i.e., similar operations are done on the elements $g[0]$ and $g[2]$, and the same is true for the odd elements $g[1]$ and $g[3]$. Furthermore we can see that there are addition and subtraction symmetries on these even and odd elements. During the calculations of the output frequencies $G[0]$ and $G[2]$, the even and odd elements are summed together. The even and odd input elements are subtracted when calculating the frequencies $G[1]$ and $G[3]$.  Finally, the odd elements in every frequency term are multiplied by a constant complex exponential $W^i_4$ where $i$ denotes the index for the frequency output, i.e., $G[i]$. 

Looking at the terms in the parentheses, we see that they are 2 point \gls{fft}. For example, consider the terms corresponding to the even input values $g[0]$ and $g[2]$. If we perform a 2 point \gls{fft} on these even terms, the lower frequency (DC value) is $g[0] + g[2]$ (see Equation \ref{eq:2ptlower}), and the higher frequency is calculated as $g[0] - g[2]$ (see Equation \ref{eq:2pthigher}). The same is true for the odd input values $g[1]$ and $g[3]$. 

We perform one more transformation on these equations.
\begin{equation}
\begin{array} {lll} 
G[0] & = & (g[0] + g[2]) + e^{\frac{-j 2 \pi 0}{4}} (g[1] + g[3])\\
G[1] & = & (g[0] - g[2]) + e^{\frac{-j 2 \pi 1}{4}} (g[1] - g[3])\\
G[2] & = & (g[0] + g[2]) - e^{\frac{-j 2 \pi 0}{4}} (g[1] + g[3])\\
G[3] & = & (g[0] - g[2]) - e^{\frac{-j 2 \pi 1}{4}} (g[1] - g[3])\\
\end{array}
\label{eq:reduced4point}
\end{equation}
The twiddle factors in the last two equations are modified from $e^{\frac{-j 2 \pi 2}{4}} = -e^{\frac{-j 2 \pi 0}{4}}$ and $e^{\frac{-j 2 \pi 3}{4}} = -e^{\frac{-j 2 \pi 1}{4}}$. This allows for a reduction in the complexity of the multiplications since we can share multiplications across two terms. 

\begin{figure}
\centering
\includegraphics[width=  .5 \textwidth]{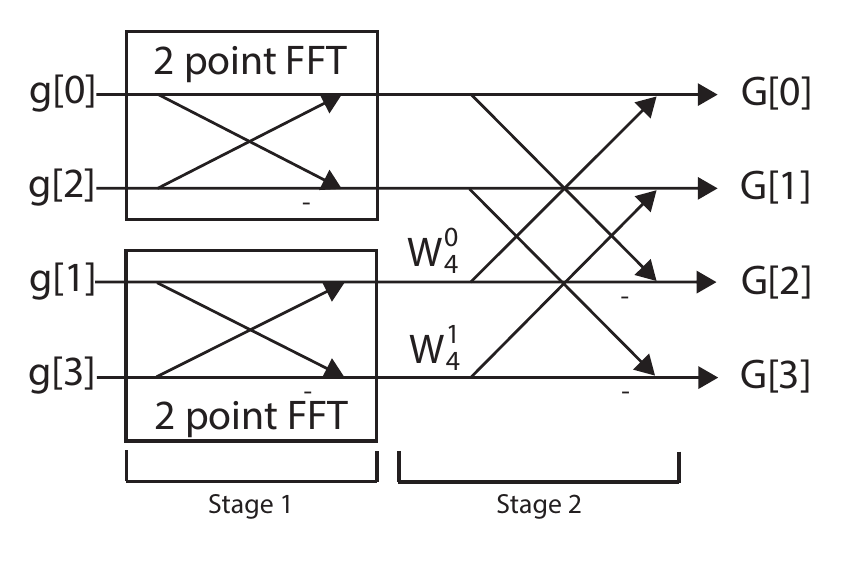}
\caption{A four point \gls{fft} divided into two stages. Stage 1 has uses two 2 point \gls{fft}s -- one 2 point \gls{fft} for the even input values and the other 2 point \gls{fft} for the odd input values. Stage 2 performs the remaining operations to complete the \gls{fft} computation as detailed in Equation \ref{eq:reduced4point}. }
\label{fig:4ptFFT}
\end{figure}

Figure \ref{fig:4ptFFT} shows the butterfly diagram for the four point \gls{fft}. We can see that the first stage is two 2 point \gls{fft} operations performed on the even (top butterfly) and odd (bottom butterfly) input values. The output of the odd 2 point \gls{fft}s are multiplied by the appropriate twiddle factor. We can use two twiddle factors for all four output terms by using the reduction shown in Equation \ref{eq:reduced4point}.

We are seeing the beginning of trend that allows the a reduction in complexity from $\mathcal{O}(n^2)$ operations for the \gls{dft} to $\mathcal{O}(n \log n)$ operations for the \gls{fft}. The key idea is building the computation through recursion. The 4 point \gls{fft} uses two 2 point \gls{fft}s. This extends to larger \gls{fft} sizes. For example, an 8 point \gls{fft} uses two 4 point \gls{fft}s, which in turn each use two 2 point \gls{fft}s (for a total of four 2 point \gls{fft}s). An 16 point \gls{fft} uses two 8 point \gls{fft}s, and so on.

\begin{exercise}
How many 2 point \gls{fft}s are used in a 32 point \gls{fft}? How many are there in a 64 point \gls{fft}? How many 4 point \gls{fft}s are required for a 64 point \gls{fft}? How about a 128 point \gls{fft}?  What is the general formula for 2 point, 4 point, and 8 point \gls{fft}s in an $N$ point \gls{fft} (where $N > 8$)?
\end{exercise}

Now let us formally derive the relationship, which provides a general way to describe the recursive structure of the \gls{fft}.  Assume that we are calculating an $N$ point \gls{fft}. The formula for calculating the frequency domain values $G[]$ given the input values $g[]$ is:
\begin{equation}
G[k] = \displaystyle\sum\limits_{n=0}^{N-1} g[n] \cdot e^{\frac{-j 2 \pi k n}{N}} \text{ for } k = 0,\dots, N-1
\label{eq:fft-full}
\end{equation}

We can divide this equation into two parts, one that sums the even components and one that sums the odd components.
\begin{equation}
G[k] = \displaystyle\sum\limits_{n=0}^{N/2-1} g[2n] \cdot e^{\frac{-j 2 \pi k (2n)}{N}} + \displaystyle\sum\limits_{n=0}^{N/2-1} g[2n+1] \cdot e^{\frac{-j 2 \pi k (2n+1)}{N}}
\label{eq:fft-split}
\end{equation}
The first part of this equation deals with the even inputs, hence the $2n$ terms in both $g[]$ and in the exponent of $e$. The second part corresponds to the odd inputs with $2n +1$ in both places. Also note that the sums now go to $N/2 -1$ in both cases which should make sense since we have divided them into two halves. 

We transform Equation \ref{eq:fft-split} to the following:
\begin{equation}
G[k] = \displaystyle\sum\limits_{n=0}^{N/2-1} g[2n] \cdot e^{\frac{-j 2 \pi k n}{N/2}} + \displaystyle\sum\limits_{n=0}^{N/2-1} g[2n+1] \cdot e^{\frac{-j 2 \pi k (2n)}{N}} \cdot e^{\frac{-j 2 \pi k}{N}}
\label{eq:fft-split-2}
\end{equation}
In the first summation (even inputs), we simply move the $2$ into the denominator so that it is now $N/2$. The second summation (odd inputs) uses the power rule to separate the $+1$ leaving two complex exponentials. We can further modify this equation to
\begin{equation}
G[k] = \displaystyle\sum\limits_{n=0}^{N/2-1} g[2n] \cdot e^{\frac{-j 2 \pi k n}{N/2}} + e^{\frac{-j 2 \pi k}{N}} \cdot \displaystyle\sum\limits_{n=0}^{N/2-1} g[2n+1] \cdot e^{\frac{-j 2 \pi k n}{N/2}} 
\label{eq:fft-split-3}
\end{equation}
Here we only modify the second summation. First we pull one of the complex exponentials outside of the summation since it does not depend upon $n$. And we also move the $2$ into the denominator as we did before in the first summation. Note that both summations now have the same complex exponential $e^{\frac{-j 2 \pi k n}{N/2}}$. Finally, we simplify this to 
\begin{equation}
G[k] = A_k + W_N^k B_k 
\label{eq:fft-split-4}
\end{equation} where $A_k$ and $B_k$ are the first and second summations, respectively. And recall that $W = e^{-j 2 \pi}$. This completely describes an N point \gls{fft} by separating even and odd terms into two summations.

For reasons that will become clear soon, let us assume that we only want to use Equation \ref{eq:fft-split-4} to calculate the first $N/2$ terms, i.e., $G[0]$ through $G[N/2 -1]$. And we will derive the remaining $N/2$ terms, i.e., those from $G[N/2]$ to $G[N-1]$ using a different equation. While this may seem counterintuitive or even foolish (why do more math than necessary?), you will see that this will allow us to take advantage of even more symmetry, and derive a pattern as we have seen in the 4 point \gls{fft}.

In order to calculate the higher frequencies $G[N/2]$ to $G[N-1]$, let us derive the same equations but this time using $k = N/2, N/2 + 1, \dots, N/2 -1$.  Thus, we wish to calculate
\begin{equation}
G[k + N/2] = \displaystyle\sum\limits_{n=0}^{N-1} g[n] \cdot e^{\frac{-j 2 \pi (k + N/2) n}{N}} \text{ for } k = 0, \dots, N/2 - 1
\label{eq:fft-upper}
\end{equation}
This is similar to Equation \ref{eq:fft-full} with different indices, i.e., we replace $k$ from Equation \ref{eq:fft-full} with $k + N/2$. Using the same set of transformations that we did previously, we can move directly to the equivalent to Equation \ref{eq:fft-split-3}, but replacing all instances of $k$ with $k + N/2$ which yields 
\begin{equation}
G[k + N/2] = \displaystyle\sum\limits_{n=0}^{N/2-1} g[2n] \cdot e^{\frac{-j 2 \pi (k + N/2) n}{N/2}} + e^{\frac{-j 2 \pi (k + N/2)}{N}} \cdot \displaystyle\sum\limits_{n=0}^{N/2-1} g[2n+1] \cdot e^{\frac{-j 2 \pi (k + N/2) n}{N/2}} 
\label{eq:fft-split-3-upper}
\end{equation}

We can reduce the complex exponential in the summations as follows:
\begin{equation}
e^{\frac{-j 2 \pi (k + N/2) n}{N/2}} = e^{\frac{-j 2 \pi k n}{N/2}} \cdot e^{\frac{-j 2 \pi (N/2) n}{N/2}} = e^{\frac{-j 2 \pi k n}{N/2}} \cdot e^{-j 2 \pi n} = e^{\frac{-j 2 \pi k n}{N/2}} \cdot 1
\label{eq:fft-split-4-upper}
\end{equation}
The first reduction uses the power rule to split the exponential. The second reduction cancels the term $N/2$ in the second exponential. The final reduction uses that fact that $n$ is a non-negative integer, and thus $e^{-j 2 \pi n}$ will always be a rotation of multiple of $2 \pi$. This means that this term is always equal to $1$. 

Now let us tackle the second complex exponential
\begin{equation}
e^{\frac{-j 2 \pi (k + N/2)}{N}} = e^{\frac{-j 2 \pi k }{N}} \cdot e^{\frac{-j 2 \pi N/2 }{N}} = e^{\frac{-j 2 \pi k }{N}} \cdot e^{-j  \pi} = - e^{\frac{-j 2 \pi k }{N}}
\label{eq:fft-split-5-upper}
\end{equation}
The first reduction splits the exponential using the power rule. The second reduction does some simplifications on the second exponential. We get the final term by realizing that $e^{-j \pi} = -1$.

By substituting Equations \ref{eq:fft-split-4-upper} and \ref{eq:fft-split-5-upper} into Equation \ref{eq:fft-split-3-upper}, we get
\begin{equation}
G[k + N/2] = \displaystyle\sum\limits_{n=0}^{N/2-1} g[2n] \cdot e^{\frac{-j 2 \pi k n}{N/2}} - e^{\frac{-j 2 \pi k}{N}} \cdot \displaystyle\sum\limits_{n=0}^{N/2-1} g[2n+1] \cdot e^{\frac{-j 2 \pi k n}{N/2}} 
\label{eq:fft-split-6-upper}
\end{equation}
Note the similarity to Equation \ref{eq:fft-split-3}. We can put it in terms of Equation \ref{eq:fft-split-4} as
\begin{equation}
G[k + N/2] = A_k - W_N^k B_k 
\label{eq:fft-split-7-upper}
\end{equation}

We can use Equations \ref{eq:fft-split-4} and \ref{eq:fft-split-7-upper} to create an $N$ point \gls{fft} from two $N/2$ point \gls{fft}s. Remember that $A_k$ corresponds to the even input values, and $B_k$ is a function of the odd input values. Equation \ref{eq:fft-split-4} covers the first $N/2$ terms, and Equation \ref{eq:fft-split-7-upper} corresponds to the higher $N/2$ frequencies. 

\begin{figure}
\centering
\includegraphics[width=  .8 \textwidth]{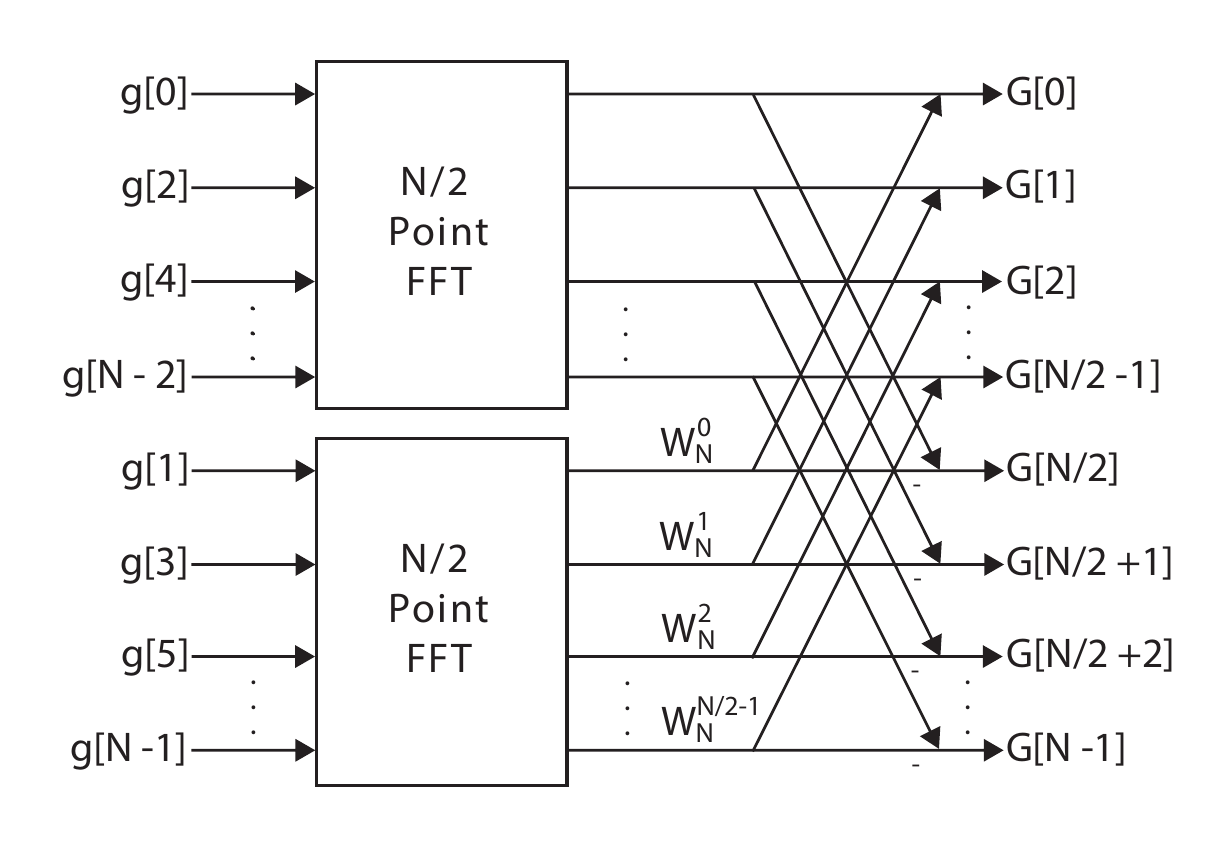}
\caption{Building an $N$ point \gls{fft} from two $N/2$ point \gls{fft}s. The upper $N/2$ point \gls{fft} is performed on the even inputs; the lower $N/2$ \gls{fft} uses the odd inputs. }
\label{fig:NptFFT}
\end{figure}

Figure \ref{fig:NptFFT} shows an $N$ point \gls{fft} derived from two $N/2$ point \gls{fft}s. $A_k$ corresponds to the top $N/2$ \gls{fft}, and $B_k$ is the bottom $N/2$ \gls{fft}.  The output terms $G[0]$ through $G[N/2-1]$ are multiplied by $W_N^0$ while the output terms $G[N/2]$ through $G[N-1]$ are multiplied by $-W_N^0$. Note that the inputs $g[]$ are divided into even and odd elements feeding into the top and bottom $n/2$ point \gls{fft}s, respectively.

We can use the general formula for creating the \gls{fft} that was just derived to recursively create the $N/2$ point \gls{fft}. That is, each of the $N/2$ point \gls{fft}s can be implemented using two $N/4$ point \gls{fft}s. And each $N/4$ point \gls{fft} uses two $N/8$ point \gls{fft}s, and so on until we reach the base case, a 2 point \gls{fft}.

\begin{figure}
\centering
\includegraphics[width=  \textwidth]{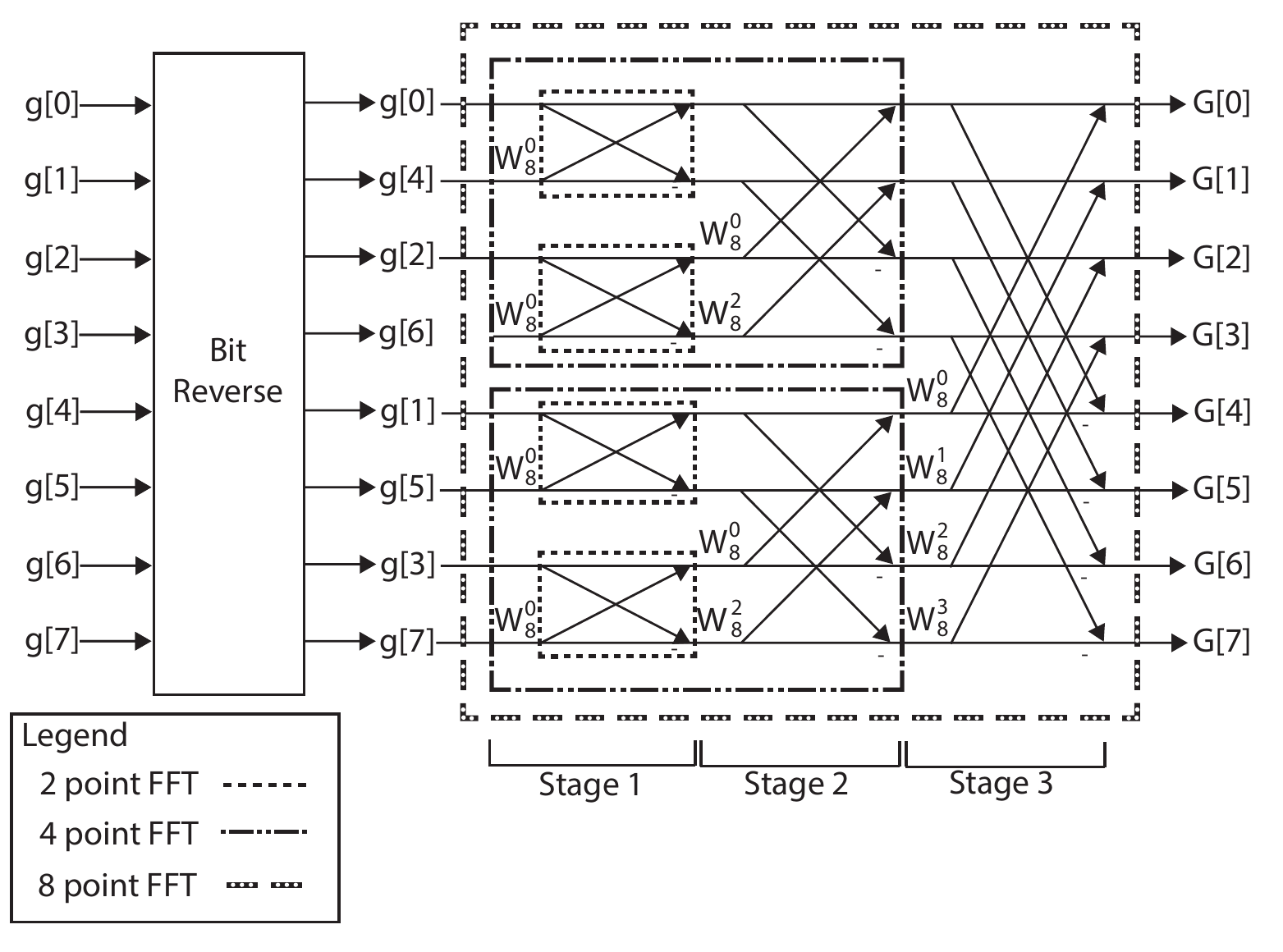}
\caption{An 8 point \gls{fft} built recursively. There are two 4 point \gls{fft}s, which each use two 2 point \gls{fft}s. The inputs must be reordered to even and odd elements twice. This results in reordering based upon the bit reversal of the indices.}
\label{fig:8ptFFT}
\end{figure}

Figure \ref{fig:8ptFFT} shows an 8 point \gls{fft} and highlights this recursive structure. The boxes with the dotted lines indicate different sizes of \gls{fft}. The outermost box indicates an 8 point \gls{fft}. This is composed by two 4 point \gls{fft}s. Each of these 4 point \gls{fft}s have two 2 point \gls{fft}s for a total of four 2 point \gls{fft}s. 

Also note that the inputs must be reordered before they are feed into the 8 point \gls{fft}. This is due to the fact that the different $N/2$ point \gls{fft}s take even and odd inputs. The upper four inputs correspond to even inputs and the lower four inputs have odd indices. However, they are reordered twice. If we separate the even and odd inputs once we have the even set $\{g[0], g[2], g[4], g[6] \}$ and the odd set $\{g[1], g[3], g[5], g[7] \}$. Now let us reorder the even set once again. In the even set $g[0]$ and $g[4]$ are the even elements, and $g[2]$ and $g[6]$ are the odd elements. Thus reordering it results in the set $\{g[0], g[4], g[2], g[6] \}$. The same can be done for the initial odd set yielding the reordered set $\{g[1], g[5], g[3], g[7] \}$. 

The final reordering is done by swapping values whose indices are in bit reversed order. Table \ref{table:bit_reverse} shows the indices and their three bit binary values. The table shows the eight indices for the 8 point \gls{fft}, and the corresponding binary value for each of those indices in the second column. The third column is the bit reversed binary value of the second column. And the last column is the decimal number corresponding the reversed binary number. 

\begin{table}[htbp]
\caption{The index, three bit binary value for that index, bit reversed binary value, and the resulting bit reversed index.}
\begin{center}
\begin{tabular}{|c|c|c|c|}
\hline
Index & Binary & Reversed & Reversed \\
 & & Binary & Index \\
\hline
0 & 000 & 000 & 0 \\
1 & 001 & 100 & 4 \\
2 & 010 & 010 & 2 \\
3 & 011 & 110 & 6 \\
4 & 100 & 001 & 1 \\
5 & 101 & 101 & 5 \\
6 & 110 & 011 & 3 \\
7 & 111 & 111 & 7 \\
\hline
\end{tabular}
\end{center}
\label{table:bit_reverse}
\end{table}

Looking at the first row, the initial index $0$, has a binary value of $000$, which when reversed remains $000$. Thus this index does not need to be swapped. Looking at Figure \ref{fig:8ptFFT} we see that this is true. $g[0]$ remains in the same location. In the second row, the index 1 has a binary value $001$. When reversed this is $100$ or $4$. Thus, the data that initially started at index 1, i.e., $g[1]$ should end up in the fourth location. And looking at index 4, we see the bit reversed value is $1$. Thus $g[1]$ and $g[4]$ are swapped.  

This bit reversal process works regardless of the input size of the \gls{fft}, assuming that the \gls{fft} is a power of two. \gls{fft} are commonly a power of two since this allows them to be recursively implemented.

\begin{exercise}
In an 32 point \gls{fft}, index 1 is swapped with which index? Which index is index 2 is swapped with?
\end{exercise}

This completes our mathematical treatment of the \gls{fft}. There are plenty of more details about the \gls{fft}, and how to optimize it. You may think that we spent too much time already discussing the finer details of the \gls{fft}; this is a book on parallel programming for FPGAs and not on digital signal processing. This highlights an important part of creating an optimum hardware implementation -- the designer must have a good understanding of the algorithm under development. Without that, it is difficult to create a good implementation. The next section deals with how to create a good \gls{fft} implementation. 

\section{Baseline Implementation}

In the remainder of this chapter, we discuss different methods to implement the Cooley-Tukey \gls{fft} \cite{cooley65} algorithm using the \VHLS tool. This is the same algorithm that we described in the previous section. We start with a common version of the code, and then describe how to restructure it to achieve a better hardware design. 

When performed sequentially, the $\mathcal{O}(n \log n)$ operations in the \gls{fft} require $\mathcal{O}(n \log n)$ time steps.   Typically, a parallel implementation will perform some portion of the \gls{fft} in parallel.  One common way of parallelizing the \gls{fft} is to organize the computation into $\log n$ stages, as shown in Figure \ref{fig:fftstages}.  The operations in each stage are dependent on the operations of the previous stage, naturally leading to a pipelining across the tasks.  Such an architecture allows $\log n$ \gls{fft}s to be computed simultaneously with a task interval determined by the architecture of each stage.  We discuss task pipelining using the \lstinline|dataflow| directive in Section \ref{sec:fft_task_pipelining}.

Each stage in the \gls{fft} also contains significant parallelism, since each butterfly computation is independent of other butterfly computations in the same stage.  In the limit, performing $n/2$ butterfly computations every clock cycle with a Task Interval of 1 can allow the entire stage to be computed with a Task Interval of 1.  When combined with a dataflow architecture, all of the parallelism in the \gls{fft} algorithm can be exploited.  Note, however that although such an architecture can be constructed, it is almost never used except for very small signals, since an entire new block of \lstinline|SIZE| samples must be provided every clock cycle to keep the pipeline fully utilized.  For instance, a 1024-point \gls{fft} of complex 32-bit floating point values, running at 250 MHz would require 1024 \text{points}*(8 {bytes}/{point})*250*$10^9$ Hz = 1Terabyte/second of data into the FPGA.   In practice, a designer must match the computation architecture to the data rate required in a system.

\begin{exercise}
Assuming a clock rate of 250 MHz and one sample received every clock cycle, approximately how many butterfly computations must be implemented to process every sample with a 1024-point \gls{fft}?  What about for a 16384-point \gls{fft}?
\end{exercise}

In the remainder of this section, we describe the optimization of an \gls{fft} with the function prototype \lstinline|void fft(DTYPE X_R[SIZE], DTYPE X_I[SIZE])| where \lstinline|DTYPE| is a user customizable data type for the representation of the input data. This may be \lstinline|int|, \lstinline|float|, or a fixed point type. For example, \lstinline|#define DTYPE int| defines \lstinline|DTYPE| as an \lstinline|int|. Note that we choose to implement the real and imaginary parts of the complex numbers in two separate arrays. The \lstinline|X_R| array holds the real input values, and the \lstinline|X_I| array holds the imaginary values. \lstinline|X_R[i]| and \lstinline|X_I[i]| hold the $i$th complex number in separate real and imaginary parts. 

\begin{aside}
There is one change in the \gls{fft} implementation that we describe in this section. Here we perform an \gls{fft} on complex numbers. The previous section uses only real numbers. While this may seem like a major change, the core ideas stay the same. The only differences are that the data has two values (corresponding to the real and imaginary part of the complex number), and the operations (add, multiply, etc.) are complex operations. 
\end{aside}

This function prototype forces an in-place implementation. That is, the output data is stored in the same array as the input data. This eliminates the need for additional arrays for the output data, which reduces the amount of memory that is required for the implementation. However, this may limit the performance due to the fact that we must read the input data and write the output data to the same arrays. Using separate arrays for the output data is reasonable if it can increase the performance. There is always a tradeoff between resource usage and performance; the same is true here. The best implementation depends upon the application requirements (e.g., high throughput, low power, size of FPGA, size of the \gls{fft}, etc.).


We start with code for an \gls{fft} that would be typical for a software implementation. Figure \ref{fig:fft_sw} shows a nested three \lstinline|for| loop structure. The outer \lstinline|for| loop, labeled \lstinline|stage_loop| implements one stage of the \gls{fft} during each iteration. There are $log_2(N)$ stages where $N$ is the number of input samples. The stages are clearly labeled in Figure \ref{fig:8ptFFT}; this 8 point \gls{fft} has $log_2(8) = 3$ stages. You can see that each stage performs the same amount of computation, or the same number of butterfly operations. In the 8 point \gls{fft}, each stage has four butterfly operations. 

\begin{figure}
\begin{lstlisting}
void fft(DTYPE X_R[SIZE], DTYPE X_I[SIZE]) {
	DTYPE temp_R; // temporary storage complex variable
	DTYPE temp_I; // temporary storage complex variable
	int i, j, k;	// loop indexes
	int i_lower;	// Index of lower point in butterfly
	int step, stage, DFTpts;
	int numBF;			// Butterfly Width
	int N2 = SIZE2; // N2=N>>1

	bit_reverse(X_R, X_I);

	step = N2;
	DTYPE a, e, c, s;

stage_loop:
	for (stage = 1; stage <= M; stage++) { // Do M stages of butterflies
		DFTpts = 1 << stage;								 // DFT = 2^stage = points in sub DFT
		numBF = DFTpts / 2;									 // Butterfly WIDTHS in sub-DFT
		k = 0;
		e = -6.283185307178 / DFTpts;
		a = 0.0;
	// Perform butterflies for j-th stage
	butterfly_loop:
		for (j = 0; j < numBF; j++) {
			c = cos(a);
			s = sin(a);
			a = a + e;
		// Compute butterflies that use same W**k
		dft_loop:
			for (i = j; i < SIZE; i += DFTpts) {
				i_lower = i + numBF; // index of lower point in butterfly
				temp_R = X_R[i_lower] * c - X_I[i_lower] * s;
				temp_I = X_I[i_lower] * c + X_R[i_lower] * s;
				X_R[i_lower] = X_R[i] - temp_R;
				X_I[i_lower] = X_I[i] - temp_I;
				X_R[i] = X_R[i] + temp_R;
				X_I[i] = X_I[i] + temp_I;
			}
			k += step;
		}
		step = step / 2;
	}
}
\end{lstlisting}
\caption{ A common implementation for the \gls{fft} using three nested \lstinline|for| loops.  While this may work well running as software on a processor, it is far from optimal for a hardware implementation.}
\label{fig:fft_sw}
\end{figure}

\begin{exercise}
For an $N$ point \gls{fft}, how many butterfly operations are there in each stage? How many total butterfly operations are there for the entire \gls{fft}?
\end{exercise}

The second \lstinline|for| loop, labeled \lstinline|butterfly_loop|, performs all of the butterfly operations for the current stage. \lstinline|butterfly_loop| has another nested \lstinline|for| loop, labeled \lstinline|dft_loop|. Each iteration of \lstinline|dft_loop| performs one butterfly operation. Remember that we are dealing with complex numbers and must perform complex additions and multiplications. 

The first line in \lstinline|dft_loop| determines the offset of the butterfly. Note that the ``width'' of the butterfly operations changes depending upon the stage. Looking at Figure \ref{fig:8ptFFT}, Stage 1 performs butterfly operations on adjacent elements, Stage 2 performs butterfly operations on elements with index differing by two, and Stage 3 performs butterfly operations on elements with index differing by four. This difference is computed and stored in the \lstinline|i_lower| variable. Notice that this offset, stored in the variable \lstinline|numBF|, is different in every stage. 

The remaining operations in \lstinline|dft_loop| perform multiplication by the twiddle factor and an addition or subtraction operation. The variables \lstinline|temp_R| and \lstinline|temp_I| hold the real and imaginary portions of the data after multiplication by the twiddle factor $W$. The variables \lstinline|c| and \lstinline|s| are the real and imaginary parts of $W$, which is calculated using the \lstinline|sin()| and \lstinline|cos()| builtin functions. We could also use the CORDIC, such as the one developed in Chapter \ref{chapter:cordic}, to have more control over the implementation. Lastly, elements of the \lstinline|X_R[]| and \lstinline|X_I[]| arrays are updated with the result of the butterfly computation.

\lstinline|dft_loop| and \lstinline|butterfly_loop| each execute a different number of times depending upon the stage. However the total number of times that the body of \lstinline|dft_loop| is executed in one stage is constant. The number of iterations for the \lstinline|butterfly for| loop depends upon the number of unique $W$ twiddle factors in that stage. Referring again to Figure \ref{fig:8ptFFT}, we can see that Stage 1 uses only one twiddle factor, in this case $W_8^0$. Stage 2 uses two unique twiddle factors and Stage 3 uses four different $W$ values. Thus, \lstinline|butterfly_loop| has only one iteration in Stage 1, 2 iterations in stage 2, and four iterations in stage 3. Similarly, the number of iterations of \lstinline|dft_loop| changes. It iterates four times for an 8 point \gls{fft} in Stage 1, two times in Stage 2, and only one time in stage 3. However in every stage, the body of \lstinline|dft_loop| is executed the same number of times in total, executing a total of four butterfly operations for each stage an 8 point \gls{fft}. 

\begin{aside}
\VHLS performs significant static analysis on each synthesized function, including computing bounds on the number of times each loop can execute.  This information comes from many sources, including variable bitwidths, ranges, and \lstinline|assert()| functions in the code. When combined with the loop II, \VHLS can compute bounds on the latency or interval of the \gls{fft} function.  In some cases (usually when loop bounds are variable or contain conditional constructs), the tool is unable to compute the latency or interval of the code and returns `'?'.  When synthesizing the code in Figure \ref{fig:fft_sw}, \VHLS may not be able to determine the number of times that \lstinline|butterfly_loop| and \lstinline|dft_loop| iterate because these loops have variable bounds.

The \lstinline|tripcount| directive enables the user to specify to the \VHLS tool more information about the number of times a loop is executed which can be used by the tool to analyze the performance of the design. It takes three optional arguments \lstinline|min|, \lstinline|max|, and \lstinline|average|. In this code, we could add a directive to \lstinline|dft_loop|. By applying this directive, the \VHLS tool can calculate bounds on the latency and interval value for the loop and the overall design.  Note that since the \VHLS tool uses the numbers that you provide, if you give the tool an incorrect tripcount then the reported task latency and task interval will be incorrect -- garbage in, garbage out.   
\end{aside}

\begin{exercise}
What is the appropriate way to use the \lstinline|trip count| directive for the \gls{fft} in Figure \ref{fig:fft_sw}? Should you set the \lstinline|max|, \lstinline|min|, and/or \lstinline|average| arguments? Would you need to modify the tripcount arguments if the size of the \gls{fft} changes?  
\end{exercise}

\section{Bit Reversal}
\label{sec:fft_bit_reversal}

We have not talked about the bit reverse function, which swaps the input data values so that we can perform an in-place \gls{fft}. This means that the inputs values are mixed, and the result is that the output data is in the correct order. We discuss that function in some detail now.

Figure \ref{fig:fft_bit_reverse} shows one possible implementation of the bit reverse function. It divides the code into two functions. The first is the bit reversal function (\lstinline|bit_reverse|), which reorders data in the given arrays so that each data is in located at a different index in the array. This function calls another function, \lstinline|reverse_bits|, which takes an input integer and returns the bit reversed value of that input.

\begin{figure}
\begin{lstlisting}
#define FFT_BITS 10			// Number of bits of FFT, i.e., log(1024)
#define SIZE 1024				// SIZE OF FFT
#define SIZE2 SIZE >> 1 // SIZE/2
#define DTYPE int

unsigned int reverse_bits(unsigned int input) {
	int i, rev = 0;
	for (i = 0; i < FFT_BITS; i++) {
		rev = (rev << 1) | (input & 1);
		input = input >> 1;
	}
	return rev;
}

void bit_reverse(DTYPE X_R[SIZE], DTYPE X_I[SIZE]) {
	unsigned int reversed;
	unsigned int i;
	DTYPE temp;

	for (i = 0; i < SIZE; i++) {
		reversed = reverse_bits(i); // Find the bit reversed index
		if (i < reversed) {
			// Swap the real values
			temp = X_R[i];
			X_R[i] = X_R[reversed];
			X_R[reversed] = temp;

			// Swap the imaginary values
			temp = X_I[i];
			X_I[i] = X_I[reversed];
			X_I[reversed] = temp;
		}
	}
}
\end{lstlisting}
\caption{ The first stage in our \gls{fft} implementation reorders the input data. This is done by swapping the value at index $i$ in the input array with the value at the bit reversed index corresponding to $i$. The function \lstinline|reverse_bits| gives the bit reversed value corresponding to the \lstinline|input| argument. And the function \lstinline|bit_reverse| swaps the values in the input array. }
\label{fig:fft_bit_reverse}
\end{figure}

Let us start with a brief overview of the \lstinline|reverse_bits| function. The function goes bit by bit through the \lstinline|input| variable and shifts it into the \lstinline|rev| variable. The \lstinline|for| loop body consists of a few bitwise operations that reorder the bits of the input. Although these operations are individually not terribly complex, the intention of this code is that the for loop is completely unrolled and \VHLS can identify that the bits of the input can simply be wired to the output.  As a result, the implementation of the \lstinline|reverse_bits| function should require no logic resources at all, but only wires.  This is a case where unrolling loops greatly simplifies the operations that must be performed.  Without unrolling the loop, the individual `or' operations must be performed sequentially.  Although this loop can be pipelined, the `or' operation would still be implemented in logic resources in the FPGA and executing the the loop would have a latency determined by the number of bits being reversed (\lstinline|\gls{fft}_BITS| in this case).

\begin{exercise}
What is the latency of the \lstinline|reverse_bits| function when no directives are applied? What is the latency when the loop is pipelined?  What is the latency when the whole function is pipelined?
\end{exercise}

\begin{aside}
It is tempting to ``blindly'' apply directives in order to achieve a better design. However, this can be counterproductive. The best designer has a deep understanding of both the application and the available optimizations and carefully considers these together to achieve the best results. 
\end{aside}

Now let us optimize the parent \lstinline|bit_reverse| function. This function has a single \lstinline{for} loop that iterates through each index of the input arrays. Note that there are two input arrays \lstinline{X_R[]} and \lstinline{X_I[]}. Since we are dealing with complex numbers, we must store both the real portion (in the array \lstinline{X_R[]}), and the imaginary portion (in the array \lstinline{X_I[]}). \lstinline{X_R[i]} and \lstinline{X_I[i]} holds the real and imaginary values of the i-th input.
In each iteration of the \lstinline{for} loop, we find the index reversed value by calling the \lstinline{reverse_bits} function. Then we swap both the real and imaginary values stored in the index \lstinline{i} and the index returned by the function \lstinline{reverse_bits}. Note that as we go through all \lstinline{SIZE} indices, we will eventually hit the reversed index for every value. Thus, the code only swaps values the first time based on the condition \lstinline{if(i < reversed)}.


\section{Task Pipelining}
\label{sec:fft_task_pipelining}

Dividing the \gls{fft} algorithm into stages enables \VHLS to generate an implementation where different stages of the algorithm are operating on different data sets. This optimization, called \gls{taskpipelining} is enabled using the \lstinline{dataflow} directive. This is a common hardware optimization, and thus is relevant across a range of applications.

\begin{figure}
\begin{lstlisting}[firstline=9]

void fft(DTYPE X_R[SIZE], DTYPE X_I[SIZE], DTYPE OUT_R[SIZE], DTYPE OUT_I[SIZE])
{
  #pragma HLS dataflow
  DTYPE Stage1_R[SIZE], Stage1_I[SIZE];
  DTYPE Stage2_R[SIZE], Stage2_I[SIZE];
  DTYPE Stage3_R[SIZE], Stage3_I[SIZE];

  bit_reverse(X_R, X_I, Stage1_R, Stage1_I);
  fft_stage_one(Stage1_R, Stage1_I, Stage2_R, Stage2_I);
  fft_stages_two(Stage2_R, Stage2_I, Stage3_R, Stage3_R);
  fft_stage_three(Stage3_R, Stage3_I, OUT_R, OUT_I);
}
\end{lstlisting}
\caption{ The code divides an 8 point \gls{fft} into four stages, each of which is a separate function. The \lstinline{bit_reverse} function is the first stages. And there are three stages for the 8 point \gls{fft}. }
\label{fig:fft_stages_code}
\end{figure}

We can naturally divide the \gls{fft} algorithm into $\log_2(N+1)$ stages where $N$ is the number of points of the \gls{fft}.  The first stage swaps each element in the input array with the element located at the bit reversed address in the array. After this bit reverse stage, we perform $\log_2(N)$ stages of butterfly operations. Each of these butterfly stages has the same computational complexity.  Figure \ref{fig:fft_stages_code} describes how to devide an 8 point \gls{fft} into four separate tasks. The code has separate functions for each of the tasks: \lstinline{bit_reverse}, \lstinline{fft_stage_one}, \lstinline{fft_stage_two}, and \lstinline{fft_stage_three}. Each stage has two input arrays and two output arrays: one for the real portion and one for the imaginary portion of the complex numbers. Assume that the \lstinline{DTYPE} is defined elsewhere, e.g., as an \lstinline{int}, \lstinline{float} or a fixed point data type.

\begin{figure}
\centering
{\scriptsize %
\executeiffilenewer{fftstages.svg}{images/fftstages.pdf}%
{inkscape -z -D --file=fftstages.svg %
--export-pdf=images/fftstages.pdf --export-latex}%
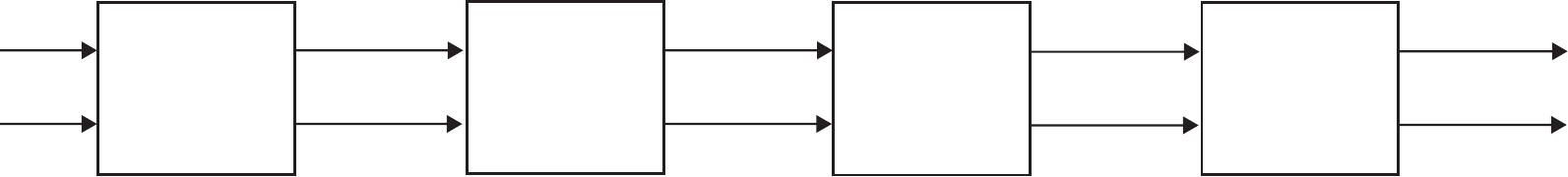%
}
\executeiffilenewer{fft_dataflow_behavior.svg}{images/fft_dataflow_behavior.pdf}%
{inkscape -z -D --file=fft_dataflow_behavior.svg %
--export-pdf=images/fft_dataflow_behavior.pdf --export-latex}%
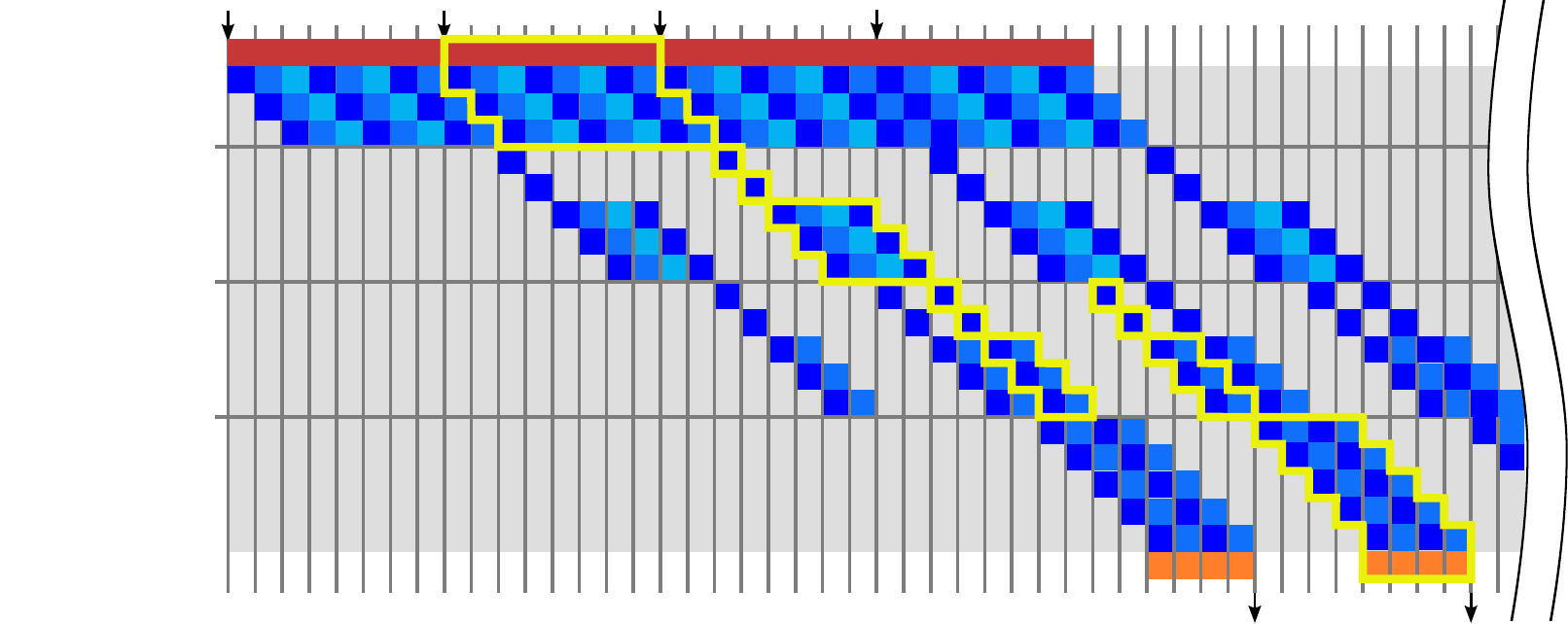%

\caption{ Dividing the \gls{fft} into different stages allows for task pipelining across each of these stages. The figure shows an example with three \gls{fft} stages (i.e., an 8 point \gls{fft}). The figure shows four 8 point \gls{fft} executing at the same time. }
\label{fig:fftstages}
\end{figure}

Refactoring the \gls{fft} code allows us to perform \gls{taskpipelining}. Figure \ref{fig:fftstages} gives an example of this. In this execution, rather than wait for the first task to complete all four four functions in the code before the second task can begin, we allow the second task to start after the first task has only completed the first function \lstinline{bit_reverse}. The first task continues to execute each stage in the pipeline in order, followed by the remaining tasks in order.  Once the pipeline is full, all four subfunctions are executing concurrently, but each one is operating on different input data. Similarly, there are four 8 point \gls{fft}s being computed simultaneously, each one executing on a different component of the hardware. This shown in the middle portion of Figure \ref{fig:fftstages}. Each of the vertical four stages represents one 8 point \gls{fft}. And the horizontal denotes increasing time. Thus, once we start the fourth 8 point \gls{fft}, we have four \gls{fft}s running simultaneously. 

The \lstinline|dataflow| directive can construct separate pipeline stages (often called \glspl{process}) from both functions and loops. The code in Figure \ref{fig:fft_stages_code} uses functions only, but we could achieve a similar result with four loops instead of four functions.  In fact, this result could be achieved by unrolling the outer \lstinline|stage_loop| in the original code either explicitly or using \lstinline|#pragma HLS unroll|.

The \lstinline{dataflow} directive and the \lstinline{pipeline} directive both generate circuits capable of pipelined execution.  The key difference is in the granularity of the pipeline. The \lstinline{pipeline} directive constructs an architecture that is efficiently pipelined at the cycle level and is characterized by the II of the pipeline.  Operators are statically scheduled and if the II is greater than one, then operations can be shared on the same operator.  The \lstinline{dataflow} directive constructs an architecture that is efficiently pipelined for operations that take a (possibly unknown) number of clock cycles, such as the the behavior of a loop operating on a block of data.  These coarse-grained operations are not statically scheduled and the behavior is controlled dyanmically by the handshake of data through the pipeline.  In the case of the \gls{fft}, each stage is an operation on a block of data (the whole array) which takes a large number of cycles.  Within each stage, loops execute individual operations on the data in a block.  Hence, this is a case where it often makes sense to use the \lstinline|dataflow| directive at the toplevel to form a coarse-grained pipeline, combined with the \lstinline|pipeline| directive within each loop to form fine-grained pipelines of the operations on each individual data element.

The \lstinline{dataflow} directive must implement memories to pass data between different processes. In the case when \VHLS can determine that processes access data in sequential order, it implements the memory using a FIFO. This requires that data is written into an array in the same order that it is read from the array.  If the is not the case, or if \VHLS can not determine if this streaming condition is met, then the memory can be implemented using a ping-pong buffer instead.   The ping-pong buffer consists of two (or more) conceptual blocks of data, each the size of the original array. One of the blocks can be written by the source process while another block is read by the destination process. The term ``ping-pong'' comes from the fact that the reading and writing to each block of data alternates in every execution of the task. That is, the source process will write to one block and then switch to the other block before beginning the next task. The destination process reads from the block that the producer is not writing to. As a result, the source and destination processes can never writing and reading from the same block at the same time. 

A ping-pong buffer requires enough memory to store each communication array at least twice.   FIFOs can often be significantly smaller, although determining a minimal size for each fifo is often a difficult design problem.  Unlike a FIFO, however, the data in a ping-pong buffer can be written to and read from in any order. Thus, FIFOs are generally the best choice when the data is produced and consumed in sequential order and ping-pong buffers are a better choice when there is not such regular data access patterns.

Using the \lstinline{dataflow} directive effectively still requires the behavior of each individual process to be optimized. Each individual process in the pipeline can still be optimized using techniques we have seen previously such as code restructuring, pipelining, and unrolling.  For example, we have already discussed some optimizations for the \lstinline{bit_reverse} function in Section \ref{sec:fft_bit_reversal}.  In general, it is important to optimize the individual tasks while considering overall toplevel performance goals. Many times it is best to start with small functions and understand how to optimize them in isolation. As a designer, it is often easier to comprehend what is going on in a small piece of code and hopefully determine the best optimizations quickly. After optimizing each individual function, then you can move up the hierarchy considering larger functions given particular implementations of low level functions, eventually reaching the toplevel function.

However, the local optimizations must be considered in the overall scope of the goals. In particular for dataflow designs the achieved interval for the overall pipeline can never be smaller than the interval of each individual process. Looking again at Figure \ref{fig:fft_stages_code}, assume that \lstinline{bit_reverse} has an interval of 8 cycles, \lstinline{fft_stage_one} takes 12 cycles, \lstinline{fft_stage_two} requires 12 cycles, and \lstinline{fft_stage_three} takes 14 cycles. When using \lstinline{dataflow}, the overal task interval is 14, determined by the maximum of all of the tasks/functions. This means that you should be careful in balancing optimizations across different processes with the goal of creating a balanced pipeline where the interval of each process is approximately the same. In this example, improving the interval of the \lstinline{bit_reverse} function cannot improve the overall interval of the \lstinline|fft| function. In fact, it might be beneficial to increase the latency of the \lstinline{bit_reverse} function, if it can be achieved with significantly fewer resources.

\section{Conclusion}
\label{sec:fft_conclusion}

The overall goal is to create the most optimal design, which is a function of your application needs. This may be to create the smallest implementation. Or the goal could be creating something that can perform the highest throughput implementation regardless of the size of the FPGA or the power/energy constraints. Or the latency of delivering the results may matter if the application has real-time constraints. All of the optimizations change these factors in different ways. 

In general, there is no one algorithm on how to optimize your design. It is a complex function of the application, design constraints, and the inherent abilities of the designer himself. Yet, it is important that the designer have a deep understanding of the application itself, the design constraints, and the abilities of the synthesis tool. 

We attempted to illustrate these bits of wisdom in this chapter. While the \gls{fft} is a well studied algorithm, with a large number of known hardware implementation tricks, it still serves as a good exemplar for high-level synthesis. We certainly did not give all of the tricks for optimization; we leave that as an exercise in Chapter \ref{chapter:ofdm} where we task the designer to create an simple orthogonal frequency-division multiplexing receiver. The core of this an \gls{fft}.  Regardless, we attempted to provide some insight into the key optimizations here, which we hope serve as a guide to how to optimize the \gls{fft} using the \VHLS tool.

First and foremost, understand the algorithm. We spent a lot of time explaining the basics of the \gls{fft}, and how it relates to the \gls{dft}. We hope that the reader understands that this is the most important part of building optimal hardware. Certainly, the designer could translate C/MATLAB/Java/Python code into \VHLS and get an working implementation. And that same designer could somewhat blindly apply directives to achieve better results. But that designer is not going to get anywhere close to optimal results without a deep understanding of the algorithm itself.

Second, we provide an introduction to  task level pipelining using the \lstinline{dataflow} directive. This is a powerful optimization that is not possible through code restructuring. I.e., the designer must use this optimization to get such a design. Thus, it is important that the designer understand its power, drawbacks, and usage.

Additionally, we give build upon some of the optimizations from previous chapters, e.g., loop unrolling and pipelining. All of these are important to get an optimized \gls{fft} hardware design. While we did not spend too much time on these optimizations, they are extremely important.

Finally, we tried to impress on the reader that these optimizations cannot be done in isolation. Sometimes the optimizations are independent, and they can be done in isolation. For example, we can focus on one of the tasks (e.g., in the \lstinline{bit_reverse} function as we did in Section \ref{sec:fft_bit_reversal}). But many times different optimizations will effect another. For example, the \lstinline{inline} directive will effect the way the pipelining of a function. And in particular, the way that we optimize tasks/functions can propagate itself up through the hierarchy of functions.  The takeaway is that it is extremely important that the designer understand the effects of the optimizations on the algorithm, both locally and globally.

\chapter{Sparse Matrix Vector Multiplication}
\glsresetall
\label{chapter:spmv}

Sparse matrix vector multiplication (SpMV) takes a sparse matrix, i.e., one in which most of its elements  are zero, and multiplies it by a vector. The vector itself may be sparse as well, but often it is dense. This is a common operation in scientific applications, economic modeling, data mining, and information retrieval. For example, it is used as an iterative method for solving sparse linear systems and eigenvalue problems. It is an operation in PageRank and it is also used in computer vision, e.g., image reconstruction.

This chapter introduces several new HLS concepts, and reinforces some previously discussed optimization. One goal of the chapter is to introduce a more complex data structure. We use a \gls{crs} representation to hold the sparse matrix. Another goal is to show how to perform testing. We build a simple structure for a testbench that can be used to help determine if the code is functionally correct. This is an important aspect of hardware design, and \VHLS makes it easy to test many aspects of the generated RTL with the same high-level C testbench. This is one of the big advantages of HLS over RTL design. We also show how you can perform C/RTL cosimulation using the testbench and \VHLS tool. This is necessary to derive the performance characteristics for the different SpMV designs. Since the execution time depends upon the number of entries in the sparse matrix, we must use input data in order to determine the clock cycles for the task interval and task latency.

\section{Background}

\begin{figure}
\centering
\includegraphics[width=  .65\textwidth]{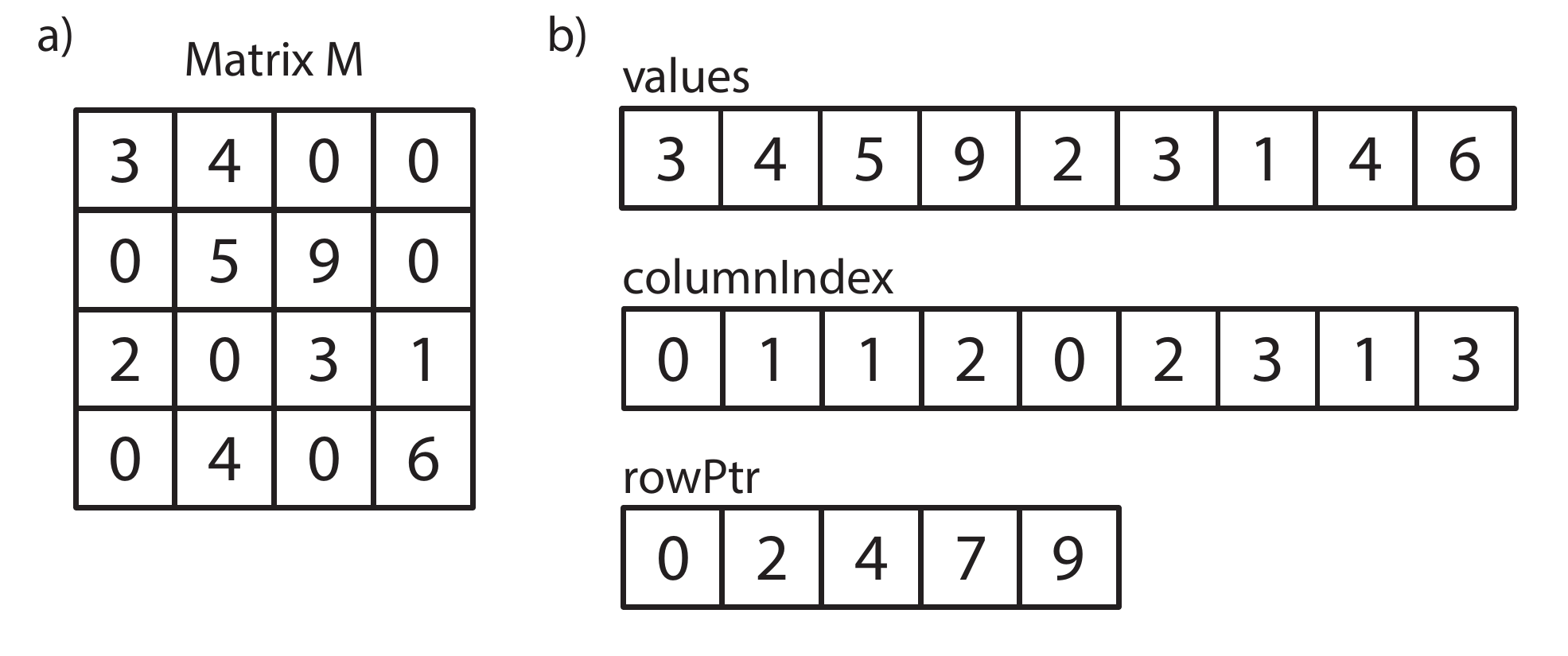}
\caption{A $4 \times 4$ matrix $\mathbf{M}$ represented in two different ways: as a `dense' matrix stored in a two-dimensional array, and as a sparse matrix stored in the compressed row storage (\gls{crs}) form, a data structure consisting of three arrays. }
\label{fig:crs}
\end{figure}

Figure~\ref{fig:crs} shows an example of a $4 \times 4$ matrix $\mathbf{M}$ represented in two different ways.  Figure ~\ref{fig:crs} a) shows the normal representation of the matrix as a two-dimensional array of 16 elements.  Each element is store in its own location in the array.  Figure ~\ref{fig:crs} b) shows the same matrix represented in \gls{crs} format.   The \gls{crs} representation is a data structure consisting of three arrays.  The \lstinline{values} array holds the value of each non-zero element in the matrix. The \lstinline{columnIndex} and \lstinline{rowPtr} arrays encode information about the location of these non-zero elements in the matrix. \lstinline{columnIndex} stores the column of each element, while \lstinline{rowPtr} contains the index in \lstinline|values| of the first element in each row. The \gls{crs} format avoids storing values in the matrix that are zero, although there is nothing to prevent a zero from being explicitly represented in the \lstinline|values| array.  In the example, however, we see that the \lstinline|values| array does not, in fact, contain any zero elements.  The tradeoff is that some additional book-keeping information (the \lstinline|columnIndex| and \lstinline|rowPtr| arrays) in order to properly interpret and manipulate the matrix.  The \gls{crs} form is commonly used when large matrices contain only a small number of non-zero elements (typically 10 percent or less), enabling these matrices to be stored with less memory and manipulated with fewer operations. However, the \gls{crs} form has no requirements about the sparsity of the matrix and can be used for any matrix. This makes it a general approach that can be used for any matrix, but not necessarily the most efficient.  The \gls{crs} form is also not the only efficient representation of sparse matrices.  Depending on the characteristics of the matrix and the types of operations to be performed, other sparse representations can also be used.

More precisely, the \gls{crs} format uses a data structure consisting of three arrays: \lstinline{values}, \lstinline{colIndex}, and \lstinline{rowPtr}. The \lstinline{values} array and the \lstinline{columnIndex} has an entry for each of the non-zero elements in the sparse matrix $\mathbf{M}$. These arrays represent the matrix $\mathbf{M}$ stored in a row-wise fashion, i.e., left to right, and top to bottom. The data in the matrix is stored in the \lstinline|values| array, while the \lstinline{columnIndex} array contains the horizontal location of the data in the matrix.  If \lstinline|values[k]| represents $M_{ij}$ then \lstinline|colIndex[k]| = j. The array \lstinline{rowPtr} has size $n+1$ for an $n$-row matrix.  \lstinline|rowPtr[k]| contains the total number of elements in all the rows in the matrix prior to row $k$, with the first element \lstinline|rowPtr[0] = 0| and the last element \lstinline|rowPtr[n]| always giving the total number of non-zero elements the matrix.  As a result, if \lstinline|values[k]| represents $M_{ij}$, then \lstinline|rowPtr[i]| $\leq k < $\lstinline|rowPtr[i+1]|.  If row \lstinline|k| contains any non-zero elements, then \lstinline|rowPtr[k]| will contain the index of the first element in the row.  Note that if there are rows in the matrix without a non-zero element, then values in the \lstinline|rowPtr| array will repeat.

Looking at Figure~\ref{fig:crs} a), we can scan the matrix in row-major order to determine the \lstinline{values} array in \gls{crs} form.  Whenever we find a non-zero element, its value is stored at the next available index $i$ in the \lstinline|values| array and its column is stored at \lstinline{columnIndex[i]}.  In addition, whenever we start scanning a new row, we store the next available index $i$ in the \lstinline|rowPtr| array.  As a result, the first element in the \lstinline|rowPtr| array is always zero. 
Looking at Figure~\ref{fig:crs} b), we can also convert the matrix back to a two-dimensional array representation.  The first step is to determine the number of elements in each row of the matrix from the \lstinline|rowPtr| array.  The number of elements in row $i$ is the difference \lstinline|rowPtr[i]-rowPtr[i+1]|.   Then the row can be reconstructed by iterating through the \lstinline|values| array starting at \lstinline|values[rowPtr[i]]|. In our example matrix, the because the first two elements of the \lstinline|rowPtr| array are $0$ and $2$, then we know that there are 2 elements in the first row, i.e., \lstinline|values[0]| and \lstinline|values[1]|.  The first non-zero element in the \lstinline{values} data structure, \lstinline|values[0]|, is $3$.  This value is in column 0, since \lstinline{columnIndex[0]} = 0. Similarly, the second non-zero value is the value 4 in column 1. The second row of the matrix has elements with $k \in [2,4)$,  the third row has elements with $k \in [4,7)$, and so on.  In this case, there are 9 non-zero entries, thus that last entry in the \lstinline{rowPtr} data structure is 9. 
 
\begin{exercise}
Given a 2-dimensional array representing a matrix, write the C code to convert the matrix to \gls{crs} form.  Write the corresponding C code to convert the matrix in \gls{crs} form back to a 2-dimensional array.
\end{exercise}

It turns out that using the \gls{crs} form, we can multiply a sparse matrix with a vector relatively efficiently without explicitly converting the matrix back to a 2-dimensional array.  In fact, for large matrices with a small number of non-zero elements, sparse matrix-vector multiply is much more efficient than the dense matrix-vector multiply we discussed in chapter \ref{chapter:dft}.  This is because we can compute the non-zero elements of the result by only looking at the non-zero elements of the operands. 

\section{Baseline Implementation}

\begin{figure}
\begin{lstlisting}
#include "spmv.h"

void spmv(int rowPtr[NUM_ROWS+1], int columnIndex[NNZ],
		DTYPE values[NNZ], DTYPE y[SIZE], DTYPE x[SIZE])
{
L1: for (int i = 0; i < NUM_ROWS; i++) {
		DTYPE y0 = 0;
	L2: for (int k = rowPtr[i]; k < rowPtr[i+1]; k++) {
#pragma HLS unroll factor=8
#pragma HLS pipeline
			y0 += values[k] * x[columnIndex[k]];
		}
		y[i] = y0;
	}
}
\end{lstlisting}
\caption{  The baseline code for sparse matrix vector (SpMV) multiplication, which performs the operation $y = \mathbf{M} \cdot x$. The variables \lstinline{rowPtr}, \lstinline{columnIndex}, and \lstinline{values}  hold $\mathbf{M}$ in \gls{crs} format. The first \lstinline{for} loop iterates across the rows while the second nested \lstinline{for} loop iterates across the columns of $\mathbf{M}$ by multiplying each non-zero element by the corresponding element in the vector \lstinline{x} which results in one element in the resulting vector \lstinline{y}. }
\label{fig:spmv_arch1}
\end{figure}

\begin{figure}
\begin{lstlisting}
#ifndef __SPMV_H__
#define __SPMV_H__

const static int SIZE = 4; // SIZE of square matrix
const static int NNZ = 9; //Number of non-zero elements
const static int NUM_ROWS = 4;// SIZE;
typedef float DTYPE;
void spmv(int rowPtr[NUM_ROWS+1], int columnIndex[NNZ],
		  DTYPE values[NNZ], DTYPE y[SIZE], DTYPE x[SIZE]);

#endif // __MATRIXMUL_H__ not defined
\end{lstlisting}
\caption{ The header file for \lstinline{spmv} function and testbench.  }
\label{fig:spmv.h}
\end{figure}

Figure \ref{fig:spmv_arch1} provides a baseline code for sparse matrix vector multiplication. The \lstinline{spmv} function has five arguments. The arguments \lstinline{rowPtr}, \lstinline{columnIndex}, and \lstinline{values} correspond to the input matrix $\mathbf{M}$ in \gls{crs} format. These are equivalent to the data structures shown in Figure \ref{fig:crs}. The argument \lstinline{y} holds the output result $y$ and the argument \lstinline{x} holds the input vector $x$ to be multiplied by the matrix.  The variable \lstinline{NUM_ROWS} indicates the number of rows in the matrix $\mathbf{M}$. The variable \lstinline{NNZ} is the number of non-zero elements in the matrix $\mathbf{M}$. Finally, the variable \lstinline{SIZE} is the number of elements in the arrays \lstinline{x} and \lstinline{y}.

The outer \lstinline{for} loop, labeled \lstinline{L1}, iterates across each row of the matrix. Multiplying this row of the matrix with the vector $x$ will produce one element of $y$.  The inner loop labeled \lstinline{L2} loop across the elements in the columns of the matrix $\mathbf{M}$. The \lstinline{L2} loop iterates \lstinline{rowPtr[i+1]} $-$ \lstinline{rowPtr[i]} times, corresponding to the number of non-zero entries in that row. For each entry, we read the value of the non-zero element of the $\mathbf{M}$ matrix from the \lstinline{values} array and multiply it by the corresponding value of the vector $x$ read from the \lstinline{x} array. That value is located at \lstinline{columnIndex[k]} since the data structure \lstinline{columnIndex} holds the column for the value \lstinline{k}. 

\begin{figure}
\begin{lstlisting}[format=none]
#include "spmv.h"
#include <stdio.h>

void matrixvector(DTYPE A[SIZE][SIZE], DTYPE *y, DTYPE *x)
{
	for (int i = 0; i < SIZE; i++) {
		DTYPE y0 = 0;
		for (int j = 0; j < SIZE; j++)
			y0 += A[i][j] * x[j];
		y[i] = y0;
	}
}

int main(){
	int fail = 0;
	DTYPE M[SIZE][SIZE] = {{3,4,0,0},{0,5,9,0},{2,0,3,1},{0,4,0,6}};
	DTYPE x[SIZE] = {1,2,3,4};
	DTYPE y_sw[SIZE];
	DTYPE values[] = {3,4,5,9,2,3,1,4,6};
	int columnIndex[] = {0,1,1,2,0,2,3,1,3};
	int rowPtr[] = {0,2,4,7,9};
	DTYPE y[SIZE];

	spmv(rowPtr, columnIndex, values, y, x);
	matrixvector(M, y_sw, x);

	for(int i = 0; i < SIZE; i++)
		if(y_sw[i] != y[i])
			fail = 1;

	if(fail == 1)
		printf("FAILED\n");
	else
		printf("PASS\n");

	return fail;
}
\end{lstlisting}
\caption{  A simple testbench for our \lstinline{spmv} function. The testbench generates one example and computes the matrix vector multiplication using a sparse (\lstinline{spmv}) and non-sparse function (\lstinline{matrixvector}).}
\label{fig:spmv_test}
\end{figure}

\section{Testbench}

Figure \ref{fig:spmv_test} shows a simple testbench for the \lstinline{spmv} function. The testbench starts by defining the \lstinline{matrixvector} function. This is a straightforward implementation of matrix vector multiplication. This does not assume a sparse matrix and does not use the \gls{crs} format. We will compare the output results from this function with the results from our \lstinline{spmv} function. 

\begin{aside}
A common testbench will implement a ``golden'' reference implementation of the function that the designer wishes to synthesis. The testbench will then compare the results of the golden reference with those generated from the code that is synthesized by the \VHLS code. A best practice for the testbench is to use alternative implementations for the golden reference and the synthesizable code. This provides more assurance that both implementations are correct. 
\end{aside}

The testbench continues in the \lstinline{main} function. Here we set the \lstinline{fail} variable equal to $0$ (later code sets this to $1$ if the output data from \lstinline{spmv} does not match that from the function \lstinline{matrixvector}). Then we define a set of variables that correspond to the matrix $\mathbf{M}$, the input vector $x$ and the output vector $y$. In case of $\mathbf{M}$, we have both the ``normal'' form and the CSR form (stored in the variables \lstinline{values}, \lstinline{columnIndex}, and \lstinline{rowPtr}). The values of the $\mathbf{M}$ matrix are the same as shown in Figure \ref{fig:crs}. We have two versions of the output vector $y$. The \lstinline{y_sw} array stores the output from the function \lstinline{matrixvector} and the \lstinline{y} array has the output from the function \lstinline{spmv}. 

After defining all of the input and output variables, we call the \lstinline{spmv} and \lstinline{matrixvector} functions using the appropriate data. The following \lstinline{for} loop compares the output results from both of the functions by comparing the elements from \lstinline{y_sw} with those in \lstinline{y}. If any of them are different, we set the \lstinline{fail} flag equal to $1$. Lastly, we print out the results of the test and then return the \lstinline{fail} variable. 

This testbench is relatively simple and probably insufficient to ensure that the implementation is correct.  Primarily, it only tests one example matrix, whereas a better testbench would test multiple matrices.  It is common, for instance, to randomly generate inputs for testing, in addition to explicitly verifying important corner-cases.   In this case, we need to make sure to vary not only the values being operated on, which which will be passed to our accelerator when it is executing, but also to vary the compile-time parameters which might be used to create different accelerators with different tradeoffs.  The key difference is that we can randomly generate multiple data values to operate on and test them all in the same execution of the program, using multiple function calls.   Compile-time parameters, on the other hand, require the code to be recompiled every time parameters change.

\begin{exercise}
Create a more sophisticated testbench which generates multiple sets of test data using a random number generator. The compile-time parameters of the sparse matrix should be modifiable (e.g., \lstinline{SIZE}, \lstinline{NNZ}, etc.).  Create an HLS synthesis script which executes the same code multiple times for different reasonable compile-time parameters.
\end{exercise}

\section{Specifying Loop Properties}

If you directly synthesize this code, you will get results for the clock period and utilization. However, you will not get the number of clock cycles either in terms of task latency or initiation interval. This is because this depends upon the input data, which is external to the \lstinline{spmv} function itself.  Primarily, the performance depends on the number of times the body of the inner loop is executed, which is equal to the number of non-zero elements in $\mathbf{M}$.  We know that the number of non-zero elements is limited by the constant \lstinline{NNZ} in the code, but it is possible to call the code with matrices of different sizes, so the actual number of iterations is data-dependent. In addition, the performance may vary depending on the location of the non-zero elements and the optimization directives utilized during synthesis. To make matters worse, the number of iterations depends on the input in a complex way and many potential inputs don't actually represent valid matrices. Thus, it is very difficult for a tool to determine the total number of clock cycles for the \lstinline{spmv} function without complex analysis and additional information. \VHLS is unable to perform this analysis.

\begin{exercise}
What are the preconditions for the spmv function to work correctly?  Prove that given these preconditions, the body of the inner loop does, in fact, execute exactly once for each non-zero element in the matrix.
\end{exercise}

There are several ways to leverage the tool to derive some performance estimates, however. One method is to provide the \VHLS tool additional information about the loop bounds. This can be done using the \lstinline{loop_tripcount} directive, which enables the designer to specify a minimum, maximum, and/or average number of iterations for each particular loop. By providing these values, the \VHLS tool is capable of providing an estimate on the number of clock cycles. 

\begin{aside}
Use the \lstinline{loop_tripcount} directive to specify minimum, maximum, and/or average number of iterations for a loop with a variable bound. This enables the \VHLS tool to provide an estimate on the number of clock cycles for the design. This does not impact the results of the synthesis; it only effects the synthesis report.
\end{aside}

\begin{exercise}
Add a \lstinline{loop_tripcount} directive to the \lstinline{spmv} function. The syntax for the pragma form of the directive is \lstinline{#pragma HLS loop_tripcount min=X, max=Y, avg=Z} where \lstinline{X}, \lstinline{Y}, and \lstinline{Z} are constant positive integers. Which loops require this directive? What happens to the synthesis report when you change the different parameters (\lstinline{min}, \lstinline{max}, and \lstinline{avg})? How does this effect the clock period? How does it change the utilization results?
\end{exercise}

The \lstinline{loop_tripcount} directive enables the designer to get a rough idea about the performance of a function. This can enable comparison between different implementations of the same function either by applying different optimization directives or by restructuring the code itself.  However, it may be difficult or impossible to determine the \lstinline{min}, \lstinline{max}, and \lstinline{avg} parameters.  It can also be difficult to provide tight bounds on the \lstinline{min} and \lstinline{max} parameters. If there is a testbench, there is another more accurate method to calculate the total number of clock cycles required for the \lstinline{spmv} function. This is done by performing C/RTL cosimulation. 

\section{C/RTL Cosimulation}

C/RTL cosimulation performs automatic testing of the \gls{rtl} designs that are generated by the \VHLS tool. It does this by executing the synthesized code together with the provided testbench. The execution is instrumented to record the input and output values for each execution of the synthesized code. The input values are converted to cycle-by-cycle \term{input vectors}.  The input vectors are used in an RTL-level simulation of the generated RTL design and the resulting \term{output vectors} are captured.  The testbench code can then be executed again replacing the synthesized code with the captured input and output values.  The testbench code can then return a zero value (indicating success) or a non-zero value (indicating failure).

The C/RTL cosimulation flow combines the cycle-accurate RTL design generated from the \VHLS tool with input values provided from the C testbench. As a result, it can generate accurate estimates of the performance of the generated RTL design which reflect any HLS optimizations, even in the presence of data-dependent behavior.  The minimum, maximum, and average latency and interval of the synthesized function are automatically extracted after simulation completes.

Note that these numbers only correspond to the clock cycles derived from the input data used by the testbench. Thus, they are only as good as the testbench itself. To put it another way, if the testbench does not exercise the function in a manner that is consistent with how it will be used upon deployment, the results will not be accurate. In addition, the input testvectors are generated with idealized timing that does not accurately model the behavior of external interfaces.  The actual performance may be lower if execution stalls waiting for input data, or if there is contention waiting for external memory access to complete.  Nevertheless, it provides a convenient method for determining clock cycles that does not require the designer to estimate the loop bounds for a variable loop.

\begin{aside}
C/RTL cosimulation provides the latency for functions with variable loop bounds. It reports the minimum, maximum, and average clock cycles for function latency and function interval. These latency values are directly dependent upon the input data from the C testbench. 
\end{aside}

\begin{exercise}
What are the minimum, maximum, and average clock cycles for the \lstinline{spmv} function latency and function interval when using the testbench provided in Figure \ref{fig:spmv_test}?
\end{exercise}

\section{Loop Optimizations and Array Partitioning}

Now that we have a method to gather all of the performance and utilization estimates from the \VHLS tool, let us consider how to best optimize the function. Pipelining, loop unrolling, and data partitioning are the most common first approaches in optimizing a design. And the typical approach is to start with the innermost loop, and then move outwards as necessary.

In this example, pipelining the inner \lstinline{L2} loop is perhaps the first and easiest optimization to consider. This overlaps the execution of the consecutive iterations of this loop, which can result is a faster overall implementation. Without pipelining, each iteration of the \lstinline{L2} loop occurs sequentially. Note that the iterations of the \lstinline{L1} loop are still done sequentially. 

\begin{figure}
\centering
\executeiffilenewer{spmv_behavior.svg}{images/spmv_behavior.pdf}%
{inkscape -z -D --file=spmv_behavior.svg %
--export-pdf=images/spmv_behavior.pdf --export-latex}%
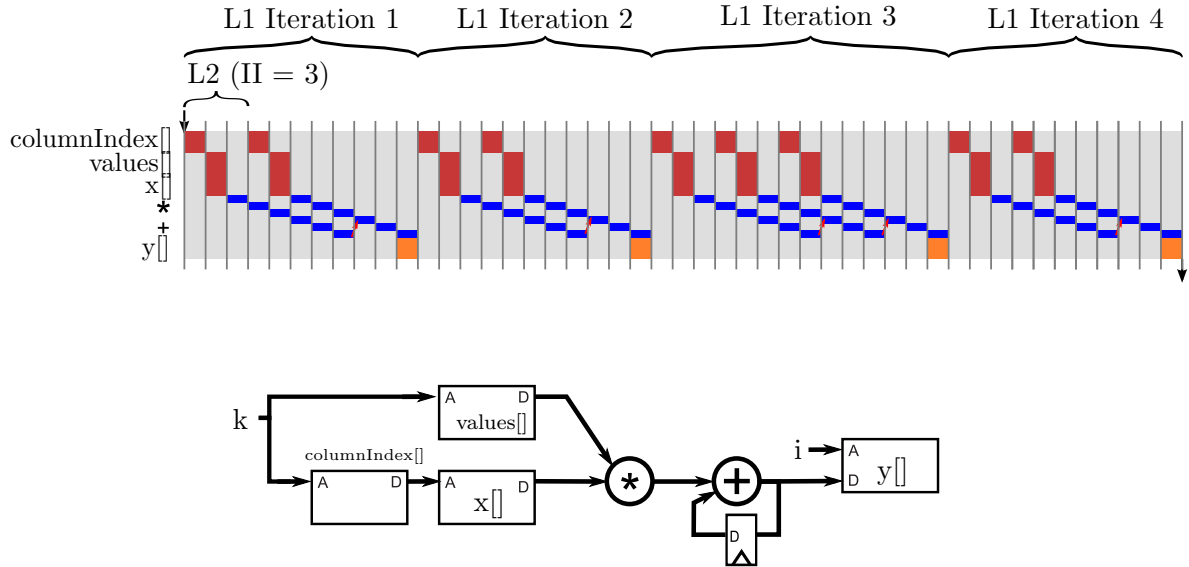%

\caption{Architecture and behavior of the \lstinline{spmv} code with a pipelined inner loop.}
\label{fig:spmv_pipeline_inner}
\end{figure}

Figure \ref{fig:spmv_pipeline_inner} illustrates the approximate manner in which the \lstinline{spmv} function executes when pipelining the \lstinline{L2 for} loop. Each iteration of the inner \lstinline{L2} loop is pipelined with II=3.  Pipelining allows multiple iterations of the inner loop from the same iteration of the outer loop execute concurrently.  In this case, the II of the inner loop is limited by a recurrence through the accumulation.  II=3 is achieved because we've assumed that the adder has a latency of 3 clock cycles.  Iterations of the outer loop are not pipelined, so the inner loop must completely finish and flush the pipeline before the next iteration of the outer \lstinline{L2} loop begins.

\begin{exercise}
Pipeline the innermost \lstinline{L2 for} loop. This can be done by adding a pipeline directive to the \lstinline{spmv} code from Figure \ref{fig:spmv_arch1}. What is the achieved initiation interval (II)? What happens to the results as you specify an II argument, and increase or decrease the target II? 
\end{exercise}

Looking at this behavior, we see that there are several factors limiting the performance of the loop.  One factor is the recurrence through the adder that limits the achieved loop II.  A second factor is that iterations of the outer loop are not pipelined.  An efficient solution for sparse matrix-vector multiply would likely come close to using each multiplier and adder every clock cycle.  This design is far from that.

In Section \ref{subsec:mvmul_implementation} we explored several design optimization techniques, including pipelining different loops, loop unrolling, and array partitioning.  Understanding the tradeoffs between these techniques can be somewhat challenging, since they are often dependent on what another.  We must often apply these techniques together with a carefully chosen goal in mind in order to get a benefit and applying one technique without applying another technique can actually make matters worse.  For instance, when performing loop unrolling, the designer must be careful to understand the effects that this has upon memory accesses. Increasing the number of operations that can execute concurrently doesn't help if performance is limited by available memory ports.  Similarly, providing more memory ports if there are insufficient operations to utilize those memory ports (or if the addresses of each memory operation can't be easily partitioned) can also incur a resource cost without increasing performance.

To see some of the complexity in applying these combinations of transforms, we encourage you to perform the following exercise:

\begin{exercise}
Synthesize the \lstinline{spmv} design using the directives specified in each of the ten cases from Table \ref{table:spmv_optimizations}. Each case has different pipeline, unroll, and partitioning directives for the different loops and arrays. These partitionings should be done across the three arrays (\lstinline{values}, \lstinline{columnIndex}, and \lstinline{x}). What sort of trends do you see? Does increasing the unroll and partitioning factors help or hurt when it comes to utilization? How about performance? Why?
\end{exercise}

\begin{table}[htbp]
  \centering
  \caption{Potential optimizations for sparse matrix-vector multiplication. }
	\begin{tabular}{*{4}{l}}
    \toprule
						& L1 		& 		L2 							 \\
    \midrule
    Case 1 & - 			& -  	 \\ 
    Case 2 & - 			& pipeline 	\\ 
    Case 3 & pipeline & - \\
    Case 4 & unroll=2 & - \\ 
    Case 5 & - 			& pipeline, unroll=2 	\\
    Case 6 & - 			& pipeline, unroll=2, cyclic=2 	\\ 
    Case 7 & - 			& pipeline, unroll=4		\\ 
    Case 8 & - 			& pipeline, unroll=4, cyclic=4   \\ 
    Case 9 & - 			& pipeline, unroll=8		\\ 
    Case 10 & - 			& pipeline, unroll=8, cyclic=8   \\ 
    Case 11 & - 			& pipeline, unroll=8, block=8   \\ 
    \bottomrule
  \end{tabular}
  \label{table:spmv_optimizations}
\end{table}

If you performed the previous exercise, you should have seen that blindly applying optimization directives may not always provide you with the expected results. It is usually more effective to consider the properties of an application under design, and to select optimizations with a particular design goal in mind. Of course, this requires some intuition behind the capabilities and limitations of a particular tool being used.  While it is certainly difficult (perhaps impossible?) to understand every detail of a complex tool like \VHLS, we can build a mental model of the most critical aspects.

One of the options we considered in cases 3 and 4 above was to increase pipeline outer loops, such as the \lstinline{L1} loop in this code, rather than inner loops.  This transformation has the effect of increasing the potential parallelism within one task.  In order to perform this optimization, the \VHLS tool must fully unroll inner loops, like the \lstinline{L2} loop in this code. If full unrolling is possible, this can reduce the cost of calculating the loop bounds and can also eliminate recurrences in the code.  However, in this code, the inner loop cannot be unrolled by \VHLS because the loop bound is not constant.

\begin{exercise}
Add a directive to pipeline the outermost \lstinline{L1} loop, i.e., implement case 3 above. What is the initiation interval (II) when you do not set a target II? What happens to the utilization? How does explicitly increasing the II change the utilization results? How does this compare to pipelining the \lstinline{L2} loop? How does this compare to the baseline design (no directives)? What is happening when you attempt to pipeline this outer loop? (hint: check the synthesis log)
\end{exercise}

Another option to increase parallelism is \gls{partial_loop_unrolling} of the inner loop, as in cases 5 through 10.  This transformation exposes more parallelism by allowing more operations from the same loop iteration to be executed concurrently.   In some cases, more operations can increase performance by enabling \VHLS to instantiate more operators when pipelining the inner loop.  However, in this case it is still difficult to improve the II of the inner loop because of the recurrence through the inner loop.  However, in this case, because we have an II greater than 1, many of those operations can be shared on the same operators. 

An partially unrolled version of the code is shown in Figure \ref{fig:spmv_unrolled}.  In this code, the L2 loop has been split into two loops labeled \lstinline|L2_1| and \lstinline|L2_2|.  The innermost \lstinline|L2_2| executes a parameterized number of times, given by the compile-time parameter \lstinline|S|.  The body of the inner loop contains the body of the original \lstinline|L2| loop, along with a condition that arises from the loop bound of the original \lstinline|L2| loop.  In this code, we now have an arbitrary number of multiply and add operations to execute in the body of the \lstinline|L2_1| loop, given by the parameter \lstinline|S|, and a single recurrence through the accumulation \lstinline|y0 += yt|.

\begin{figure}
\centering
\executeiffilenewer{spmv_partial_unroll.svg}{images/spmv_partial_unroll.pdf}%
{inkscape -z -D --file=spmv_partial_unroll.svg %
--export-pdf=images/spmv_partial_unroll.pdf --export-latex}%
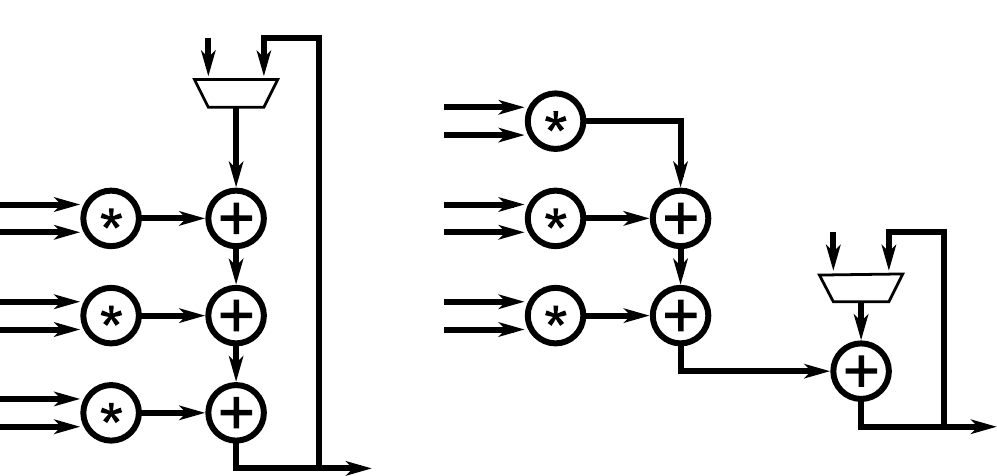%

\caption{Two different partially unrolled versions of an accumulation.  The version on the left has a recurrence with three additions, whereas the version on right only has one addition in the recurrence.}
\label{fig:spmv_partial_unroll}
\end{figure}

Note that the code in Figure \ref{fig:spmv_unrolled} is slightly different from the code that is generated from automatic loop unrolling.  Automatic loop unrolling duplicates operations, but must also preserve the order of each operation (additions in this case).  This results in a long chain of operation dependencies in the inner loop shown on the left side of Figure \ref{fig:spmv_partial_unroll}.  Reordering the operations results in operation dependencies show on the right side of the figure.  In this case, only the final accumulation results in a recurrence.   When using floating-point data types, this reordering of operations can slightly change the behavior of the program, so \VHLS does not apply this kind of operation reordering automatically.  

\begin{figure}
\begin{lstlisting}
#include "spmv.h"

const static int S = 7;

void spmv(int rowPtr[NUM_ROWS+1], int columnIndex[NNZ],
          DTYPE values[NNZ], DTYPE y[SIZE], DTYPE x[SIZE])
{
  L1: for (int i = 0; i < NUM_ROWS; i++) {
	  DTYPE y0 = 0;
    L2_1: for (int k = rowPtr[i]; k < rowPtr[i+1]; k += S) {
#pragma HLS pipeline II=S
		  DTYPE yt = values[k] * x[columnIndex[k]];
	  L2_2: for(int j = 1; j < S; j++) {
			  if(k+j < rowPtr[i+1]) {
				  yt += values[k+j] * x[columnIndex[k+j]];
			  }
		  }
		  y0 += yt;
	  }
    y[i] = y0;
  }
}
\end{lstlisting}
\caption{A partially unrolled version of the \lstinline|spmv| code from Figure \ref{fig:spmv_arch1}.}
\label{fig:spmv_unrolled}
\end{figure}

\begin{figure}
\centering
\executeiffilenewer{spmv_unrolled_behavior.svg}{images/spmv_unrolled_behavior.pdf}%
{inkscape -z -D --file=spmv_unrolled_behavior.svg %
--export-pdf=images/spmv_unrolled_behavior.pdf --export-latex}%
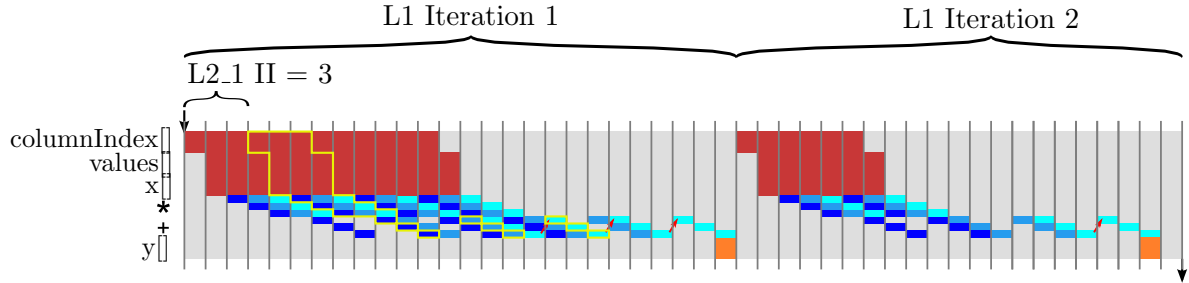%

\caption{Architecture and behavior of the \lstinline{spmv} code based on the partially unrolled and pipelined inner loop shown in Figure \ref{fig:spmv_unrolled}.}
\label{fig:spmv_unrolled_behavior}
\end{figure}

A possible implementation of this design is shown in Figure \ref{fig:spmv_unrolled_behavior}.  In this case, \lstinline|S = 3| to match the best achievable II where there is a latency of 3 through the adder.  In this case, we see that all the operations have been successfully shared on a single multiplier and adder.  Comparing this behavior to the behavior in Figure \ref{fig:spmv_pipeline_inner}, we see that there are some disadvantages.  In particular, the depth of the pipeline of the inner loop is much longer, which implies that the number of cycles to flush the pipeline to start a new iterations of the outer \lstinline|L1| loop is much larger.  Processing of the non-zero elements in a row also occurs in blocks of size \lstinline|S|.  A row with 3 elements takes exactly the same time to compute as a row with one element.  The remaining operations which are still scheduled in the loop pipeline must still `execute' even though their results are discarded.  In order to rigorously compare the characteristics of the two designs, we need to understand the expected number of non-zero elements in each row of the matrix.  Fewer non-zero elements in each line would favor the first implementation, while more non-zero elements in each line would favor the second implementation.

Notice that there is, to some extent a chicken-and-egg problem here.   We need to know the target device and clock period to determine the number of pipeline stages required for the adder to meet timing.  Only after we know the number of pipeline stages (perhaps by running with \lstinline|S=1| and investigating the \VHLS logs to identify the adder recurrence) can we select an appropriate version of the parameter \lstinline|S| that achieves II=1.  Once we've determined \lstinline|S|, we can run \gls{cosimulation} to determine the achieved performance on a set of benchmark test data.  Because of the variable loop bounds, the achieved performance is data-dependent so we might have to explore different values of \lstinline|S| to determine the value that maximizes performance.  Changing the target device or clock period might affect all of these decisions!  Although it may seem like high-level synthesis provides little assistance in solving this problem, it's still much faster (and possible to easily script) compared to evaluating each new version with a new \gls{rtl} design that must be verified!  

\begin{exercise}
The behavior in Figure \ref{fig:spmv_unrolled_behavior} is achieved when \lstinline|S| is the same as the number of pipeline stages for the adder.   What happens to the behavior when \lstinline|S| is set larger?  What happens to the behavior when is it set smaller?  What happens when the target II is smaller than \lstinline|S|?  What happens when the target II is larger?
\end{exercise}

\section{Conclusion}
In this chapter, we looked at sparse matrix-vector multiplication (SpMV). This continues our study of matrix operations. This operation is particularly interesting because it uses a unique data structure. In order to reduce the amount of storage, the matrix is stored in a compressed row storage format. This requires a design that uses some indirect references to find the appropriate entry in the matrix. 

This chapter is the first to discuss at length the testing and simulation abilities of the \VHLS tool. We provide a simple testbench for SpMV and describe how it can be integrated into the HLS work-flow. Additionally, we describe the C/RTL cosimulation features of the \VHLS tool. This is particularly important for us in order to get precise performance results. The task interval and task latency depends upon the input data. The less sparse the matrix, the more computation that must be performed. The cosimulation provides a precise trace of execution using the given testbench. This allows the tool to compute the clock cycles to include in the performance results.  Finally, we discuss optimizing the code using loop optimizations and array partitioning. 

\chapter{Matrix Multiplication}
\glsresetall
\label{chapter:matrix_multiplication}

This chapter looks at a bit more complex design -- matrix multiplication. We consider two different versions. We start with a ``straightforward'' implementation, i.e., one that takes two matrices as inputs and outputs the result of their multiplication. We call this complete matrix multiplication. Then, we look at a block matrix multiplication. Here the input matrices are feed into the function in portions, and the function computes partial results.

\section{Background}

Matrix multiplication is a binary operation that combines two matrices into a third. The operation itself can be described as a linear operation on the vectors that compose the two matrices. The most common form of matrix multiplication is call the \term{matrix product}. The matrix product $\mathbf{AB}$ creates an $n \times p$ matrix when matrix $\mathbf{A}$ has dimensions $n \times m$ and matrix $\mathbf{B}$ has dimensions $m \times p$. 

More precisely, we define the following: \begin{equation}
\mathbf{A} =
 \begin{bmatrix}
  A_{11} & A_{12}  & \cdots & A_{1m} \\
  A_{21} & A_{22}  & \cdots & A_{2m} \\
  \vdots  & \vdots  &\ddots & \vdots  \\
    A_{n1} & A_{n2}  & \cdots & A_{nm} \\
 \end{bmatrix},  \quad
\mathbf{B} =
 \begin{bmatrix}
  B_{11} & B_{12}  & \cdots & B_{1p} \\
  B_{21} & B_{22}  & \cdots & B_{2p} \\
  \vdots  & \vdots   &\ddots & \vdots  \\
  B_{m1} & B_{m2}  & \cdots & B_{mp} \\
 \end{bmatrix}
\end{equation} 

\begin{equation}
\mathbf{AB} = \begin{bmatrix}
 \left(\mathbf{AB}\right)_{11} & \left(\mathbf{AB}\right)_{12} & \cdots & \left(\mathbf{AB}\right)_{1p} \\
 \left(\mathbf{AB}\right)_{21} & \left(\mathbf{AB}\right)_{22} & \cdots & \left(\mathbf{AB}\right)_{2p} \\
\vdots & \vdots & \ddots & \vdots \\
 \left(\mathbf{AB}\right)_{n1} & \left(\mathbf{AB}\right)_{n2} & \cdots & \left(\mathbf{AB}\right)_{np} \\
\end{bmatrix}
\end{equation} where the operation $\left(\mathbf{AB}\right)_{ij}$ is defined as $\left(\mathbf{A}\mathbf{B}\right)_{ij} = \sum_{k=1}^m A_{ik}B_{kj}$.

Now we provide a simple example. Let
\begin{equation}
\mathbf{A} =
 \begin{bmatrix}
 \label{eq:ABmatrix}
  A_{11} & A_{12}  &  A_{13} \\
  A_{21} & A_{22} & A_{23} \\
 \end{bmatrix},  \quad
\mathbf{B} =
 \begin{bmatrix}
  B_{11} & B_{12}  \\
  B_{21} & B_{22}   \\
   B_{31} & B_{32}  \\
 \end{bmatrix}
\end{equation}
The result of the matrix product is
\begin{equation}\label{eq:ABmatrix_product}\mathbf{AB} = \begin{bmatrix}
 A_{11}B_{11} + A_{12}B_{21} + A_{13}B_{31} & A_{11}B_{12} + A_{12}B_{22} + A_{13}B_{32} \\
 A_{21}B_{11} + A_{22}B_{21} + A_{23}B_{31} & A_{21}B_{12} + A_{22}B_{22} + A_{23}B_{32}\\
\end{bmatrix}
\end{equation}

Matrix multiplication is a fundamental operation in numerical algorithms. Computing the product between large matrices can take a significant amount of time. Therefore, it is critically important part of many of problems in numerical computing. Fundamentally, matrices represent linear transforms between vector spaces; matrix multiplication provides way to compose the linear transforms.  Applications include linearly changing coordinates (e.g., translation, rotation in graphics), high dimensional problems in statistical physics (e.g., transfer-matrix method), and graph operations (e.g., determining if a path exists from one vertex to another). Thus it is a well studied problem, and there are many algorithms that aim to increase its performance, and reduce the memory usage.

\begin{figure}
\begin{lstlisting}[firstline=5]

void matrixmul(int A[N][M], int B[M][P], int AB[N][P]) {
  #pragma HLS ARRAY_RESHAPE variable=A complete dim=2
  #pragma HLS ARRAY_RESHAPE variable=B complete dim=1
  /* for each row and column of AB */
  row: for(int i = 0; i < N; ++i) {
    col: for(int j = 0; j < P; ++j) {
      #pragma HLS PIPELINE II=1
      /* compute (AB)i,j */
      int ABij = 0;
    product: for(int k = 0; k < M; ++k) {
        ABij += A[i][k] * B[k][j];
      }
      AB[i][j] = ABij;
    }
  }
}
\end{lstlisting}
\caption{A common three \lstinline{for} loop structure for matrix multiplication. The outer \lstinline{for} loops, labeled \lstinline{rows} and \lstinline{cols}, iterate across the rows and columns of the output matrix $\mathbf{AB}$.  The innermost loop, labeled \lstinline{product} multiplies the appropriate elements of one row of $\mathbf{A}$ and one column of $\mathbf{B}$ and accumulates them until it has the result for the element in $\mathbf{AB}$ .  }
\label{fig:matrixmultiplication_sw}
\end{figure}

\section{Complete Matrix Multiplication}

We start our optimization process with perhaps the most common method to compute a matrix multiplication -- using three nested \lstinline{for} loops. Figure \ref{fig:matrixmultiplication_sw} provides the code for such an implementation.  The outer \lstinline{for} loops, labeled \lstinline{rows} and \lstinline{cols}, iterate across the rows and columns of the output matrix $\mathbf{AB}$. The innermost \lstinline{for} loop computes a dot product of one row of $\mathbf{A}$ and one column of $\mathbf{B}$. Each dot product is a completely independent set of computations that results in one element of $\mathbf{AB}$.  Conceptually, we are performing \lstinline{P} matrix-vector multiplications, one for each column of $\mathbf{B}$.

In this case, we've applied a \lstinline{pipeline} directive to the \lstinline{col} loop with a target initiation interval of 1. The result is that the innermost \lstinline{for} loop is fully unrolled, and we expect the resulting circuit include roughly $M$ multiply-add operators and to have an interval of roughly $N*P$ cycles.  As discussed in Chapter \ref{chapter:dft}, this is only one reasonable choice.  We could choose to place the \lstinline{pipeline} directive in different locations in the function with the goal of achieving different resource-throughput tradeoffs. For instance, placing the same directive at the top of the function (outside all of the \lstinline{for} loops) will result in all of the loops being completely unrolled, which would take roughly $N*M*P$ multiply-add operators and would have an interval of 1 cycle. Placing it inside the \lstinline{row} loop would result in roughly $M*P$ multiply-add operators and an interval of roughly $N$ cycles. These design points are relatively easy to achieve, given the corresponding array partitioning.  It's also possible to pipeline the innermost loop with the goal of achieving a design with only one multiply-add operator, although achieving an II=1 implementation at high clock frequencies can be difficult because of the recurrence involved in the accumulation of variable \lstinline{ABij}.  We can also partially unroll different loops to achieve yet more design points.  The fundamental tradeoff here is between the complexity of the resulting architecture, i.e., the number of multiply-add operators, and performance, i.e., the number of cycles that the hardware is busy.  In an ideal world, each doubling of the resource usage should result in exactly half the number of clock cycles being required, although in practice such `perfect scaling' is difficult to achieve.

\begin{exercise}
Change the location of the \lstinline{pipeline} directive. How does the location effect the resource usage? How does it change the performance? Which alternative provides the best performance in terms of function interval? Which provides the smallest resource usage? Where do you think is the best place for the directive? Would increasing the size of the matrices change your decision?
\end{exercise}

\begin{figure}
\centering
\executeiffilenewer{matmul_array_reshape.svg}{images/matmul_array_reshape.pdf}%
{inkscape -z -D --file=matmul_array_reshape.svg %
--export-pdf=images/matmul_array_reshape.pdf --export-latex}%
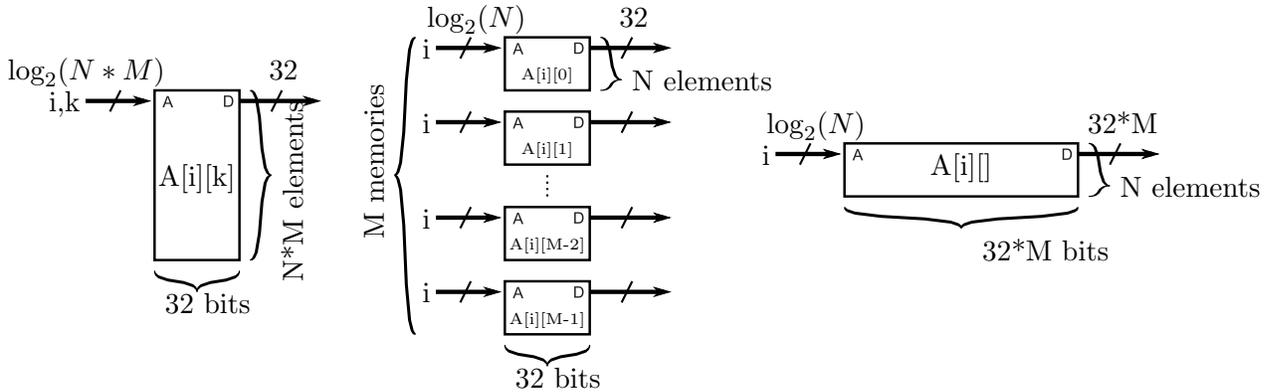%

\caption{Three different implementations of a two-dimensional array.  On the left is the original array consisting of $N*M$ elements.  In the middle, the array has been transformed using the \lstinline{array_partition} directive, resulting in $M$ memories, each with $N$ elements.  On the right, the array has been transformed using the \lstinline{array_reshape} directive, resulting in one memory with $N$ locations and each location contains $M$ elements of the original array.}
\label{fig:matmul_array_reshape}
\end{figure}

Executing large numbers of operations every cycle requires being able to supply all of the required operands and to store the results of each operation.   Previously we have used the \lstinline{array_partition} directive to increase the number of accesses that can be performed on each memory.  As long as the partition of the array that each memory access can be determined at compile time, then array partitioning is a simple and efficient way to increase the number of memory accesses that can be performed each clock cycle.  In this case, we use the slightly different \lstinline{array_reshape} directive to perform array partitioning.  This directive not only partitions the address space of the memory into separate memory blocks, but then recombines the memory blocks into a single memory.  This transformation increases the data width of the memory used to store the array, but doesn't change the overall number of bits being stored.  The difference is shown in Figure \ref{fig:matmul_array_reshape}.

Both \lstinline{array_reshape} and \lstinline{array_partition} increase the number of array elements that can be read each clock cycle.  They also support the same options, enabling \lstinline{cyclic} and \lstinline{block} partitions or partitioning along different dimensions of a multi-dimensional array.  In the case of \lstinline{array_reshape}, the elements must each have the same address in the transformed array, whereas with \lstinline{array_partition}, the addresses in the transformed array can be unrelated.  Although it may seem like one would always want to use \lstinline{array_partition} because it is more flexible, it makes each individual memory smaller, which can sometimes result in inefficient memory usage.  The \lstinline{array_reshape} directive results in larger memory blocks, which can sometimes be mapped more efficiently into primitive FPGA resources.   In particular, the smallest granularity of \gls{bram} blocks in Xilinx Virtex Ultrascale+ devices is 18 Kbits with several different supported combination of depths and widths.  When the partitions of an array become smaller than around 18Kbits, then BRAMs are no longer used efficiently.  If we start with an original array which is a 4-bit array with dimensions [1024][4], this array can fit in a single \gls{bram} resource configured as a 4Kbit x 4 memory.  Partitioning this array completely in the second dimension would result in 4 1Kbit x 4 memories, each of which are much smaller than one one \gls{bram} resource.  Reshaping the array instead using the \lstinline{array_reshape} directive results in a memory which is 1Kbit x 16, a supported \gls{bram} configuration.

\begin{aside}
Note that this partitioning is effective because the partitioned dimension (dimension 2 of $\mathbf{A}$ or dimension 1 of $\mathbf{B}$) is indexed by a constant.   In addition, the non-partitioned dimension is indexed by the same (variable) value.  When partitioning multi-dimensional arrays, this is a good rule of thumb to identify which dimension should be partitioned.
\end{aside}

\begin{exercise}
Remove the \lstinline{array_reshape} directives. How does this effect the performance? How does it change the resource usage? Does it make sense to use any other \lstinline{array_reshape} directives (with different arguments) on these arrays?  In this case, how does the result differ if you use the \lstinline{array_reshape} directive instead?
\end{exercise}

The size of the arrays can have a substantial effect on the optimizations that you wish to perform. Some applications might use very small matrices, say $2 \times 2$ or $4 \times 4$. In this cases, it may be desirable to implement a design with the absolute highest performance, which is generally achieved by applying the \lstinline{pipeline} directive on the entire function.   As the size of the arrays increase, in the range of $32 \times 32$, this approach quickly become infeasible because of the resource limits available in a single device.  There will simply not be enough DSP resources to implement that many multiplications every clock cycle or enough external bandwidth to get get data on and off the chip.  Many FPGA designs are often coupled to the data rates of other components in a system, such as an Analog to Digital (A/D) converter, or the symbol rate in a communication system.  In these designs it is common to instead apply the \lstinline|pipeline| directive on inner loops with the goal of matching the interval of the computation with the data rate in a system.  In such cases, we often need to explore different resource-throughput tradeoffs by moving the \lstinline{pipeline} directive into an inner loop or partially unrolling loops. When  
 dealing with very large matrices containing thousands or millions of elements, we often need to take into account more complex architectural considerations.  The next section discusses a common approach to scaling matrix multiply to larger designs, called \term{blocking} or \term{tiling}.

\begin{exercise}
Optimize your design for the $128 \times 128$ matrix multiplication. Then start increasing the size of the matrices by a factor of two (to $512 \times 512$, $1024 \times 1024$, $2048 \times 2048$, etc. How does this effect the resource usage and performance? How about the runtime of the tool? What is the best way to optimize for large large matrix sizes?
\end{exercise}

\section{Block Matrix Multiplication}

A \term{block matrix} is interpreted as being partitioned into different submatrices. This can be visualized by drawing different horizontal and vertical lines across the elements of the matrix. The resulting ``blocks'' can be viewed as submatrices of the original matrix.  Alternatively, we can view the original matrix as a matrix of blocks.  This naturally leads to many hierarchical algorithms in Linear Algebra where we compute matrix operations, such as matrix multiply, on large block matrices by decomposing them into smaller matrix operations on the blocks themselves.

For instance, when we talk about the matrix multiplication operation between matrix $\mathbf{A}$ and $\mathbf{B}$ in Equations \ref{eq:ABmatrix} and \ref{eq:ABmatrix_product}, we might normally think of each element of the matrices $\mathbf{A}_{11}$ or $\mathbf{B}_{23}$ as a single number or perhaps a complex number.  Alternatively, we can consider each element in these matrix operations as a block of the original matrix.  In this case, as long as the sizes of the individual blocks are compatible, we simply have to perform the correct matrix operations instead of the original scalar operations.  For instance, to compute $\mathbf{AB}_{11}$, we would need to compute two matrix products and two matrix sums to compute $\mathbf{A}_{11}\mathbf{B}_{11} + \mathbf{A}_{12}\mathbf{B}_{21} + \mathbf{A}_{13}\mathbf{B}_{31}$.

Matrix blocking turns out to be a very useful technique for a number of reasons.  One reason is that blocking is an easy way to find more structure in the algorithm that we can explore.  In fact, some of the optimizations that we have already seen as loop transformations, such as loop unrolling, can be viewed as specific simple forms of blocking.  Another reason is that we can choose to block a matrix according to the natural structure of the matrix.  If a matrix has a large block of zeros, then many individual products may be zero.  If we want to skip these individual products then this can be difficult in a statically schedule pipeline, whereas it may be easier to skip a large block of zeros.  Many matrices are \term{block-diagonal} where the blocks on the diagonal are non-zero and blocks off the diagonal are zero.  Yet another reason is that the blocked decomposition results in lots of smaller problems operating on smaller sets of data.  This increases the data locality of a computation.  In processor systems, it is common to choose block sizes that conveniently match the memory hierarchy of a processor or the natural size of the vector data types supported by the processor.  Similarly, in FPGAs we can choose the blocking sizes to match the available on-chip memory size or to the number of multiply-add operators that we can budget to support.

Until now, we have assumed that accelerators always have all of their data available before the start of a task.  However in designs dealing with large datasets, such as large matrices, this can sometimes be an unneeded constraint.  Since it is unlikely that our accelerator will be able to process all input data immediately, we can build an accelerator that receives input data only right before it is needed.  This allows an accelerator to more efficiently use the available on-chip memory.  We call this a \term{streaming architecture} since we transfer the input data (and potentially the output data) one portion at a time rather than all at once.

Streaming architectures are common in many applications.  In some cases this is because of a conscious design choice that we make to decompose a large computation into multiple smaller computations.  For instance, we may design a matrix multiplication system that reads and processes one block of data at a time from external memory.  In other cases, we might process a stream of data because the data is being sampled in real time from the physical world, for instance, from an A/D converter.  In other cases, the data we are processing may simply be created in sequence from a previous computation or accelerator.  In fact, we've already seen an instance of this in Section \ref{sec:fft_task_pipelining}.

One potential advantage of streaming is a reduction in the memory that we can use to store the input and output data. The assumption here is that we can operate on the data in portions, create partial results, and then we are done with that data, thus we do not need to store it. When the next data arrives, we can overwrite the old data resulting in smaller memories.

In the following, we develop a streaming architecture for matrix multiplication. We divide the input arrays $\mathbf{A}$ and $\mathbf{B}$ into blocks, which are a contiguous set of rows and columns, respectively. Using these blocks, we compute a portion of the product $\mathbf{AB}$. Then we stream the next set of blocks, compute another portion of $\mathbf{AB}$ until the entire matrix multiplication is complete.

\begin{figure}
\centering
\includegraphics[width=  .85\textwidth]{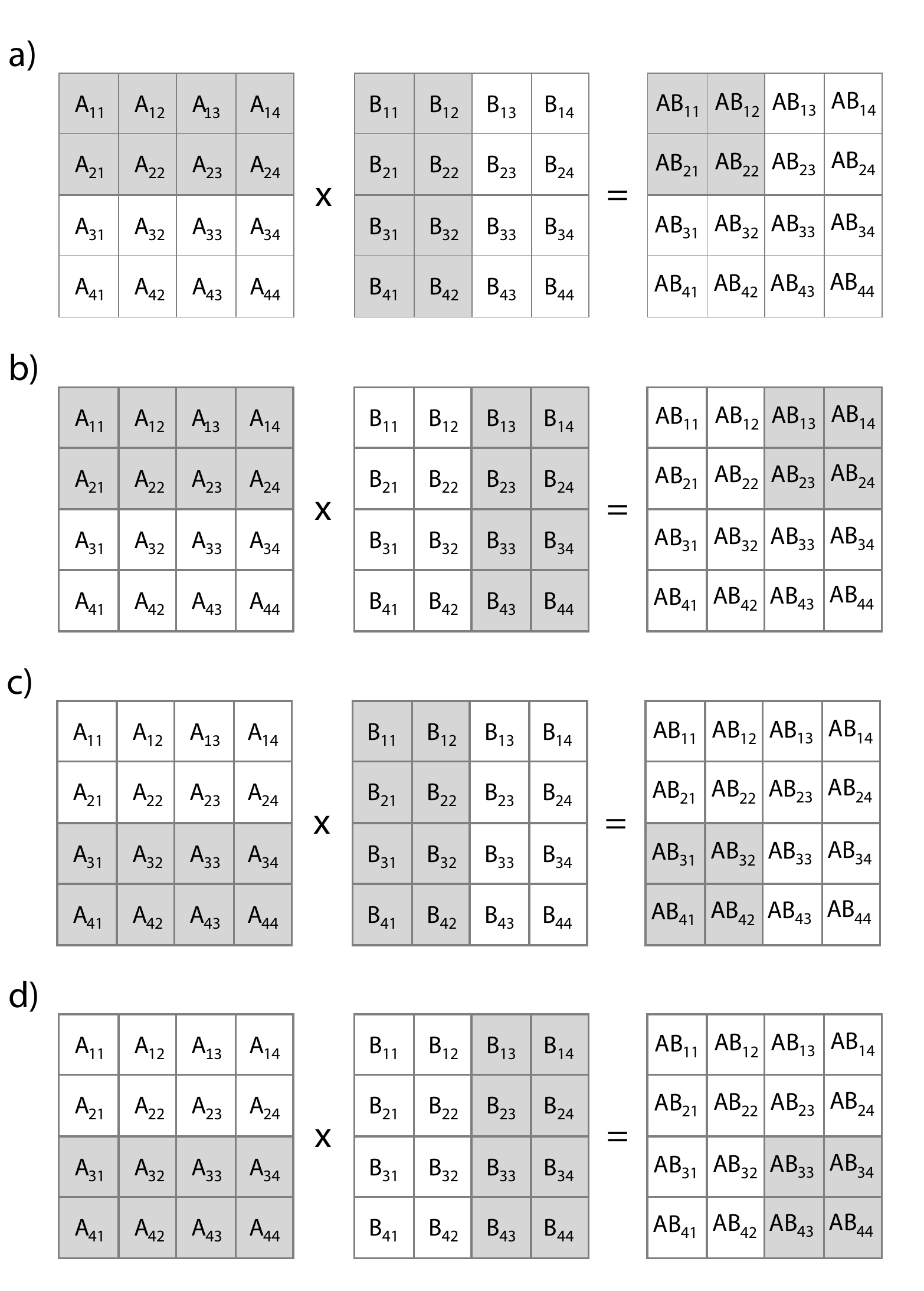}
\caption{ One possible blocked decomposition of the matrix multiplication of two $4 \times 4$ matrices. The entire $\mathbf{AB}$ product is decomposed into four matrix multiply operations operating on a $2 \times 4$ block of $\mathbf{A}$ and a $4 \times 2$ block of $\mathbf{B}$.}
\label{fig:blockmm}
\end{figure}

Figure \ref{fig:blockmm} provides the a description of the streaming architecture that we create. Our architecture has a variable \lstinline{BLOCK_SIZE} that indicates the number of rows that we take from the $\mathbf{A}$ matrix on each execution, the number of columns taken from the $\mathbf{B}$ matrix, and the \lstinline{BLOCK_SIZE} $\times$ \lstinline{BLOCK_SIZE} result matrix corresponding to the data that we compute each time for the $\mathbf{AB}$ matrix. 

The example in Figure \ref{fig:blockmm} uses a \lstinline{BLOCK_SIZE = 2}. Thus we take two rows from $\mathbf{A}$, and two columns from $\mathbf{B}$  on each execution of the streaming architecture that we define. The result of each call to the \lstinline{blockmatmul} function is a $2 \times 2$ matrix for the $\mathbf{AB}$ architecture.

Since we are dealing with $4 \times 4$ matrices in the example, we need to do this process four times. Each time we get a $2 \times 2$ set of results for the $\mathbf{AB}$ matrix. The figure shows a progression of the rows and columns that we send. In Figure \ref{fig:blockmm} a) we send the first two rows of $\mathbf{A}$ and the first two columns of $\mathbf{B}$. The function will compute a $2 \times 2$ matrix corresponding to the first two elements in the rows and columns of the resulting matrix $\mathbf{AB}$. 

In Figure \ref{fig:blockmm} b), we use again the first two rows of $\mathbf{A}$, but this time we send the last two columns of $\mathbf{B}$. We do not need to resend the data from the rows of $\mathbf{A}$ since they are the same as the previous data from the previous execution. And we get the results for the $2 \times 2$ matrix corresponding to the data in ``upper left'' corner of $\mathbf{AB}$. 

Figure \ref{fig:blockmm} c) sends different data for both the $\mathbf{A}$ and $\mathbf{B}$ matrices. This time we send the last two rows of $\mathbf{A}$ and the first two columns of $\mathbf{B}$. The results from this computation provide the ``lower left'' corner of the $\mathbf{AB}$ matrix.

The final execution of the streaming block matrix multiply, shown in Figure \ref{fig:blockmm} d), uses the same last two rows of the $\mathbf{A}$ matrix from the previous iteration. And it sends the last two columns of the $\mathbf{B}$ matrix. The result provides the elements in the ``lower right'' corner of the $\mathbf{AB}$ matrix. 

\begin{figure}
\begin{lstlisting}
#ifndef _BLOCK_MM_H_
#define _BLOCK_MM_H_
#include "hls_stream.h"
#include <iostream>
#include <iomanip>
#include <vector>
using namespace std;

typedef int DTYPE;
const int SIZE = 8;
const int BLOCK_SIZE = 4;

typedef struct { DTYPE a[BLOCK_SIZE]; } blockvec;

typedef struct { DTYPE out[BLOCK_SIZE][BLOCK_SIZE]; } blockmat;

void blockmatmul(hls::stream<blockvec> &Arows, hls::stream<blockvec> &Bcols,
								 blockmat & ABpartial, DTYPE iteration);
#endif
\end{lstlisting}
\caption{  The header file for the block matrix multiplication architecture. The file defines the data types used within the function, the key constants, and the \lstinline{blockmatmul} function interface.   }
\label{fig:block_mm_h}
\end{figure}

Before we show the code for the block matrix multiplication, we define some data types that we will use. Figure \ref{fig:block_mm_h} shows the header file for the project. We create a custom data type \lstinline{DTYPE} that specifies the type of data that we will multiply in the $\mathbf{A}$ and $\mathbf{B}$ matrices, and the corresponding $\mathbf{AB}$ matrix. This is currently set to an \lstinline{int} data type.

\begin{aside}
It is good coding practice to use a custom data type in your designs. This allows you to easily change the data type, and to have one source of information so that you do not have errors when changing the data type in the future design iterations. And it is quite common to change the data type over the course of the design. For example, first you may start out with a \lstinline{float} or \lstinline{double} type while you get a functionally correct design. This also provides a baseline for error since later you will likely change your design to use fixed point data type.  Fixed point data can reduce the number of resources, and increase the performance potentially at the cost of a reduction in the precision of the resulting data. You will likely try many different fixed point types until you find the right tradeoff between accuracy/error, performance, and resource usage.
\end{aside}

\lstinline{SIZE} determines the number of rows and columns in the matrices to be multiplied. We limit this to square matrices although handling arbitrary matrix sizes could be done by changing the code in a number of different places. We leave that as an exercise for the reader.

\begin{exercise}
Change the code to allow it to handle matrices of arbitrary size. 
\end{exercise}

The \lstinline{BLOCK_SIZE} variable defines the number of rows from $\mathbf{A}$ and the number of columns from $\mathbf{B}$ that we operate upon in each execution. This also defines how much data that we stream at one time into the function. The output data that we receive from the function at each execution is an \lstinline{BLOCK_SIZE} $\times$ \lstinline{BLOCK_SIZE} portion of the $\mathbf{AB}$ matrix. 

The \lstinline{blockvec} data type is used to transfer the \lstinline{BLOCK_SIZE} rows of $\mathbf{A}$ and columns of $\mathbf{B}$ to the function on each execution. The \lstinline{blockmat} data type is where we store the partial results for the $\mathbf{AB}$ matrix. 

Finally, the \lstinline{blockmat} data type is a structure consisting of an \lstinline{BLOCK_SIZE} $\times$ \lstinline{BLOCK_SIZE} array. This holds the resulting values from one execution of the \lstinline{matrmatmul} function. 

The function prototype itself takes the two inputs which are both of the type \lstinline{hls::stream<blockvec> &}. These are a sequence of \lstinline{blockvec} data. Remember that a \lstinline{blockvec} is a data type that consists of an array with \lstinline{BLOCK_SIZE} elements. 

The \lstinline{hls::stream<>} template class is one way in \VHLS of creating a FIFO data structure that works well in simulation and synthesis. The samples are sent in sequential order using the \lstinline{write()} function, and retrieved using the \lstinline{read()} function. This library was developed since streaming is a common methodology for passing data in hardware design, yet this same operation can be modeled in many different ways using the C programming language, for instance, by using arrays. In particular, it can be difficult for the \VHLS tool to infer streaming behaviors when dealing complex access patterns or multi-dimensional arrays. The built-in stream library enables the programmer to explicitly specify the order of stream accesses, avoiding any limitations of this inference. 

\begin{aside}
The \lstinline{hls::stream} class must always be passed by reference between functions, e.g., as we have done in the \lstinline{blockmatmul} function in Figure \ref{fig:block_mm}.
\end{aside}

\begin{figure}
\begin{lstlisting}[format=none]
#include "block_mm.h"
void blockmatmul(hls::stream<blockvec> &Arows, hls::stream<blockvec> &Bcols,
        blockmat &ABpartial, int it) {
  #pragma HLS DATAFLOW
  int counter = it % (SIZE/BLOCK_SIZE);
  static DTYPE A[BLOCK_SIZE][SIZE];
  if(counter == 0){ //only load the A rows when necessary
    loadA: for(int i = 0; i < SIZE; i++) {
      blockvec tempA = Arows.read();
      for(int j = 0; j < BLOCK_SIZE; j++) {
        #pragma HLS PIPELINE II=1
        A[j][i] = tempA.a[j];
      }
    }
  }
  DTYPE AB[BLOCK_SIZE][BLOCK_SIZE] = { 0 };
  partialsum: for(int k=0; k < SIZE; k++) {
    blockvec tempB = Bcols.read();
    for(int i = 0; i < BLOCK_SIZE; i++) {
      for(int j = 0; j < BLOCK_SIZE; j++) {
        AB[i][j] = AB[i][j] +  A[i][k] * tempB.a[j];
      }
    }
  }
  writeoutput: for(int i = 0; i < BLOCK_SIZE; i++) {
    for(int j = 0; j < BLOCK_SIZE; j++) {
      ABpartial.out[i][j] = AB[i][j];
    }
  }
}
\end{lstlisting}
\caption{  The \lstinline{blockmatmul} function takes a \lstinline{BLOCK_SIZE} set of rows from $\mathbf{A}$ matrix, a \lstinline{BLOCK_SIZE} set of columns from the $\mathbf{B}$ matrix, and creates a \lstinline{BLOCK_SIZE} $\times$ \lstinline{BLOCK_SIZE} partial result for the $\mathbf{AB}$ matrix. The first part of the code (denoted by the label \lstinline{loadA}) stores the rows from $\mathbf{A}$ into a local memory, the second part in the nested \lstinline{partialsum for} performs the computation for the partial results, and the final part (with the \lstinline{writeoutput} label) takes these results and puts them the proper form to return from the function.}
\label{fig:block_mm}
\end{figure}

The code for executing one part of the streaming block matrix multiplication is shown in Figure \ref{fig:block_mm}. The code has three portions denoted by the labels \lstinline{loadA}, \lstinline{partialsum}, and \lstinline{writeoutput}. 

The first part of the code, denoted by the \lstinline{loadA} label, is only executed on certain conditions, more precisely when \lstinline{it 

Remember that in each execution of this \lstinline{blockmatmul} function we send \lstinline{BLOCK_SIZE} rows from the $\mathbf{A}$ matrix and \lstinline{BLOCK_SIZE} columns from the $\mathbf{B}$ matrix. We send multiple \lstinline{BLOCK_SIZE} of columns for each \lstinline{BLOCK_SIZE} of rows from $\mathbf{A}$. The variable \lstinline{it} keeps track of the number of times that we have called the \lstinline{blockmatmul} function. Thus, we do a check on each execution of the function to determine if we need to load the rows from $\mathbf{A}$. When we do not, this saves us some time. When it is executed, it simply pulls data from the \lstinline{Arows} stream and puts it into a static local two-dimensional matrix \lstinline{A[BLOCK_SIZE][SIZE]}. 

Fully understanding this code requires some explanation about the \lstinline{stream} class, and how we are using it. The \lstinline{stream} variable \lstinline{Arows} has elements of the type \lstinline{blockvec}. A \lstinline{blockvec} is a matrix of size \lstinline{BLOCK_SIZE}. We use this in the following manner; each element in the \lstinline{Arows} stream has an array that holds one element from each of the \lstinline{BLOCK_SIZE} rows of the $\mathbf{A}$ matrix. Thus, in each call the the \lstinline{blockmatmul} function, the \lstinline{Arows} stream will have \lstinline{SIZE} elements in it, each of those holding one of each of the \lstinline{BLOCK_SIZE} rows. The statement \lstinline{tempA = Arows.read()} takes one element from the \lstinline{Arows} stream. Then we load each of these elements into the appropriate index in the local \lstinline{A} matrix.

\begin{aside}
The \lstinline{stream} class overloads the \lstinline{<<} operator to be equivalent to the \lstinline{read()} function. Thus, the statements \lstinline{tempA = Arows.read()} and \lstinline{tempA << Arows} perform the same operation.
\end{aside}

The next part of the computation calculates the partial sums. This is the bulk of the computation in the \lstinline{blockmatmul} function. 

The \lstinline{Bcols} stream variable is utilized in a very similar manner to the \lstinline{Arows} variable. However, instead of storing rows of $\mathbf{A}$, it stores the data corresponding the columns of $\mathbf{B}$ that the current execution of the function is computing upon. Every call of the \lstinline{blockmatmul} function will provide new data for the columns of the $\mathbf{B}$ matrix. Thus, we do not need to conditionally load this data as we do with the $\mathbf{A}$ matrix. The function itself works in a very similar manner to that from the \lstinline{matmul} in Figure \ref{fig:matrixmultiplication_sw} except that we are only calculating \lstinline{BLOCK_SIZE} $\times$ \lstinline{BLOCK_SIZE} results from the $\mathbf{AB}$ matrix. Thus we only have to iterate across \lstinline{BLOCK_SIZE} rows of $\mathbf{A}$ and \lstinline{BLOCK_SIZE} columns of $\mathbf{B}$. But each row and column has \lstinline{SIZE} elements, hence the bounds on the outer \lstinline{for} loop. 

The final portion of the function moves the data from the local \lstinline{AB} array, which has dimensions of \lstinline{BLOCK_SIZE} $\times$ \lstinline{BLOCK_SIZE}; this holds the partial results of the $\mathbf{AB}$ output matrix. 

Of the three parts of the function, the middle part, which calculates the partial sum, requires the most computation. By inspecting the code, we can see that this part has three nested \lstinline{for} loops with a total of \lstinline{SIZE} $\times$ \lstinline{BLOCK_SIZE} $\times$ \lstinline{BLOCK_SIZE} iterations. The first part has \lstinline{SIZE} $\times$ \lstinline{BLOCK_SIZE} iterations; and the last part has \lstinline{BLOCK_SIZE} $\times$ \lstinline{BLOCK_SIZE} iterations. Thus, we should focus our optimizations on the middle part, i.e., the \lstinline{partialsum} nested \lstinline{for} loops.

The common starting point for optimizations of nested \lstinline{for} loops is to pipeline the innermost \lstinline{for} loop. Then, if that does not require too many resources, the designer can move the \lstinline{pipeline} directive into higher level \lstinline{for} loops. Whether the resulting design consume too many resource depends upon the specified \lstinline{BLOCK_SIZE}; if this is small, then it is likely worth moving the \lstinline{pipeline} directive. It may even be worthwhile to move it inside the outermost \lstinline{for} loop. This will unroll the two inner \lstinline{for} loops and thus very likely increase the resource usage by a substantial amount. However, it will increase the performance.

\begin{exercise}
How does changing the \lstinline{BLOCK_SIZE} effect the performance and resource usage? How about changing the \lstinline{SIZE} constant? How does moving the \lstinline{pipeline} directive across the three different nested \lstinline{for} loops in the \lstinline{partialsum} portion of the function change the performance and resource usage?
\end{exercise}

The \lstinline{dataflow} directive at the start of the function creates a pipeline across the portions of the function, i.e., the \lstinline{loadA for} loop, the \lstinline{partialsum} nested \lstinline{for} loop, and the \lstinline{writeoutput for} loop. Using this directive will decrease the interval of the \lstinline{blockmatmul} function. However, this is limited by the largest interval of all three of the portions of the code. That is, the maximum interval for the \lstinline{blockmatmul} function -- let us call it interval(\lstinline{blockmatmul}) -- is greater than or equal to the the interval of the three parts which are defined as interval(\lstinline{loadA}), interval(\lstinline{partialsum}), and interval(\lstinline{writeoutput}). More formally, 

\begin{align}
\label{eq:interval}
interval(\texttt{blockmatmul}) \ge \max(&interval(\texttt{loadA}), interval(\texttt{partialsum}), \nonumber \\
 & interval(\texttt{writeoutput}))
 \end{align}

We need to keep Equation \ref{eq:interval} in mind as we optimize the \lstinline{blockmatmul} function. For example, assume that interval(\lstinline{partialsum}) is much larger than the other two portions of the function. Any performance optimizations that minimize interval(\lstinline{loadA}) and interval(\lstinline{writeoutput}) are useless since the function interval, i.e., interval(\lstinline{blockmatmul}) would not decrease. Thus, the designer should focus any performance optimization effort to decrease interval(\lstinline{partialsum}), i.e., target performance optimizations on those three nested \lstinline{for} loops. 

It is important to note that this only applies to performance optimizations. The designer can (and should) optimize the resource usage of these other two parts. In fact, they are ripe for such optimizations since reducing the resource usage often increases the interval and/or latency. In this case, it is ok to increase the interval as it will not effect the overall performance of the \lstinline{blockmatmul} function. In fact, the ideal case is to optimize all three parts of the function such that they all have the same interval, assuming that we can easily tradeoff between the interval and resource usage (which is not always the case). 

The testbench for the \lstinline{blockmatmul} function is shown in Figures \ref{fig:block_mm_init} and \ref{fig:block_mm_final}. We split it across two figures to make it more readable since it is a longer piece of code. Up until this point, we have not shown the testbenches. We show this testbench for several reasons. First, it provides insight into how the \lstinline{blockmatmul} function works. In particular, it partitions the input matrices into blocks and feeds them into the \lstinline{blockmatmul} function in a block by block manner. Second, it gives a complex usage scenario for using \lstinline{stream} template for simulation. Finally, it gives the reader an idea about how to properly design testbenches.

The \lstinline{matmatmul_sw} function is a simple three \lstinline{for} loop implementation of matrix multiplication. It takes two two-dimensional matrices as inputs, and outputs a single two-dimensional matrix. It is very similar to what we have seen in the \lstinline{matrixmul} function in Figure \ref{fig:matrixmultiplication_sw}. We use this to compare our results from the blocked matrix multiplication hardware version.

Let us focus on the first half of the testbench shown in Figure \ref{fig:block_mm_init}. The beginning block of code initializes variables of the rest of the function. The variable \lstinline{fail} keeps track of whether the matrix multiplication was done correctly. We will check this later in the function. The variables \lstinline{strm_matrix1} and \lstinline{strm_matrix2} are \lstinline{hls:stream<>} variables that hold the rows and columns of the $\mathbf{A}$ and $\mathbf{B}$ matrices, respectively. Each element of these \lstinline{stream} variables is a \lstinline{<blockvec>}. Referring back at the \lstinline{block_mm.h} file in Figure \ref{fig:block_mm_h}, we recall that a \lstinline{blockvec} is defined as an array of data; we will use each \lstinline{blockvec} to store one row or column of data. 

\begin{aside}
The \lstinline{stream} variable resides in the \lstinline{hls} namespace. Thus, we can use that namespace and forgo the \lstinline{hls::stream} and instead simply use \lstinline{stream}. However, the preferred usage is to keep the \lstinline{hls::} in front of the \lstinline{stream} to insure code readers that the stream is relevant to \VHLS and not C construct from another library. Also, it avoids having to deal with any potential conflicts that may occur by introducing a new namespace.
\end{aside}

The next definitions in this beginning block of code are the variables \lstinline{strm_matrix1_element} and \lstinline{strm_matrix2_element}. These two variables are used as placeholders to populate each \lstinline{blockvec} variable that we write into the \lstinline{strm_matrix1} and \lstinline{strm_matrix2} stream variables. The \lstinline{block_out} variable is used to store the output results from the \lstinline{blockmatmul} function. Note that this variable uses the data type \lstinline{blockmat} which is a two-dimensional array of \lstinline{BLOCK_SIZE} $\times$ \lstinline{BLOCK_SIZE} as defined in the \lstinline{block_mm.h} header file (see Figure \ref{fig:block_mm_h}). The final definitions are \lstinline{A},  \lstinline{B},  \lstinline{matrix_swout},  and\lstinline{matrix_hwout}. These are all \lstinline{SIZE} $\times$ \lstinline{SIZE} two-dimensional arrays with the \lstinline{DTYPE} data type. 

\begin{aside}
You can name the streams using an initializer. This is good practice as it gives better error messages. Without the name, the error message provides a generic reference to the stream with the data type. If you have multiple stream declarations with the same data type, then you will have to figure out which stream the error is referring to. Naming the stream variable is done by giving the variable an argument which is the name, e.g., \lstinline{hls::stream<blockvec> strm_matrix1("strm_matrix1");}.
\end{aside}

The next set of nested \lstinline{initmatrices for} loops sets the values of the four two-dimensional arrays \lstinline{A}, \lstinline{B}, \lstinline{matrix_swout}, and \lstinline{matrix_hwout}. The variables \lstinline{A} and \lstinline{B} are input matrices. These are initialized to a random value between [0, 512). We picked the number 512 for no particular reason other than it can fit any 9 bit value. Keep in mind that while the \lstinline{DTYPE} is set as an \lstinline{int}, and thus has significantly more range than [0,512), we often move to fixed point values with much smaller ranges later in the design optimization process. The \lstinline{matrix_swout} and \lstinline{matrix_hwout} are both initialized to $0$. These are filled in later by calls to the functions \lstinline{matmatmul_sw} and \lstinline{blockmatmul}.

\begin{figure}
\begin{lstlisting}
#include "block_mm.h"
#include <stdlib.h>
using namespace std;

void matmatmul_sw(DTYPE A[SIZE][SIZE], DTYPE B[SIZE][SIZE],
      DTYPE out[SIZE][SIZE]){
 DTYPE sum = 0;
 for(int i = 0; i < SIZE; i++){
  for(int j = 0;j<SIZE; j++){
   sum = 0;
   for(int k = 0; k < SIZE; k++){
    sum = sum + A[i][k] * B[k][j];
   }
   out[i][j] = sum;
  }
 }
}

int main() {
 int fail = 0;
 hls::stream<blockvec> strm_matrix1("strm_matrix1");
 hls::stream<blockvec> strm_matrix2("strm_matrix2");
 blockvec strm_matrix1_element, strm_matrix2_element;
 blockmat block_out;
 DTYPE A[SIZE][SIZE], B[SIZE][SIZE];
 DTYPE matrix_swout[SIZE][SIZE], matrix_hwout[SIZE][SIZE];

 initmatrices: for(int i = 0; i < SIZE; i++){
  for(int j = 0; j < SIZE; j++){
   A[i][j] = rand() % 512;
   B[i][j] = rand() % 512;
   matrix_swout[i][j] = 0;
   matrix_hwout[i][j] = 0;
  }
 }

 //the remainder of this testbench is displayed in the next figure
\end{lstlisting}
\caption{  The first part of the testbench for block matrix multiplication. The function is split across two figures since it is too long to display on one page. The rest of the testbench is in Figure \ref{fig:block_mm_final}. This has a ``software'' version of matrix multiplication, and variable declarations and initializations. }
\label{fig:block_mm_init}
\end{figure}

The second part of the testbench is continued in Figure \ref{fig:block_mm_final}. This has the last portion of the code from the \lstinline{main} function. 

\begin{figure}
\begin{lstlisting}[format=none]
// The beginning of the testbench is shown in the previous figure
int main() {
  int row, col, it = 0;
  for(int it1 = 0; it1 < SIZE; it1 = it1 + BLOCK_SIZE) {
    for(int it2 = 0; it2 < SIZE; it2 = it2 + BLOCK_SIZE) {
      row = it1; //row + BLOCK_SIZE * factor_row;
      col = it2; //col + BLOCK_SIZE * factor_col;
      
      for(int k = 0; k < SIZE; k++) {
        for(int i = 0; i < BLOCK_SIZE; i++) {
          if(it % (SIZE/BLOCK_SIZE) == 0) strm_matrix1_element.a[i] = A[row+i][k];
          strm_matrix2_element.a[i] = B[k][col+i];
        }
        if(it % (SIZE/BLOCK_SIZE) == 0) strm_matrix1.write(strm_matrix1_element);
        strm_matrix2.write(strm_matrix2_element);
      }
      blockmatmul(strm_matrix1, strm_matrix2, block_out, it);
      
      for(int i = 0; i < BLOCK_SIZE; i++)
        for(int j = 0; j < BLOCK_SIZE; j++)
          matrix_hwout[row+i][col+j] = block_out.out[i][j];
      it = it + 1;
    }
  }
  
  matmatmul_sw(A, B, matrix_swout);
  
  for(int i = 0; i<SIZE; i++)
    for(int j = 0; j<SIZE; j++)
      if(matrix_swout[i][j] != matrix_hwout[i][j]) { fail=1; }
  
  if(fail==1) cout << "failed" << endl;
  else cout << "passed" << endl;
  
  return 0;
}
\end{lstlisting}
\caption{  The second portion of the block matrix multiply testbench. The first part is shown in Figure \ref{fig:block_mm_init}. This shows the computation required to stream the data to the \lstinline{blockmatmul} function, and the code that tests that this function matches a simpler three \lstinline{for} loop implementation. }
\label{fig:block_mm_final}
\end{figure}

The first part of this figure has a complex set of nested \lstinline{for} loops. The overall goal of the computation in these \lstinline{for} loops is to set up the data from the input matrices $\mathbf{A}$ and $\mathbf{B}$ so that it can be streamed to the \lstinline{blockmatmul} function. Then the results of the \lstinline{blockmatmul} function are stored in the \lstinline{matrix_hwout} array. 

The outer two \lstinline{for} loops are used to step across the input arrays in a blocked manner. You can see that these both iterate by a step of \lstinline{BLOCK_SIZE}. The next two \lstinline{for} loops write rows from $\mathbf{A}$ into \lstinline{strm_matrix1_element} and the columns from $\mathbf{B}$ into \lstinline{strm_matrix2_element}. It does this in an element by element fashion by using the variable \lstinline{k} to access the individual values from the rows (columns) and write these into the one dimensional array for each of these ``elements''. Remember that both \lstinline{strm_matrix1_element} and \lstinline{strm_matrix2_element} have the datatype \lstinline{blockvec}, which is a one dimensional array of size \lstinline{BLOCK_SIZE}. It is meant to hold \lstinline{BLOCK_SIZE} elements from each row or column. The inner \lstinline{for} loop iterates \lstinline{BLOCK_SIZE} times. The \lstinline{strm_matrix1} and \lstinline{strm_matrix2} stream variables are written to \lstinline{SIZE} times. That means that has a buffer of the entire row (or column) and each element in the buffer holds \lstinline{BLOCK_SIZE} values. 

\begin{aside}
The \lstinline{stream} class overloads the \lstinline{>>} operator to be equivalent to the \lstinline{write(data)} function. This is similar to overloading the \lstinline{read()} function to the \lstinline{<<} operator. Thus, the statements \lstinline{strm_matrix1.write(strm_matrix1_element)} and \lstinline{strm_matrix1_element >> strm_matrix1} perform the same operation.
\end{aside}

The final part of this portion of the code to highlight is the \lstinline{if} statements. These are correspond to the values $\mathbf{A}$ matrix. Essentially, these are there so that we do not constantly write the same values to \lstinline{strm_matrix1}. Recall that the values from the $\mathbf{A}$ matrix are used across several calls to the \lstinline{blockmatmul} function. See Figure \ref{fig:blockmm} for a discussion on this. These \lstinline{if} statements are placed there to highlight the fact that you should not continually write the same data over and over. This is important because the internal code of the \lstinline{blockmatmul} only does a read of this data when it is necessary. So if we continued to write this consistently, the code would not function correctly do to the fact that this stream is written to more than it is read from.

Now that the input data, the testbench calls the \lstinline{blockmatmul} function. After the function call, it receives the partial computed results in the \lstinline{block_out} variable. The next two \lstinline{for} loops but these results into the appropriate locations in the \lstinline{matrix_hwout} array. 

After this complex set of \lstinline{for} loops, the block matrix multiplication is complete. And the testbench continues to insure that the code is written correctly. It does this by comparing the results from the multiple calls to the \lstinline{blockmatmul} function to results that were computed in the \lstinline{matmatmul_sw}, which is a much simpler version of matrix matrix multiplication. After this function call, the testbench iterates through both two-dimensional matrices \lstinline{matrix_hwout} and \lstinline{matrix_swout} and makes sure that all of the elements are equivalent. If there is one or more element that is not equal, it sets the \lstinline{fail} flag equal to 1. The testbench completes by printing out \lstinline{failed} or \lstinline{passed}.

\begin{aside}
It is important that note that you cannot directly compare the performance of the function \lstinline{blockmatmul} with that of code for matrix multiplication, such as the code in Figure \ref{fig:matrixmultiplication_sw}. This is because it takes multiple calls to the \lstinline{blockmatmul} function in order to perform the entire matrix multiplication. It is important to always compare apples to apples.

\begin{exercise}
Derive a function to determine the number of times that \lstinline{blockmatmul} must be called in order to complete the entire matrix multiplication. This function should be generic, e.g., it should not be assume a specific value of \lstinline{BLOCK_SIZE} or size of the matrix (i.e., \lstinline{SIZE}).
\end{exercise}

\begin{exercise}
Compare the resource usage of block matrix multiplication versus matrix multiplication. How do the resources change as the size of the matrices increases? Does the block size play a role in the resource usage? What are the general trends, if any?
\end{exercise}

\begin{exercise}
Compare the performance of block matrix multiplication versus matrix multiplication. How does the performance change as the size of the matrices increases? Does the block size play a role in the performance? Pick two architectures with similar resource usage. How does the performance for those architectures compare?
\end{exercise}

\end{aside}

\section{Conclusion}

Block matrix multiplication provides a different way to compute matrix multiplication. It computes partial results of the result matrix by streaming a subset of the input matrices to a function. This function is then computed multiple times in order to complete the entire matrix multiplication computation.


\chapter{Prefix Sum and Histogram}
\glsresetall
\label{chapter:prefixsum}

\section{Prefix Sum}
\label{sec:prefixSum}

Prefix sum is a common kernel used in many applications, e.g., recurrence relations, 
compaction problems, string comparison, polynomial evaluation, histogram, radix sort, and quick sort \cite{blelloch1990prefix}. Prefix sum requires restructuring in order to create an efficient FPGA design.

The prefix sum is the cumulative sum of a sequence of numbers. Given a sequence of inputs $in_n$, the prefix sum $out_n$ is the summation of the first $n$ inputs, namely $out_n = in_0 + in_1 + in_2 + \cdots + in_{n-1} + in_n$. The following shows the computation for the first four elements of the output sequence $out$.
\begin{align*} 
out_0 & = in_0 &\\
out_1 & = in_0 + in_1 &\\
out_2 & = in_0 + in_1 + in_2 \\
out_3 & = in_0 + in_1 + in_2 + in_3 \\
\cdots
\end{align*}

Of course, in practice we don't want to store and recompute the sum of all of the previous inputs, so the prefix sum is often computed by the recurrence equation:
\begin{equation}
out_n = out_{n-1} + in_n
\end{equation}

The disadvantage of the recurrence equation is that we must compute $out_{n-1}$ before computing $out_n$, which fundamentally limits the parallelism and throughput that this computation can be performed.  In contrast, the original equations have obvious parallelism where each output can be computed independently at the expense of a significant amount of redundant computation.  C code implementing the recurrence equation is shown in Figure \ref{fig:prefixsumSW}. Ideally, we'd like to achieve $II=1$ for the loop in the code, but this can be challenging even for such simple code.  Implementing this code with \VHLS results in behavior like that shown in Figure \ref{fig:prefixsumSW}.
\begin{figure}
\begin{minipage}{.5\textwidth}
\begin{lstlisting}
#define SIZE 128
void prefixsum(int in[SIZE], int out[SIZE]) {
	out[0]=in[0];
    for(int i = 1; i < SIZE; i++) {
#pragma HLS PIPELINE
        out[i] = out[i-1] + in[i];
    }
}
\end{lstlisting}
\end{minipage}
\begin{minipage}{.5\textwidth}
\centering
\executeiffilenewer{prefixsumBO_behavior.svg}{images/prefixsumBO_behavior.pdf}%
{inkscape -z -D --file=prefixsumBO_behavior.svg %
--export-pdf=images/prefixsumBO_behavior.pdf --export-latex}%
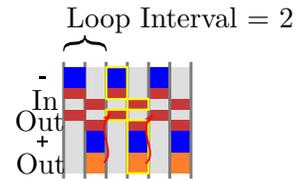%

\end{minipage}
\caption{ Code for implementing prefix sum, and its accompanying behavior. }
\label{fig:prefixsumSW}
\end{figure}

The way this code is written, each output is written into the output memory \lstinline|out[]| and then in the next iteration is read back out of the memory again.  Since the memory read is has a latency of one, data read from memory cannot be processed until the following clock cycle.  As a result, such a design can only achieve a loop II of 2.  In this case there is a relatively easy way to rewrite the code: we can simply perform the accumulation on a separate local variable, rather than reading the previous value back from the array.  Avoiding extra memory accesses in favor of register storage is often advantageous in processor code, but in HLS designs it is often more significant since other operations are rarely a performance bottleneck.  Code that does this is shown in Figure \ref{fig:prefixsum_optimized}.

\begin{figure}
\begin{minipage}{.5\textwidth}
\begin{lstlisting}
#define SIZE 128
void prefixsum(int in[SIZE], int out[SIZE]) {
  int A = in[0];
  for(int i=0; i < SIZE; i++) {
    #pragma HLS PIPELINE
    A = A + in[i];
    out[i] = A;
  }
}
\end{lstlisting}
\end{minipage}
\begin{minipage}{.5\textwidth}
\centering
\executeiffilenewer{prefixsum_optimized_behavior.svg}{images/prefixsum_optimized_behavior.pdf}%
{inkscape -z -D --file=prefixsum_optimized_behavior.svg %
--export-pdf=images/prefixsum_optimized_behavior.pdf --export-latex}%
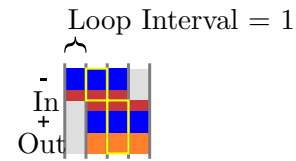%

\end{minipage}
\caption{ Code for implementing an optimized prefix sum, and its accompanying behavior. }
\label{fig:prefixsum_optimized}
\end{figure}

You might ask why the compiler is not able to optimize the memory loads and stores automatically in order to improve the II of the design.  It turns out that \VHLS is capable of optimizing loads and stores to array, but only for reads and writes within the scope of a single basic block.  You can see this if we unroll the loop, as shown in Figure \ref{fig:prefixsum_unrolled}.  Note that we also have to add appropriate \lstinline{array_partition} s in order to be able to read and write multiple values at the interfaces.   In this case, \VHLS is able to eliminate most of the read operations of the \lstinline{out[]} array within the body of the loop, but we still only achieve a loop II of 2.  In this case the first load in the body of the loop is still not able to be removed.  We can, however, rewrite the code manually to use a local variable rather than read from the \lstinline{out[]} array.

\begin{figure}
\begin{minipage}{.5\textwidth}
\begin{lstlisting}
#define SIZE 128
void prefixsum(int in[SIZE], int out[SIZE]) {
  #pragma HLS ARRAY_PARTITION variable=out cyclic factor=4 dim=1
  #pragma HLS ARRAY_PARTITION variable=in cyclic factor=4 dim=1
  int A = in[0];
  for(int i=0; i < SIZE; i++) {
    #pragma HLS UNROLL factor=4
    #pragma HLS PIPELINE
    out[i] = out[i-1] + in[i];
  }
}
\end{lstlisting}
\end{minipage}
\begin{minipage}{.5\textwidth}
\raggedleft
\executeiffilenewer{prefixsum_unrolled_behavior.svg}{images/prefixsum_unrolled_behavior.pdf}%
{inkscape -z -D --file=prefixsum_unrolled_behavior.svg %
--export-pdf=images/prefixsum_unrolled_behavior.pdf --export-latex}%
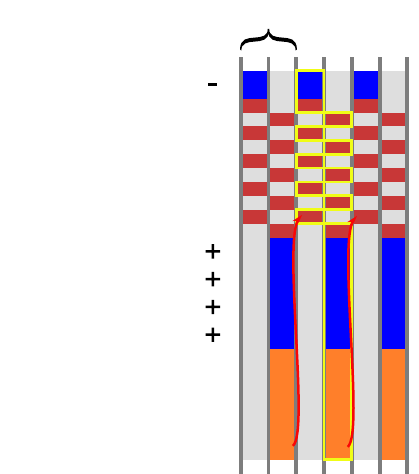%

\end{minipage}
\caption{ Optimizing the prefixsum code using \lstinline{unroll}, \lstinline{pipeline}, and \lstinline{array_partition} directives. }
\label{fig:prefixsum_unrolled}
\end{figure}



Ideally, when we unroll the inner loop, we the perform more operations per clock and reduce the interval to compute the function. If we unroll by a factor of two, then the performance doubles. A factor of four would increase the performance by factor four, i.e., the performance scales in a linear manner as it is unrolled. While this is mostly the case, as we unroll the inner loop there are often some aspects of the design that don't change.  Under most circumstances, such as when the iteration space of loops execute for a long time, these aspects represent a small fixed overhead which doesn't contribute significantly to the performance of the overall function.  However, as the number of loop iterations decreases, these fixed portions of the design have more impact.  The largest fixed component of a pipelined loop is the depth of the pipeline itself.  The control logic generate by \VHLS for a pipelined loop requires the pipeline to completely flush before code after the loop can execute.

\begin{exercise}
Unroll the \lstinline{for} loop corresponding to the prefix sum code in Figure \ref{fig:prefixsum_optimized} by different factors in combination with array partitioning to achieve a loop II of 1.  How does the \lstinline{prefixsum} function latency change? What are the trends in the resource usages? Why do you think you are seeing these trends? What happens when the loop becomes fully unrolled?
\end{exercise}

\begin{figure}
\centering
\includegraphics[width=  .7\textwidth]{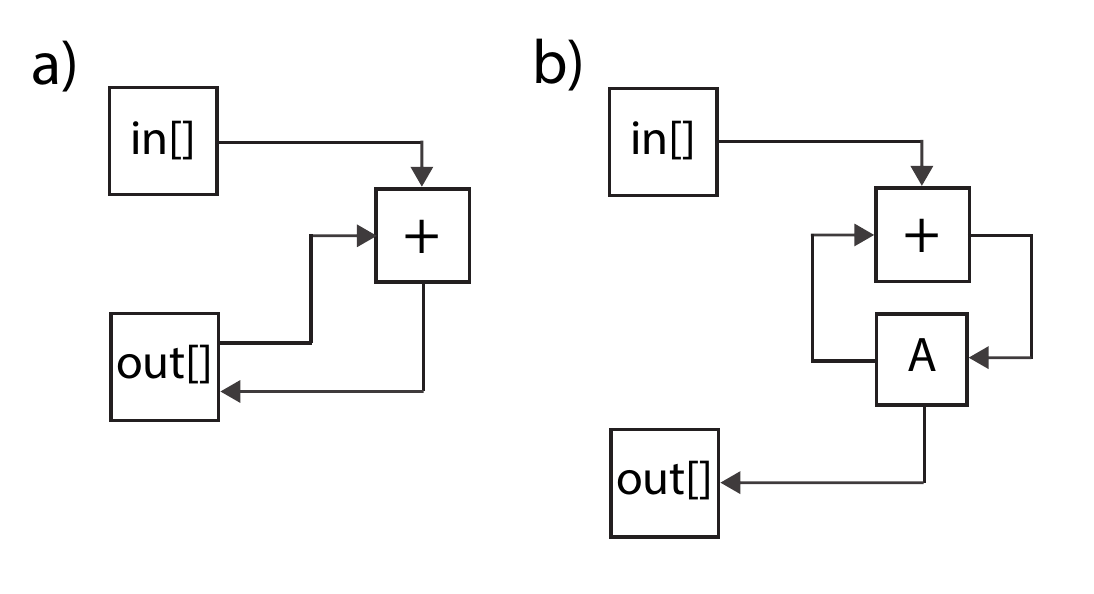}
\caption{ Part a) displays an architecture corresponding to the code in Figure~\ref{fig:prefixsumSW}. The dependence on the \lstinline{out[]} array can prevent achieving a loop II of 1. Computing the recurrence with a local variable, as shown in the code in Figure~\ref{fig:prefixsum_optimized}, is able to reduce the latency in the recurrence and achieve a loop II of 1.}
\label{fig:architecture_prefixsum}
\end{figure}

Figure~\ref{fig:architecture_prefixsum} shows the hardware architecture resulting from synthesizing the code from Figure \ref{fig:prefixsumSW} and Figure~\ref{fig:prefixsum_optimized}.   In part a), we can see that the `loop' in the circuit includes the output memory that stores the \lstinline{out[]} array, whereas in part b), the loop in the circuit includes only a register that stores the accumulated value and the output memory is only written.  Simplifying recurrences and eliminating unnecessary memory accesses is a common practice in optimizing HLS code.


\note{Could add a whole part about doing reduction. Probably not necessary.}
\note{Should have something here about what to do with floating point accumulation.  This is fundamentally more problematic than what's above (which is relatively easily handled by improving store->load optimization.}

The goal of this section is to show that even a small changes in the code can sometimes have a significant effect on the hardware design. Some changes may not necessarily be intuitive, but can be identified with feedback from the tool.

\section{Histogram}
\label{sec:histogram}

A Histogram models the probability distribution of a discrete signal. Given a sequence of discrete input values, the histogram counts the number of times each value appears in the sequence.  When normalized by the total number of input values, the histogram becomes the probability distribution function of the sequence.  Creating a histogram is a common function used in image processing, signal processing, database processing, and many other domains.  In many cases, it is common to quantize high-precision input data into a smaller number of intervals or \term{bins} as part of the histogram computation.  For the purpose of this section, we will skip the actual process by which this is done and focus on what happens with the binned data.

\begin{figure}
\centering
\includegraphics[width=  .7\textwidth]{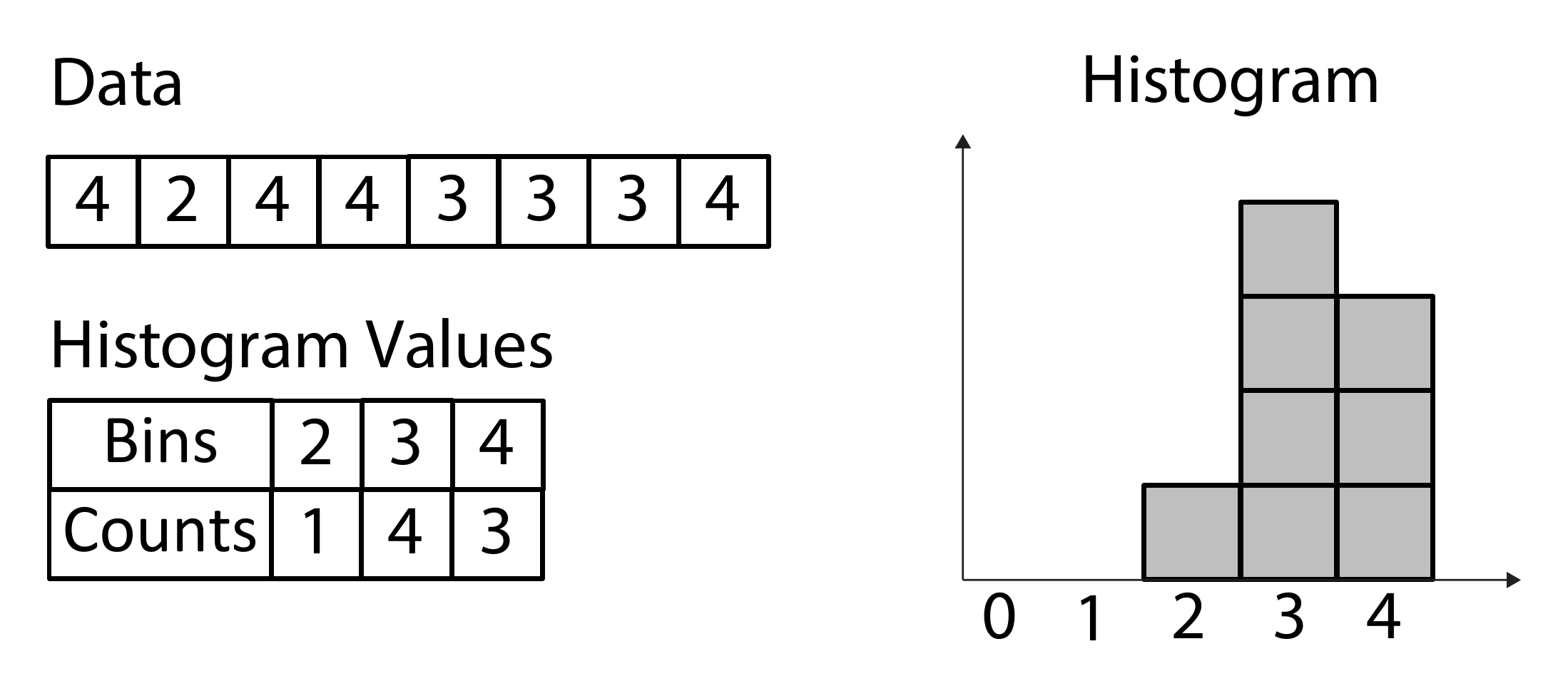}
\caption{ An example of a histogram.  }
\label{fig:histogram_introd}
\end{figure}

Figure~\ref{fig:histogram_introd} provides a simple example of a histogram.  The data set consists of a sequence of binned values, in this case represented by an integer in $[0,4]$. The corresponding histogram, consisting of a count for each bin, is shown below along with a graphical representation of the histogram, where the height of each bar corresponding to the count of each separate value.  Figure~\ref{fig:histogramSW} shows baseline code for the \lstinline{histogram} function.

\begin{figure}
\begin{lstlisting}
void histogram(int in[INPUT_SIZE], int hist[VALUE_SIZE]) {
  int val;
  for(int i = 0; i < INPUT_SIZE; i++) {
    #pragma HLS PIPELINE
    val = in[i];
    hist[val] = hist[val] + 1;
  }
}
\end{lstlisting}
\caption{ Original code for calculating the histogram. The \lstinline{for} loop iterates across the input array and increments the corresponding element of the \lstinline{hist} array. }
\label{fig:histogramSW}
\end{figure}

The code ends up looking very similar to the prefix sum in the previous section.  The difference is that the prefix sum is essentially only performing one accumulation, while in the \lstinline|histogram| function we compute one accumulation for each bin.  The other difference is that in the prefix sum we added the input value each time, in this case we only add 1.  When pipelining the inner loops using the \lstinline|pipeline| directive, we return to the same problem as with the code in Figure \ref{fig:prefixsumSW}, where we can only achieve a loop II of 2 due to the recurrence through the memory.   This is due to the fact that we are reading from the \lstinline{hist} array and writing to the same array in every iteration of the loop.  
Figure~\ref{fig:architecture_histogram} shows the hardware architecture for the code in Figure~\ref{fig:histogramSW}. You can see that the \lstinline{hist} array has a read and write operation. The \lstinline{val} variable is used as the index into the \lstinline{hist} array, and the variable at that index is read out, incremented, and written back into the same location.

\begin{figure}
\centering
\includegraphics[width=  .5\textwidth]{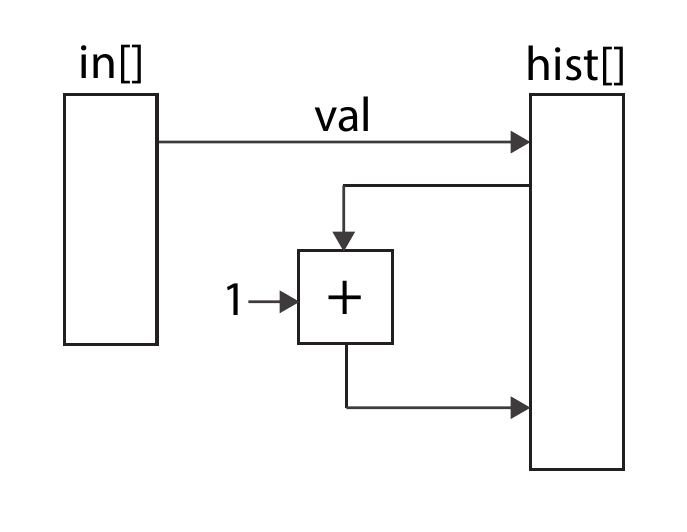}
\caption{ An architecture resulting from the code in Figure \ref{fig:histogramSW}. The \lstinline{val} data from the \lstinline{in} array is used to index into the \lstinline{hist} array. This data is incremented and stored back into the same location.}
\label{fig:architecture_histogram}
\end{figure}

\section{Histogram Optimization and False Dependencies}
Let's look deeper at the recurrence here.  In the first iteration of the loop, we read the \lstinline{hist} array at some location $x_0$ and write back to the same location $x_0$.  The read operation has a latency of one clock cycle, so the write has to happen in the following clock.  Then in the next iteration of the loop, we read at another location $x_1$.  Both $x_0$ and $x_1$ are dependent on the input and could take any value, so we consider the worst case when generating the circuit.  In this case, if $x_0 == x_1$, then the read at location $x_1$ cannot begin until the previous write has completed.  As a result, we must alternate between reads and writes.  

It turns out that we must alternate between reads and writes as long as $x_0$ and $x_1$ are independent.  What if they are {\em not} actually independent?  For instance, we might know that the source of data never produces two consecutive pieces of data that actually have the same bin.  What do we do now?  If we could give this extra information to the HLS tool, then it would be able to read at location $x_1$ while writing at location $x_0$ because it could guarantee that they are different addresses.   In \VHLS, this is done using the \lstinline|dependence| directive.

The modified code is shown in Figure \ref{fig:histogram_dependence}.  Here we've explicitly documented (informally) that the function has some preconditions.  In the code, we've added an \lstinline|assert()| call which checks the second precondition.
in \VHLS, this assertion is enabled during simulation to ensure that the simulation testvectors meet the required precondition.  The \lstinline|dependence| directive captures the effect of this precondition on the circuit, generated by the tool.  Namely, it indicates to \VHLS that reads and writes to the \lstinline|hist| array are dependent only in a particular way.  In this case, \lstinline|inter|-iteration dependencies consisting of a read operation after a write operation (RAW) have a distance of 2.  In this case a distance of $n$ would indicate that read operations in iteration $i+n$ only depend on write operations in iteration $i$.  In this case, we assert that \lstinline|in[i+1] != in[i]|, but it could be the case that \lstinline|in[i+2] == in[i]| so the correct distance is 2.

\begin{figure}
\begin{lstlisting}
#include <assert.h>
#include "histogram.h"

// Precondition: hist[] is initialized with zeros.
// Precondition: for all x, in[x] != in[x+1]
void histogram(int in[INPUT_SIZE], int hist[VALUE_SIZE]) {
#pragma HLS DEPENDENCE variable=hist inter RAW distance=2
    int val;
    int old = -1;
    for(int i = 0; i < INPUT_SIZE; i++) {
#pragma HLS PIPELINE
        val = in[i];
        assert(old != val);
        hist[val] = hist[val] + 1;
        old = val;
    }
}
\end{lstlisting}
\caption{ An alternative function for computing a histogram.  By restricting the inputs and indicating this restriction to \VHLS via the \lstinline|dependence| directive, II=1 can be achieved without significantly altering the code. }
\label{fig:histogram_dependence}
\end{figure}

\begin{exercise}
In Figure \ref{fig:histogram_dependence}, we added a precondition to the code, checked it using an assertion, and indicated the effect of the precondition to the tool using the \lstinline|dependence| directive.  What happens if your testbench violates this precondition?  What happens if you remove the \lstinline|assert()| call?  Does \VHLS still check the precondition?   What happens if the precondition is not consistent with the \lstinline|dependence| directive?
\end{exercise}

Unfortunately, the \lstinline{dependence} directive doesn't really help us if we are unwilling to accept the additional precondition.  It's also clear that we can't directly apply same optimization as with the \lstinline|prefixsum| function, since we might need to use all of the values stored in the \lstinline|hist| array.  Another alternative is implement the \lstinline|hist| array with a different technology, for instance we could partition the \lstinline|hist| array completely resulting in the array being implemented with \gls{ff} resources.  Since the data written into a \gls{ff} on one clock cycle is available immediately on the next clock cycle, this solves the recurrence problem and can be a good solution when a small number of bins are involved.  The architecture resulting from such a design is shown in Figure \ref{fig:histogram_partitioned}.  However, it tends to be a poor solution when a large number of bins are required.  Commonly histograms are constructed with hundreds to thousands of bins and for large data sets can require many bits of precision to count all of the inputs.   This results in a large number of FF resources and a large mux, which also requires logic resources.  Storing large histograms in \gls{bram} is usually a much better solution.

\begin{figure}
\centering
\note{fixme!}
\includegraphics[width=  .5\textwidth]{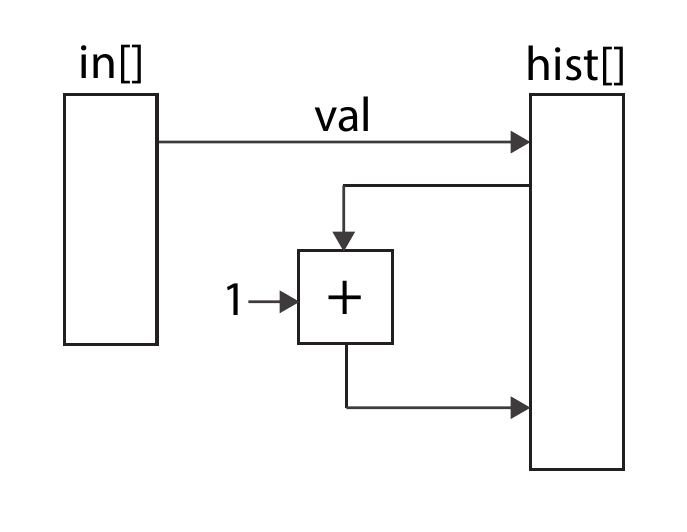}
\caption{ An architecture resulting from the code in Figure \ref{fig:histogramSW} when the \lstinline|hist| array is completely partitioned.}
\label{fig:histogram_partitioned}
\end{figure}

Returning to the code in Figure~\ref{fig:histogramSW}, we see that there are really two separate cases that the architecture must be able to handle.  One case is when the input contains consecutive values in the same bin.  In this case, we'd like to use a simple register to perform the accumulation with a minimal amount of delay.  The second case is when the input does not contain consecutive values in the same bin, in which case we need to read, modify, and write back the result to the memory.  In this case, we can guarantee that the read operation of the \lstinline|hist| array can not be affected by the previous write operation.   We've seen that both of these cases can be implemented separately, perhaps we can combine them into a single design.  The code to accomplish this is shown in Figure~\ref{fig:histogramOpt1}. This code uses a local variable \lstinline|old| to store the bin from the previous iteration and another local variable \lstinline{accu} to store the count for that bin.  Each time through the loop we check to see if we are looking at the same bin as the previous iteration.  If so, then we can simply increment \lstinline|accu|.  If not, then we need to store the value in \lstinline|accu| in the \lstinline|hist| array and increment the correct value in the \lstinline|hist| array instead.  In either case, we update \lstinline|old| and \lstinline|accu| to contain the current correct values.  The architecture corresponding to this code is shown in Figure \ref{fig:architecture_histogram_restructured}.  

In this code, we still need a \lstinline|dependence| directive, just as in Figure \ref{fig:histogram_dependence}, however the form is slightly different.  In this case the read and write accesses are to two different addresses in the same loop iteration.  Both of these addresses are dependent on the input data and so could point to any individual element of the \lstinline|hist| array.  Because of this, \VHLS assumes that both of these accesses could access the same location and as a result schedules the read and write operations to the array in alternating cycles, resulting in a loop II of 2.  However, looking at the code we can readily see that \lstinline|hist[old]| and \lstinline|hist[val]| can never access the same location because they are in the \lstinline|else| branch of the conditional \lstinline|if(old == val)|.  Within one iteration (an \lstinline|intra|-dependence) a read operation after a write operation (\lstinline|RAW|) can never occur and hence is a \lstinline|false| dependence.  In this case we are not using the \lstinline|dependence| directive to inform the tool about a precondition of the function, but instead about a property of the code itself. 

\begin{figure}
\begin{lstlisting}
#include "histogram.h"
void histogram(int in[INPUT_SIZE], int hist[VALUE_SIZE]) {
  int acc = 0;
  int i, val;
  int old = in[0];
  #pragma HLS DEPENDENCE variable=hist intra RAW false
  for(i = 0; i < INPUT_SIZE; i++) {
    #pragma HLS PIPELINE II=1
    val = in[i];
    if(old == val) {
      acc = acc + 1;
    } else {
      hist[old] = acc;
      acc = hist[val] + 1;
    }
    old = val;
  }
  hist[old] = acc;
}
\end{lstlisting}
\caption{ Removing the read after write dependency from the \lstinline{for} loop. This requires an \lstinline{if/else} structure that may seem like it is adding unnecessary complexity to the design. However, it allows for more effective pipelining despite the fact that the datapath is more complicated. }
\label{fig:histogramOpt1}
\end{figure}

\begin{exercise}
Synthesize the code from Figure \ref{fig:histogramSW} and Figure \ref{fig:histogramOpt1}. What is the initiation interval (II) in each case? What happens when you remove the \lstinline{dependence} directive from the code in Figure \ref{fig:histogramOpt1}? How does the loop interval change in both cases? What about the resource usage?
\end{exercise}

\begin{aside}
For the code in Figure \ref{fig:histogramOpt1}, you might question why a tool like \VHLS cannot determine this property.  In fact, while in some simple cases like this one better code analysis could propagate the \lstinline|if| condition property into each branch, we must accept that there are some pieces of code where properties of memory accesses are actually undecidable.  The highest performance in such cases will only be achieved in a static schedule with the addition of user information.  Several recent research works have looked to improve this by introducing some dynamic control logic into the design\cite{winterstein13dynamic, liu17elasticflow, dai17dynamic}.
\end{aside}

A pictorial description of the restructured code from Figure \ref{fig:histogramOpt1} is shown in Figure \ref{fig:architecture_histogram_restructured}. Not all of the operations are shown here, but the major idea of the function is there. You can see the two separate \lstinline{if} and \lstinline{else} regions (denoted by dotted lines). The \lstinline{acc} variable is replicated twice in order to make the drawing more readable; the actual design will only have one register for that variable.  The figure shows the two separate datapaths for the \lstinline{if} and the \lstinline{else} clause with the computation corresponding to the \lstinline{if} clause on the top and the \lstinline{else} clause datapath on the bottom. 

\begin{figure}
\centering
\includegraphics[width=  .6\textwidth]{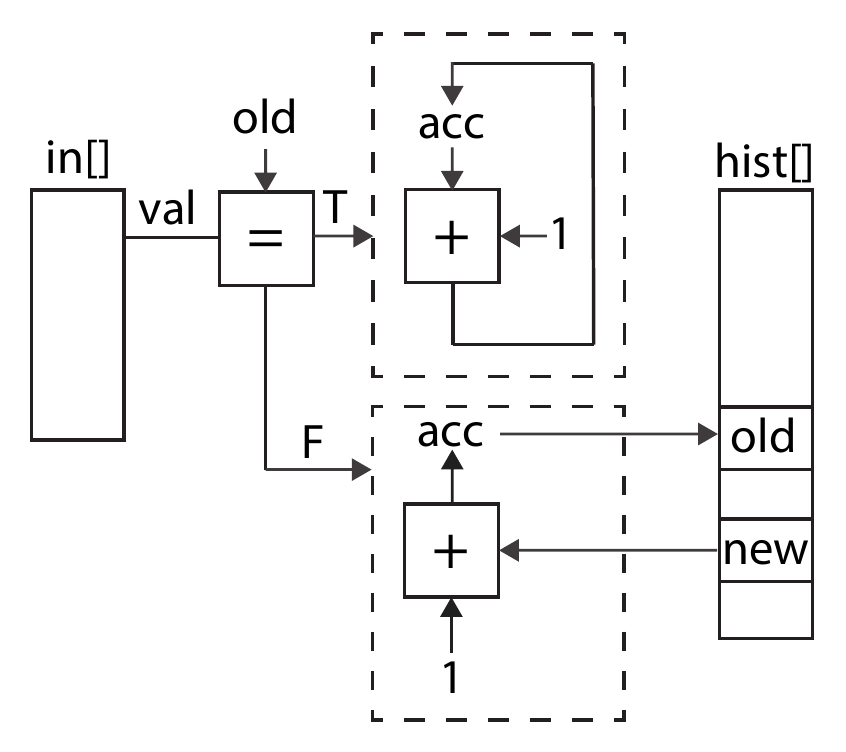}
\caption{ A depiction of the datapath corresponding to the code in Figure \ref{fig:histogramOpt1}. There are two separate portions corresponding to the \lstinline{if} and \lstinline{else} clauses. The figure shows the important portions of the computation, and leaves out some minor details. }
\label{fig:architecture_histogram_restructured}
\end{figure}

\section{Increasing Histogram Performance}

With some effort, we've achieved a design with a loop II of 1.  Previously we have seen how further reducing the execution time of a design can be achieved by partial unrolling of the inner loop.  However, with the \lstinline|histogram| function this is somewhat difficult for several reasons.  One reason is the challenging recurrence, unless we can break up the input data in some fashion, the computation of one iteration of the loop must be completed with the computation of the next iteration of the loop. A second reason is that with a loop II of 1, the circuit performs a read and a write of the \lstinline|hist| array each clock cycle, occupying both ports of a \gls{bram} resource in the FPGA.   Previously we have considered \gls{arraypartitioning} to increase the number of memory ports for accessing an array, but there's not a particularly obvious way to partition the \lstinline|hist| array since the access order depends on the input data.

All is not lost, however, as there is a way we can expose more parallelism by decomposing the histogram computation into two stages.  In the first stage, we divide the input data into a number of separate partitions.  The histogram for each partition can be computed independently using a separate instance, often called a \gls{pe}, of the histogram solution we've developed previously.  In the second stage, the individual histograms are combined to generate the histogram of the complete data sets.  This partitioning (or mapping) and merging (or reducing) process is very similar to that adopted by the MapReduce framework \cite{dean08mapreduce} and is a common pattern for parallel computation.  The map-reduce pattern is applicable whenever there is recurrence which includes a commutative and associative operation, such as addition in this case. This idea is shown in Figure~\ref{fig:architecture_histogram_parallel}. 

\begin{figure}
\centering
\includegraphics[width=  .9\textwidth]{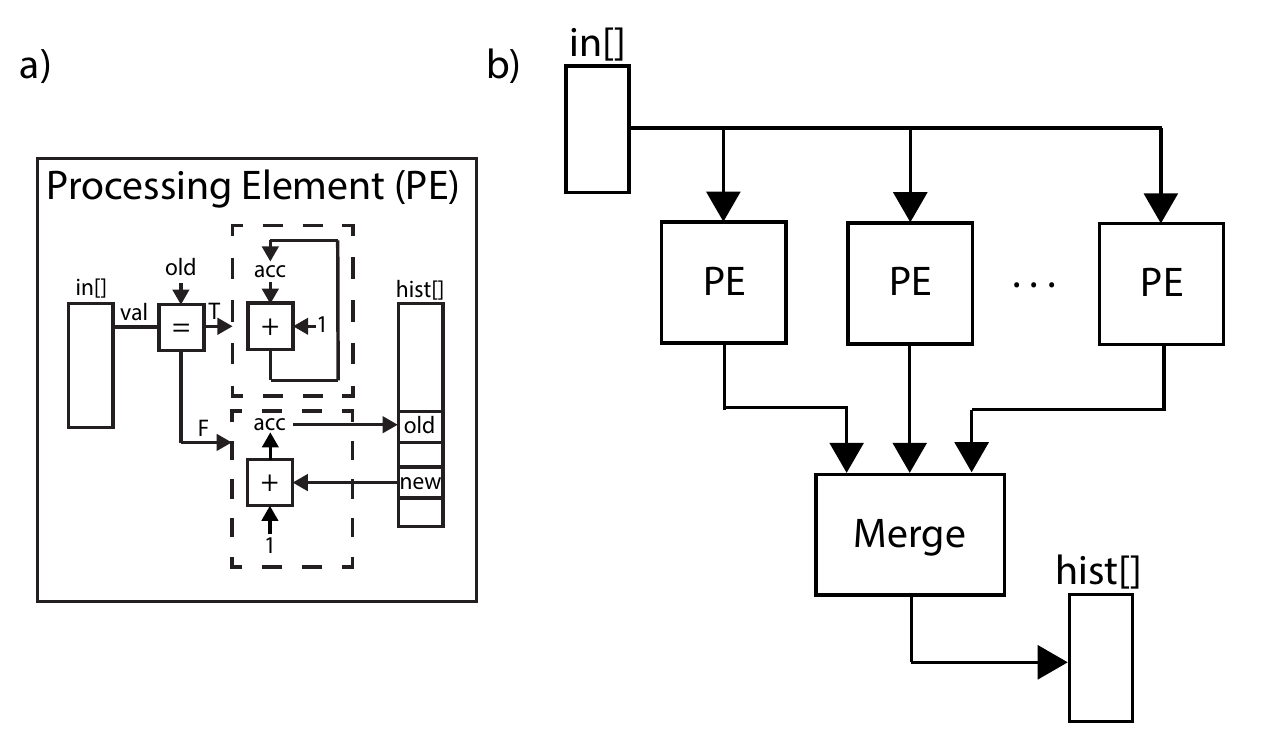}
\caption{The histogram computation implemented using a map-reduce pattern.  The processing element (PE) in Part a) is the same architecture as shown in Figure \ref{fig:architecture_histogram_restructured}. The \lstinline|in| array is partitioned and each partition is processed by a separate \gls{pe}. The merge block combines the individual histograms to create the final histogram.  }
\label{fig:architecture_histogram_parallel}
\end{figure}

The code for implementing this architecture is shown in Figure \ref{fig:histogram_parallel}. The \lstinline{histogram_map} function implements the `map' portion of the map-reduce pattern and will be instantiated multiple times. The code is very similar to the code in Figure \ref{fig:histogramOpt1}. The main difference is that we have added the additional code to initialize the \lstinline|hist| array.  The \lstinline{histogram_map} function takes an input array \lstinline{in} which will contain a partition of the data being processed and computes the histogram of that partition in the \lstinline|hist| array.   The \lstinline{histogram_reduce} function implements the `reduce' portion of the pattern.  It takes as input a number of partial histograms and combines them into complete histogram by adding together the count for each histogram bin.  In our code example in Figure \ref{fig:histogram_parallel}, we have only two processing elements, thus the merge has two input arrays \lstinline{hist1} and \lstinline{hist2}. This can easily be extended to handle more processing elements.

The new \lstinline{histogram} function takes as an input two partitions of the input data, stored in the \lstinline{inputA} and \lstinline{inputB} arrays.  It computes the histogram of each partition using the \lstinline {histogram_map} function, which are then stored in the \lstinline{hist1} and \lstinline{hist2} arrays. These are feed into the \lstinline{histogram_reduce} function which combines them and stores the result in the \lstinline{hist} array, which is the final output of the top level function \lstinline{histogram}. 

\begin{figure}
\begin{lstlisting}
#include "histogram_parallel.h"
void histogram_map(int in[INPUT_SIZE/2], int hist[VALUE_SIZE]) {
#pragma HLS DEPENDENCE variable=hist intra RAW false
    for(int i = 0; i < VALUE_SIZE; i++) {
#pragma HLS PIPELINE II=1
        hist[i] = 0;
    }
  int old = in[0];
  int acc = 0;
  for(int i = 0; i < INPUT_SIZE/2; i++) {
#pragma HLS PIPELINE II=1
    int val = in[i];
    if(old == val) {
      acc = acc + 1;
    } else {
      hist[old] = acc;
      acc = hist[val] + 1;
    }
    old = val;
  }
  hist[old] = acc;
}

void histogram_reduce(int hist1[VALUE_SIZE], int hist2[VALUE_SIZE], int output[VALUE_SIZE]) {
  for(int i = 0; i < VALUE_SIZE; i++) {
#pragma HLS PIPELINE II=1
    output[i] = hist1[i] + hist2[i];
  }
}

//Top level function
void histogram(int inputA[INPUT_SIZE/2], int inputB[INPUT_SIZE/2], int hist[VALUE_SIZE]){
  #pragma HLS DATAFLOW
  int hist1[VALUE_SIZE];
  int hist2[VALUE_SIZE];

  histogram_map(inputA, hist1);
  histogram_map(inputB, hist2);
  histogram_reduce(hist1, hist2, hist);
}
\end{lstlisting}
\caption{ Another implementation of histogram that uses task level parallelism and pipelining. The histogram operation is split into two sub tasks, which are executed in the two \lstinline{histogram_map} functions. These results are combined in the final histogram result using the \lstinline{histogram_reduce} function. The \lstinline{histogram} function is the top level function that connects these three functions together. }
\label{fig:histogram_parallel}
\end{figure}

\begin{exercise}
Modify the code in Figure \ref{fig:histogram_parallel} to support a parameterizable number \lstinline|NUM_PE| of \glspl{pe}? Hint: You'll need to combine some of the arrays into a single array that is correctly partitioned and add some loops that depend on \lstinline|NUM_PE|.  What happens to the throughput and task interval as you vary the number of \glspl{pe}? 
\end{exercise}

We use the \lstinline{dataflow} directive in the \lstinline{histogram} function in order to enable a design with \gls{taskpipelining}.   In this case there are three processes: two instances of the \lstinline{partial_histogram} function and one instance of the \lstinline{histogram_reduce} function. Within a single task, the two \lstinline{partial_histogram} processes can execute concurrently since they work on independent data, while the \lstinline{histogram_reduce} function must execute after since it uses the results from the  \lstinline{partial_histogram} processes. Thus, the \lstinline{dataflow} directive essentially creates a two stage task pipeline with the \lstinline{partial_histogram} functions in the first stage and the \lstinline{histogram_reduce} function in the second stage.   As with any dataflow design, the interval of the entire \lstinline{histogram} function depends upon the maximum initiation interval of the two stages. The two \lstinline{partial_histogram} functions in the first stage are the same and will have the same interval ($II_\mathrm{histogram\_map}$). The \lstinline{histogram_reduce} function will have another interval ($II_\mathrm{histogram\_reduce}$). The interval of the toplevel \lstinline{histogram} function $II_\mathrm{histogram}$ is then $\max (II_\mathrm{histogram\_map}, II_\mathrm{histogram\_reduce})$.

\begin{exercise}
What happens when you add or change the locations of the \lstinline{pipeline} directives? For example, is it beneficial to add a \lstinline{pipeline} directive to the \lstinline{for} loop in the \lstinline{histogram_reduce} function? What is the result of moving the \lstinline{pipeline} directive into the \lstinline{histogram_map} function, i.e., hoisting it outside of the \lstinline{for} loop where it currently resides? 
\end{exercise}

The goal of this section was to walk through the optimization the histogram computation, another small but important kernel of many applications. The key takeaway is that there are often limits to what tools can understand about our programs.  In some cases we must take care in how we write the code and in other cases we must actually give the tool more information about the code or the environment that the code is executing in.  In particular, properties about memory access patterns often critically affect the ability of HLS to generate correct and efficient hardware.  In \VHLS, these properties can be expressed using the \lstinline|dependence| directive.  Sometimes these optimizations might even be counter-intuitive, such as the addition of the \lstinline{if/else} control structure in \ref{fig:histogramOpt1}.  In other cases optimizations might require some creativity, as in applying the map-reduce pattern in Figures \ref{fig:architecture_histogram_parallel} and \ref{fig:histogram_parallel}).  

\section{Conclusion}
In this section, we've looked at the prefix sum and histogram kernels.  Although these functions seem different, they both contain recurrences through a memory access.  These recurrences can limit throughput if the memory access is not pipelined.  In both cases, by rewriting the code we can remove the recurrence.  In the case of the prefix sum, this is much easier since the access patterns are deterministic.  In the case of the histogram we must rewrite the code to address the recurrence or ensure that recurrence never happens in practice.   In either case we needed a way to describe to \VHLS information about the environment or about the code itself that the tool was unable to determine for itself.  This information is captured in the \lstinline{dependence} directive.  Lastly, we looked at ways of parallelizing both algorithms yet further, so that they could process a number of data samples each clock cycle.

\chapter{Video Systems}
\glsresetall
\label{chapter:video}

\section{Background}
\label{chapVideo}
Video Processing is a common application for FPGAs.  One reason is that common video data rates match well the clock frequencies that can achieved with modern FPGAs.  For instance, the common High-Definition TV format known as FullHD or 1080P60 video requires
 $1920 \unit{pixels}{line} * 1080 \unit{lines}{frame} * 60 \unit{frames}{second} = 124,416,000 \unit{pixels}{second}$.  

When encoded in a digital video stream, these pixels are transmitted along with some blank pixels at 148.5 MHz, and can be processed in a pipelined FPGA circuit at that frequency.  Higher data rates can also be achieved by processing multiple samples per clock cycle.  Details on how digital video is transmitted will come in Section \ref{sec:video:formats}.  Another reason is that video is mostly processed in \term{scanline order} line-by-line from the top left pixel to the lower right pixel, as shown in Figure \ref{fig:scanlineOrder}.  This predictable order allows highly specialized memory architectures to be constructed in FPGA circuits to efficiently process video without excess storage.  Details on these architectures will come in Section \ref{sec:video:buffering}

\begin{figure}
\centering
\executeiffilenewer{videoScanlineOrder.svg}{images/videoScanlineOrder.pdf}%
{inkscape -z -D --file=videoScanlineOrder.svg %
--export-pdf=images/videoScanlineOrder.pdf --export-latex}%
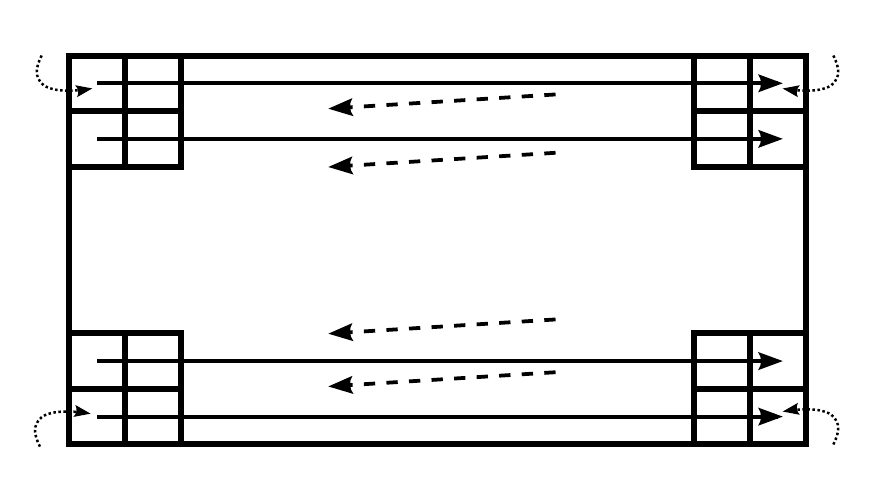%

\caption{Scanline processing order for video frames.}\label{fig:scanlineOrder}
\end{figure}

Video processing is also a good target application for HLS.  Firstly, video processing is typically tolerant to processing latency.  Many applications can tolerate several frames of processing delay, although some applications may limit the overall delay to less than one frame.  As a result, highly pipelined implementations can be generated from throughput and clock constraints in HLS with little concern for processing latency.   Secondly, video algorithms are often highly non-standardized and developed based on the personal taste or intuition of an algorithm expert.  This leads them to be developed in a high-level language where they can be quickly developed and simulated on sequences of interest.  It is not uncommon for FullHD video processing algorithms to run at 60 frames per second in an FPGA system, one frame per second in synthesizable C/C++ code running on a development laptop, but only one frame per hour (or slower) in an RTL simulator.  Lastly, video processing algorithms are often easily expressed in a nested-loop programming style that is amenable to HLS.  This means that many video algorithms can be synthesized into an FPGA circuit directly from the C/C++ code that an algorithm developer would write for prototyping purposes anyway

\subsection{Representing Video Pixels}
Many video input and output systems are optimized around the way that the human vision system perceives light.  One aspect of this is that the cones in the eye, which sense color, are sensitive primarily to red, green, and blue light.  Other colors are perceived as combinations of red, green, and blue light. As a result, video cameras and displays mimic the capabilities of the human vision system and are primarily sensitive or capable of displaying red, green, and blue light and pixels are often represented in the RGB colorspace as a combination of red, green, and blue components.  Most commonly each component is represented with 8 bits for a total of 24 bits per pixel, although other combinations are possible, such as 10 or even 12 bits per pixel in high-end systems. 

A second aspect is that the human visual system interprets brightness with somewhat higher resolution than color. Hence, within a video processing system it is common to convert from the RGB colorspace to the YUV colorspace, which describes pixels as a combination of Luminance (Y) and Chrominance (U and V).  This allows the color information contained in the U and V components to be represented independently of the brightness information in the Y component. One common video format, known as YUV422, represents two horizontally adjacent pixels with two Y values, one U value and one V value.  This format essentially includes a simple form of video compression called \term{chroma subsampling}.  Another common video format, YUV420, represents four pixels in a square with 4 Y values, one U value and one V value, further reducing the amount of data required.  Video compression is commonly performed on data in the YUV colorspace.

A third aspect is that the rods and codes in eye are more sensitive to green light than red or blue light and that the brain primarily interprets brightness primarily from green light. As a result, solid-state sensors and displays commonly use a mosaic pattern, such as the \term{Bayer} pattern\cite{bayer76} which consists of 2 green pixels for every red or blue pixel.  The end result is that higher resolution images can be produced for the same number of pixel elements, reducing the manufacturing cost of sensors and displays.

\begin{aside}
Video systems have been engineered around the human visual system for many years.  The earliest black and white video cameras were primarily sensitive to blue-green light to match the eye's sensitivity to brightness in that color range.  However, they were unfortunately not very sensitive to red light, as a result red colors (such as in makeup) didn't look right on camera.  The solution was decidedly low-tech: actors wore garish green and blue makeup.
\end{aside}

\subsection{Digital Video Formats}
\label{sec:video:formats}

In addition to representing individual pixels, digital video formats must also encode the organization of pixels into video frames.  In many cases, this is done with synchronization or \term{sync} signals that indicate the start and stop of the video frame in an otherwise continuous sequence of pixels.  In some standards (such as the Digital Video Interface or \term{DVI}) sync signals are represented as physically separate wires.  In other standards (such as the Digital Television Standard BTIR 601/656) the start and stop of the sync signal is represented by special pixel values that don't otherwise occur in the video signal.

Each line of video (scanned from left to right) is separated by a \term{Horizontal Sync Pulse}.  The horizontal sync is active for a number of cycles between each video line.  In addition, there are a small number of pixels around the pulse where the horizontal sync is not active, but there are not active video pixels.  These regions before and after the horizontal sync pulse are called the \term{Horizontal Front Porch} and \term{Horizontal Back Porch}, respectively.  Similarly, each frame of video (scanned from top to bottom) is separated by a \term{Vertical Sync Pulse}.  The vertical sync is active for a number of lines between each video frame.  Note that the vertical sync signal only changes at the start of the horizontal sync signal.  There are also usually corresponding \term{Vertical Front Porch} and \term{Vertical Back Porch} areas consisting of video lines where the vertical sync is not active, but there are not active video pixels either.  In addition, most digital video formats include a Data Enable signal that indicates the active video pixels.  Together, all of the video pixels that aren't active are called the \term{Horizontal Blanking Interval}
 and \term{Vertical Blanking Interval}.  These signals are shown graphically in Figure \ref{fig:video_syncs}.

\begin{figure}
\centering
\executeiffilenewer{video_syncs.svg}{images/video_syncs.pdf}%
{inkscape -z -D --file=video_syncs.svg %
--export-pdf=images/video_syncs.pdf --export-latex}%
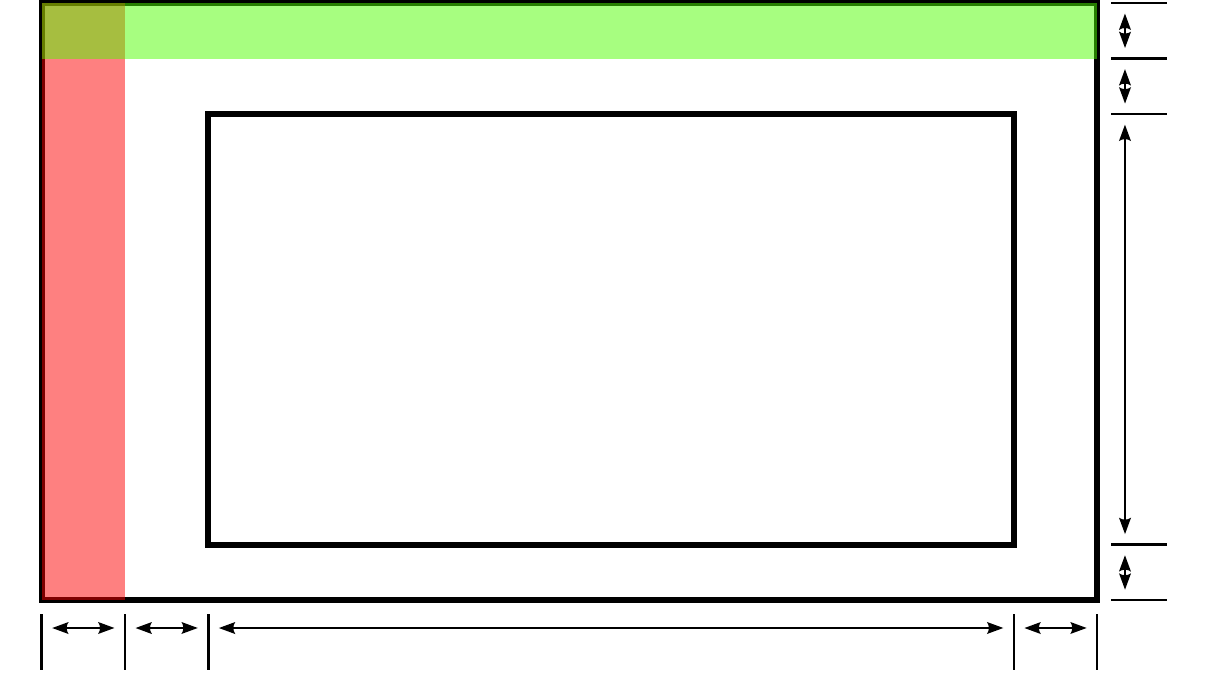%

\caption{Typical synchronization signals in a 1080P60 high definition video signal.}\label{fig:video_syncs}
\end{figure}

\begin{aside}
The format of digital video signals is, in many ways, an artifact of the original analog television standards, such as NTSC in the United States and PAL in many European countries.  Since the hardware for analog scanning of Cathode Ray Tubes contained  circuits with limited slew rates, the horizontal and vertical sync intervals allowed time for the scan recover to the beginning of the next line.  These sync signals were represented by a large negative value in the video signal.  In addition, since televisions were not able to effectively display pixels close to the strong sync signal, the front porches and the back porches were introduced to increase the amount of the picture that could be shown.  Even then, many televisions were designed with \term{overscan}, where up to 20\% of the pixels at the edge the frame were not visible.
\end{aside}

The typical 1080P60 video frame shown in Figure \ref{fig:video_syncs} contains a total of $2200*1125$ data samples.  At 60 frames per second, this corresponds to an overall sample rate of 148.5 Million samples per second.   This is quite a bit higher than the average rate of active video pixels in a frame, $1920*1080*60 = 124.4$ Million pixels per second. Most modern FPGAs can comfortably process at this clock rate, often leading to 1 sample-per-clock cycle architectures.  Systems that use higher resolutions, such as 4K by 2K for digital cinema, or higher frame rates, such as 120 or even 240 frames per second often require more the one sample to be processed per clock cycle. Remember that such architectures can often be generated by unrolling loops in HLS (see Section \ref{sec:filterThroughputTradeoffs}).  Similarly, when processing lower resolutions or frame rates, processing each sample over multiple clocks may be preferable, enabling operator sharing.  Such architectures can often be generated by increasing the II of loops. 
\note{Insert something here about handling syncs.}

\begin{figure}
\begin{lstlisting}
#include "video_common.h"

unsigned char rescale(unsigned char val, unsigned char offset, unsigned char scale) {
	return ((val - offset) * scale) >> 4;
}

rgb_pixel rescale_pixel(rgb_pixel p, unsigned char offset, unsigned char scale) {
#pragma HLS pipeline
    p.R = rescale(p.R, offset, scale);
    p.G = rescale(p.G, offset, scale);
    p.B = rescale(p.B, offset, scale);
    return p;
}

void video_filter_rescale(rgb_pixel pixel_in[MAX_HEIGHT][MAX_WIDTH],
													rgb_pixel pixel_out[MAX_HEIGHT][MAX_WIDTH],
													unsigned char min, unsigned char max) {
#pragma HLS interface ap_hs port = pixel_out
#pragma HLS interface ap_hs port = pixel_in
row_loop:
	for (int row = 0; row < MAX_WIDTH; row++) {
	col_loop:
		for (int col = 0; col < MAX_HEIGHT; col++) {
#pragma HLS pipeline
			rgb_pixel p = pixel_in[row][col];
			p = rescale_pixel(p,min,max);
			pixel_out[row][col] = p;
		}
	}
}
\end{lstlisting}
\caption{Code implementing a simple video filter.}\label{fig:videoSimple}
\end{figure}

For instance, the code shown in Figure \ref{fig:videoSimple} illustrates a simple video processing application that processes one sample per clock cycle with the loop implemented at II=1.  The code is written with a nested loop over the pixels in the image, following the scanline order shown in \ref{fig:scanlineOrder}.  An II=3 design could share the rescale function computed for each component, enabling reduced area usage.  Unrolling the inner loop by a factor of 2 and partitioning the input and the output arrays by an appropriate factor of 2 could enable processing 2 pixels every clock cycle in an II=1 design.  This case is relatively straightforward, since the processing of each component and of individual pixels is independent.  More complicated functions might not benefit from resource sharing, or might not be able to process more than one pixel simultaneously.

\begin{exercise}
A high-speed computer vision application processes small video frames of 200 * 180 pixels at 10000 frames per second.  This application uses a high speed sensor interfaced directly to the FPGA and requires no sync signals.  How many samples per clock cycle would you attempt to process?  Is this a good FPGA application?  Write the nested loop structure to implement this structure using HLS.  
\end{exercise}

\subsection{Video Processing System Architectures}
\label{sec:video:architectures}
Up to this point, we have focused on building video processing applications without concern for how they are integrated into a system.  In many cases, such as the example code in Figure \ref{fig:video:2Dfilter_linebuffer_extended}, the bulk of the processing occurs within a loop over the pixels and can process one pixel per clock when the loop is active.  In this section we will discuss some possibilities for system integration.

By default, \VHLS generates a simple memory interface for interface arrays.  This interface consists of address and data signals and a write enable signal in the case of a write interface.  Each read or write of data is associated with a new address and the expected latency through the memory is fixed.  It is simple to integrate such an interface with on-chip memories created from Block RAM resources as shown in Figure \ref{fig:video:BRAM_interface}.  However Block RAM resources are generally a poor choice for storing video data because of the large size of each frame, which would quickly exhaust the Block RAM resources even in large expensive devices.

\begin{figure}
\centering
\framebox{%
\executeiffilenewer{video_BRAM_interface.svg}{images/video_BRAM_interface.pdf}%
{inkscape -z -D --file=video_BRAM_interface.svg %
--export-pdf=images/video_BRAM_interface.pdf --export-latex}%
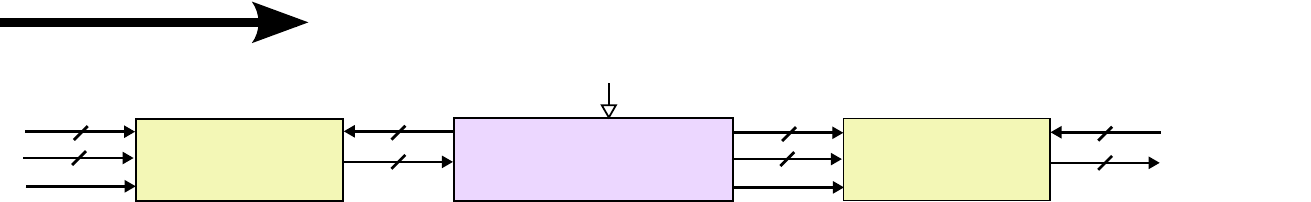%
}
\begin{lstlisting}
void video_filter(rgb_pixel pixel_in[MAX_HEIGHT][MAX_WIDTH],
				rgb_pixel pixel_out[MAX_HEIGHT][MAX_WIDTH]) {
#pragma HLS interface ap_memory port = pixel_out // The default
#pragma HLS interface ap_memory port = pixel_in // The default
\end{lstlisting}
\caption{Integration of a video design with BRAM interfaces.}\label{fig:video:BRAM_interface}
\end{figure}

\begin{exercise}
For 1920x1080 frames with 24 bits per pixel, how many Block RAM resources are required to store each video frame?  How many frames can be stored in the Block RAM of the FPGA you have available?
\end{exercise}

A better choice for most video systems is to store video frames in external memory, typically some form of double-data-rate (DDR) memory.  Typical system integration with external memory is shown in Figure \ref{fig:video:DDR_interface}.   An FPGA component known as \term{external memory controller} implements the external DDR interface and provides a standardized interface to other FPGA components through a common interface, such as the ARM AXI4 slave interface \cite{ARMAXI4}.   FPGA components typically implement a complementary master interface which can be directly connected to the slave interface of the external memory controller or connected through specialized \term{AXI4 interconnect} components.  The AXI4 interconnect allows multiple multiple master components to access a number of slave components.  This architecture abstracts the details of the external memory, allowing different external memory components and standards to be used interchangeably without modifying other FPGA components.   

\begin{figure}
\centering
\framebox{%
\executeiffilenewer{video_DDR_interface.svg}{images/video_DDR_interface.pdf}%
{inkscape -z -D --file=video_DDR_interface.svg %
--export-pdf=images/video_DDR_interface.pdf --export-latex}%
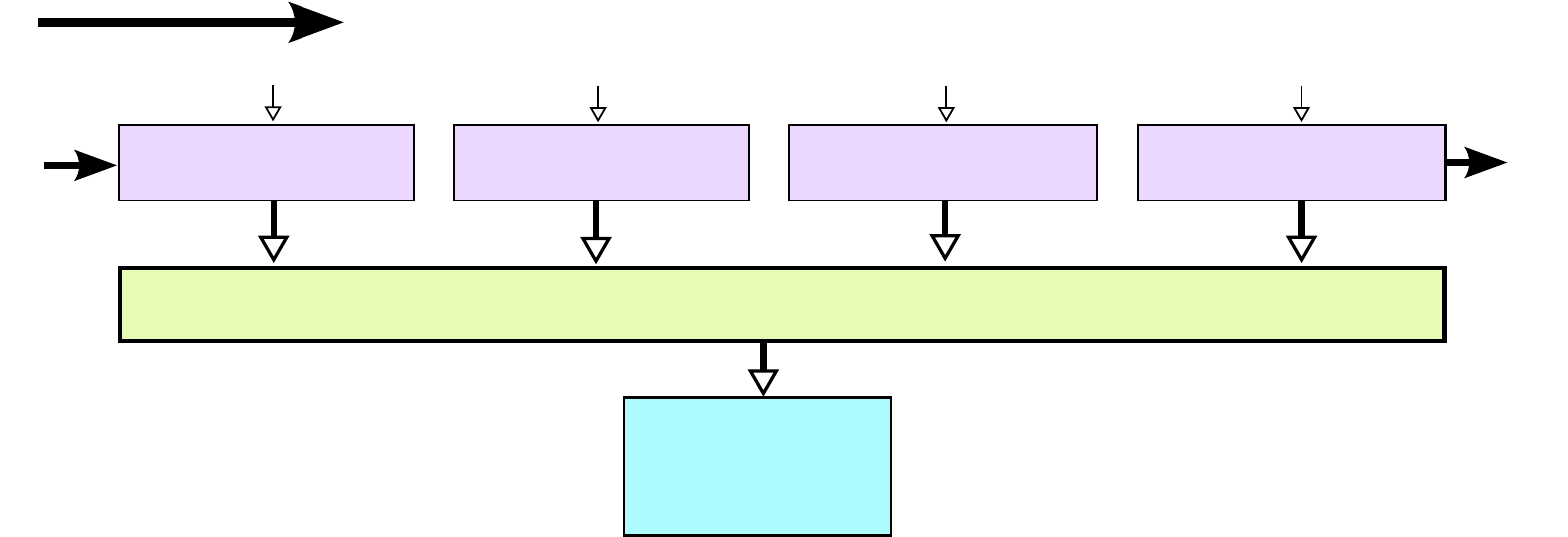%
}
\begin{lstlisting}
void video_filter(pixel_t pixel_in[MAX_HEIGHT][MAX_WIDTH],
				pixel_t pixel_out[MAX_HEIGHT][MAX_WIDTH]) {
#pragma HLS interface m_axi port = pixel_out
#pragma HLS interface m_axi port = pixel_in
\end{lstlisting}
\caption{Integration of a video design with external memory interfaces.}\label{fig:video:DDR_interface}
\end{figure}

Although most processor systems are built with caches and require them for high performance processing, it is typical to implement FPGA-based video processing systems as shown in \ref{fig:video:DDR_interface} without on-chip caches.  In a processor system, the cache provides low-latency access to previously accessed data and improves the bandwidth of access to external memory by always reading or writing complete cache lines. Some processors also use more complex mechanisms, such as prefetching and speculative reads in order to reduce external memory latency and increase external memory bandwidth.  For most FPGA-based video processing systems simpler techniques leveraging line buffers and window buffers are sufficient to avoid fetching any data from external memory more than once, due to the predictable access patterns of most video algorithms.  Additionally, \VHLS is capable of scheduling address transactions sufficiently early to avoid stalling computation due to external memory latency and is capable of statically inferring burst accesses from consecutive memory accesses.

\begin{figure}
\centering
\framebox{%
\executeiffilenewer{video_DDR_DMA_interface.svg}{images/video_DDR_DMA_interface.pdf}%
{inkscape -z -D --file=video_DDR_DMA_interface.svg %
--export-pdf=images/video_DDR_DMA_interface.pdf --export-latex}%
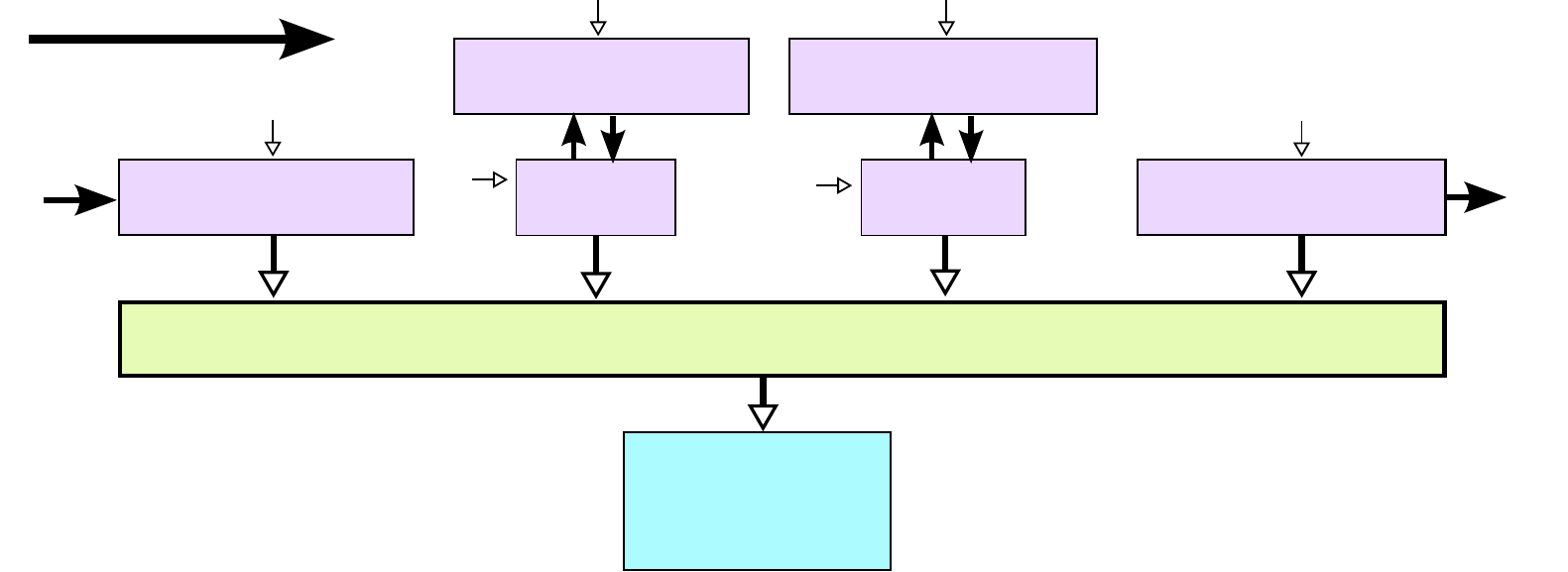%
}
\begin{lstlisting}
void video_filter(pixel_t pixel_in[MAX_HEIGHT][MAX_WIDTH],
				pixel_t pixel_out[MAX_HEIGHT][MAX_WIDTH]) {
#pragma HLS interface s_axi port = pixel_out
#pragma HLS interface s_axi port = pixel_in
\end{lstlisting}
\caption{Integration of a video design with external memory interfaces through a DMA component.}\label{fig:video:DDR_DMA_interface}
\end{figure}

An alternative external memory architecture is shown in Figure \ref{fig:video:DDR_DMA_interface}.  In this architecture, an accelerator is connected to an external Direct Memory Access (DMA) component that performs the details of generating addresses to the memory controller.  The DMA provides a stream of data to the accelerator for processing and consumes the data produced by the accelerator and writes it back to memory.  In \VHLS, there are multiple coding styles that can generate a streaming interface, as shown in Figure \ref{fig:video:stream_interfaces}.  One possibility is to model the streaming interfaces as arrays.  In this case, the C code is very similar to the code seen previously, but different interface directives are used.  An alternative is to model the streaming interface explicitly, using the \code{hls::stream<>} class.  In either case, some care must be taken that the order of data generated by the DMA engine is the same as the order in which the data is accessed in the C code.

\begin{figure}
\centering
\begin{lstlisting}
void video_filter(pixel_t pixel_in[MAX_HEIGHT][MAX_WIDTH],
									pixel_t pixel_out[MAX_HEIGHT][MAX_WIDTH]) {
#pragma HLS interface ap_hs port = pixel_out
#pragma HLS interface ap_hs port = pixel_in
\end{lstlisting}
\begin{lstlisting}
void video_filter(hls::stream<pixel_t> &pixel_in,
									hls::stream<pixel_t> &pixel_out) {
\end{lstlisting}

\caption{Coding styles for modelling streaming interfaces in HLS.}\label{fig:video:stream_interfaces}
\end{figure}

One advantage of streaming interfaces is that they allow multiple accelerators to be composed in a design without the need to store intermediate values in external memory.  In some cases, FPGA systems can be built without external memory at all, by processing pixels as they are received on an input interface (such as HDMI) and sending them directly to an output interface, as shown in Figure \ref{fig:video:streaming_interface}.  Such designs typically have accelerator throughput requirements that must achieved in order to meet the strict real-time constraints at the external interfaces.   Having at least one frame buffer in the system provides more flexibility to build complex algorithms that may be hard to construct with guaranteed throughput.  A frame buffer can also simplify building systems where the input and output pixel rates are different or potentially unrelated (such as a system that receives an arbitrary input video format and outputs an different arbitrary format).

\begin{figure}
\centering
\framebox{%
\executeiffilenewer{video_streaming_interface.svg}{images/video_streaming_interface.pdf}%
{inkscape -z -D --file=video_streaming_interface.svg %
--export-pdf=images/video_streaming_interface.pdf --export-latex}%
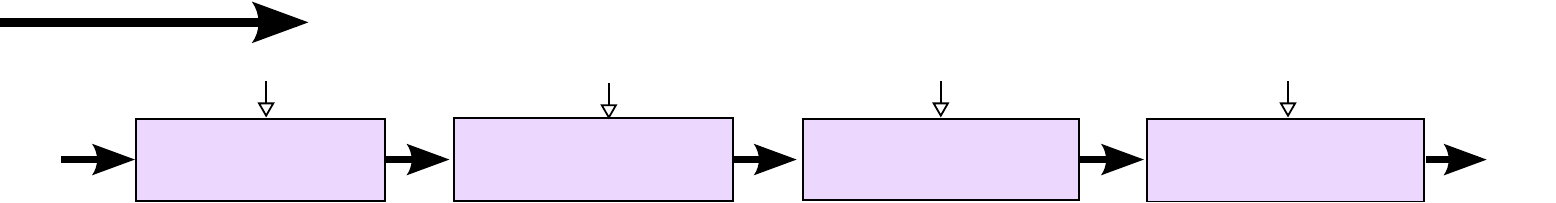%
}
\begin{lstlisting}
void video_filter(pixel_t pixel_in[MAX_HEIGHT][MAX_WIDTH],
				pixel_t pixel_out[MAX_HEIGHT][MAX_WIDTH]) {
#pragma HLS interface ap_hs port = pixel_out
#pragma HLS interface ap_hs port = pixel_in
\end{lstlisting}
\caption{Integration of a video design with streaming interfaces.}\label{fig:video:streaming_interface}
\end{figure}

\section{Implementation}

When actually processing video in a system it is common to factor out the system integration aspects from the implementation of video processing algorithms.  For the remainder of this chapter we will assume that input pixels are arriving in a stream of pixels and must be processed in scanline order.  The actual means by which this happens is largely unimportant, as long as HLS meets the required performance goals. 

\subsection{Line Buffers and Frame Buffers}
\label{sec:video:buffering}

Video processing algorithms typically compute an output pixel or value from a nearby region of input pixels, often called a \term{window}.   Conceptually, the window scans across the input image, selecting a region of pixels that can be used to compute the corresponding output pixel.  For instance, Figure \ref{fig:video:2Dfilter} shows code that implements a 2-dimensional filter on a video frame.  This code reads a window of data from the input video frame (stored in an array) before computing each output pixel.  

\begin{figure}
\begin{lstlisting}[format=none, firstline=3]
rgb_pixel filter(rgb_pixel window[3][3]) {
	const char h[3][3] = {{1, 2, 1}, {2, 4, 2}, {1, 2, 1}};
	int r = 0, b = 0, g = 0;
i_loop: for (int i = 0; i < 3; i++) {
	j_loop: for (int j = 0; j < 3; j++) {
			r += window[i][j].R * h[i][j];
			g += window[i][j].G * h[i][j];
			b += window[i][j].B * h[i][j];
		}
	}
	rgb_pixel output;
	output.R = r / 16;
	output.G = g / 16;
	output.B = b / 16;
	return output;
}

void video_2dfilter(rgb_pixel pixel_in[MAX_HEIGHT][MAX_WIDTH],
					rgb_pixel pixel_out[MAX_HEIGHT][MAX_WIDTH]) {
	rgb_pixel window[3][3];
row_loop: for (int row = 0; row < MAX_HEIGHT; row++) {
	col_loop: for (int col = 0; col < MAX_WIDTH; col++) {
#pragma HLS pipeline
			for (int i = 0; i < 3; i++) {
				for (int j = 0; j < 3; j++) {
					int wi = row + i - 1;
					int wj = col + j - 1;
					if (wi < 0 || wi >= MAX_HEIGHT || wj < 0 || wj >= MAX_WIDTH) {
						window[i][j].R = 0;
						window[i][j].G = 0;
						window[i][j].B = 0;
					} else
						window[i][j] = pixel_in[wi][wj];
				}
			}
			if (row == 0 || col == 0 || row == (MAX_HEIGHT - 1) || col == (MAX_WIDTH - 1)) {
				pixel_out[row][col].R = 0;
				pixel_out[row][col].G = 0;
				pixel_out[row][col].B = 0;
			} else
				pixel_out[row][col] = filter(window);
		}
	}
}

\end{lstlisting}
\caption{Code implementing a 2D filter without an explicit line buffer.}\label{fig:video:2Dfilter}
\end{figure}

\begin{exercise}
In Figure \ref{fig:video:2Dfilter}, there is the code \lstinline{int wi = row+i-1; int wj = col+j-1;}.  Explain why these expressions include a '-1'.  Hint: Would the number change if the filter were 7x7 instead of 3x3?
\end{exercise}

Note that in this code, multiple reads of \lstinline|pixel_in| must occur to populate the \lstinline|window| memory and compute one output pixel.  If only one read can be performed per cycle, then this code is limited in the pixel rate that it can support.  Essentially this is a 2-Dimensional version of the one-tap-per-cycle filter from Figure \ref{fig:FIR}.  In addition, the interfacing options are limited, because the input is not read in normal scan-line order.  (This topic will be dealt with in more detail in Section \ref{sec:video:architectures}.

A key observation about adjacent windows is that they often overlap, implying a high locality of reference.  This means that pixels from the input image can be buffered locally or cached and accessed multiple times.  By refactoring the code to read each input pixel exactly once and store the result in a local memory, a better result can be achieved.  In video systems, the local buffer is also called a \term{line buffer}, since it typically stores several lines of video around the window.   Line buffers are typically implemented in \gls{bram} resources, while window buffers are implemented using \gls{ff} resources.  Refactored code using a line buffer is shown in Figure \ref{fig:video:2Dfilter_linebuffer}.
Note that for an NxN image filter, only N-1 lines need to be stored in line buffers.  

\begin{figure}
\begin{lstlisting}[format=none,firstline=20]
void video_2dfilter_linebuffer(rgb_pixel pixel_in[MAX_HEIGHT][MAX_WIDTH],
							   rgb_pixel pixel_out[MAX_HEIGHT][MAX_WIDTH]) {
#pragma HLS interface ap_hs port=pixel_out
#pragma HLS interface ap_hs port=pixel_in
	rgb_pixel window[3][3];
	rgb_pixel line_buffer[2][MAX_WIDTH];
#pragma HLS array_partition variable=line_buffer complete dim=1
row_loop: for (int row = 0; row < MAX_HEIGHT; row++) {
	col_loop: for (int col = 0; col < MAX_WIDTH; col++) {
#pragma HLS pipeline
			for(int i = 0; i < 3; i++) {
				window[i][0] = window[i][1];
				window[i][1] = window[i][2];
			}

			window[0][2] = (line_buffer[0][col]);
			window[1][2] = (line_buffer[0][col] = line_buffer[1][col]);
			window[2][2] = (line_buffer[1][col] = pixel_in[row][col]);

			if (row == 0 || col == 0 ||
				row == (MAX_HEIGHT - 1) ||
				col == (MAX_WIDTH - 1)) {
				pixel_out[row][col].R = 0;
				pixel_out[row][col].G = 0;
				pixel_out[row][col].B = 0;
			} else {
				pixel_out[row][col] = filter(window);
			}
		}
	}
}
\end{lstlisting}
\caption{Code implementing a 2D filter with an explicit line buffer.}\label{fig:video:2Dfilter_linebuffer}
\end{figure}

The line buffer and window buffer memories implemented from the code in Figure \ref{fig:video:2Dfilter_linebuffer} are shown in Figure \ref{fig:video:video_buffers}.    Each time through the loop, the window is shifted and filled with one pixel coming from the input and two pixels coming from the line buffer.  Additionally, the input pixel is shifted into the line buffer in preparation to repeat the process on the next line.  Note that in order to process one pixel each clock cycle, most elements of the window buffer must be read from and written to every clock cycle. In addition, after the 'i' loop is unrolled, each array index to the \lstinline|window| array is a constant.  In this case, \VHLS will convert each element of the array into a scalar variable (a process called \term{scalarization}).  Most of the elements of the \lstinline|window| array will be subsequently implemented as Flip Flops.  Similarly, each row of the \lstinline|line_buffer| is accessed twice (being read once and written once).  The code explicitly directs each row of the \lstinline|line_buffer| array to be partitioned into a separate memory.  For most interesting values of \lstinline|MAX_WIDTH| the resulting memories will be implemented as one or more Block RAMs.  Note that each Block RAM can support two independent accesses per clock cycle.

\begin{figure}
\centering
\executeiffilenewer{video_buffers.svg}{images/video_buffers.pdf}%
{inkscape -z -D --file=video_buffers.svg %
--export-pdf=images/video_buffers.pdf --export-latex}%
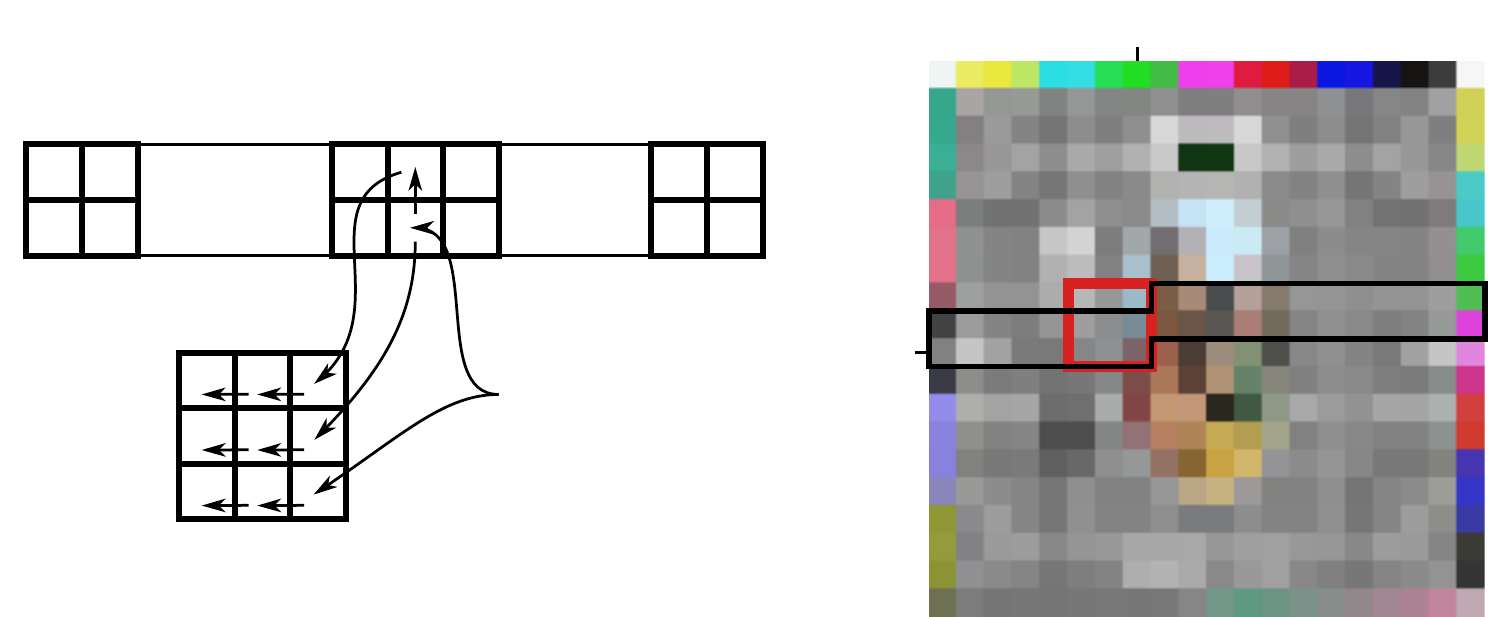%

\caption{ Memories implemented by the code in Figure \ref{fig:video:2Dfilter_linebuffer}.  These memories store a portion of the input image shown in the diagram on the right at the end of a particular iteration of the loop.  The pixels outlined in black are stored in the line buffer and the pixels outlined in red are stored in the window buffer.}\label{fig:video:video_buffers}
\label{fig:dft-visualization}
\end{figure}


\begin{aside}
Line buffers are a special case of a more general concept known as a \term{reuse buffer}, which is often used in stencil-style computations.  High-level synthesis of reuse buffers and line buffers from code like Figure \ref{fig:video:2Dfilter} is an area of active research.  See, for instance \cite{bayliss12sdram}\cite{hegarty2016rigel}.
\end{aside}
\begin{aside}
\VHLS includes \lstinline|hls::line_buffer<>| and \lstinline|hls::window_buffer<>| classes that simplify the management of window buffers and line buffers.  
\end{aside}

\begin{exercise}
For a 3x3 image filter, operating on 1920x1080 images with 4 bytes per pixel, How many FPGA Block RAMs are necessary to store each video line?
\end{exercise}

\subsection{Causal Filters}

The filter implemented in Figure \ref{fig:video:2Dfilter_linebuffer} reads a single input pixel and produces a single output pixel each clock cycle, however the behavior is not quite the same as the code in Figure \ref{fig:video:2Dfilter}.  The output is computed from the window of previously read pixels, which is 'up and to the left' of the pixel being produced. As a result, the output image is shifted 'down and to the right' relative to the input image.  The situation is analogous to the concept of causal and non-causal filters in signal processing.  Most signal processing theory focuses on causal filters because only causal filters are practical for time sampled signals (e.g. where x[n] = x(n*T) and y[n] = y(n*T)).

\begin{aside}
A \term{causal} filter $h[n]$ is a filter where $\forall k < 0, h[k] = 0$.
A finite filter $h[n]$ which is not causal can be converted to a causal filter $\hat{h}[n]$ by delaying the taps of the filter so that $\hat{h}[n] = h[n-D]$. The output of the new filter $\hat{y} = x \otimes \hat{h}$ is the same as a delayed output of the old filter $y = x \otimes h$.  Specifically,  $\hat{y}[n] = y[n-D]$.
\end{aside}

\begin{exercise}
Prove the fact in the previous aside using the definition of the convolution for

$y = x \otimes h$: y[n] = $\sum\limits_{k=-\infty}^\infty x[k] *h[n-k]$
\end{exercise}

For the purposes of this book, most variables aren't time-sampled signals and the times that individual inputs and outputs are created may be determined during the synthesis process.  For systems involving time-sampled signals, we treat timing constraints as a constraint during the HLS implementation process.  As long as the required task latency is achieved, then the design is correct.

In most video processing algorithms, the spatial shift introduced in the code above is undesirable and needs to be eliminated.  Although there are many ways to write code that solves this problem, a common way is known as \term{extending the iteration domain}.  In this technique, the loop bounds are increased by a small amount so that the first input pixel is read on the first loop iteration, but the first output pixel is not written until later in the iteration space.  A modified version of the filter code is shown in Figure \ref{fig:video:2Dfilter_linebuffer_extended}.  The behavior of this code is shown in Figure \ref{fig:video:timelines}, relative to the original linebuffer code in Figure \ref{fig:video:2Dfilter_linebuffer}.  After implementation with HLS, we see that the data dependencies are satisfied in exactly the same way and that the implemented circuit is, in fact, implementable.

\begin{figure}
\begin{lstlisting}[format=none,firstline=21]

void video_2dfilter_linebuffer_extended(
										rgb_pixel pixel_in[MAX_HEIGHT][MAX_WIDTH],
										rgb_pixel pixel_out[MAX_HEIGHT][MAX_WIDTH]) {
#pragma HLS interface ap_hs port=pixel_out
#pragma HLS interface ap_hs port=pixel_in
	rgb_pixel window[3][3];
	rgb_pixel line_buffer[2][MAX_WIDTH];
#pragma HLS array_partition variable=line_buffer complete dim=1
row_loop: for(int row = 0; row < MAX_HEIGHT+1; row++) {
	col_loop: for(int col = 0; col < MAX_WIDTH+1; col++) {
#pragma HLS pipeline II=1
			rgb_pixel pixel;
			if(row < MAX_HEIGHT && col < MAX_WIDTH) {
				pixel = pixel_in[row][col];
			}

			for(int i = 0; i < 3; i++) {
				window[i][0] = window[i][1];
				window[i][1] = window[i][2];
			}

			if(col < MAX_WIDTH) {
				window[0][2] = (line_buffer[0][col]);
				window[1][2] = (line_buffer[0][col] = line_buffer[1][col]);
				window[2][2] = (line_buffer[1][col] = pixel);
			}
			if(row >= 1 && col >= 1) {
				int outrow = row-1;
				int outcol = col-1;
				if(outrow == 0 || outcol == 0 ||
				   outrow == (MAX_HEIGHT-1) || outcol == (MAX_WIDTH-1)) {
					pixel_out[outrow][outcol].R = 0;
					pixel_out[outrow][outcol].G = 0;
					pixel_out[outrow][outcol].B = 0;
				} else {
					pixel_out[outrow][outcol] = filter(window);
				}
			}
		}
	}
}
\end{lstlisting}
\caption{Code implementing a 2D filter with an explicit line buffer.  The iteration space is extended by 1 to allow the filter to be implemented without a spatial shift.}\label{fig:video:2Dfilter_linebuffer_extended}
\end{figure}

\begin{figure}
\centering
\executeiffilenewer{filter2d_results_withshifting.svg}{images/filter2d_results_withshifting.pdf}%
{inkscape -z -D --file=filter2d_results_withshifting.svg %
--export-pdf=images/filter2d_results_withshifting.pdf --export-latex}%
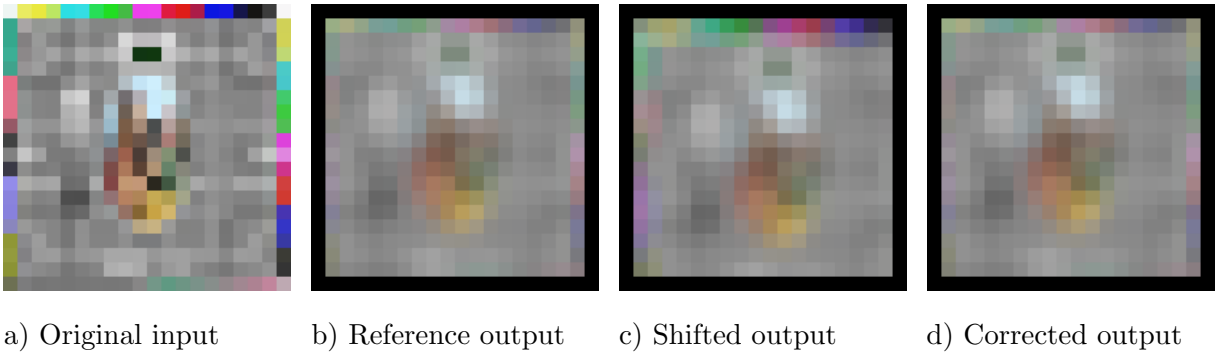%

\caption{Results of different filter implementations. The reference output in image b is produced by the code in Figure \ref{fig:video:2Dfilter}.  The shifted output in image c is produced by the code in Figure \ref{fig:video:2Dfilter_linebuffer}.  The output in image d is produced by the code in Figure \ref{fig:video:2Dfilter_linebuffer_extended} and is identical to image b.}\label{fig:video:filter2D_results_withshifting}
\end{figure}

\begin{figure}
\centering
\executeiffilenewer{video_timelines.svg}{images/video_timelines.pdf}%
{inkscape -z -D --file=video_timelines.svg %
--export-pdf=images/video_timelines.pdf --export-latex}%
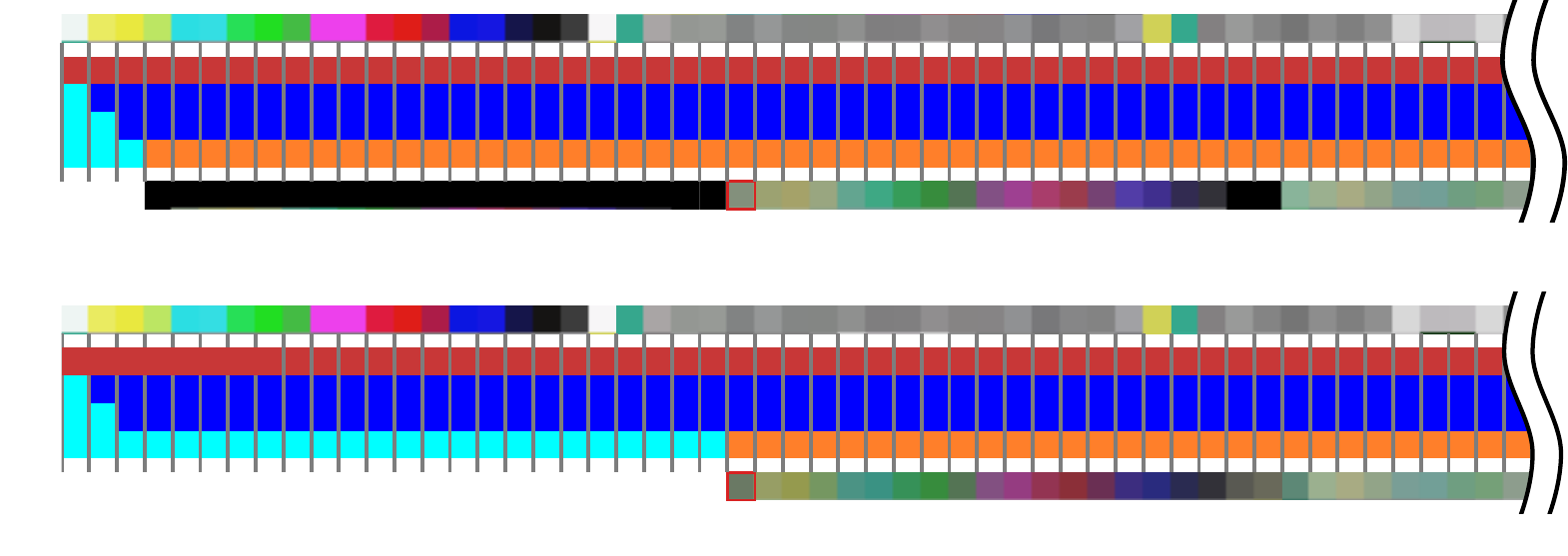%

\caption{Timeline of execution of code implemented with line buffers.  The top timeline shows the behavior of the code in  Figure \ref{fig:video:2Dfilter_linebuffer}.  The bottom timeline shows the behavior of the code in  Figure \ref{fig:video:2Dfilter_linebuffer_extended}.  The pixel marked in red is output on the same cycle in both implementations.  In the first case it is interpreted to be the second pixel of the second line and in the second case, it is interpreted as the first pixel of the first line.}\label{fig:video:timelines}
\end{figure}

\subsection{Boundary Conditions}

In most cases, the processing window contains a region of the input image.  However, near the boundary of the input image, the filter may extend beyond the boundary of the input image.  Depending on the requirements of different applications, there are many different ways of accounting for the behavior of the filter near the boundary.  Perhaps the simplest way to account for the boundary condition is to compute a smaller output image that avoids requiring the values of input pixels outside of the input image.  However, in applications where the output image size is fixed, such as Digital Television, this approach is generally unacceptable.  In addition, if a sequence of filters is required, dealing with a large number images with slightly different sizes can be somewhat cumbersome.  The code in Figure \ref{fig:video:2Dfilter_linebuffer} creates an output with the same size as the input by padding the smaller output image with a known value (in this case, the color black).   Alternatively, the missing values can be synthesized, typically in one of several ways.
\begin{itemize}
\item Missing input values can be filled with a constant
\item Missing input values can be filled from the boundary pixel of the input image.
\item Missing input values can be reconstructed by reflecting pixels from the interior of the input image.
\end{itemize}
Of course, more complicated and typically more computationally intensive schemes are also used.

\begin{figure}
\centering
\executeiffilenewer{filter2d_results_boundary_conditions.svg}{images/filter2d_results_boundary_conditions.pdf}%
{inkscape -z -D --file=filter2d_results_boundary_conditions.svg %
--export-pdf=images/filter2d_results_boundary_conditions.pdf --export-latex}%
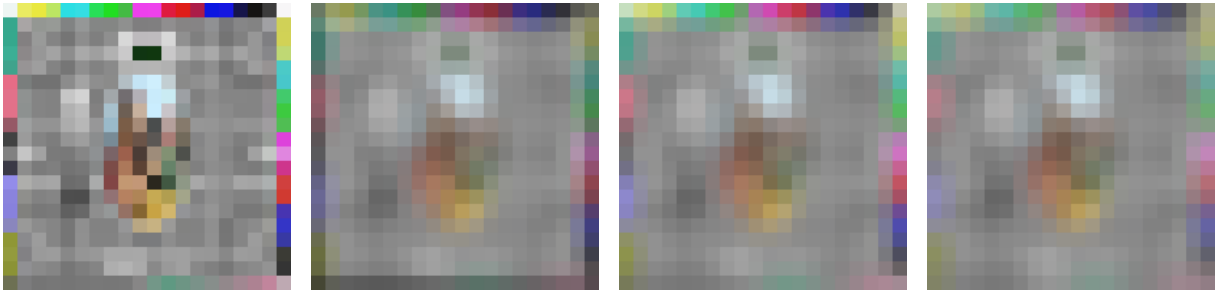%

\caption{Examples of the effect of different kinds of boundary conditions.}\label{fig:video:boundary_conditions}
\end{figure}

One way of writing code to handle boundary conditions is shown in Figure \ref{fig:video:boundaryConditionExtendBad}.  This code computes an offset address into the window buffer for each pixel in the window buffer.  However, there is a significant disadvantage in this code since each read from the window buffer is at a variable address.  This variable read results in multiplexers before the filter is computed.  For an N-by-N tap filter, there will be approximately N*N multiplexers with N inputs each.  For simple filters, the cost of these multiplexers (and the logic required to compute the correct indexes) can dominate the cost of computing the filter.

\begin{figure}
\begin{lstlisting}[format=none,firstline=22]
void video_2dfilter_linebuffer_extended_constant(
	rgb_pixel pixel_in[MAX_HEIGHT][MAX_WIDTH], rgb_pixel pixel_out[MAX_HEIGHT][MAX_WIDTH]) {
#pragma HLS interface ap_hs port=pixel_out
#pragma HLS interface ap_hs port=pixel_in
	rgb_pixel window[3][3];
	rgb_pixel line_buffer[2][MAX_WIDTH];
#pragma HLS array_partition variable=line_buffer complete dim=1
row_loop: for(int row = 0; row < MAX_HEIGHT+1; row++) {
	col_loop: for(int col = 0; col < MAX_WIDTH+1; col++) {
#pragma HLS pipeline II=1
			rgb_pixel pixel;
			if(row < MAX_HEIGHT && col < MAX_WIDTH) {
				pixel = pixel_in[row][col];
			}
			for(int i = 0; i < 3; i++) {
				window[i][0] = window[i][1];
				window[i][1] = window[i][2];
			}
			if(col < MAX_WIDTH) {
				window[0][2] = (line_buffer[0][col]);
				window[1][2] = (line_buffer[0][col] = line_buffer[1][col]);
				window[2][2] = (line_buffer[1][col] = pixel);
			}
			if(row >= 1 && col >= 1) {
				int outrow = row-1;
				int outcol = col-1;
				rgb_pixel window2[3][3];
				for (int i = 0; i < 3; i++) {
					for (int j = 0; j < 3; j++) {
						int wi, wj;
						if (i < 1 - outrow) wi = 1 - outrow;
						else if (i >= MAX_HEIGHT - outrow + 1) wi = MAX_HEIGHT - outrow;
						else wi = i;
						if (j < 1 - outcol) wj = 1 - outcol;
						else if (j >= MAX_WIDTH - outcol + 1) wj = MAX_WIDTH - outcol;
						else wj = j;
						window2[i][j] = window[wi][wj];
					}
				}
				pixel_out[outrow][outcol] = filter(window2);
			}
		}
	}
}
\end{lstlisting}
\caption{Code implementing a 2D filter with an explicit line buffer and constant extension to handle the boundary condition.   Although correct, this code is relatively expensive to implement.}\label{fig:video:boundaryConditionExtendBad}
\end{figure}

An alternative technique is to handle the boundary condition when data is written into the window buffer and to shift the window buffer in a regular pattern.  In this case, there are only N multiplexers, instead of N*N, resulting in significantly lower resource usage.  

\begin{exercise}
Modify the code in Figure \ref{fig:video:boundaryConditionExtendBad} to read from the window buffer using constant addresses.  How many hardware resources did you save?
\end{exercise}


\section{Conclusion}

Video processing is a common FPGA application and are highly amenable to HLS implementation.  A key aspect of most video processing algorithms is a high degree of data-locality enabling either streaming implementations or applications with local buffering and a minimum amount of external memory access.


\chapter{Sorting Algorithms}
\glsresetall
\label{chapter:sorting}

\section{Introduction}
\label{sec:sorting_introduction}

Sorting is a common algorithm in many systems.  It is a core algorithm for many data structures because sorted data can be efficiently searched in $\mathcal{O}(\log n)$ time using binary search.  For example, given an sequence of elements:
\begin{equation}
A = \{1,4,17,23,31,45,74,76\}
\end{equation}
We can find whether a number exists in this set of data without comparing it against all 8 elements.  Because the data is sorted we can start by picking an element in the middle of the array and checking to see whether our number is greater than or less than the element.  For example, if we were looking to find the element 45, we could start by comparing with $A(4) = 31$.  Since $45 > 31$ we can eliminate $A(0..4)$ from further comparison and only consider $A(5..7)$. 
\begin{equation}
\setlength{\fboxsep}{1pt}
A = \{\colorbox{black}{1},\colorbox{black}{4},\colorbox{black}{17},\colorbox{black}{23},\colorbox{black}{31},45,74,76\}
\end{equation}
Next if we compare with $A(6) = 74$, we see that $45 < 74$, so we can eliminate all but $A(5)$ from consideration. 
\begin{equation}
\setlength{\fboxsep}{1pt}
A = \{\colorbox{black}{1},\colorbox{black}{4},\colorbox{black}{17},\colorbox{black}{23},\colorbox{black}{31},45,\colorbox{black}{74},\colorbox{black}{76}\}
\end{equation}
 One additional comparison with $A(5)$ allows us to determine that 45 is indeed contained in the sequence.

In our example, the sequence $A$ could represent a variety of different concepts.  $A$ could represent a mathematical set and searching for an element in the sequence could tell us whether a value exists in the set.  $A$ could also represent only a portion of the data, often called a \term{key}, which is useful for indexing the rest of the information.  For instance, the key could be a person's name.  After searching based on the key then we know the position where the rest of the data about the person, such as their birthdate, is stored.  In yet other cases, the key could be something more abstract, such as a cryptographic hash of some data or a key.  In this case, the order that the data is stored is likely to be randomized, but we can still find it if we know the right cryptographic hash to search for.  In each case, the fundamental operations of required for sorting and searching are pretty much the same, primarily we need to have the ability to compare two different values.  For the remainder of this chapter we will mostly ignore these differences. 

There are a wide variety of different sorting techniques that have been studied in processor systems\cite{knuth1998art}.  These different algorithms vary in terms of their fundamental $\mathcal{O}()$ complexity, but many require $\mathcal{O}(n \log n)$ comparisons to sort $n$ elements.  Conceptually, this is intuitive since we could search for the right location to insert a new value in a sorted data set with $\mathcal{O}(\log n)$ comparisons using binary search. To insert $n$ elements this process would need to be repeated $n$ times, once for each element.

In practice, the cost of inserting an element can be significant, depending on the data structure being sorted.  In processor systems a variety of factors influence overall performance, such as memory locality when processing large data sets or their ability to easily parallelize across multiple cores.  In HLS, we have similar considerations where it is common to trade off increased resource usage for reduced processing time.  In many cases, this might require a variety of algorithms and implementation techniques in order to obtain the best design.  The best techniques for making these tradeoffs are an area of active research\cite{marcelino2008sorting,mueller2012sorting,matai2016sorting}.

Characteristics other than performance also affect the choice of sorting algorithms.  For instance, we might consider:
\begin{itemize}
\item \textbf{Stability:} A sort is a stable if when two items in the input data have the same key, then they will appear in the same order on the output. For example, we might sort a set of records containing people's names and ages using the ages as the sort key. In the input data, John appears before Jane, and both are 25 years old. A \gls{stable_sort} will ensure that John and Jane remain in the same order after sorting.  
\item \textbf{Online:} The algorithm allows for data to be sorted as it is received. This can be particularly valuable when data is not accessible when the sort starts or must be read in sequence from external storage. 
\item \textbf{In-place:} A list with \lstinline{n} elements can be sorted using \lstinline{n} memory elements. Some algorithms require additional storage during the sorting process.
\item \textbf{Adaptive:} It is efficient for data that is already relatively sorted. For example, if data is already sorted, then some algorithms might run faster, e.g. in linear $\mathcal{O}(n)$ time.
\end{itemize}

\section{Insertion Sort}
\label{sec:insertion_sort}

Insertion sort is one of the basic sorting algorithms. It works by iteratively placing the items of an array into sorted order and builds the sorted array one element at a time.  Each iteration selects an unsorted element and places it in the appropriate order within the previously sorted elements. It then moves onto the next element. This continues until all of the elements are considered, and the entire array is sorted.

To make this more formal, assume that we are given an input array \lstinline{A} that should be put into sorted order.  The base case is the first element of that array \lstinline{A[0]}, which by default is a sorted subarray (since it is only one element). The next step is to consider element \lstinline{A[1]}, and place it into the sorted subarray such that the new subarray (with two elements) is also sorted. We continue this process for each element \lstinline{A[i]} until we have iterated across all of the elements of \lstinline{A}. At each step, we take the new element \lstinline{A[i]} and insert it into the proper location such that the subarray \lstinline{A[0..i-1]} remains sorted. Figure \ref{eq:insertion_sort} gives a step by step view of insertion sort operating on an array.
\begin{figure}
\centering
\setlength{\fboxsep}{1pt}
\{\hspace*\fboxsep\underline{3},\hspace*\fboxsep{} 2, 5, 4, 1\} \\
\{\colorbox{black!10}{3,} \underline{2}, 5, 4, 1\} \\
\{\colorbox{black!10}{2, 3,} \underline{5}, 4, 1\} \\
\{\colorbox{black!10}{2, 3, 5,} \underline{4}, 1\} \\
\{\colorbox{black!10}{2, 3, 4, 5,} \underline{1}\} \\
\{\colorbox{black!10}{1, 2, 3, 4, 5}\} \\
\caption{The Insertion Sort algorithm operating on an array.  The initial array is shown at the top.  In each step of the algorithm, the underlined algorithm is considered and placed into sorted order of the elements to it's left. At each stage, the shaded elements are in sorted order. }
\label{eq:insertion_sort}
\end{figure}
The first line is trivial. We consider only the first value \lstinline{3} which makes a subarray with one element. Any subarray with one element is in sorted order. The second line places the second value \lstinline{2} into the sorted subarray. The end result is that the value \lstinline{2} is placed into the first element of the sorted subarray, which shifts the previous sorted element \lstinline{3} to the right. The third line moves the third entry of the initial array into its appropriate place in the sorted subarray. In this case, since \lstinline{A[2] = 5}, it is already in its correct location. Thus, nothing needs to happen. The fourth line considers the value \lstinline{4}. This is moved into its appropriate place, shifting \lstinline{5} to the right. Finally, the fifth line considers the placement of the value \lstinline{1}. This is placed into the first location of the array, and all of the previous sorted values are shifted by one location.

Insertion sort is a stable, online, in-place, adaptive sorting algorithm.  Because of these properties, insertion sort is often preferred when sorting small arrays or as a base case in a recursive sorting algorithm. For example, more complex algorithms might decompose a large data set into a number of smaller arrays and then these small arrays will be sorted using insertion sort. The result is that formed by combining the sorted arrays.

\subsection{Basic Insertion Sort Implementation}
\label{sec:basic_insertion_sort}

\begin{figure}
\begin{lstlisting}
#include "insertion_sort.h"
void insertion_sort(DTYPE A[SIZE]) {
 L1:
    for(int i = 1; i < SIZE; i++) {
      DTYPE item = A[i];
	  int j = i;
      DTYPE t = A[j-1];
    L2:
      while(j > 0 && A[j-1] > item && j > 0) {
          #pragma HLS pipeline II=1
		  A[j] = A[j-1];
		  j--;
	  }
	  A[j] = item;
  }
}
\end{lstlisting}
\caption{  The complete code for insertion sort. The outer \lstinline{for} loop iterates across the elements one at a time. The inner \lstinline{while} loop moves the current element into sorted place.  }
\label{fig:insertion_sort.cpp}
\end{figure}

Figure \ref{fig:insertion_sort.cpp} shows basic C code for insertion sort. The outer loop, labeled \lstinline{L1}, iterates from elements \lstinline{A[1]} to \lstinline{A[SIZE - 1]} where \lstinline{SIZE} denotes the number of elements in the array \lstinline{A}. We do not need to start at element \lstinline{A[0]} since any one element is already in sorted order. Each iteration of the \lstinline{L1} loop starts by copying the current element that we wish to insert into the sorted subarray (i.e., \lstinline{A[i]}) into the \lstinline{item} variable and then executes the inner \lstinline{L2} loop. The inner loop walks down the sorted portion of \lstinline{A[]} looking for the appropriate location to place the value \lstinline{index}. The inner loop executes as long as it has not arrived at the end of the array (the condition \lstinline{j > 0}) and the array elements are greater than the item being inserted (the condition \lstinline{A[j-1] > index}). As long as the loop condition is satisfied, elements of the sorted subarray are shifted by one element (the statement \lstinline{A[j] = A[j-1])}. This will make room for the insertion of \lstinline{index} when we eventually find its correct location. When the loop exits, we have found the correct location for \lstinline{index} and store it there. After the completion of iteration \lstinline{i}, the elements from \lstinline{A[0]} to \lstinline{A[i]} are in sorted order.

The code in Figure \ref{fig:insertion_sort.cpp} is an straightforward implementation without any optimizations. We can optimize it using different \VHLS directives, such as \lstinline{pipeline}, \lstinline{unroll}, and \lstinline{array_partition}. The simplest optimization would be to pipeline the inner loop, by applying the \lstinline{pipeline} directive to the body of the inner loop.  In this case, even though the inner loop is accessing different elements of \lstinline{A[]} there are no data dependencies in these array accesses, so we could expect to achieve a loop II of 1. The resulting accelerator would perform roughly $N^2/4$ data comparisons\cite{sedgewickalgorithmsinC} on average and have a latency of roughly $N^2/4$ cycles, since it performs one comparison per clock cycle.  In actuality, an accelerator generated from \VHLS will have slightly higher latency to account for the sequential execution of the outer loop.   In order to achieve higher performance, we could also attempt to move the \lstinline{pipeline} directive to the outer \lstinline{L1} loop or to the function body itself.  We could also combine these options with partial loop unrolling.  Some of these options are shown in Table \ref{table:insertion_sort:options}.

\begin{table}[htbp]
  \centering
  \caption{Possible options to optimize the basic \lstinline{insertion_sort} function in Figure \ref{fig:insertion_sort.cpp} through directives.}
  \label{tbl:sw_insertionsort_vs_hardwareA}
	\tabcolsep=0.2cm
	 \centering 
	 \begin{tabularx}{400pt}{cXccc}
    \toprule
     & Directives & II &  Period & Slices   \\
    \midrule
1 & \lstinline|L2: pipeline II=1|  & ? & ? & ?  \\ \midrule
2 & \lstinline|L2: pipeline II=1|& ?  & ? & ?   \\
 &\lstinline|L2: unroll factor=2|&&&\\
 &\lstinline|array_partition variable=A cyclic factor=2| &&&\\ \midrule
3 & \lstinline|L1: pipeline II=1| & ? & ? & ?      \\ \midrule
4 & \lstinline|L1: pipeline II=1| & ?  & ? & ?   \\
& \lstinline|L1: unroll factor=2| &&&\\
& \lstinline|array_partition variable=A complete| &&& \\ \midrule
5 & \lstinline|function pipeline II=1| & ? & ? & ?     \\
& \lstinline|array_partition variable=A complete| &&& \\ 
    \bottomrule
    \end{tabularx}
  \label{table:case-studies}
\end{table}

\begin{exercise}
Explore the options in Table \ref{table:case-studies}.  Synthesize each of these designs and determine the initiation interval (II), clock period, and required number of slices.  Which options are successful in improving latency and/or throughput?  What would happen if you combined the directives from multiple rows into one design?
\end{exercise}

Unfortunately, although this code seems similar to other nested loop programs we've looked at previously, it does have some aspects that can make it difficult to optimize.  Even Option 1, which simply attempts to pipeline the inner loop, can fail to achieve II=1.  Although there are no significant data recurrences, there is a recurrence in the control path that affects whether or not the pipeline can execute.  In this case, the code must read \lstinline{A[i-1]} in order to determine whether the loop should actually execute.  \VHLS incorporates this read into the loop pipeline, but if the read of \lstinline{A} is pipelined then the loop exit check cannot be made in the first stage of the pipeline.  This is an example of a recurrence which includes the HLS-generated control logic for the loop.  If such recurrences exist, then \VHLS will issue a message indicating that the loop exit condition cannot be scheduled in the first II clock cycles.  This situation can also occur a \lstinline{break} or \lstinline{continue} statement exists under a complex control condition.  One solution is to explicitly speculate the read of \lstinline{A[i-1]} out of the loop.  This enables the loop exit check to be scheduled with II=1 at the expense of an additional array access on the last iteration.  This code is shown in Figure \ref{fig:insertion_sort_relaxed.cpp}.

\begin{figure}
\begin{lstlisting}
#include "insertion_sort.h"
void insertion_sort(DTYPE A[SIZE]) {
 L1:
    for(i = 1; i < SIZE; i++) {
        DTYPE item = A[i];
        j = i;
        DTYPE t = A[j-1];
    L2:
        while(j > 0 && t > item) {
#pragma HLS pipeline II=1
            A[j] = t;
            t = A[j-2];
            j--;
        }
        A[j] = item;
    }
}
\end{lstlisting}
\caption{Refactored insertion sort for Option 1 in Table \ref{table:case-studies}.}
\label{fig:insertion_sort_relaxed.cpp}
\end{figure}

Option 2 in the Table unrolls the inner loop by a factor of 2, attempting to perform two shift operations every clock cycle.  This could potentially reduce the latency to compute \lstinline{insertion_sort} by a factor of two.  Unfortunately, \VHLS cannot achieve a loop II of 1 for this code using \gls{arraypartitioning}, because each array access cannot be assigned to a different memory partition.

\note{We should show manual code for how to address this.}

\begin{aside}
In \VHLS, the \lstinline{array_partition} directive results in implementing a number of completely separate memories.  For instance, \lstinline{array_partition variable=A cyclic factor=4} would result in generating four separate memories from array \lstinline{A[]}, each of which contains a portion of the array contents.  We can often think that this optimization provides four times the number of memory accesses each clock, but this is only the case if each memory access can be assigned to exactly one of the memory partitions.   For example an array access \lstinline{A[i]} for unknown \lstinline{i} could reference data stored in any partition, while an array access \lstinline{A[4*i+2]} would only access data in the third partition.  More complex logic, often called \term{memory banking}, can resolve a number of independent accesses \lstinline{A[i]}, \lstinline{A[i+1]}, \lstinline{A[i+6]}, \lstinline{A[i+7]} and perform these accesses in the same clock cycle.  Memory banking requires additional crossbar logic to route these simultaneous accesses in a circuit, since \lstinline{i} can be arbitrary.  At compile time, we can guarantee that the constant offsets of these accesses will hit in different banks, but the actual banks cannot be determined until \lstinline{i} is known.  Yet more complex logic could implement stalling logic, enabling a set of unrelated accesses to complete in a single clock cycle if they hit in different banks.  If the accesses happen to all hit in the same bank, then the stalling logic can delay the progress of the circuit for a number of clocks until all of the accesses have completed.  Lastly, multiport architectures have designed that can allow a number of accesses guaranteed completion every clock cycle\cite{ug574,Abdelhadi2014multiport,Laforest2014multiport} by replicating data across normal memories with one or two physical ports. 
\end{aside}

Option 3 in the table also fails to achieve significant improvement.  Since the inner \lstinline{L2} loop does not have a statically computable loop bound \VHLS is unable to construct a pipeline from the body of the \lstinline{L1} loop.  Unfortunately, this is a case where exploring the design space of interesting alternatives requires code restructuring in addition to the use of directives.  Finding the best code restructuring requires not only understanding the algorithm, but also having a sense of the architecture that will be generated by the HLS process\cite{george2014hardware, matai2014enabling}. For example, we discuss one such code restructuring for insertion sort in the following section. You will see that that code is significantly different from the code in Figure \ref{fig:insertion_sort.cpp}.

In the following, we attempt to demonstrate several concepts. First, writing efficient high-level synthesis code requires that the designer must understand hardware concepts like unrolling and partitioning. Second, the designer must be able to diagnose any throughput problems, which requires substantial knowledge about both the application and the hardware implementation of that design. Third, and most importantly, in order to achieve the best results, i.e. high performance and low-area, it is often required to rewrite the code in a manner that will create an efficient hardware architecture.  This can be very different from code that results in the most efficient software.

\subsection{Parallelising Insertion Sort}

In order significantly increase the performance of insertion sort, we'd like to get to the point where we can insert a new element every clock cycle.  When inserting the last element into the sorted list, this might require shifting all of the elements in the array.  For the code in Figure \ref{fig:insertion_sort.cpp}, this means that the inner while loop could actually execute over all of the elements in the array.  Intuitively, we realize that inserting a new element into the sorted list every clock cycle requires enough hardware operators to perform a comparison on every element of the array in a single clock cycle.  To enable pipelining of the outer loop, we can convert the inner \lstinline{L2} loop with variable loop bounds into a fixed-bound loop, enabling it to be unrolled and integrated into the \lstinline{L1} loop pipeline.  Code that does this is shown in Figure \ref{fig:insertion_sort_parallel.cpp}.

\begin{figure}
\begin{lstlisting}
#include "insertion_sort_parallel.h"
#include "assert.h"
void insertion_sort_parallel(DTYPE A[SIZE], DTYPE B[SIZE]) {
#pragma HLS array_partition variable=B complete
 L1:
    for(int i = 0; i < SIZE; i++) {
#pragma HLS pipeline II=1
        DTYPE item = A[i];
    L2:
        for(int j = SIZE-1; j >= 0; j--) {
            DTYPE t;
            if(j > i) {
                t = B[j];
            } else if(j > 0 && B[j-1] > item) {
                t = B[j-1];
            } else {
                t = item;
                if (j > 0)
                    item = B[j-1];
            }
            B[j] = t;
        }
    }
}
\end{lstlisting}
\caption{Refactored insertion sort for Option 3 in Table \ref{table:case-studies}.}
\label{fig:insertion_sort_parallel.cpp}
\end{figure}

This code contains the exit condition of the original loop (\lstinline{L2} in Figure \ref{fig:insertion_sort.cpp}) as an \lstinline{if} condition in the body of the new \lstinline{L2} loop.  The other branches of the conditional are added to handle the expanded iteration space of the loop, essentially performing no operations when the original loop would not be executing.  In addition, the loop now contains the final assignment of \lstinline{item} in the array, rather than performing the assignment outside of the loop.  When the inner loop is unrolled, remember that \lstinline{j} will become a constant in all of the unrolled instances of the loop.  As a result, each read and write from \lstinline{B[]} will be performed at a constant index and comparisons between \lstinline{j} and a constant will be completely optimized away.  The \lstinline{item} variable, on the other hand, is possibly assigned in every copy of the inner loop.  During compilation, \VHLS creates a separate copy of this variable for each separate possible assignment and each possible assignment results in multiplexer in the implemented circuit.

\begin{aside}
The conversion of a single variable into multiple versions is a common internal transformation used by compilers.  The resulting internal representation is called \gls{ssa}.  To merge values coming from different points in the code, the \gls{ssa} internal representation include artificial `phi-functions' represented by the greek letter $\phi$.  These phi-functions often result in multiplexers in the circuit generated by \VHLS and you'll probably find the related resources in the tool reports if you look carefully.
\end{aside}

The parallelized insertion sort in Figure \ref{fig:insertion_sort_parallel.cpp} essentially results in a number of copies of the body of the inner loop.   The contents of this inner loop consists of a few multiplexers, a comparator to determine the smallest of two elements, and a register to store an element of \lstinline{B[]}.  Incidentally, each stage might also include pipeline registers, if needed, to ensure that the resulting circuit runs at an effective clock frequency.  We'll call the contents of the inner loop a \gls{sorting_cell}.  The whole insertion sort function consists of a one-dimensional array of sorting cells, combined with a small amount of additional logic to feed data in the input and capture the output at the right time, in this case after \lstinline{SIZE} elements have been processed by the outer loop.  This array of sorting cells has an interesting property where each sorting cell only communicates with it's neighboring sorting cells, rather than with all cells.  Designs like this are called \glspl{systolic_array} and are a common technique for parallelizing algorithms. In many cases, including sorting, systolic array implementations can naturally arise when we unroll inner loops, as long as the communication between different loop iterations is limited.   We call this type of design an implicit systolic array.

\subsection{Explicit Systolic Array For Insertion Sort}
\label{sec:insertion_cells}
Systolic arrays have been well researched and many parallel algorithms are published as systolic arrays.  In particular, the idea of using a linear array of sorting cells to implement insertion sort is well understood\cite{ortiz2011streaming, bednara2000tradeoff, marcelino2008sorting, arcas2014empirical}. However, rather than being described as an unrolled loop, systolic arrays are often described as components communicating by streams of data. This section describes an alternative coding style based on explicit streams and the \lstinline{dataflow} directive that can be a more intuitive way to describe a systolic array.
  
\begin{figure}
\centering
\includegraphics[width= \textwidth]{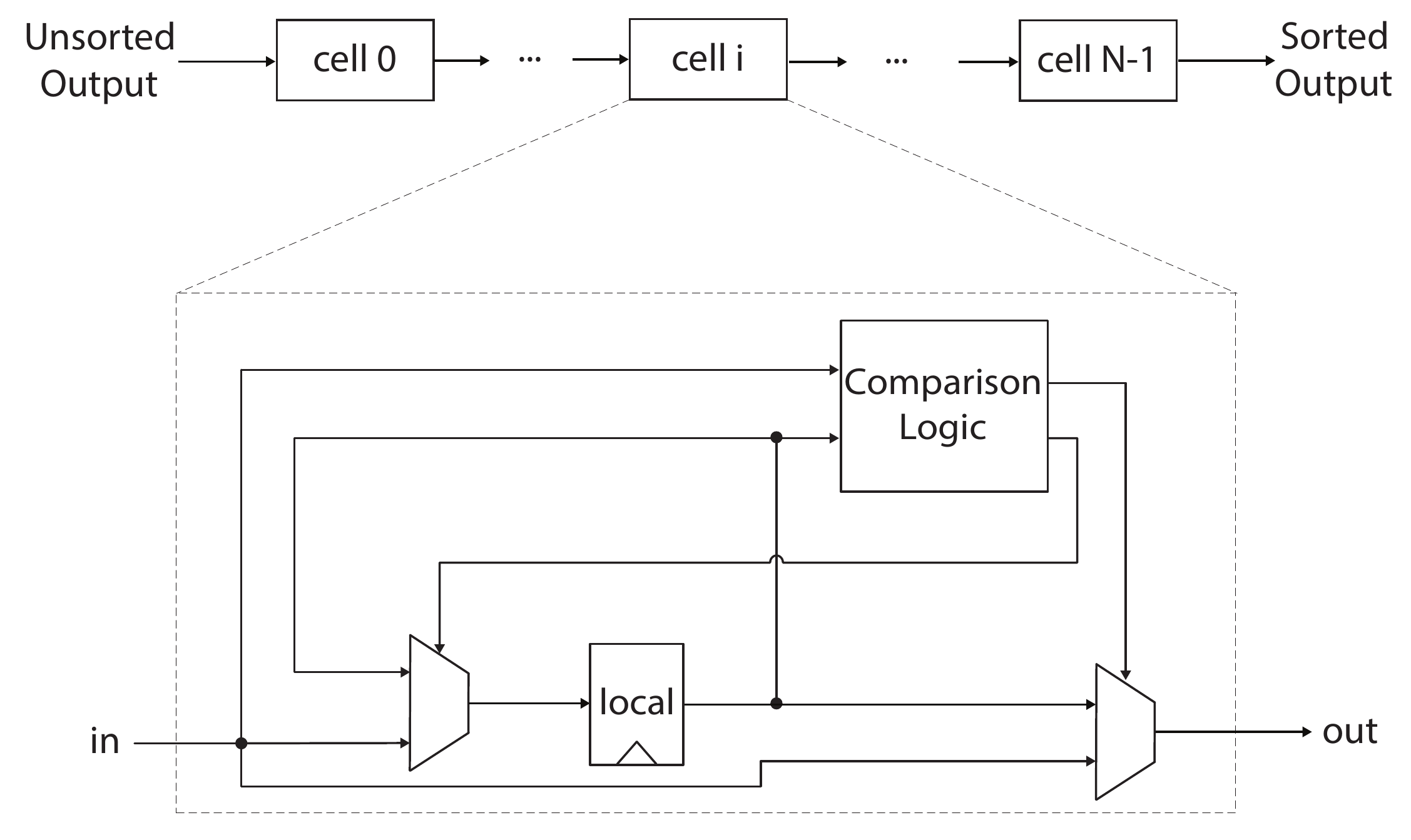}
\caption{ The architecture of one insertion cell. Each cell holds exactly one value in the register \lstinline{local}. On each execution it receives an input value \lstinline{in}, compares that to its \lstinline{local} value, and writes the smaller of the two to \lstinline{output}. Sorting \lstinline{N} values requires \lstinline{N} cells. }
\label{fig:insertion_cell}
\end{figure}

Figure~\ref{fig:insertion_cell} shows a \gls{systolic_array} implementing insertion sort. Each cell is identical and compares its input (\lstinline{in}) with the value in current register \lstinline{local}. The smaller value is passed to the output \lstinline{out}, while the larger value is stored back in \lstinline{local}. In other words, \lstinline{out = min(in,local)}.  The output of cell $i$ is passed as input to the next cell $i+1$ in the linear array. As a new input arrives it will be compared against elements stored in the array until it finds it's correct place.  If a new input is larger than all of the  values in the array, then the sorted values will shift one cell to the right.  If a new input is smaller than all of the values in the array, then it will propagate through the array, and eventually become stored in the furthest right cell. After all the data has moved through the array, the smallest data element will be sorted in cell $N-1$ and can be read from the output. 

\begin{figure}
\begin{lstlisting}[firstline=3, lastline=15]
void cell0(hls::stream<DTYPE> & in, hls::stream<DTYPE> & out)
{
	static DTYPE local = 0;
	DTYPE in_copy = in.read();
	if(in_copy > local) {
		out.write(local);
		local = in_copy;
	}
	else
	{
		out.write(in_copy);
	}
}
\end{lstlisting}
\caption{  The \VHLS C code corresponding to one insertion cell \lstinline{cell0}. The other cells have the exact same code except with a different function name (\lstinline{cell1}, \lstinline{cell2}, etc.). The code performs the same functionality as shown in the architectural diagram in Figure \ref{fig:insertion_cell}. It uses the \lstinline{hls:stream} interface for the input and output variables. \lstinline{hls::stream} provides a convenient method to create FIFOs that work both for synthesis and simulation. }
\label{fig:insertion_cell_sort.cpp}
\end{figure}

The code for one insertion cell is shown in Figure \ref{fig:insertion_cell_sort.cpp}. The code uses a streaming interface by declaring the input and output variables as an \lstinline{hls::stream} type. \lstinline{DTYPE} is a type parameter enabling different types to be operated on. The \lstinline{local} variable stores one element of the array being sorted. It is \lstinline{static} because we want to preserve its value across multiple function calls. This does create a problem since we must replicate this \lstinline{cell} function \lstinline{N} times.  Using the same function (e.g., calling the same \lstinline{cell} function N times) would cause a problem since each cell must have a separate static variable.  One static variable cannot be shared across N functions.



Linking the insertion cells together is straightforward. The function \lstinline{insertion_cell_sort} in Figure \ref{fig:partial_insertion_cell_sort.cpp} shows the code for sorting eight elements. Expanding this to a larger number of elements simply requires replicating more \lstinline{cell} functions, additional \lstinline{hls::stream} variables, instantiating these functions, and connecting their input and output function arguments in the appropriate manner.

\begin{figure}
\begin{lstlisting}
void insertion_cell_sort(hls::stream<DTYPE> & in, hls::stream<DTYPE> & out)
{
	#pragma HLS DATAFLOW
	hls::stream<DTYPE> out0("out0_stream");
	hls::stream<DTYPE> out1("out1_stream");
	hls::stream<DTYPE> out2("out2_stream");
	hls::stream<DTYPE> out3("out3_stream");
	hls::stream<DTYPE> out4("out4_stream");
	hls::stream<DTYPE> out5("out5_stream");
	hls::stream<DTYPE> out6("out6_stream");

	// Function calls;
	cell0(in, out0);
	cell1(out0, out1);
	cell2(out1, out2);
	cell3(out2, out3);
	cell4(out3, out4);
	cell5(out4, out5);
	cell6(out5, out6);
	cell7(out6, out);
}
\end{lstlisting}
\caption{  Insertion cell sorting for eight elements. The function takes an input \lstinline{hls::stream} and outputs the elements in sorted order one at a time through the variable \lstinline{out}. The order starts with the smallest element first, and then continues on to increasingly larger elements. }
\label{fig:partial_insertion_cell_sort.cpp}
\end{figure}

\begin{exercise}
The existing implementation of \lstinline{insertion_cell_sort} outputs the data starting with the smallest element and continues outputting increasingly larger elements. What changes are required in order to reverse this order, i.e., to output the largest element first, and then the decreasingly smaller outputs? 
\end{exercise}

The \lstinline{insertion_cell_sort} function must be called multiple times in order to sort the entire data. Each call to \lstinline{insertion_cell_sort} provides one data element to the array for sorting. The first call places the input data in its appropriate place. Where would this data be placed? To answer that question, we should point out that we make an assumption on the input data since we are initializing the \lstinline{local} variable to \lstinline{0}. This is done in all of the \lstinline{cell} functions.

\begin{exercise}
Initializing the \lstinline{static} variable \lstinline{local} to \lstinline{0} makes an assumption on the type of data that will be provided for sorting. What is this assumption? In other words, what is the range of values for the input data that will be properly handled? What would happen if input data from outside of this range was given to the \lstinline{insertion_cell_sort} function? Is there a better way to initialize the \lstinline{local} variable?
\end{exercise}

After making eight calls to the \lstinline{insertion_cell_sort} function, all of the data will be put into the eight \lstinline{local} variables in each of the \lstinline{cell} functions. 

\begin{exercise}
How many times must we call the \lstinline{insertion_cell_sort} function in order to get the first sorted element? How many calls are necessary in order to all of the sorted data? What if we increased the array to \lstinline{N} cells? Can you generalize these two functions (number of times to call \lstinline{insertion_cell_sort} to load \lstinline{N} elements, and the number of calls required to output all \lstinline{N} elements)?
\end{exercise}

To achieve a task-pipelined architecture consisting of the eight \lstinline{cell} functions, the code specifies the \lstinline{dataflow} directive. Each execution of the toplevel function processes a new input sample and inserts it into the sorted sequence.  The actual insertion is pipelined, with one pipeline stage for each call to the \lstinline{cell} function. With eight calls to the \lstinline{cell} function, we can only sort sequences with eight values, but this can be extended almost arbitrarily since the code contains no recurrences between the cells.

\begin{exercise}
How many cycles are required to sort an entire array? We define the sorting to be completed when all of the sorted data is output from the array of \lstinline{cell} function inside of the \lstinline{insertion_cell_sort}, i.e., all eight elements have been output from argument \lstinline{out} in the function \lstinline{insertion_cell_sort}? How does the cycle count change if we remove the \lstinline{dataflow} directive? How does change the resource utilization?
\end{exercise}

Figure \ref{fig:insertion_cell_sort_test.cpp} shows the code for the testbench. The testbench generates random data to be sorted in \lstinline{input[]}. This array is sorted by calling the \lstinline{insertion_cell_sort()} function multiple times, with the result appearing in \lstinline{cell_output[]}. Next, the same data is sorted in place using the \lstinline{insertion_sort} function from Figure \ref{fig:insertion_sort.cpp}. Finally, the testbench compares the results of these two sort implementations. The testbench passes if the sorted order from both implementations is the same.

\begin{figure}
\footnotesize
\begin{lstlisting}[format=none]
#include "insertion_cell_sort.h"
#include <iostream>
#include <stdlib.h>

const static int DEBUG=1;
const static int MAX_NUMBER=1000;
int main () {
	int fail = 0;
	DTYPE input[SIZE];
	DTYPE cell_output[SIZE] = {0};
	hls::stream<DTYPE> in, out;

	//generate random data to sort
	if(DEBUG) std::cout << "Random Input Data\n";
	srand(20); //change me if you want different numbers
	for(int i = 0; i < SIZE; i++) {
		input[i] = rand() % MAX_NUMBER + 1;
		if(DEBUG) std::cout << input[i] << "\t";
	}

	//process the data through the insertion_cell_sort function
	for(int i = 0; i < SIZE*2; i++) {
		if(i < SIZE) {
			//feed in the SIZE elements to be sorted
			in.write(input[i]);
			insertion_cell_sort(in, out);
			cell_output[i] = out.read();
		} else {
			//then send in dummy data to flush pipeline
			in.write(MAX_NUMBER);
			insertion_cell_sort(in, out);
			cell_output[i-SIZE] = out.read();
		}
	}

	//sort the data using the insertion_sort function
	insertion_sort(input);

	//compare the results of insertion_sort to insertion_cell_sort; fail if they differ
	if(DEBUG) std::cout << "\nSorted Output\n";
	for(int i = 0; i < SIZE; i++) {
		if(DEBUG) std::cout << cell_output[i] << "\t";
	}
	for(int i = 0; i < SIZE; i++) {
		if(input[i] != cell_output[i]) {
			fail = 1;
			std::cout << "golden=" << input[i] << "hw=" << cell_output[i] << "\n";
		}
	}

	if(fail == 0) std::cout << "PASS\n";
	else std::cout << "FAIL\n";
	return fail;
}
\end{lstlisting}
\caption{  The testbench for the \lstinline{insertion_cell_sort} function. }
\label{fig:insertion_cell_sort_test.cpp}
\end{figure}

In the testbench, the \lstinline{SIZE} constant is set to the number of elements being sorted. This would be \lstinline{8} in the running example throughout this chapter. The \lstinline{DEBUG} constant is used to provided output detailing the execution of the testbench. This should be set to a non-zero value if you wish to see the debugging data, and \lstinline{0} if you want the output to be quiet. The input data is randomly generated using the \lstinline{rand()} function and stored in the \lstinline{input[]} array. You can change the data by modifying the argument in the call to \lstinline{srand()}.  The argument sets the seed for the random number generator to a specific value to ensure the generation of the same sequence of random numbers on each execution.  Changing this value to a different integer will result in a different, but predictable, random sequence.

Note again that the testbench calls the \lstinline{insertion_cell_sort()} function a total of \lstinline{SIZE*2} times. Each of the first \lstinline{SIZE} function calls feeds a single input element into function, but produce no useful output. The next \lstinline{SIZE} calls provide dummy input data and produce one sorted element each function call.  The data is produced one at a time starting with the smallest element.

\begin{exercise}
After executing a sort, the code in Figure \ref{fig:insertion_cell_sort.cpp} leaves the \lstinline{local} variable in a different state than when it started.  Unfortunately, this means that we can only sort one array! In most cases, we not only want to sort more than one array, but we'd like to process each array back to back with no bubbles in the pipeline.   Modify the code and the testbench to enable this and demonstrate that your modified code can sort multiple arrays.
\end{exercise}

\section{Merge Sort}
\label{sec:sort:merge}
Merge sort is a stable, divide and conquer algorithm invented by John von Neumann in 1945 \cite{knuth1998art}. The basic idea of the merge sort algorithm is that combining two sorted arrays into a larger sorted array is a relatively simple operation, which can be completed in $\mathcal{O}(n)$ time.    Conceptually, we divide the array into two subarrays, sort each subarray, and then combine the sorted subarrays into the final result. 

\begin{figure}
\centering
\setlength{\fboxsep}{1pt}
\begin{tabular} {l l l } 
width = 1, A[] = \{\colorbox{black!10}{3},\colorbox{black!10}{7},\colorbox{black!10}{6},\colorbox{black!10}{4},\colorbox{black!10}{5},\colorbox{black!10}{8},\colorbox{black!10}{2},\colorbox{black!10}{1}\} \\
width = 2, A[] = \{\colorbox{black!10}{3,\hspace*{2\fboxsep}7},\colorbox{black!10}{4,\hspace*{2\fboxsep}6},\colorbox{black!10}{5,\hspace*{2\fboxsep}8},\colorbox{black!10}{1,\hspace*{2\fboxsep}2}\} \\
width = 4, A[] = \{\colorbox{black!10}{3,\hspace*{2\fboxsep}4,\hspace*{2\fboxsep}7,\hspace*{2\fboxsep}6},\colorbox{black!10}{1,\hspace*{2\fboxsep}2,\hspace*{2\fboxsep}5,\hspace*{2\fboxsep}8}\} \\
width = 8, A[] = \{\colorbox{black!10}{1,\hspace*{2\fboxsep}2,\hspace*{2\fboxsep}3,\hspace*{2\fboxsep}4,\hspace*{2\fboxsep}5,\hspace*{2\fboxsep}6,\hspace*{2\fboxsep}7,\hspace*{2\fboxsep}8}\} \\
\\
\end{tabular}
\caption{The Merge Sort algorithm operating on an array.  The initial state where each element is considered to be a sorted subarray of length one is shown at the top.   At each step of the algorithm, subarrays are merged placing the shaded elements are in sorted order.}
\label{fig:merge_sort_behavior}
\end{figure}

Since we focus on sorting arrays of data, it turns out that the `divide' operation is actually trivial.  It requires no arithmetic or data movement and we can simply consider each individual element of the input array as a trivially sorted subarray.  All of the computation is involved with merging subarrays into larger sorted subarrays.  With other data representations, such as a linked-list, dividing an input array into subarrays can require traversing the data structure.   The merge sort process is shown in Figure \ref{fig:merge_sort_behavior}.  

\begin{figure}
\centering
\setlength{\fboxsep}{0pt}
\begin{tabular} {l l l } 
in1[] = \{\underline{3}, 4, 6, 7\} & in2[] = \{\underline{1}, 2, 5, 8\} & out[] = \{ \} \\
in1[] = \{\underline{3}, 4, 6, 7\} & in2[] = \{\colorbox{black}{1}, \underline{2}, 5, 8\}  & out[] = \{1\} \\
in1[] = \{\underline{3}, 4, 6, 7\} & in2[] = \{\colorbox{black}{1}, \colorbox{black}{2}, \underline{5}, 8\} & out[] = \{1, 2\} \\
in1[] = \{\colorbox{black}{3}, \underline{4}, 6, 7\} & in2[] = \{\colorbox{black}{1}, \colorbox{black}{2}, \underline{5}, 8\} & out[] = \{1, 2, 3\} \\
in1[] = \{\colorbox{black}{3}, \colorbox{black}{4}, \underline{6}, 7\} & in2[] = \{\colorbox{black}{1}, \colorbox{black}{2}, \underline{5}, 8\}  & out[] = \{1, 2, 3, 4\} \\
in1[] = \{\colorbox{black}{3}, \colorbox{black}{4}, \underline{6}, 7\} & in2[] = \{\colorbox{black}{1}, \colorbox{black}{2}, \colorbox{black}{5}, \underline{8}\}  & out[] = \{1, 2, 3, 4, 5\} \\
in1[] = \{\colorbox{black}{3}, \colorbox{black}{4}, \colorbox{black}{6}, \underline{7}\}  & in2[] = \{\colorbox{black}{1}, \colorbox{black}{2}, \colorbox{black}{5}, \underline{8}\}  &out[] = \{1, 2, 3, 4, 5, 6\} \\
in1[] = \{\colorbox{black}{3}, \colorbox{black}{4}, \colorbox{black}{6}, \colorbox{black}{7}\}  & in2[] = \{\colorbox{black}{1}, \colorbox{black}{2}, \colorbox{black}{5}, \underline{8}\}  &out[] = \{1, 2, 3, 4, 5, 6, 7\} \\
in1[] = \{\colorbox{black}{3}, \colorbox{black}{4}, \colorbox{black}{6}, \colorbox{black}{7}\}  & in2[] = \{\colorbox{black}{1}, \colorbox{black}{2}, \colorbox{black}{5}, \colorbox{black}{8}\} &out[] = \{1, 2, 3, 4, 5, 6, 7, 8\} \\
\end{tabular}
\caption{The process of merging two sorted arrays.  The initial state is shown at the top.  In each step of the algorithm, the underlined elements are considered and one is placed into sorted order in the output array.}
\label{fig:merge_behavior}
\end{figure}

The process of combining two sorted arrays into one larger sorted array is sometimes called the ``two finger algorithm''.  Figure \ref{fig:merge_behavior} describes the process using two sorted input arrays, named \lstinline{in1[]} and \lstinline{in2[]}. These are merged into a sorted output array, named \lstinline{out[]}.

The process starts with a ``finger'' pointing to the first element of each array.  This conceptual finger is just an index into the array storing the data.  As the algorithm proceeds, the fingers will point to different elements of the arrays. We underline these elements in order to show where the fingers are placed.  To begin with, the fingers point to the first element of each array, elements \lstinline{3} and \lstinline{1} in arrays \lstinline{in1[]} and \lstinline{in2[]}, respectively.

The first line of Figure \ref{fig:merge_behavior} shows the initial state. There are four elements in each of the two input arrays and zero elements in the output array. We compare the first two elements of the input arrays, and move the smaller of these two to the output array. In this case, we compare \lstinline{3} and \lstinline{1}, and move \lstinline{1} into \lstinline{out[]}. This reduces the number of elements in \lstinline{in2[]}, and our ``finger'' moves to the next element in \lstinline{in2[]} which is the next smallest element since the array is sorted. Once again, we compare the two elements from each of the input arrays, and move the smaller of the two elements to \lstinline{out[]}. In this case, we compare \lstinline{3} and \lstinline{2}, and move the element from \lstinline{in2[]} to \lstinline{out[]}. This process continues until all of the elements in one of the arrays is empty. In that case, we copy the remaining elements from the non-empty array into the output array.

Although the merge sort algorithm is a common textbook example of recursive function calls, most high-level synthesis tools do not support recursion or only support it in a limited manner. Thus, we will focus on a non-recursive implementation of the algorithm.  The code might look substantially different from what you are used to, but the core of the algorithm is exactly the same.

\subsection{Basic Merge Sort}

Figure \ref{fig:merge_sort.cpp} shows basic code implementing a non-recursive merge sort.  This code sorts an array by performing roughly $N \log N$ comparisons, but requires a temporary array to store the partially sorted data. The code starts by considering each element of the array as a length one sorted subarray.  Each iteration of the outer loop merges pairs of sorted subarrays into larger sorted subarrays.  After the first iteration we have sorted subarrays with maximum size 2, after the second iteration the sorted subarrays have maximum size 4, then 8 and so on.  Note that if the size of the input array is not a power of two, then it is possible that we might end up with some subarrays which are smaller than the maximum size.

\begin{figure}
\begin{lstlisting}
#include "merge_sort.h"
#include "assert.h"

// subarray1 is in[i1..i2-1], subarray2 is in[i2..i3-1], result is in out[i1..i3-1]
void merge(DTYPE in[SIZE], int i1, int i2, int i3, DTYPE out[SIZE]) {
    int f1 = i1, f2 = i2;
    // Foreach element that needs to be sorted...
    for(int index = i1; index < i3; index++) {
        // Select the smallest available element.
        if((f1 < i2 && in[f1] <= in[f2]) || f2 == i3) {
            out[index] = in[f1];
            f1++;
        } else {
            assert(f2 < i3);
            out[index] = in[f2];
            f2++;
        }
    }
}

void merge_sort(DTYPE A[SIZE]) {
    DTYPE temp[SIZE];
    // Each time through the loop, we try to merge sorted subarrays of width elements
    // into a sorted subarray of 2*width elements.
 stage:
    for (int width = 1; width < SIZE; width = 2 * width) {
    merge_arrays:
        for (int i1 = 0; i1 < SIZE; i1 = i1 + 2 * width) {
            // Try to merge two sorted subarrays:
            // A[i1..i1+width-1] and A[i1+width..i1+2*width-1] to temp[i1..2*width-1]
            int i2 = i1 + width;
            int i3 = i1 + 2*width;
            if(i2 >= SIZE) i2 = SIZE;
            if(i3 >= SIZE) i3 = SIZE;
            merge(A, i1, i2, i3, temp);
        }

        // Copy temp[] back to A[] for next iteration
    copy:
        for(int i = 0; i < SIZE; i++) {
            A[i] = temp[i];
        }
    }
}


\end{lstlisting}
\caption{  A non-recursive implementation of merge sort. The \lstinline{merge_sort()} function iteratively merges subarrays until the entire array has been sorted. }
\label{fig:merge_sort.cpp}
\end{figure}

The sorting process starts in the \lstinline{merge_sort()} function. The function primarily operates on the input \lstinline{A[]} array, but leverages internal storage in \lstinline{temp[]}. The size of both of these arrays is determined by the \lstinline{SIZE} parameter. The parameter \lstinline{DTYPE} determines the type of data being sorted.

The computation of the function consists of two nested \lstinline{for} loops. The outer \lstinline{stage} loop keeps track of the number of elements in each sorted subarray in the \lstinline{width} variable. The function starts by considering each element as a separate subarray; hence \lstinline{width} is initialized as \lstinline{1}. Every iteration of the \lstinline{stage} loop results in the generation of longer sorted subarrays. These new subarrays potentially have twice as many elements, which is why \lstinline{width} doubles on each iteration of the \lstinline{stage} loop. The \lstinline{stage} loop terminates when \lstinline{width} is greater than or equal to \lstinline{SIZE}, indicating that all elements of \lstinline{A[]} are in a single sorted subarray.

Each iteration of the inner \lstinline{for} loop, labeled \lstinline{merge_arrays} merges two consecutive subarrays. These subarrays each consist of up to \lstinline{width} elements, starting at index \lstinline{i1} and \lstinline{i2}.  These two subarrays are merged and copied into a single subarray stored in \lstinline{temp[]} using the \lstinline{merge()} function.  The main complexity here is dealing with the boundary condition at the end of the loop if \lstinline{SIZE} is not exactly a power of two.  In this case the subarrays might contain less than \lstinline{width} elements. After merging subarrays into \lstinline{temp[]}, the final loop copies the data back into \lstinline{A[]} for the next iteration of the loop.

\begin{exercise}
The code in Figure \label{fig:merge_sort.cpp} attempts to handle a wide variety of values for the parameter \lstinline{SIZE}.  What values are allowed?  When the \lstinline{merge()} function is called, whare are the possible relationships between \lstinline{i1}, \lstinline{i2}, and \lstinline{i3}?   If we restrict the allowed values of the parameter \lstinline{SIZE} can the code be simplified?  What is the affect on the resulting HLS generated circuit? 
\end{exercise}

The \lstinline{merge()} function performs the ``two finger'' algorithm on two subarrays within the \lstinline{in[]} array.  The function takes input in the \lstinline{in[]} array and produces output in the \lstinline{out[]} array. The function also takes as input variables \lstinline{i1}, \lstinline{i2}, and \lstinline{i3} which describe the extent of the two subarrays to be merged.  One subarray starts at index \lstinline{i1} and includes all the elements before \lstinline{i2} and the second subarray starts at index \lstinline{i2} and includes the elements up to \lstinline{i3}.  The merged output subarray will be stored from index \lstinline{i1} up to \lstinline{i3} in \lstinline{out[]}. 

The \lstinline{merge()} function consists of a single loop which iterates over the elements being stored into \lstinline{out[]}.  Each iteration places an element into its correctly sorted location in \lstinline{out[]}. The variables \lstinline{f1} and \lstinline{f2} within the function correspond to the position of the fingers for each subarray.  The \lstinline{if} condition selects the smaller of \lstinline{in[f1]} or \lstinline{in[f2]} to copy to the next sorted position in 
\lstinline{out[]}.  However, the \lstinline{if} condition is more complex since it has to deal with several special cases.  One case is where \lstinline{f1 == i2} and we have run out of elements to consider for \lstinline{in[f1]}, in which case we must select \lstinline{in[f2]} as the smallest element.  Alternatively, if \lstinline{f2 == i3}, then we  have run out of elements to consider for \lstinline{in[f2]}, in which case we must select \lstinline{in[f1]} as the smallest element.

\begin{exercise}
What happens to the \lstinline{in[]} array over the course of the computation? Describe the state of the elements in the \lstinline{in[]} array after each iteration of the outer \lstinline{for} loop. What is the final order of the elements in the \lstinline{in[]} array when \lstinline{merge_sort()} returns?
\end{exercise}

\begin{exercise}
The performance report after synthesis may not be able to determine the number of cycles for the latency and interval. Why is that the case? What are appropriate  \lstinline{min}, \lstinline{max}, and \lstinline{avg} values to provide in a \lstinline{loop_tripcount} directive(s)?
\end{exercise}

The code is not particularly optimized for any particular \gls{hls} implementation. The best starting point for optimization is by adding directives. Given that we have several nested \lstinline{for} loops, we generally look first at optimizing the inner loops.  Optimizations of inner loops are usually much more significant than optimizations of outer loops, which are executed relatively rarely. By this point, you should be familiar with the common \lstinline{pipeline} and \lstinline{unroll} directives for loop optimization. 

\begin{exercise}
Perform different optimizations using the \lstinline{pipeline} and \lstinline{unroll} directives on the \lstinline{for} loops. What provides the best performance? Which gives the best tradeoff between resource utilization and performance?  What aspects of the code prevent higher performance?  Are these aspects fundamental to the algorithm, or only because of the way the algorithm is captured in the code?
\end{exercise}

Pipelining and unrolling can be hindered by resource constraints; in particular, we must carefully consider the number of memory ports for the arrays. The arrays in this code seem relatively straightforward as they are both one-dimensional. Yet, the designer must carefully consider the access patterns to insure that performance optimizations match with the resource constraints. 

\begin{exercise}
Optimize the code using loop optimizations and array partitioning, i.e., create a set of designs using the \lstinline{array_partition}, \lstinline{pipeline}, and \lstinline{unroll} directives. Were you able to achieve better results than by using the \lstinline{pipeline} and \lstinline{unroll} directives alone? What was the best strategy for performing design space exploration using these directives? What was your best design in terms of performance? What was the best design that provides a good tradeoff between resource utilization and performance?
\end{exercise}

Many times the best designs are only possible by performing code restructuring. Although \VHLS provides many directives to enable common code optimizations, it is impractical to provide directives for every optimization.  Sometimes we must resort to rewriting the code in addition to providing directives in order to achieve a design that meets our requirements.  In the next section we describe one way to significantly restructure the merge sort code in order to increase throughput.

\subsection{Restructured Merge Sort}
Looking first at the inner loop of the \lstinline{merge()} function, you might have found that it was difficult to achieve a loop II of 1.  One challenge is that there are actually four reads of \lstinline{in[]}, but only at two different addresses.   The HLS tool must recognize that these reads are redundant, since \gls{bram} memories can only support two accesses each clock.  However because these reads are in different basic blocks, it is more difficult for a compiler to eliminate the redundant loads.  By eliminating the redundant reads, the compiler needs to do less optimization to achieve a loop II of 1.  The restructured code is shown in Figure \ref{fig:merge_sort_restructured.cpp}.  In addition, there is a recurrence through the \lstinline{f1} and \lstinline{f2} variables.  These variables are incremented in one of the branches of the \lstinline{if} conditional, which must be used in the next iteration to determine which locations in \lstinline{in[]} to compare and subsequently which branch of the conditional to take.  Because the floating point comparison is relatively complex, this recurrence can also limit the achievable II and clock period.

\begin{figure}
\begin{lstlisting}[format=none, lastline=23]
#include "merge_sort.h"
#include "assert.h"

// subarray1 is in[ii..i2-1]; subarray2 is in[i2..i3-1]
// sorted merge is stored in out[i1..i3-1]
void merge(DTYPE in[SIZE], int i1, int i2, int i3, DTYPE out[SIZE]) {
    int f1 = i1, f2 = i2;
    // Foreach element that needs to be sorted...
    for(int index = i1; index < i3; index++) {
#pragma HLS pipeline II=1
        DTYPE t1 = in[f1];
        DTYPE t2 = in[f2];
        // Select the smallest available element.
        if((f1 < i2 && t1 <= t2) || f2 == i3) {
            out[index] = t1;
            f1++;
        } else {
            assert(f2 < i3);
            out[index] = t2;
            f2++;
        }
    }
}
\end{lstlisting}
\caption{Restructured code for the \lstinline{merge()} function, which can achieve a loop II of 1 in \VHLS.}
\label{fig:merge_sort_restructured.cpp}
\end{figure}

The behavior of this code is shown in Figure \ref{fig:merge_sort_restructured_behavior}.  Although the inner loop achieves a loop II of 1, this inner loop often has a very small number of loop iterations.  When the inner loop finishes, the pipeline must empty before code executing after the pipeline can execute.  Although the loop pipeline is relatively short in this case, the bubble caused by the loop completing is significant, since the number of iterations is also small.  Unfortunately, because of the limits of static loop analysis, the performance of this particular code is somewhat hard to visualize.  In this case, the number of iterations of the inner loop is data dependent.

\begin{figure}
\centering
\executeiffilenewer{merge_sort_restructured_behavior.svg}{images/merge_sort_restructured_behavior.pdf}%
{inkscape -z -D --file=merge_sort_restructured_behavior.svg %
--export-pdf=images/merge_sort_restructured_behavior.pdf --export-latex}%
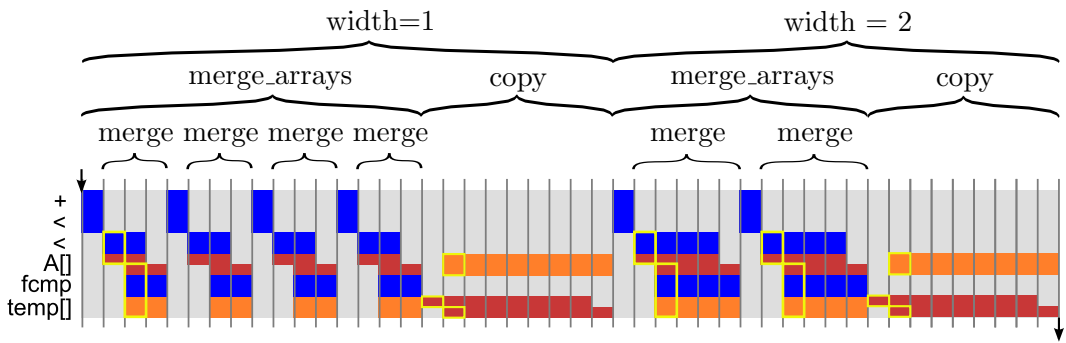%

\caption{Behavior of the restructured code in Figure \ref{fig:merge_sort_restructured.cpp}.}
\label{fig:merge_sort_restructured_behavior}
\end{figure}

A common approach is to flatten loop nests like these into a single loop, reducing the number of times that the pipeline must flush when exiting a loop.  In fact, \VHLS will automatically flatten perfect loop nests.  In this case, however, since the code does not contain a perfect loop nest, we can resort to flattening the loops manually.  Code resulting from manually flattening the \lstinline{merge_arrays} loop with the loop inside the \lstinline{merge()} function is shown in Figure \ref{fig:merge_sort_loop_merged.cpp}.  Note that one advantage of this code is that the \lstinline{merge_arrays} loop also has a constant number of loop iterations, making understanding performance much easier.

\begin{figure}
\begin{lstlisting}[format=none]
#include "merge_sort.h"
#include "assert.h"

void merge_sort(DTYPE A[SIZE]) {
    DTYPE temp[SIZE];
 stage:
    for (int width = 1; width < SIZE; width = 2 * width) {
        int f1 = 0;
        int f2 = width;
        int i2 = width;
        int i3 = 2*width;
        if(i2 >= SIZE) i2 = SIZE;
        if(i3 >= SIZE) i3 = SIZE;
    merge_arrays:
        for (int i = 0; i < SIZE; i++) {
#pragma HLS pipeline II=1
            DTYPE t1 = A[f1];
            DTYPE t2 = A[f2];
            if((f1 < i2 && t1 <= t2) || f2 == i3) {
                temp[i] = t1;
                f1++;
            } else {
                assert(f2 < i3);
                temp[i] = t2;
                f2++;
            }
            if(f1 == i2 && f2 == i3) {
                f1 = i3;
                i2 += 2*width;
                i3 += 2*width;
                if(i2 >= SIZE) i2 = SIZE;
                if(i3 >= SIZE) i3 = SIZE;
                f2 = i2;
            }
        }

    copy:
        for(int i = 0; i < SIZE; i++) {
#pragma HLS pipeline II=1
            A[i] = temp[i];
        }
    }
}

\end{lstlisting}
\caption{Restructured code for Merge Sort which can achieve a loop II of 1 with fewer pipeline bubbles in \VHLS.}
\label{fig:merge_sort_loop_merged.cpp}
\end{figure}

\begin{aside}
Estimate the performance of the code in Figure \ref{fig:merge_sort_loop_merged.cpp}.  Even though the inner loops have achieved a loop II of 1, is the design using hardware efficiently?  Is there a way to further reduce the latency of the \lstinline{merge_sort()} function to the point where it is using approximately $N \log N$ clock cycles?
\end{aside}

So far we have focused on optimizing the \lstinline{merge_sort()} function to reduce the latency of the computation without significantly increasing the number of resources.   As a result, the accelerator is becoming more efficient.  However, after achieving a reasonably efficient design, the only way to further reduce latency and/or increase throughput is increase parallelism.  Previously we have seen ways of unrolling inner loops and partitioning arrays as a way to perform more work each clock cycle.  An alternative way to increase parallelism is leverage pipelining.  In addition to operator-level pipelinling we can also look for coarser-granularity task-level pipelining.

With Merge Sort it turns out that we can make a dataflow pipeline out of each iteration of the \lstinline{stage} loop, assuming that we have a fixed size array to sort.  In \VHLS this implementation can be conceptually achieved by unrolling the \lstinline{stage} loop and using the \lstinline{dataflow} directive.  Each instance of the \lstinline{merge_arrays} loop then becomes an independent process which can operate on a different set of data.  The resulting architecture is shown in Figure \ref{fig:restructured_mergesort_dataflow}.  

\begin{figure}
\centering
\executeiffilenewer{restructured_mergesort_dataflow.svg}{images/restructured_mergesort_dataflow.pdf}%
{inkscape -z -D --file=restructured_mergesort_dataflow.svg %
--export-pdf=images/restructured_mergesort_dataflow.pdf --export-latex}%
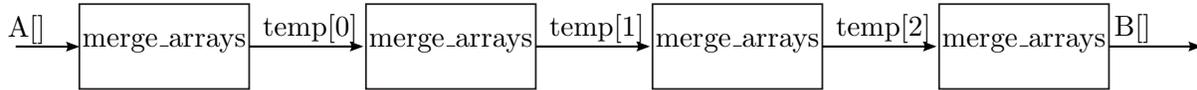%

\caption{Dataflow pipelined architecture for implementing 4 stages of Merge Sort.  This architecture can sort up to 16 elements.}
\label{fig:restructured_mergesort_dataflow}
\end{figure}

\begin{figure}
\footnotesize
\begin{lstlisting}
#include "merge_sort_parallel.h"
#include "assert.h"

void merge_arrays(DTYPE in[SIZE], int width, DTYPE out[SIZE]) {
  int f1 = 0;
  int f2 = width;
  int i2 = width;
  int i3 = 2*width;
  if(i2 >= SIZE) i2 = SIZE;
  if(i3 >= SIZE) i3 = SIZE;
 merge_arrays:
  for (int i = 0; i < SIZE; i++) {
#pragma HLS pipeline II=1
      DTYPE t1 = in[f1];
      DTYPE t2 = in[f2];
	if((f1 < i2 && t1 <= t2) || f2 == i3) {
	  out[i] = t1;
	  f1++;
	} else {
	  assert(f2 < i3);
	  out[i] = t2;
	  f2++;
	}
	if(f1 == i2 && f2 == i3) {
      f1 = i3;
	  i2 += 2*width;
	  i3 += 2*width;
	  if(i2 >= SIZE) i2 = SIZE;
	  if(i3 >= SIZE) i3 = SIZE;
      f2 = i2;
 	}
  }
}

void merge_sort_parallel(DTYPE A[SIZE], DTYPE B[SIZE]) {
#pragma HLS dataflow

	DTYPE temp[STAGES-1][SIZE];
#pragma HLS array_partition variable=temp complete dim=1
	int width = 1;

	merge_arrays(A, width, temp[0]);
	width *= 2;

	for (int stage = 1; stage < STAGES-1; stage++) {
#pragma HLS unroll
		merge_arrays(temp[stage-1], width, temp[stage]);
		width *= 2;
	}

	merge_arrays(temp[STAGES-2], width, B);
}
\end{lstlisting}
\caption{Restructured code for Merge Sort which can be implemented as a dataflow pipeline by \VHLS.}
\label{fig:merge_sort_parallel.cpp}
\end{figure}

The code to implement this is shown in Figure \ref{fig:merge_sort_parallel.cpp}.  This code is similar in many ways to the original code, but has several important differences.  One key difference is that the \lstinline{merge_arrays} loop has been extracted into a function, making it easier simpler to rewrite the toplevel code. A second key difference is that the output of \lstinline{merge_sort_parallel()} is produced in a separate array than the input, enabling \VHLS to build an architecture that pipelined architecture processing using the \lstinline{dataflow} directive.  Additionally, the \lstinline{temp[][]} array is now used to model the dataflow ping-pong channels between the processes implemented by the \lstinline{merge_arrays()} function, making extra array copies unnecessary.  This two-dimensional array is partitioned in the first dimension, making it easy to represent a parameterized number of channels.

The \lstinline{merge_sort_parallel()} function consists of \lstinline{STAGES} calls to the \lstinline{merge_arrays()} function.  Each The first call reads from the input and writes to \lstinline{temp[0]}.  The loop performs the intermediate stages  and writes to the remaining partitions of \lstinline{temp[]}.  The final call writes to the output array \lstinline{B[]}.  The code is parameterized in terms of \lstinline{SIZE} and \lstinline{STAGES} and supports array lengths 4 or greater.

\begin{aside}
Estimate the performance of the code in Figure \ref{fig:merge_sort_parallel.cpp}.  What is the interval and latency of the implementation?  How much memory is required to support that processing rate?   Is all of the memory necessary?
\end{aside}

\section{Conclusion}

This chapter introduced a number of basic sorting algorithms with fundamentally different algorithmic tradeoffs.  Insertion sort operating on an array performs on average $N^2/4$ comparisons, but requires much fewer (approximately $N$ comparisons) when operating on sorted data. We showed several ways of parallelizing Insertion sort which can increase performance at the expense of using more hardware resources.  Because of bubbles in the statically scheduled pipeline, we end up needing roughly $N$ comparators with an interval of $N$ cycles to perform insertion sort.  Merge sort operating on an array performs generally fewer comparisons, roughly $N \log N$, but requires additional memory to store partially sorted results.  The more complex loop structure of Merge Sort means that we required additional refactoring to reach an efficient solution that took roughly $N \log N$ cycles with one comparator.  We also showed a task-pipelined implementation which performs $\log N$ comparisons every clock cycle with an interval of $N$ cycles to perform merge sort.  In contrast with insertion sort, this requires much fewer comparisons to achieve the same interval, but requires more memory (in the form of dataflow channels) and has much higher latency.

In practice, many FPGA-based implementations of sorting will have to address these fundamental tradeoffs in order to achieve the best integration with other aspects of a system. Another sort with a fundamentally different tradeoff is the Radix Sort, which focuses on data that exists within a bounded range, unlike the more general sorting algorithms in this chapter which only require comparison between items.  Radix sort will be implemented as part of the Huffman Coding in Chapter \ref{chapter:huffman}.

Parallel implementations of sorting algorithms are often described as sorting networks.  Sorting networks are sometimes described as \glspl{systolic_array} and other times as pipelines.  We can often gain inspiration for HLS designs by looking investigating these existing parallel implementations and then capturing them as C code.  In \VHLS, these networks can be described using either loop pipelines, dataflow pipelines, or a combination of both.
 


\chapter{Huffman Encoding}
\glsresetall
\label{chapter:huffman}

\section{Background}

Lossless data compression is a key ingredient for efficient data storage, and Huffman coding is amongst the most popular algorithms for variable length coding \cite{huffman1952method}. Given a set of data symbols and their frequencies of occurrence, Huffman coding generates codewords in a way that assigns shorter codes to more frequent symbols to minimize the average code length. Since it guarantees optimality, Huffman coding has been widely adopted for various applications \cite{flannery1992numerical}. In modern multi-stage compression designs, it often functions as a back-end of the system to boost compression performance after a domain-specific front-end as in GZIP \cite{deutsch1996deflate}, JPEG \cite{pennebaker1992jpeg}, and MP3 \cite{sherigar2004huffman}. Although arithmetic encoding \cite{witten1987arithmetic}  (a generalized version of Huffman encoding which translates an entire message into a single number) can achieve better compression for most scenarios, Huffman coding has often been the algorithm of choice for many systems because of patent concerns with arithmetic encoding \cite{langdon1990arithmetic}.

Canonical Huffman coding has two main benefits over traditional Huffman coding. In basic Huffman coding, the encoder passes the complete Huffman tree structure to the decoder. Therefore, the decoder must traverse the tree to decode every encoded symbol. On the other hand, canonical Huffman coding only transfers the number of bits for each symbol to the decoder, and the decoder reconstructs the codeword for each symbol. This makes the decoder more efficient both in memory usage and computation requirements. Thus, we focus on canonical Huffman coding.

In basic Huffman coding, the decoder decompresses the data by traversing the Huffman tree from the root until it hits the leaf node. This has two major drawbacks: it requires storing the entire Huffman tree which increases memory usage. Furthermore, traversing the tree for each symbol is computationally expensive. Canonical Huffman encoding addresses these two issues by creating codes using a standardized canonical format.  The benefit of using a canonical encoding is that we only need to transmit the length of each Huffman codeword.  A Canonical Huffman code has two additional properties.  Firstly, longer length codes have a higher numeric value than the same length prefix of shorter codes.  Secondly, codes with the same length increase by one as the symbol value increases. This means if we know the starting symbol for each code length, we can easily reconstruct the canonical Huffman code.  The Huffman tree is essentially equivalent to a `sorted' version of the original Huffman tree so that longer codewords are on the right-most branch of the tree and all of the nodes at the same level of the tree are sorted in order of the symbols.  

\begin{figure}
\centering
\includegraphics[width= \textwidth]{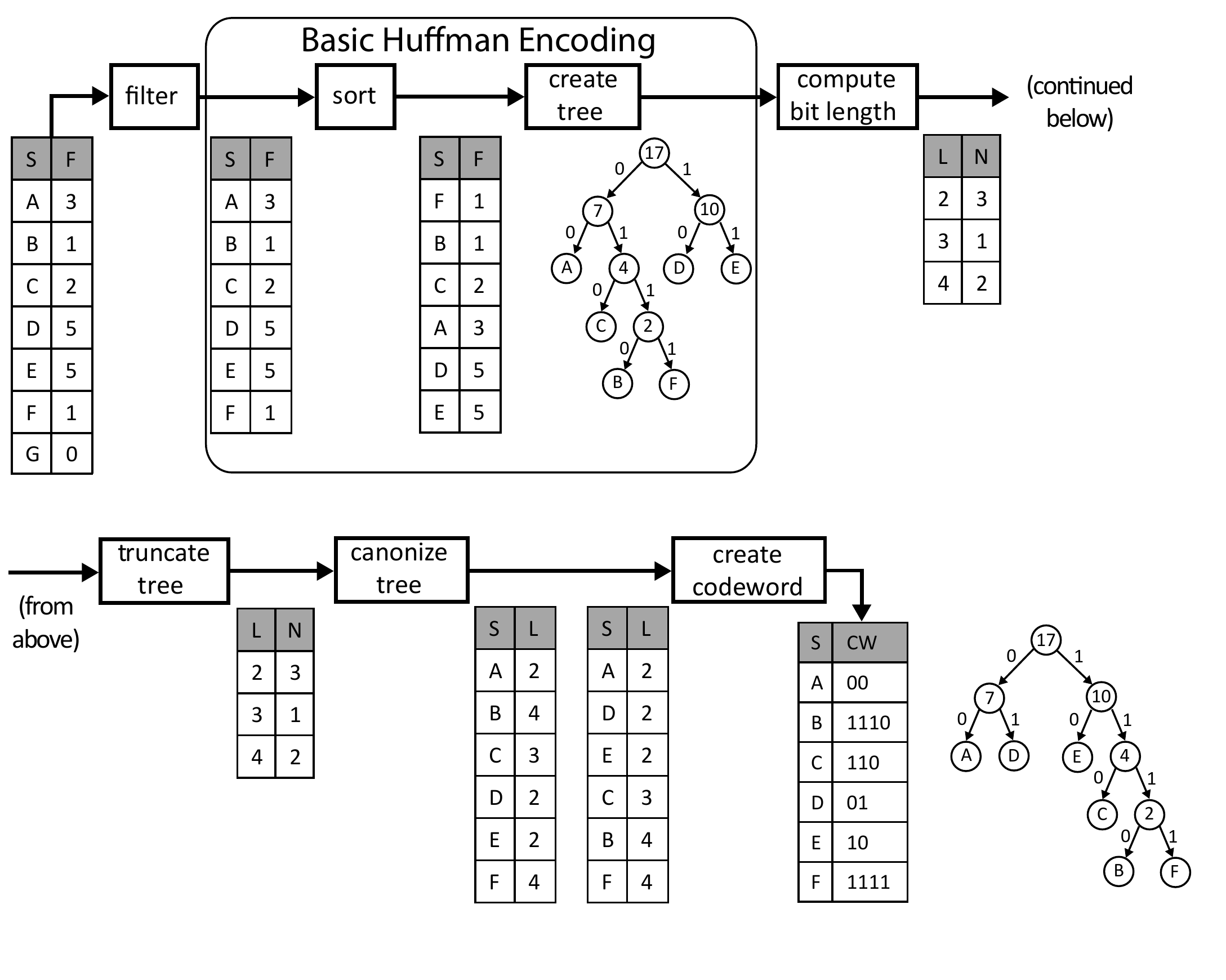}
\caption{ The Canonical Huffman Encoding process. The symbols are filtered and sorted, and used to build a Huffman tree. Instead of passing the entire tree to the decoder (as is done in ``basic'' Huffman coding), the encoding is done such that only the length of the symbols in the tree is required by the decoder. Note that the final canonical tree is different from the initial tree created near the beginning of the process.}
\label{fig:canonical_huffman_flow}
\end{figure}

Figure~\ref{fig:canonical_huffman_flow} shows the process of creating a canonical Huffman code. The \lstinline{filter} module only passes symbols with non-zero frequencies. The \lstinline{sort} module rearranges the symbols in ascending order based upon their frequencies. Next, the \lstinline{create tree} module builds the Huffman tree using three steps: 1) it uses the two minimum frequency nodes as an initial sub-tree and generates a new parent node by summing their frequencies; 2) it adds the new intermediate node to the list and sorts them again; and 3) it selects the two minimum elements from the list and repeats these steps until one element remains. The result is a Huffman tree where each leaf node in the tree represents a symbol that can be coded and each internal node is labeled with the frequency of the nodes in that sub-tree.  By associating the left and right edges in the tree with bits $0$ and $1$, we can determine the unique codeword for each symbol based on the path to reach it from the root node. For example, the codeword for \sym{A} is $00$ and codeword for \sym{B} is $1110$. This completes the basic Huffman encoding process, but does not necessarily create the canonical Huffman tree.

To create the canonical Huffman tree, we perform several additional transformations.   First, the \lstinline{compute bit len} module calculates the bit length of each codeword and then counts the frequency of each length. The result is a histogram of the codeword lengths (see Section \ref{sec:histogram}).  In the example case, we have three symbols (\sym{A},\sym{D},\sym{E}) with the code length of 2. Therefore, the computed histogram maps contains value $3$ in location $2$. Next, the \lstinline{truncate tree} module rebalances the Huffman tree in order to avoid excessively long codewords. This can improve decoder speed at the cost of a slight increase in encoding time. This is not necessary in the example in Figure \ref{fig:canonical_huffman_flow}. We set the maximum height of the tree to 27.  Lastly, the \lstinline{canonize tree} module creates two sorted tables. The first table contains symbols and lengths sorted by symbol. The second table contains symbols and lengths sorted by lengths. These tables simplify the creation of the canonical Huffman codewords for each symbol.

The \lstinline{create codeword} module creates a table of canonical Huffman codewords by traversing the sorted tables. Beginning with the first codeword in the sorted table, it is assigned the all-zero codeword with the appropriate length.  Each following symbol with the same bit length is assigned the following codeword, which is formed by simply adding $1$ to the previous code word.  In our example, symbols \sym{A}, \sym{D}, and \sym{E} all have bit length $l = 2$ and are assigned the codewords $\sym{A} = 00$, $\sym{D} = 01$, and $\sym{E} = 10$.  Note that the symbols are considered in alphabetical order, which is necessary to make the tree canonical.  This process continues until we get to a codeword that requires a larger length, in which case we not only increment the previous codeword, but also shift left to generate a codeword of the correct length.  In the example, the next symbol is \sym{C} with a length of 3, which receives the codeword $\sym{C} = (10 + 1) << 1 = 11 << 1 = 110$. Continuing on, the next symbol is \sym{B} with a length of 4.  Once again we increment and shift by one. Thus the codeword for $\sym{B} = (110 + 1) << 1 = 1110$. The final codeword for symbol $\sym{F} = 1110 + 1 = 1111$. We explain this in more detail in Chapter \ref{sec:create_codewords}.  


The creation of a canonical Huffman code includes many complex and inherently sequential computations. For example, the \lstinline{create tree} module needs to track the correct order of the created sub trees, requiring careful memory management. Additionally, there is very limited parallelism that can be exploited. In the following, we discuss the hardware architecture and the implementation of the canonical Huffman encoding design using \VHLS. 

\begin{figure}
\begin{lstlisting}[format=none, lastline=43]
#include "huffman.h"
#include "assert.h"
void huffman_encoding(
    /* input */ Symbol symbol_histogram[INPUT_SYMBOL_SIZE],
    /* output */ PackedCodewordAndLength encoding[INPUT_SYMBOL_SIZE],
    /* output */ int *num_nonzero_symbols) {
    #pragma HLS DATAFLOW

    Symbol filtered[INPUT_SYMBOL_SIZE];
    Symbol sorted[INPUT_SYMBOL_SIZE];
    Symbol sorted_copy1[INPUT_SYMBOL_SIZE];
    Symbol sorted_copy2[INPUT_SYMBOL_SIZE];
    ap_uint<SYMBOL_BITS> parent[INPUT_SYMBOL_SIZE-1];
    ap_uint<SYMBOL_BITS> left[INPUT_SYMBOL_SIZE-1];
    ap_uint<SYMBOL_BITS> right[INPUT_SYMBOL_SIZE-1];
    int n;

    filter(symbol_histogram, filtered, &n);
    sort(filtered, n, sorted);

    ap_uint<SYMBOL_BITS> length_histogram[TREE_DEPTH];
    ap_uint<SYMBOL_BITS> truncated_length_histogram1[TREE_DEPTH];
    ap_uint<SYMBOL_BITS> truncated_length_histogram2[TREE_DEPTH];
    CodewordLength symbol_bits[INPUT_SYMBOL_SIZE];

    int previous_frequency = -1;
 copy_sorted:
    for(int i = 0; i < n; i++) {
        sorted_copy1[i].value = sorted[i].value;
        sorted_copy1[i].frequency = sorted[i].frequency;
        sorted_copy2[i].value = sorted[i].value;
        sorted_copy2[i].frequency = sorted[i].frequency;
        // std::cout << sorted[i].value << " " << sorted[i].frequency << "\n";
        assert(previous_frequency <= (int)sorted[i].frequency);
        previous_frequency = sorted[i].frequency;
    }

    create_tree(sorted_copy1, n, parent, left, right);
    compute_bit_length(parent, left, right, n, length_histogram);

#ifndef __SYNTHESIS__
    // Check the result of computing the tree histogram
    int codewords_in_tree = 0;
\end{lstlisting}
\end{figure}
\begin{figure}
\begin{lstlisting}[format=none,firstline=44]
 merge_bit_length:
    for(int i = 0; i < TREE_DEPTH; i++) {
        #pragma HLS PIPELINE II=1
        if(length_histogram[i] > 0)
            std::cout << length_histogram[i] << " codewords with length " << i << "\n";
        codewords_in_tree += length_histogram[i];
    }
    assert(codewords_in_tree == n);
#endif

        truncate_tree(length_histogram, truncated_length_histogram1, truncated_length_histogram2);
    canonize_tree(sorted_copy2, n, truncated_length_histogram1, symbol_bits);
    create_codeword(symbol_bits, truncated_length_histogram2, encoding);

    *num_nonzero_symbols = n;
}
\end{lstlisting}
\caption{  The ``top'' \lstinline{huffman_encoding} function. It defines the arrays and variables  between the various subfunctions.  These are described graphically in Figures \ref{fig:canonical_huffman_flow} and \ref{fig:che_dataflow}.  }
\label{fig:huffman_encoding.cpp}
\end{figure}

Figure \ref{fig:huffman_encoding.cpp} shows the entire ``top'' \lstinline{huffman_encoding} function. This sets up the arrays and other variables that are passed between the various subfunctions. And it instantiates these functions. 

There is some additional copying of data that may seem unnecessary. This is due to our use of the \lstinline{dataflow} directive. This imparts some restrictions on the flow of the variables between the subfunctions. In particular, there are some strict rules on producer and consumer relationships of data between the parts of the function. This requires that we replicate some of the data. For example, we create two copies of the arrays \lstinline{parent}, \lstinline{left} and \lstinline{right}. We also do the same with the array \lstinline{truncated_bit_length}. The former is done in a \lstinline{for} loop in the top \lstinline{huffman_encoding} function; the latter is done inside of the \lstinline{canonize_tree} function.

\begin{aside}
The \lstinline{dataflow} directive imposes restrictions on the flow of information in the function. Many of the restrictions enforce a strict producer and consumer relationship between the subfunctions.  One such restriction is that an array should be written to by only one function and it should be read by only one function. i.e., it should only serve as an output from one function and an input to another function. If multiple functions read from the same array, \VHLS will synthesize the code but will issue a warning and not use a dataflow pipelined architecture.  As a result, using dataflow mode often requires replicating data into multiple arrays. A similar problem occurs if a function attempts to read from and write to an array which is also accessed by another function.  In this case it is necessary to maintain an additional internal copy of the data inside the function.  We will discuss both of these requirements and how to adhere to them as we go through the code in the remainder of this chapter.
\end{aside}

\section{Implementation}
The canonical Huffman encoding process is naturally divided into subfunctions. Thus, we can work on each of these subfunctions on a one-by-one basis. Before we do that, we should consider the interface for each of these functions. 

Figure \ref{fig:che_dataflow} shows the functions and their input and output data. For the sake of simplicity, it only shows the interfaces with arrays, which, since they are large, we can assume are stored in block rams (BRAMs). Before we describe the functions and their inputs and outputs, we need to discuss the constants, custom data types, and the function interface that are defined in \lstinline{huffman.h}. Figure \ref{fig:huffman_h} shows the contents of this file. 

\begin{figure}
\begin{lstlisting}[lastline=40]
#include "ap_int.h"

// input number of symbols
const static int INPUT_SYMBOL_SIZE = 256;

// upper bound on codeword length during tree construction
const static int TREE_DEPTH = 64;

// maximum codeword tree length after rebalancing
const static int MAX_CODEWORD_LENGTH = 27;

// Should be log2(INPUT_SYMBOL_SIZE)
const static int SYMBOL_BITS = 10;

// Should be log2(TREE_DEPTH)
const static int TREE_DEPTH_BITS = 6;

// number of bits needed to record MAX_CODEWORD_LENGTH value
// Should be log2(MAX_CODEWORD_LENGTH)
const static int CODEWORD_LENGTH_BITS = 5;

// A marker for internal nodes
const static ap_uint<SYMBOL_BITS> INTERNAL_NODE = -1;

typedef ap_uint<MAX_CODEWORD_LENGTH> Codeword;
typedef ap_uint<MAX_CODEWORD_LENGTH + CODEWORD_LENGTH_BITS> PackedCodewordAndLength;
typedef ap_uint<CODEWORD_LENGTH_BITS> CodewordLength;
typedef ap_uint<32> Frequency;

struct Symbol {
	 ap_uint<SYMBOL_BITS> value;
	 ap_uint<32> frequency;
};

void huffman_encoding (
	Symbol in[INPUT_SYMBOL_SIZE],
	PackedCodewordAndLength encoding[INPUT_SYMBOL_SIZE],
	int *num_nonzero_symbols
);
\end{lstlisting}
\caption{  The parameters, custom data type, and function interface for the top level function \lstinline{huffman_encoding}.  }
\label{fig:huffman_h}
\end{figure}

The \lstinline{INPUT_SYMBOL_SIZE} parameter specifies the maximum number of symbols that will be given as input for encoding.  In this case, we've set it to $256$, enabling the encoding of 8-bit ASCII data. The \lstinline{TREE_DEPTH} parameter specifies the upper bound for the length of an individual codeword during the initial Huffman tree generation. The \lstinline{CODEWORD_LENGTH} parameter specifies the target tree height when the Huffman tree is rebalanced in the function \lstinline{truncate_tree}. Finally, the \lstinline{CODEWORD_LENGTH_BITS} constant determines the number of bits required to encode a codeword length. This is equal to $\log_2 \lceil $\lstinline{CODEWORD_LENGTH}$\rceil$, which in this case is $5$.

We create a custom data type \lstinline{Symbol} to hold the data corresponding the input values and their frequencies. This datatype is used in the \lstinline{filter}, \lstinline{sort}, and other functions in the encoding process that require access to such information. The data type has two fields \lstinline{value} and \lstinline{frequency}.  In this case we've assumed that the block of data being encoded contains no more than $2^{32}$ symbols.

Finally, the \lstinline{huffman.h} file has the \lstinline{huffman_encoding} function interface. This is the specified top level function for the \VHLS tool. It has three arguments. The first argument is an array of \lstinline{Symbols} of size \lstinline{INPUT_SYMBOL_SIZE}.  This array represents a histogram of the frequencies of the data in the block being encoded. The next two arguments are outputs. The \lstinline{encoding} argument outputs the codeword for each possible symbol. The \lstinline{num_nonzero_symbols} argument is the number of non-zero symbols from the input data. This is the same as the number of symbols that remain after the \lstinline{filter} operation. 

\begin{figure}
\centering
\includegraphics[width= \textwidth]{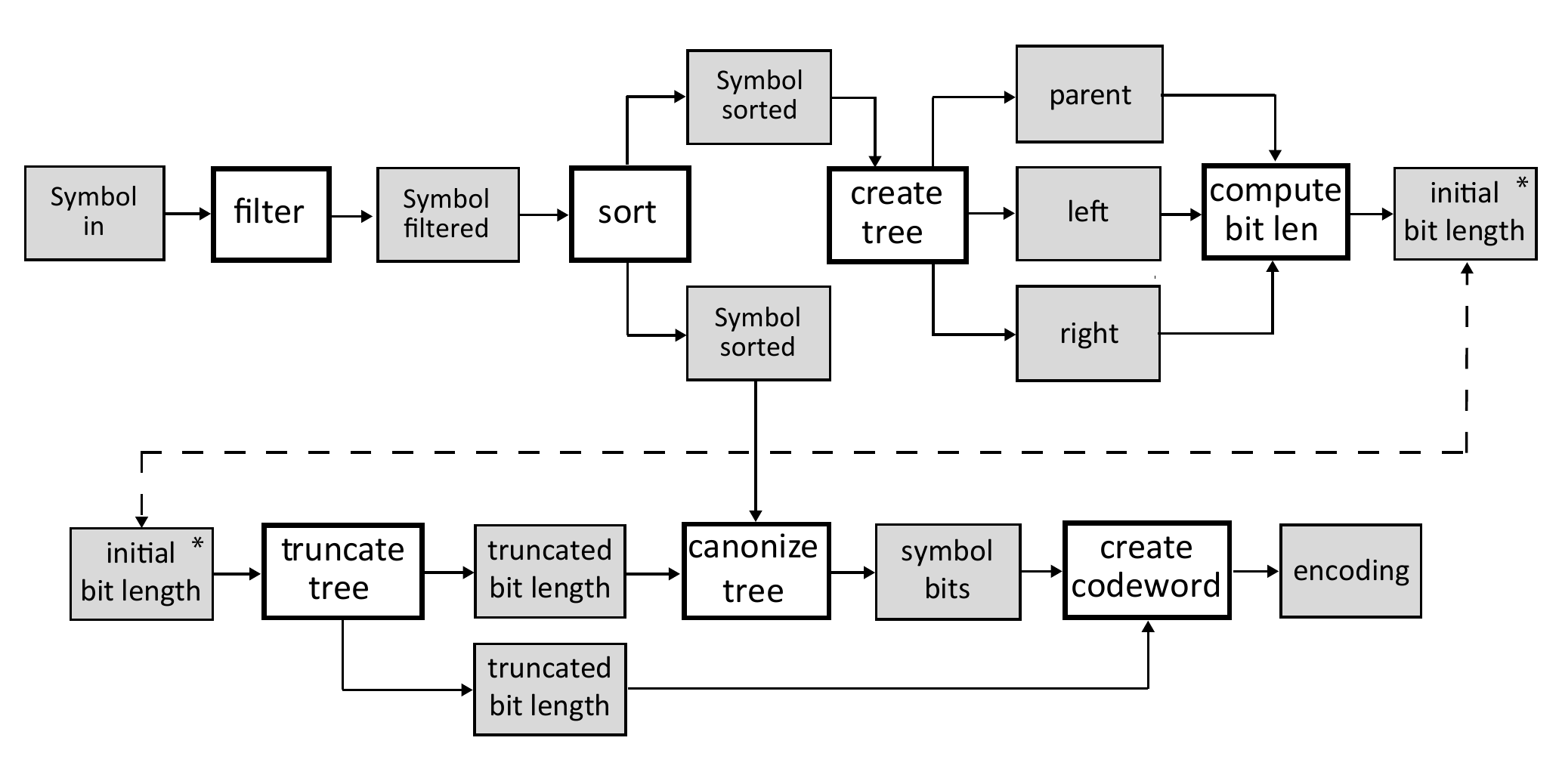}
\caption{ The block diagram for our hardware implementation of canonical Huffman encoding. The gray blocks represent the significant input and output data that is generated and consumed by the different subfunctions. The white blocks correspond to the functions (computational cores). Note that the array initial bit length appears twice to allow the figure to be more clear.}
\label{fig:che_dataflow}
\end{figure}

The input to the system is an array of \lstinline{Symbol}. This holds the symbol value and frequencies in the array \lstinline{in}. Each symbol holds a 10-bit \lstinline{value} and a 32-bit \lstinline{frequency}. The size of this array is set as the constant \lstinline{INPUT_SYMBOL_SIZE} which is 256 in our example. The \lstinline{filter} module reads from the \lstinline{in} array and writes its output to the \lstinline{filtered} array. This is an array of \lstinline{Symbols} which holds the number of non-zero elements which is the input to the \lstinline{sort} module. The \lstinline{sort} module writes the symbols sorted by frequency into two different arrays -- one is used for the \lstinline{create tree} module and the other for the \lstinline{canonize tree} module. The \lstinline{create tree} module creates a Huffman tree from the sorted array and stores it into three arrays (\lstinline{parent}, \lstinline{left}, and \lstinline{right}); these arrays hold all the info for each node of the Huffman tree. Using the Huffman tree information, the \lstinline{compute bit len} module calculates the bit length of each symbol and stores this information to a \lstinline{initial bit len} array. We set the maximum number of entries to 64, covering up to maximum 64-bit frequency number, which is sufficient for most applications given that our Huffman tree creation rebalances its height. The \lstinline{truncate tree} module rebalances the tree height and copies the bit length information of each codeword into two separate \lstinline{truncated bit length} arrays. They each have the exact same information, but they must be copied to ensure that the \VHLS tool can perform functional pipelining; we will talk about that in more detail later. The \lstinline{canonize tree} module walks through each symbol from the \lstinline{sort} module and assigns the appropriate bit length using the \lstinline{truncated bit length} array. The output of the \lstinline{canonize} module is an array that contains the bit lengths for the codeword of each symbol. Finally, the \lstinline{create codeword} module generates the canonical codewords for each symbol.

\subsection{Filter}
 
The first function for the Huffman encoding process is \lstinline{filter}, which is shown in Figure \ref{fig:huffman_filter.cpp}. This function takes as input a \lstinline{Symbol} array. The output is another \lstinline{Symbol} array that is a subset of the input array \lstinline{in}. The \lstinline{filter} function removes any entry with a frequency equal to $0$.  The function itself simply iterates across the \lstinline{in} array, storing each element to the \lstinline{out} array if its \lstinline{frequency} field is non-zero. In addition, the function counts the number of non-zero entries to the output. This is passed as the output argument \lstinline{n}, enabling further functions to only process the `useful' data.

\begin{figure}
\begin{lstlisting}
#include "huffman.h"
// Postcondition: out[x].frequency > 0
void filter(
            /* input  */ Symbol in[INPUT_SYMBOL_SIZE],
            /* output */ Symbol out[INPUT_SYMBOL_SIZE],
            /* output */ int *n) {
#pragma HLS INLINE off
    ap_uint<SYMBOL_BITS> j = 0;
    for(int i = 0; i < INPUT_SYMBOL_SIZE; i++) {
#pragma HLS pipeline II=1
        if(in[i].frequency != 0) {
            out[j].frequency = in[i].frequency;
            out[j].value = in[i].value;
            j++;
        }
    }
    *n = j;
}
\end{lstlisting}
\caption{  The \lstinline{filter} function iterates across the input array \lstinline{in} and add any \lstinline{Symbol} entry with a non-zero \lstinline{frequency} field to the output array \lstinline{out}. Additionally, it records the number of non-zero frequency elements and passes that in the output argument \lstinline{n}.  }
\label{fig:huffman_filter.cpp}
\end{figure}

\begin{aside}
\VHLS can decide to automatically inline functions in order to generate a more efficient architecture. Most often, this happens for small functions.  The directive \lstinline{inline} allows the user to explicitly specify whether or not \VHLS should inline particular functions. In this case, \lstinline{INLINE off} ensures that this function will not be inlined and will appear as a module in the generated \gls{rtl} design. In this case, disabling inlining allows us to get a performance and resource usage for this function and to ensure that it will be implemented as a process in the toplevel dataflow design.
\end{aside}


\begin{figure}
\begin{lstlisting}[format=none, lastline=39]
#include "huffman.h"
#include "assert.h"
const unsigned int RADIX = 16;
const unsigned int BITS_PER_LOOP = 4; // should be log2(RADIX)
typedef ap_uint<BITS_PER_LOOP> Digit;

void sort(
    /* input */ Symbol in[INPUT_SYMBOL_SIZE],
    /* input */ int num_symbols,
    /* output */ Symbol out[INPUT_SYMBOL_SIZE]) {
    Symbol previous_sorting[INPUT_SYMBOL_SIZE], sorting[INPUT_SYMBOL_SIZE];
    ap_uint<SYMBOL_BITS> digit_histogram[RADIX], digit_location[RADIX];
#pragma HLS ARRAY_PARTITION variable=digit_location complete dim=1
#pragma HLS ARRAY_PARTITION variable=digit_histogram complete dim=1
    Digit current_digit[INPUT_SYMBOL_SIZE];

    assert(num_symbols >= 0);
    assert(num_symbols <= INPUT_SYMBOL_SIZE);
 copy_in_to_sorting:
    for(int j = 0; j < num_symbols; j++) {
#pragma HLS PIPELINE II=1
        sorting[j] = in[j];
    }

 radix_sort:
    for(int shift = 0; shift < 32; shift += BITS_PER_LOOP) {
    init_histogram:
        for(int i = 0; i < RADIX; i++) {
#pragma HLS pipeline II=1
            digit_histogram[i] = 0;
        }

    compute_histogram:
        for(int j = 0; j < num_symbols; j++) {
#pragma HLS PIPELINE II=1
            Digit digit = (sorting[j].frequency >> shift) & (RADIX - 1); // Extrract a digit
            current_digit[j] = digit;  // Store the current digit for each symbol
            digit_histogram[digit]++;
            previous_sorting[j] = sorting[j]; // Save the current sorted order of symbols
\end{lstlisting}
\end{figure}
\begin{figure}
\begin{lstlisting}[format=none,firstline=40]
        }

        digit_location[0] = 0;
    find_digit_location:
        for(int i = 1; i < RADIX; i++)
#pragma HLS PIPELINE II=1
            digit_location[i] = digit_location[i-1] + digit_histogram[i-1];

    re_sort:
        for(int j = 0; j < num_symbols; j++) {
#pragma HLS PIPELINE II=1
            Digit digit = current_digit[j];
            sorting[digit_location[digit]] = previous_sorting[j]; // Move symbol to new sorted location
            out[digit_location[digit]] = previous_sorting[j]; // Also copy to output
            digit_location[digit]++; // Update digit_location
        }
    }
}
\end{lstlisting}
\caption{  The \lstinline{sort} function employs a radix sort on the input symbols based upon their frequency values. }
\label{fig:huffman_sort.cpp}
\end{figure}

\subsection{Sort}

The \lstinline{sort} function, shown in Figure \ref{fig:huffman_sort.cpp}, orders the input symbols based on their \lstinline{frequency} values. The function itself consists of two \lstinline{for} loops, labeled \lstinline{copy_in_to_sorting} and \lstinline{radix_sort}. 

The \lstinline{copy_in_to_sorting} loop moves input data from the \lstinline{in} array into the \lstinline{sorting} array.  This ensures that the \lstinline{in} array is read-only to meet the requirements of the \lstinline{dataflow} directive used at the toplevel. The \lstinline{sorting} function reads and writes to the \lstinline{sorting} array throughout its execution. Even for simple loops like this, it is important to use the \lstinline{pipeline} directive to generate the most efficient result and accurate performance estimates.

The \lstinline{radix_sort} loop implements the core radix-sorting algorithm.   In general, radix sorting algorithms sort data by considering one digit or group of bits at a time.  The size of each digit determines the \term{radix} of the sort.  Our algorithm considers 4 bits at a time of the 32-bit \lstinline{Symbol.frequency} variable.  Hence we are using radix $r=2^{4}=16$ sort.  For each 4-bit digit in the 32-bit number, we perform a counting sort. The \lstinline{radix_sort} loop performs these 8 counting sort operations, iterating to $32$ in steps of $4$.   Radix-sorting algorithms can also operate from left to right (least significant digit first) or right to left (most significant digit first).   The algorithm implemented here works from least significant digit to most significant digit.   In the code, the radix can be configured by setting the \lstinline{RADIX} and \lstinline{BITS_PER_LOOP} parameters. 

\begin{exercise}
What would happen if we increased or decreased the radix? How would this effect the number of counting sort operations that are performed? How would this change the resource usage, e.g., the size of the arrays?
\end{exercise}

The code stores the current state of the sort in \lstinline{sorting[]} and \lstinline{previous_sorting[]}.  Each iteration of \lstinline{radix_sort_loop}, the current value of \lstinline{sorting[]} is copied to \lstinline{previous_sorting[]} and then the values are sorted as they are copied back into \lstinline{sorting[]}.  The \lstinline{digit_histogram[]} and \lstinline{digit_location[]} arrays are used in \lstinline{radix_sort_loop} to implement the counting sort on a particular digit. The two \lstinline{array_partition} s declare that these two arrays should be completely partitioned into registers. These arrays are small and used frequently, thus this does not use many resources and can provide performance benefits. Finally, \lstinline{current_digit[]} stores the digit being sorted for each item in the current iteration of the radix sort.

This code also contains two \lstinline{assert()} calls which check assumptions about the input \lstinline{num_symbols}.  Since this variable determines the number of valid elements in the \lstinline{in} array, it must be bounded by the size of that array.  Such assertions are good defensive programming practice in general to ensure that the assumptions of this function are met.  In \VHLS they serve an additional purpose as well.  Since \lstinline{num_symbols} determines the number of times that many of the internal loops execute, \VHLS can infer the tripcount of the loop based on these assertions.  In addition, \VHLS also uses these assertions to minimize the bitwidth of the variables used in the implemented circuit.

\begin{aside}
Previously we've seen the \lstinline{loop_tripcount} directive used to give \VHLS information about the tripcount of loops.   Using \lstinline{assert()} statements serves many of the same purposes, with some advantages and disadvantages.  One advantage of using \lstinline{assert()} statements is that they are checked during simulation and this information can be used to further optimize the circuit.  In contrast, the \lstinline{loop_tripcount} directive only affects performance analysis and is not used for optimization.  On the other hand, \lstinline{assert()} statements can only be used to give bounds on variable values, but can't be used to set expected or average values, which can only be done through the \lstinline{loop_tripcount} directive.  In most cases, it is recommended to first provide worst case bounds through \lstinline{assert()} statements, and then if necessary also add \lstinline{loop_tripcount} directives.
\end{aside}

The body of the \lstinline{radix_sort} loop is divided into four subloops, labeled \lstinline{init_histogram}, \lstinline{compute_histogram}, \lstinline{find_digit_location}, and \lstinline{re_sort}.  \lstinline{init_histogram} and \lstinline{compute_histogram} loops combine to compute the histogram of the input, based on the current digit being considered.  This produces a count of the number of each times each digit occurs in \lstinline{digit_histogram[]}.  The \lstinline{compute_histogram} loop also stores the current digit being sorted for each symbol in \lstinline{current_digit[]}.   Next, the \lstinline{find_digit_location} loop computes a prefix sum of the resulting histogram values, placing the result in \lstinline{digit_location[]}.  In the context of the counting sort, \lstinline{digit_location[]} contains the location of the first symbol with each digit in the newly sorted array.  Lastly, the \lstinline{re_sort} loop reorders the symbols based upon these results, placing each element in its correct place in the newly sorted array.    It uses the key stored in \lstinline{current_digit[]} to select the right location from \lstinline{digit_location[]}. This location is incremeted each time through the \lstinline{re_sort} loop to place the next element with the same digit in the next location in the sorted array.  Overall, each iteration through the \lstinline{radix_sort} loop implements a counting sort on one digit.  The counting sort is a \gls{stable_sort}, so that elements with the same digit remain in the same order.  After stable-sorting based on each digit, the array is returned in the correct final order.

We have previous discussed the histogram and prefix sum algorithms in Chapter \ref{sec:histogram} and \ref{sec:prefixSum}.  In this case, with simple code and complete partitioning of \lstinline{digit_histogram[]} and \lstinline{digit_location[]}, we can achieve a loop II of 1 to compute the histogram and prefix sum, since the number of bins is relatively small.  The optimization of the \lstinline{re_sort} loop is similar.  Since the only recurrence is through the relatively small \lstinline{digit_location[]} array, achieving a loop II of 1 is also straightforward.  Note that this approach works primarily because we've configured \lstinline{RADIX} to be relatively small.  With larger values of \lstinline{RADIX}, it would be preferable to implement \lstinline{digit_histogram[]} and \lstinline{digit_location[]} as memories, which might require additional optimization to achieve a loop II of 1.

Another alternative that may make sense in the context of this code is to combine complete partitioning of \lstinline{digit_histogram[]} and \lstinline{digit_location[]} with complete unrolling of the \lstinline{init_histogram} and \lstinline{find_digit_location} loops.   These loops access each location in these small arrays and perform operations with a minimal amount of logic.  In this case, although unrolling loops would likely result in replicating the circuit for each loop body, fewer resources would be required to implement this circuit since the array accesses would be at constant indexes.  However, for larger values of the \lstinline{BITS_PER_LOOP} parameter this change becomes prohibitive, since each additional bit doubles the \lstinline{RADIX} parameter, doubling the cost of these unrolled loops.  This is a somewhat common situation with parameterized code where different optimizations make sense with different parameter values.

\begin{exercise}
What happens to the performance and utilization results when you perform the optimizations on the prefix sum and histogram loops as specified in Chapter \ref{sec:histogram} and \ref{sec:prefixSum}?  Are these optimizations necessary in this case?
\end{exercise}

\begin{exercise}
Is the \lstinline{re_sort for} loop able to achieve the specified initiation interval of one cycle? Why or why not?
\end{exercise}

\begin{exercise}
For a large dataset ($n > 256$), what is the approximate latency, in terms of $n$, of the code in Figure \ref{fig:huffman_sort.cpp}.  What portions of the code dominate the number of cycles?  How would this change as the \lstinline{RADIX} parameter changes?
\end{exercise}

Note that the \lstinline{re_sort} loop not only stores the sorted arrray in \lstinline{sorting[]} but also stores the sorted array in \lstinline{out[]}.  While this may seem redundant, we need to ensure that \lstinline{out[]} is only written to in order to obey the requirements of the toplevel \lstinline{dataflow} directive.  In this case, \lstinline{out[]} will be overwritten multiple times with partially sorted results, but only the final result will be passed on the following function.

\begin{aside}
The \lstinline{dataflow} directive has several requirements in order to perform the task level pipelining optimization. One of them is the need for single producer and consumer of data between the tasks. Since we would like to perform task level pipelining for the Huffman encoding process as shown in Figure \ref{fig:che_dataflow}, we must insure that each of these tasks follow this requirement. In the case of this \lstinline{sort} function, which is one of the tasks, it must only consume (read from but not write to) the input argument data and only produce (write to but not read from) the output argument data. In order to met this requirement, we create the internal array \lstinline{sorting}, which is read from and written to throughout the function. We copy the input data from the argument \lstinline{in} at the beginning of the function and write the final results to the output argument \lstinline{out} at the end of the function. This insures that we follow the producer/consumer requirements for the \lstinline{dataflow} directive.
\end{aside}

\subsection{Create Tree}
The next function in the Huffman encoding process forms the binary tree representing the Huffman code. This is implemented in the \lstinline{create_tree} function shown in Figure \ref{fig:huffman_create_tree.cpp}. \lstinline{in[]} contains \lstinline{num_symbols} \lstinline{Symbol} elements, sorted from lowest to highest frequency. The function creates a binary tree of those symbols which is stored into three output arrays named \lstinline{parent}, \lstinline{left}, and \lstinline{right}.  The \lstinline{left} and \lstinline{right} arrays represent the left and right children of each intermediate node in the tree.  If the child is a leaf node, then the corresponding element of the \lstinline{left} or \lstinline{right} array will contain the symbol value of the child, otherwise it contains the special flag \lstinline{INTERNAL_NODE}.   Similarly, the \lstinline{parent} array holds the index of the parent node of each intermediate node.  The parent of the root node of the tree is defined to be index zero.  The tree is also ordered, in the sense that a parent  always has a higher index than its children.  As a result, we can efficiently implement bottom-up and top-down traversals of the tree. 

Figure \ref{fig:huffman_create_tree} shows an example of these data structures. Six symbols sorted by their frequencies are stored in the \lstinline{in} array. The resulting Huffman tree is stored in three arrays \lstinline{parent}, \lstinline{left}, and \lstinline{right}.  In addition, the frequency of each intermediate node is stored in the \lstinline{frequency} array.  We directly denote the node numbers for the \lstinline{left} and \lstinline{right} arrays (e.g., \lstinline{n0}, \lstinline{n1}, etc.) for the sake of illustration. These will hold a special internal node value in reality. 

\begin{aside}
While it may be odd to think of storing a complex data structure in a tree like this, it is actually very common in embedded programming where data allocation is not allowed\cite{misra2012}.  In fact, the C library implementations of \lstinline{malloc()} and \lstinline{free()} often implement low-level memory management in this way to enable small allocations to be created from larger memory allocations, usually called \term{pages}, returned from the operating system.  This enables the operating system to efficiently manage large allocations of memory efficiently and to coordinate virtual memory using the processor page table and disk storage which usually handle large blocks of data.  4 Kilo-bytes is a typical size for these pages.  For more ideas about implementing data structures using arrays, see \cite{sedgewickalgorithmsinC}.
\end{aside}

\begin{figure}
\centering
\includegraphics[width= \textwidth]{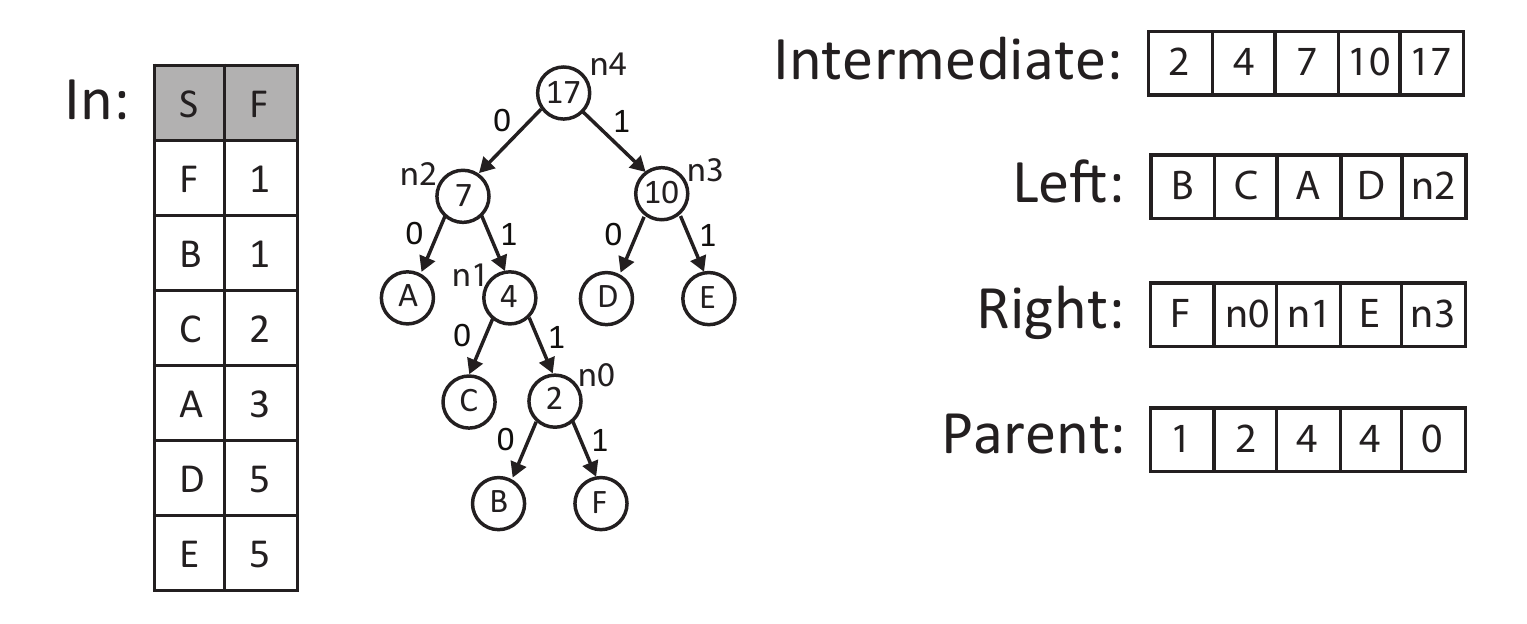}
\caption{ The \lstinline{Symbol} array \lstinline{in} is used to create the Huffman tree. The tree is shown graphically along with the corresponding values for the four arrays used to represent the tree (\lstinline{intermediate}, \lstinline{left}, \lstinline{right}, and \lstinline{parent}).   }
\label{fig:huffman_create_tree}
\end{figure}

In the Huffman tree, each symbol is associated with a leaf node in the tree.  Intermediate nodes in the tree are created by grouping the two symbols with the smallest frequency and using them as the left and right nodes of a new intermediate node. That intermediate node has a frequency which is the sum of the frequencies of each child node. This process continues by iteratively creating intermediate nodes from the two nodes with the smallest frequencies, which may include other intermediate nodes or leaf nodes.  The tree building process completes when all of the intermediate nodes have been incorporated into the binary tree.

There are many ways to represent this process in code.  For instance, we might explicitly create an array representing every node in the tree which is sorted by frequency.  In this case selecting nodes to add to the tree is simple, since they will always be in the same locations of the sorted array.  On the other hand, inserting a newly created node into the list is relatively complex because the array must again be sorted, which would require moving elements around.  Alternatively, we might add pointer-like array indexes to the data structure in our array to enable the data to be logically sorted without actually moving the data around.  This would reduce data copying, but would increase the cost of accessing each element and require extra storage.  Many of the normal algorithmic tradeoffs in the design of data structures apply in the context of HLS just as well as they apply to processors.

In this case, however, we can make some additional simplifying observations.  The most important observation is that new intermediate nodes are always created in order of frequency.  We might create an intermediate node with a frequency that is less than the frequency of some leaf node, but we will never create an intermediate node with a frequency less than an already created intermediate node.  This suggests that we can maintain a sorted data structure by storing the nodes in two separate arrays: a sorted array of symbols and a sorted array of intermediate nodes.  As we `use' the lowest frequency elements of each list, we only need to append to the end of the list of intermediate nodes.  There is a small extra complexity because we might need to remove zero, one, or two elements from either array, but this turns out to be much less complex than resorting the node array.

\begin{aside}
Conceptually this algorithm is very similar to the mergesort algorithm discussed in Section \ref{sec:sort:merge}.   The key difference is what operation is done as elements are removed from the sorted arrays.  In mergesort, the least element is simply inserted at the appropriate position in the array.  In this case, the two least elements are identified and then merged into a new tree node.
\end{aside}

\begin{figure}
\begin{lstlisting}[format=none, lastline=42]
#include "huffman.h"
#include "assert.h"
void create_tree (
    /* input */ Symbol in[INPUT_SYMBOL_SIZE],
    /* input */ int num_symbols,
    /* output */ ap_uint<SYMBOL_BITS> parent[INPUT_SYMBOL_SIZE-1],
    /* output */ ap_uint<SYMBOL_BITS> left[INPUT_SYMBOL_SIZE-1],
    /* output */ ap_uint<SYMBOL_BITS> right[INPUT_SYMBOL_SIZE-1]) {
    Frequency frequency[INPUT_SYMBOL_SIZE-1];
    ap_uint<SYMBOL_BITS> tree_count = 0;  // Number of intermediate nodes assigned a parent.
    ap_uint<SYMBOL_BITS> in_count = 0;    // Number of inputs consumed.

    assert(num_symbols > 0);
    assert(num_symbols <= INPUT_SYMBOL_SIZE);
    for(int i = 0; i < (num_symbols-1); i++) {
#pragma HLS PIPELINE II=5
        Frequency node_freq = 0;

        // There are two cases.
        // Case 1: remove a Symbol from in[]
        // Case 2: remove an element from intermediate[]
        // We do this twice, once for the left and once for the right of the new intermediate node.
        assert(in_count < num_symbols || tree_count < i);
        Frequency intermediate_freq = frequency[tree_count];
        Symbol s = in[in_count];
        if((in_count < num_symbols && s.frequency <= intermediate_freq) || tree_count == i) {
            // Pick symbol from in[].
            left[i] = s.value; // Set input symbol as left node
            node_freq = s.frequency; // Add symbol frequency to total node frequency
            in_count++; // Move to the next input symbol
        } else {
            // Pick internal node without a parent.
            left[i] = INTERNAL_NODE; // Set symbol to indicate an internal node
            node_freq = frequency[tree_count]; // Add child node frequency
            parent[tree_count] = i; // Set this node as child's parent
            tree_count++; // Go to next parentless internal node
        }

        assert(in_count < num_symbols || tree_count < i);
        intermediate_freq = frequency[tree_count];
        s = in[in_count];
        if((in_count < num_symbols && s.frequency <= intermediate_freq) || tree_count == i) {
\end{lstlisting}
\end{figure}
\begin{figure}
\begin{lstlisting}[format=none, firstline=43, lastline=86]
            // Pick symbol from in[].
            right[i] = s.value;
            frequency[i] = node_freq + s.frequency;
            in_count++;
        } else {
            // Pick internal node without a parent.
            right[i] = INTERNAL_NODE;
            frequency[i] = node_freq + intermediate_freq;
            parent[tree_count] = i;
            tree_count++;
        }
        // Verify that nodes in the tree are sorted by frequency
        assert(i == 0 || frequency[i] >= frequency[i-1]);
    }

    parent[tree_count] = 0; //Set parent of last node (root) to 0
}
\end{lstlisting}
\end{figure}
\begin{figure}
\begin{lstlisting}[format=none, firstline=87]

\end{lstlisting}
\caption{  The complete code for Huffman tree creation. The code takes as input the sorted \lstinline{Symbol} array \lstinline{in}, the number of elements in that array \lstinline{n}, and outputs the Huffman tree in the three arrays \lstinline{left}, \lstinline{right}, and \lstinline{parent}. }
\label{fig:huffman_create_tree.cpp}
\end{figure}

This code to implement the \lstinline{create_tree} function is shown in Figure \ref{fig:huffman_create_tree.cpp}.  The first block of code defines the local variables that we use in the function.  \lstinline{frequency[]} stores the frequencies for each intermediate node as it is created.   \lstinline{in_count} tracks which symbols have been given a parent node in the tree, while \lstinline{tree_count} tracks which newly created intermediate nodes have been given a parent node.  Each iteration through the main loop creates a new intermediate node without a parent, so all of the intermediate nodes between \lstinline{tree_count} and \lstinline{i} have not yet been assigned a parent in the tree.

The main loop contains two similar blocks of code.  Each block compares the frequency of the next available symbol \lstinline{in[in_count].frequency} with the frequency of the next available intermediate node \lstinline{frequency[tree_count]}.  It then selects the lowest frequency of the two to be incorporated as the leaf of a new intermediate node.  The first block does this for the left child of the new node, storing in \lstinline{left[i]}, while the second block selects the right child of the new node, storing in \lstinline{right[i]}.  In both cases, we need to be careful to ensure that the comparison is meaningful.  In the first iteration of the loop, \lstinline{tree_count == 0} and \lstinline{i == 0}, so there is no valid intermediate node to be considered and we must always select an input symbol.  During the final iterations of the loop, it is likely that all of the input symbols will have been consumed, so \lstinline{in_count == num_symbols} and we must always consume an intermediate node.

The number of iterations of the loop depends on the input \lstinline{num_symbols} in an interesting way.  Since each input symbol becomes a leaf node in the binary tree, we know that there will be exactly \lstinline{num_symbols-1} intermediate nodes to be created, since this is a basic property of a binary tree.  At the end of the loop we will have created \lstinline{num_symbols-1} new nodes, each of which has two children. \lstinline{num_symbols} of these children will be input symbols and \lstinline{num_symbols-2} will be intermediate nodes.  There will be one intermediate node remaining as the root of the tree without a parent.  This last node is artificially assigned a parent index of zero in the last line of code.  This completes the building of the Huffman tree.

In the tree, the children of an intermediate node can be either a symbol or a intermediate node.  In creating the huffman tree, this information isn't very important, although it will be important later when we traverse the tree later.  To store this difference
a special value \lstinline{INTERNAL_NODE} is stored in \lstinline{left[]} and \lstinline{right[]} if the corresponding child is an internal node.   Note that this storage essentially requires one more bit to represent in the array.  As a result, the \lstinline{left[]} and \lstinline{right[]} arrays are one bit larger than you might expect.

\begin{exercise}
For a large dataset ($n > 256$), what is the approximate latency, in terms of $n$, of the code in Figure \ref{fig:huffman_create_tree.cpp}?  What portions of the code dominate the number of cycles?
\end{exercise}

\subsection{Compute Bit Length}

The \lstinline{compute_bit_length} function determines the depth in the tree for each symbol.  The depth is important because it determines the number of bits used to encode each symbol. Computing the depth of each node in the tree is done using the recurrence:
\begin{equation}
\begin{array}{rrcl}
&\mathrm{depth}(\mathrm{root}) &=& 0 \\
\forall n != \mathrm{root}, &\mathrm{depth}(n) &=& \mathrm{depth}(\mathrm{parent}(n)+1)\\
\forall n, &\mathrm{child\_depth}(n) &=& \mathrm{depth}(n)+1
\end{array}
\end{equation}

This recurrence can be computed by traversing the tree starting at the root node and exploring each internal node in order.  As we traverse each internal node, we can compute the depth of the node and the corresponding depth (incremented by one) of any child nodes.  It turns out that we don't actually care about the depth of the internal nodes, only about the depth of the child nodes.  As a result, the code actually computes the recurrence:
\begin{equation}
\begin{array}{rrcl}
&\mathrm{child\_depth}(\mathrm{root}) &=& 1 \\
\forall n != \mathrm{root}, &\mathrm{child\_depth}(n) &=& \mathrm{child\_depth}(\mathrm{parent}(n)+1)
\end{array}
\end{equation}

The code for this function is shown in Figure \ref{fig:huffman_compute_bit_length.cpp}.  The input arguments to the function represent a Huffman tree in \lstinline{parent[]}, \lstinline{left[]}, and \lstinline{right[]}.   \lstinline{num_symbols} contains the number of input symbols, which is one more than the number of intermediate nodes in the tree.  The output \lstinline{length_histogram[]}.  Each element of that array stores the number of symbols with the given depth. Thus, if there are five symbols with depth three, then \lstinline{length_histogram[3] = 5}. 
\begin{figure}
\begin{lstlisting}[format=none]
#include "huffman.h"
#include "assert.h"
void compute_bit_length (
    /* input */ ap_uint<SYMBOL_BITS> parent[INPUT_SYMBOL_SIZE-1],
    /* input */ ap_uint<SYMBOL_BITS> left[INPUT_SYMBOL_SIZE-1],
    /* input */ ap_uint<SYMBOL_BITS> right[INPUT_SYMBOL_SIZE-1],
    /* input */ int num_symbols,
    /* output */ ap_uint<SYMBOL_BITS> length_histogram[TREE_DEPTH]) {
    assert(num_symbols > 0);
    assert(num_symbols <= INPUT_SYMBOL_SIZE);
    ap_uint<TREE_DEPTH_BITS> child_depth[INPUT_SYMBOL_SIZE-1];
    ap_uint<SYMBOL_BITS> internal_length_histogram[TREE_DEPTH];
 init_histogram:
    for(int i = 0; i < TREE_DEPTH; i++) {
        #pragma HLS pipeline II=1
        internal_length_histogram[i] = 0;
    }

    child_depth[num_symbols-2] = 1; // Depth of the root node is 1.

traverse_tree:
    for(int i = num_symbols-3; i >= 0; i--) {
#pragma HLS pipeline II=3
        ap_uint<TREE_DEPTH_BITS> length = child_depth[parent[i]] + 1;
        child_depth[i] = length;
        if(left[i] != INTERNAL_NODE || right[i] != INTERNAL_NODE){
            int children;
            if(left[i] != INTERNAL_NODE && right[i] != INTERNAL_NODE) {
                // Both the children of the original node were symbols
                children = 2;
            } else {
                // One child of the original node was a symbol
                children = 1;
            }
            ap_uint<SYMBOL_BITS> count = internal_length_histogram[length];
            count += children;
            internal_length_histogram[length] = count;
            length_histogram[length] = count;
        }
    }
}
\end{lstlisting}
\caption{The complete code for determining the number of symbols at each bit length.}
\label{fig:huffman_compute_bit_length.cpp}
\end{figure}

\lstinline{child_depth[]} stores the depth of each internal node while the tree is being traversed.  After the depth of each internal node is determined in the \lstinline{traverse_tree} loop, \lstinline{length_histogram[]} is updated.  \lstinline{internal_length_histogram[]} is used to ensure that our function adheres the requirements for the \lstinline{dataflow} directive, where the output array \lstinline{length_histogram[]} is never read.  The \lstinline{init_histogram} loop initializes these two arrays.

\begin{exercise}
The \lstinline{init_histogram} loop has a \lstinline{pipeline} directive with \lstinline{II = 1}. Is it possible to meet this II? What happens if we increase the II to something larger? What happens if we do not apply this directive?
\end{exercise}

Internal nodes in the tree are traversed from the root node, which has the largest index, down to index zero.  Since the array of nodes were created in bottom-up order, this reverse order results in a top-down traversal of the tree enabling the computation of the recurrence for each node in a single pass through the nodes.  For each node, we determine the depth of its children.  Then if the node actually does have any children which are symbols, we figure out how many children and update the histogram accordingly.  Child nodes which are internal nodes are represented by the special value \lstinline{INTERNAL_NODE}.

\begin{exercise}
For a large dataset ($n > 256$), what is the approximate latency, in terms of $n$, of the code in Figure \ref{fig:huffman_compute_bit_length.cpp}?  What portions of the code dominate the number of cycles?
\end{exercise}

\begin{exercise}
This code has several recurrences.  For example, one recurrence occurs because of the histogram computation.  In this case, the loop is synthesized with an II of 3.  What happens if you target a lower II in the \lstinline{pipeline} directive?  Can you rewrite the code to eliminate the recurrences and achieve a lower II?
\end{exercise}

\subsection{Truncate Tree}
\label{sec:huffman_truncate_tree}

The next part of the Huffman encoding process reorganizes nodes with a depth that is larger than that specified in \lstinline{MAX_CODEWORD_LENGTH}. This is done by finding any symbols with a greater depth, and moving them to a level that is smaller than that specified target.   Interestingly, this can be done entirely by manipulating the histogram of symbol depths, as long as the histogram is modified in a way that is consistent with the same modifications on the original tree.


The input histogram is contained in \lstinline{input_length_histogram}, which was derived by the \lstinline{compute_bit_length()} function described in the previous section. There are two identical output arrays \lstinline{truncated_length_histogram1} and \lstinline{truncated_length_histogram2}. These arrays are passed to two separate functions later in the process (\lstinline{canonize_tree} and \lstinline{create_codewords}), and thus we must have two arrays to adhere to the single producer, single consumer constraint of the \lstinline{dataflow} directive.

\begin{figure}
\begin{lstlisting}[lastline=39]
#include "huffman.h"
#include "assert.h"
void truncate_tree(
    /* input */ ap_uint<SYMBOL_BITS> input_length_histogram[TREE_DEPTH],
    /* output */ ap_uint<SYMBOL_BITS> output_length_histogram1[TREE_DEPTH],
    /* output */ ap_uint<SYMBOL_BITS> output_length_histogram2[TREE_DEPTH]
) {
    // Copy into temporary storage to maintain dataflow properties
 copy_input:
    for(int i = 0; i < TREE_DEPTH; i++) {
        output_length_histogram1[i] = input_length_histogram[i];
    }

    ap_uint<SYMBOL_BITS> j = MAX_CODEWORD_LENGTH;
 move_nodes:
    for(int i = TREE_DEPTH - 1; i > MAX_CODEWORD_LENGTH; i--) {
        // Look to see if there is any nodes at lengths greater than target depth
    reorder:
        while(output_length_histogram1[i] != 0) {
#pragma HLS LOOP_TRIPCOUNT min=3 max=3 avg=3
            if (j == MAX_CODEWORD_LENGTH) {
                // Find deepest leaf with codeword length < target depth
                do {
#pragma HLS LOOP_TRIPCOUNT min=1 max=1 avg=1
                    j--;
                } while(output_length_histogram1[j] == 0);
            }

            // Move leaf with depth i to depth j+1.
            output_length_histogram1[j] -= 1; // The node at level j is no longer a leaf.
            output_length_histogram1[j+1] += 2; // Two new leaf nodes are attached at level j+1.
            output_length_histogram1[i-1] += 1; // The leaf node at level i+1 gets attached here.
            output_length_histogram1[i] -= 2; // Two leaf nodes have been lost from level i.

            // now deepest leaf with codeword length < target length
            // is at level (j+1) unless j+1 == target length
            j++;
        }
    }
\end{lstlisting}
\end{figure}
\begin{figure}
\begin{lstlisting}[format=none, firstline=40]

    // Copy the output to meet dataflow requirements and check the validity
    unsigned int limit = 1;
 copy_output:
    for(int i = 0; i < TREE_DEPTH; i++) {
        output_length_histogram2[i] = output_length_histogram1[i];
        assert(output_length_histogram1[i] >= 0);
        assert(output_length_histogram1[i] <= limit);
        limit *= 2;
    }
}
\end{lstlisting}
\caption{The complete code for rearranging the Huffman tree such that the depth of any node is under the target specified by the parameter \lstinline{MAX_CODEWORD_LENGTH}. }
\label{fig:huffman_truncate_tree.cpp}
\end{figure}

The code is shown in Figure \ref{fig:huffman_truncate_tree.cpp}. The \lstinline{copy_input} loop copies the data from the input array \lstinline{input_length_histogram}. The \lstinline{move_nodes} loop contains the bulk of the processing to modify the histogram.  Lastly, the \lstinline{input_length_histogram} function copies the internal result to other output at the end of the function. 

\begin{exercise}
The \lstinline{copy_in for} loop is not optimized. What happens to the latency and initiation interval of the \lstinline{truncate_tree} function if we use a \lstinline{pipeline} or \lstinline{unroll} directive on this loop. What happens to the overall latency and initiation interval of the design (i.e., the \lstinline{huffman_encoding} function)?
\end{exercise}

The function continues in the second \lstinline{move_nodes for} loop, which performs the bulk of the computation. This \lstinline{for} loop starts by iterating through the \lstinline{truncated_length_histogram} array from the largest index (\lstinline{TREE_DEPTH} - the specified maximum depth for a tree). This continues down through the array until there is a non-zero element or \lstinline{i} reaches the \lstinline{MAX_CODEWORD_LENGTH}. If we do not find a non-zero element, that means the initial input Huffman tree does not have any nodes with a depth larger than the target depth. In other words, we can exit this function without performing any truncation. If there is a value larger than the target depth, then the function continues by reorganizing the tree so that all of the nodes have depth smaller than the target depth. This is done by the operations in the \lstinline{reorder while} loop. When there are nodes to move, the \lstinline{move_nodes for} loop goes through them from those with the largest depth, and continues to smaller depths until all nodes are rearranged with a depth smaller than the target. Each iteration of this \lstinline{move_nodes for} loops works on moving nodes from one depth at a time.

The \lstinline{reorder while} loop moves one node in each iteration. The first \lstinline{if} statement is used to find the leaf node with the largest depth. We will then alter this node by making it an intermediate node, and adding it and the leaf node with a depth larger than than target as children. This \lstinline{if} clause has a \lstinline{do/while} loop that iterates downward from the target looking for a non-zero entry in the \lstinline{truncated_length_histogram} array. It works in a similar manner as the beginning of the \lstinline{move_nodes for} loop. When it has found the deepest leaf node less than the target depth, it stops. The depth of this node is stored in \lstinline{j}.

Now we have a node with a depth \lstinline{i} larger than the target, and a node with a depth smaller than the target stored in \lstinline{j}. We move the node from depth \lstinline{i} and a node from \lstinline{j} into child nodes at depth \lstinline{j + 1}. Therefore, we add two symbols to \lstinline{truncated_length_histogram[j+1]}. We are making a new intermediate node a depth \lstinline{j} thus, we subtract a symbol from that level. We move the other leaf node from depth \lstinline{i} to depth \lstinline{i - 1}. And we subtract two from \lstinline{truncated_length_histogram[i]} since one of the nodes went to level \lstinline{j + 1} and the other when to level \lstinline{i - 1}. These operations are performed in the four statements on the array \lstinline{truncated_length_histogram}. Since we added a symbol to level \lstinline{j + 1}, we update \lstinline{j}, which holds the highest level under the target level, and then we repeat the procedure. This is done until there are no additional symbols with a depth larger than the target.

The function completes by creating an additional copy of the new bit lengths. This is done by storing the updated bit lengths in the array \lstinline{truncated_length_histogram1} into the array \lstinline{truncated_length_histogram2}. We will pass these two arrays to the final two functions in the \lstinline{huffman_encoding} top function; we need two arrays to insure that the constraints of the \lstinline{dataflow} directive are met.

\subsection{Canonize Tree}
\label{sec:huffman_canonize_tree}

The next step in the encoding process is to determine the number of bits for each of the symbols. We do this in the \lstinline{canonize_tree} function shown in Figure \ref{fig:huffman_canonize_tree.cpp}. The function takes as input an array of symbols in sorted order, the total number of symbols (\lstinline{num_symbols}), and a histogram of lengths describing the Huffman tree. The output \lstinline{symbol_bits[]} contains the number of encoded bits used for each symbol. Thus, if the symbol with value \lstinline{0x0A} is encoded in 4 bits, then \lstinline{symbol_bits[10] = 4}. 

\begin{figure}
\begin{lstlisting}[firstline=1]
#include "huffman.h"
#include "assert.h"
void canonize_tree(
    /* input */ Symbol sorted[INPUT_SYMBOL_SIZE],
    /* input */ ap_uint<SYMBOL_BITS> num_symbols,
    /* input */ ap_uint<SYMBOL_BITS> codeword_length_histogram[TREE_DEPTH],
    /* output */ CodewordLength symbol_bits[INPUT_SYMBOL_SIZE] ) {
    assert(num_symbols <= INPUT_SYMBOL_SIZE);

 init_bits:
    for(int i = 0; i < INPUT_SYMBOL_SIZE; i++) {
        symbol_bits[i] = 0;
    }
    
    ap_uint<SYMBOL_BITS> length = TREE_DEPTH;
    ap_uint<SYMBOL_BITS> count = 0;

    // Iterate across the symbols from lowest frequency to highest
    // Assign them largest bit length to smallest
 process_symbols:
    for(int k = 0; k < num_symbols; k++) {
        if (count == 0) {
            //find the next non-zero bit length
            do {
#pragma HLS LOOP_TRIPCOUNT min=1 avg=1 max=2
                length--;
                // n is the number of symbols with encoded length i
                count = codeword_length_histogram[length];
            }
            while (count == 0);
        }
        symbol_bits[sorted[k].value] = length; //assign symbol k to have length bits
        count--; //keep assigning i bits until we have counted off n symbols
    }
}
\end{lstlisting}
\caption{ The complete code for canonizing the Huffman tree, which determins the number of bits for each symbol. }
\label{fig:huffman_canonize_tree.cpp}
\end{figure}

The canonization process consists of two loops, labeled \lstinline{init_bits} and \lstinline{process_symbols}.  The \lstinline{init_bits} loop executes first, initializing \lstinline{symbol_bits[]} array to \lstinline{0}. The \lstinline{process_symbols} loop then processes the symbols in sorted order from smallest frequency to largest frequency. Naturally, the least frequent symbols are assigned the longest codes while the most frequent symbols are assigned the shortest code.  Each time through the \lstinline{process_symbols} loop, we assign the length of one symbol.  The length of the symbol is determined by the the inner \lstinline{do/while} loop, which steps through the histogram of lengths. This loop finds the largest bit length that has not yet had codewords assigned and stores the number of codewords in that length in \lstinline{count}. Each time through the outer loop, \lstinline{count} is decremented until we run out of codewords.  When \lstinline{count} becomes zero, the inner \lstinline{do/while} loop executes again to find a length with codewords to assign.

\begin{figure}
\begin{lstlisting}
int k = 0;
process_symbols:
for(length = TREE_DEPTH; length >= 0; length--) {
    count = codeword_length_histogram[length];
    for(i = 0; i < count; i++) {
#pragma HLS pipeline II=1
        symbol_bits[sorted[k++].value] = length;
    }
 }
\end{lstlisting}
\caption{ Alternate loop structure for the \lstinline{process_symbols} loop in Figure \ref{fig:huffman_canonize_tree.cpp}. }
\label{fig:huffman_canonize_alternate}
\end{figure}

Note that the \lstinline{process_symbols} loop cannot be pipelined because the inner \lstinline{do/while} loop cannot be unrolled.  This is somewhat awkward as the inner loop will usually execute exactly once, stepping to the next length in the histogram.  Only in somewhat rare cases will the inner loop need to execute more than once if we happen to get to a length which does not have any codewords assigned.  In this case, there's not too much of a loss since all the operations in the loop are simple operations that are unlikely to be pipelined, with the exception of the memory operations.  There are other ways to structure this loop, however, which {\em can} be pipelined.  One possibility is to use an outer \lstinline{for} loop to iterate over \lstinline{codeword_length_histogram[]} and an inner loop to count each symbol, as shown in Figure \ref{fig:huffman_canonize_alternate}.

\begin{exercise}
Implement the code in Figure \ref{fig:huffman_canonize_tree.cpp} and the alternate code structure in Figure \ref{fig:huffman_canonize_alternate}.  Which results in higher performance?  Which coding style is more natural to you?
\end{exercise}

\subsection{Create Codeword}
\label{sec:create_codewords}

The final step in the encoding process is to create the codeword for each symbol.  This process simply assigns each symbol in order according to the properties of a Canonical Huffman code.  The first property is that longer length codes have a higher numeric value than the same length prefix of shorter codes.  The second property is that codes with the same length increase by one as the symbol value increases.  In order to achieve these properties while keeping the code simple, it is useful to determine the first codeword of each length.  If we know the number of codewords of each length given by $\mathrm{codeword\_length\_histogram}$, then this can be found using the following recurrence:
\begin{equation}
\begin{array} {rrcl}
&\mathrm{first\_codeword}(1) &=& 0 \\
\forall i > 1, &\mathrm{first\_codeword}(i) &=& (\mathrm{first\_codeword}(i-1) + \mathrm{codeword\_length\_histogram}(i-1)) << 1
\label{eq:first_codeword_recurrence}
\end{array}
\end{equation}  
Essentially, rather than actually assigning the codewords one after another, this recurrence allocates all the codewords first.  This allows us to actually assign the codewords in order of symbol value without being concerned about also ordering them by length or frequency.

In addition to assigning codewords to symbols, we also need to format the codewords so that they can be easily used for encoding and decoding.  Systems that use Huffman encoding often store codewords in bit-reversed order. This can make the decoding process easier since the bits are stored in the same order that the tree is traversed during decoding, from root node to leaf node. 

The code implementing the \lstinline{create_codewords} function is shown in Figure \ref{fig:huffman_create_codeword.cpp}. \lstinline{symbol_bits[]} contains the length of the codeword for each symbol and \lstinline{codeword_length_histogram[]} contains the number of codewords with each length. The output \lstinline{encoding[]} represents the encoding for each symbol. Each element consists of the actual codeword and the length of each codeword packed together.  The maximum length of a codeword is given by the \lstinline{MAX_CODEWORD_LENGTH} parameter.  In turn, this determines the number of bits required to hold the codeword, which is given by  \lstinline{CODEWORD_LENGTH_BITS}. The \lstinline{CODEWORD_LENGTH_BITS} least significant bits of each element in the \lstinline{encoding} array contains the same value received from the input array \lstinline{symbol_bits}. The high order \lstinline{MAX_CODEWORD_LENGTH} bits of each \lstinline{encoding} element contains the actual codeword.  Using 27 bits for \lstinline{MAX_CODEWORD_LENGTH} resulting in \lstinline{CODEWORD_LENGTH_BITS} of 5 is a particularly useful combination, since each element of \lstinline{encoding[]} fits in a single 32-bit word.

The code consists primarily of two loops, labeled \lstinline{first_codewords} and \lstinline{assign_codewords}.  The \lstinline{first_codewords} loop finds the first codeword with each length, implementing the recurrence in Equation \ref{eq:first_codeword_recurrence}.   The \lstinline{assign_codewords} loop finally associates each symbol with a codeword.  The codeword is found using the length of each codeword and indexing into the correct element of \lstinline{first_codeword[]}.  The main complexity of this code is in the bit reversal process, which is based on the \lstinline{bit_reverse32} function. We have talked about this function previously in the FFT chapter (see Chapter \ref{sec:fft_bit_reversal}), so we will not discuss it here again. After reversing the bits in the codeword, the next statement removes the least significant '0' bits leaving only the bit-reversed codeword. The bit-reversed codeword is then packed in the high-order bits together with the length of the symbol in the low-order bits and stored in \lstinline{encoding[]}.  Lastly, the value in \lstinline{first_codeword[]} is incremented.

\begin{figure}
\begin{lstlisting}[firstline=1]
#include "huffman.h"
#include "assert.h"
#include <iostream>
void create_codeword(
  /* input */ CodewordLength symbol_bits[INPUT_SYMBOL_SIZE],
  /* input */ ap_uint<SYMBOL_BITS> codeword_length_histogram[TREE_DEPTH],
  /* output */ PackedCodewordAndLength encoding[INPUT_SYMBOL_SIZE]
) {
    Codeword first_codeword[MAX_CODEWORD_LENGTH];

    // Computes the initial codeword value for a symbol with bit length i
    first_codeword[0] = 0;
 first_codewords:
    for(int i = 1; i < MAX_CODEWORD_LENGTH; i++) {
#pragma HLS PIPELINE II=1
        first_codeword[i] = (first_codeword[i-1] + codeword_length_histogram[i-1]) << 1;
        Codeword c = first_codeword[i];
        //        std::cout << c.to_string(2) << " with length " << i << "\n";
    }

 assign_codewords:
  for (int i = 0; i < INPUT_SYMBOL_SIZE; ++i) {
#pragma HLS PIPELINE II=5
      CodewordLength length = symbol_bits[i];
      //if symbol has 0 bits, it doesn't need to be encoded
  make_codeword:
      if(length != 0) {
          //          std::cout << first_codeword[length].to_string(2) << "\n";
          Codeword out_reversed = first_codeword[length];
          out_reversed.reverse();
          out_reversed = out_reversed >> (MAX_CODEWORD_LENGTH - length);
          // std::cout << out_reversed.to_string(2) << "\n";
          encoding[i] = (out_reversed << CODEWORD_LENGTH_BITS) + length;
          first_codeword[length]++;
      } else {
          encoding[i] = 0;
      }
  }
}
\end{lstlisting}
\caption{ The complete code for generating the canonical Huffman codewords for each of the symbols. The codewords can be computed with knowledge of number of bits that each symbol uses (stored in the input array \lstinline{symbol_bits[]}). Additionally, we have another input array \lstinline{codeword_length_histogram[]} which stores at each entry the number of symbols with codewords at that bit length. The output is the code word for each symbol stored in the \lstinline{encoding[]} array. }
\label{fig:huffman_create_codeword.cpp}
\end{figure}

\begin{exercise}
In the code in Figure \ref{fig:huffman_create_codeword.cpp}, the inputs actually contain some redundant information.  In particular, we could compute the number of symbols for each bit length stored in \lstinline{codeword_length_histogram[]} from the length of each codeword \lstinline{symbol_bits[]} using a histogram computation.  Instead, in this code we've chosen to reuse the histogram originally computed in the \lstinline{truncate_tree()} function.  Instead we could save the storage by recomputing the histogram.  Do you think this is a good tradeoff?   How many resources are required to compute the histogram in this function?  How many resources are required to communicate the histogram through the pipeline?
\end{exercise}

\begin{exercise}
Estimate the latency of the code in Figure \ref{fig:huffman_create_codeword.cpp}
\end{exercise}

Let us now go through our running example and show how this is used to derive the initial codewords. In the example, the symbols \sym{A}, \sym{D}, and \sym{E} have two bits for their encoding; symbol \sym{C} has three bits; and symbols \sym{B} and \sym{F} have four bits. Thus, we have:
\begin{equation}
\begin{array} {lcl} 
\mathrm{bit\_length}(1) & = & 0 \\
\mathrm{bit\_length}(2) & = & 3 \\
\mathrm{bit\_length}(3) & = & 1 \\
\mathrm{bit\_length}(4) & = & 2 \\
\end{array}
\label{eq:bit_lengths}
\end{equation}

Using Equation \ref{eq:first_codeword_recurrence} to calculate the values of \lstinline{first_codeword}, we determine:
\begin{equation}
\begin{array} {lllll} 
\mathrm{first\_codeword}(1) & = & 0 & = & \mathrm{0b0}\\
\mathrm{first\_codeword}(2) & = & (0 + 0) << 1 &=& \mathrm{0b00} \\
\mathrm{first\_codeword}(3) & = & (0 + 3) << 1 &=& 6 = \mathrm{0b110} \\
\mathrm{first\_codeword}(4) & = & (6 + 1) << 1 &=& 14 = \mathrm{0b1110} \\
\end{array}
\label{eq:symbols}
\end{equation}

Once we have determined these values, then consider each symbol in order from smallest to largest. For each symbol, we determine the length of its codeword and assign the next codeword of the appropriate length.  In the running example, we consider symbols \sym{A}, \sym{B}, \sym{C}, \sym{D}, \sym{E}, and \sym{F} in alphabetical order. The symbol \sym{A} has two bits for its encoding. We perform a lookup into \lstinline{first_codeword[2] = 0}. Thus we assign the codeword for \sym{A} to \lstinline{0b00}. We increment the value at \lstinline{first_codeword[2]} to $1$. The symbol \sym{B} has four bits. Since \lstinline{first_codeword[4] = 14 = 0b1110}, it gets assigned the codeword \lstinline{0b1110}. Symbol \sym{C} has three bits. The value of \lstinline{first_codeword[3] = 6 = 0b110}, thus it gets the codeword \lstinline{110}. Symbol \sym{D} has two bits so it gets \lstinline{first_codeword[2] = 1 = 0b01}; remember that we incremented this value after we assigned the codeword to symbol \sym{A}. Symbol \sym{E} has two bits so it gets the codeword \lstinline{0b01 + 1 = 0b10}. And F has four bits so it gets the codeword \lstinline{0b1110 + 1 = 0b1111}.

The final codewords for all of the symbols are:
\begin{equation}
\begin{array} {lll} 
\sym{A} & \rightarrow & 00 \\
\sym{B} & \rightarrow & 1110\\
\sym{C} & \rightarrow  & 110 \\
\sym{D} & \rightarrow & 01 \\
\sym{E} & \rightarrow & 10\\
\sym{F} & \rightarrow & 1111\\
\end{array}
\label{eq:codewords}
\end{equation}

\subsection{Testbench}
\label{sec:huffman_testbench}

The final part of the code is the testbench. This is shown in Figure \ref{fig:huffman_encoding_test.cpp}. The general structure is to read the input frequency values from a file, process them using the \lstinline{huffman_encoding} function, and compare the resulting codewords with an existing golden reference that is stored in a file. 

\begin{figure}
\begin{lstlisting}[format=none, firstline=1,lastline=42]
#include "huffman.h"
#include <stdio.h>
#include <stdlib.h>

void file_to_array(const char *filename, ap_uint<16> *&array, int array_length) {
    printf("Start reading file [%s]\n", filename);
    FILE *file = fopen(filename, "r");
    if(file == NULL) {
        printf("Cannot find the input file\n");
        exit(1);
    }

    int     file_value = 0;
    int     count = 0;
    array = (ap_uint<16> *) malloc(array_length*sizeof(ap_uint<16>));

    while(1) {
        int eof_check = fscanf(file, "%x", &file_value);
        if(eof_check == EOF) break;
        else {
            array[count++] = (ap_uint<16>) file_value ;
        }
    }
    fclose(file);

    if(count != array_length) exit(1);
}

int main() {
    printf("Starting canonical Huffman encoding testbench\n");
    FILE *output_file;
    int return_val = 0;
    ap_uint<16> *frequencies = NULL;
    file_to_array("huffman.random256.txt", frequencies, INPUT_SYMBOL_SIZE);

    Symbol in[INPUT_SYMBOL_SIZE];
    for (int i = 0 ; i <  INPUT_SYMBOL_SIZE; i++) {
        in[i].frequency = frequencies[i];
        in[i].value = i;
    }

    int num_nonzero_symbols;
\end{lstlisting}
\end{figure}
\begin{figure}
\lstinputlisting[format=none, firstline=43]{examples/huffman_encoding_test.cpp}
\caption{ The complete code for the canonical Huffman encoding testbench. The code initializes the \lstinline{in} array with data from an input file. It passes that into the top \lstinline{huffman_encoding} function. Then it stores the resulting codewords into a file, and compares that with another golden reference file. It prints out the results of the comparison, and returns the appropriate value.}
\label{fig:huffman_encoding_test.cpp}
\end{figure}

The \lstinline{main()} function starts by setting up the variables required to read the frequencies from a file (in this case the file is \texttt{huffman.random256.txt}) and puts them into \lstinline{in[]}. This is done in the \lstinline{file_to_array} function, which takes as input the \lstinline{filename} for the input data and the length of the data (\lstinline{array_length}), and stores the entries in that file into \lstinline{array[]} variable. This file contains the frequency of each symbol. Frequency values are stored in symbol order, thus the first value of the file represents the frequency of symbol '0', and so on. 

The \lstinline{main()} function continues by initializing \lstinline{in[]} using the frequencies from the file. It then calls the top \lstinline{huffman_encoding} function. The function returns the encoded symbol values in \lstinline{encoding[]}. Since the result of processing should be a prefix code, we check that the properties of a prefix code are actually satisfied.  The result is then compared to the codewords stored in a golden reference, which is stored in the file \texttt{huffman.random256.gold}. We do this by writing the result to a file named \texttt{random256.out} and performing a file comparison using the \texttt{diff} tool. The \texttt{diff} tool returns '0' if the files are identical and non-zero if the files are different. Thus, the \lstinline{if} condition occurs when the files are different, and the \lstinline{else} condition is executed when the files are the same. In both cases, we print out a message and set the \lstinline{return_val} to the appropriate value. This return value is used by the \VHLS tool during cosimulation to check the correctness of results.  The return value should be '0' if it passes, and non-zero if it does not pass.

\section{Conclusion}

Huffman Coding is a common type of data compression used in many applications.  While encoding and decoding using a Huffman code are relatively simple operations, generating the Huffman code itself can be a computationally challenging problem.  In many systems it is advantageous to have relatively small blocks of data, implying that new Huffman codes must be created often, making it worthwhile to accelerate.

Compared to other algorithms we've studied in this book, creating a Huffman code contains a number of steps with radically different code structures.  Some are relatively easy to parallelize, while others are more challenging.  Some portions of the algorithm naturally have higher $\mathcal{O}(n)$ complexity, meaning that they must be more heavily parallelized to achieve a balanced pipeline.  However, using the \lstinline{dataflow} directive in \VHLS, these different code structures can be linked together relatively easily


\bibliographystyle{plainnat}
\bibliography{all} 
\printnoidxglossaries

\end{document}